\documentclass[useAMS,usenatbib]{mn2e}
\usepackage{graphicx}
\usepackage{amssymb, amsmath}
\usepackage{times}
\usepackage{xcolor}
\usepackage{lscape}
\usepackage{hyperref}
\usepackage{url}
\usepackage{lineno}

\def\apj{\rm{ApJ}}                 
\def\apjs{\rm{ApJS}}             
\def\mnras{\rm{MNRAS}}     
 \newcommand{\commentH}[1]{\relax}
 \newcommand{\commentM}[1]{\relax}

  \newcommand{\change}[1]{\relax #1}
\begin{document}
\title[Theoretical Systematics in Anisotropic BAO Analysis from the completed SDSS-III/BOSS]{The clustering of galaxies in the completed SDSS-III Baryon Oscillation Spectroscopic Survey: theoretical systematics and Baryon Acoustic Oscillations in the galaxy correlation function}
\author[Vargas-Magana et al.]{\parbox{\textwidth}{
\Large
Mariana Vargas-Maga\~na$^{1,2,3}$\thanks{E-mail:mmaganav@fisica.unam.mx}, 
Shirley Ho$^{2,3,4,5}$, 
Antonio J. Cuesta$^{6}$, 
Ross O'Connell$^{2,3}$,  
Ashley J. Ross$^{7}$,  
Daniel J. Eisenstein$^{8}$,
Will J. Percival$^{9}$, 		 
Jan Niklas Grieb$^{10,11}$,
Ariel G. S\'anchez$^{10}$,
Jeremy L. Tinker$^{12}$, 
Rita Tojeiro$^{13}$,	 
Florian Beutler$^{9}$,
Chia-Hsun Chuang$^{14,15}$,	 
Francisco-Shu Kitaura$^{14,15,5}$,
Francisco Prada$^{14,16,17}$,
Sergio A. Rodr\'iguez-Torres$^{14,16,17}$,
Graziano Rossi$^{18}$,	 
Hee-Jong Seo$^{19}$,
Joel R. Brownstein$^{20}$,
Matthew Olmstead$^{21}$,
Daniel Thomas$^{9}$ 
}\vspace*{4pt} \\ 
\scriptsize $^{1}$ Instituto de F\'isica, Universidad Nacional Aut\'onoma de M\'exico, Apdo. Postal 20-364, M\'exico.\vspace*{-2pt} \\ 
\scriptsize $^{2}$ Departments of Physics, Carnegie Mellon University, 5000 Forbes Ave., Pittsburgh, PA 15217\vspace*{-2pt} \\ 
\scriptsize $^{3}$ McWilliams Center for Cosmology, Carnegie Mellon University, 5000 Forbes Ave., Pittsburgh, PA 15217 \vspace*{-2pt} \\ 
\scriptsize $^{4}$ Lawrence Berkeley National Lab, 1 Cyclotron Rd, Berkeley CA 94720, USA\vspace*{-2pt} \\  
\scriptsize $^{5}$ Departments of Physics and Astronomy, University of California, Berkeley, CA 94720, USA\vspace*{-2pt} \\  
\scriptsize $^{6}$ Institut de Ci{\`e}ncies del Cosmos (ICCUB), Universitat de Barcelona (IEEC-UB), Mart{\'\i} i Franqu{\`e}s 1, E08028 Barcelona, Spain\vspace*{-2pt} \\ 
\scriptsize $^{7}$ Center for Cosmology and AstroParticle Physics, The Ohio State University, Columbus, OH 43210, USA\vspace*{-2pt}\\
\scriptsize $^{8}$ Harvard-Smithsonian Center for Astrophysics, 60 Garden St., Cambridge, MA 02138, USA\vspace*{-2pt} \\ 
\scriptsize $^{9}$ Institute of Cosmology \& Gravitation, Dennis Sciama Building, University of Portsmouth, Portsmouth, PO1 3FX, UK\vspace*{-2pt} \\ 
\scriptsize $^{10}$ Universit\"ats-Sternwarte M\"unchen,Ludwig-Maximilians-Universit\"at M\"unchen, Scheinerstra\ss{}e 1, 81679, M\"unchen, Germany\vspace*{-2pt}\\
\scriptsize $^{11}$ Max-Planck-Institut f\"ur extraterrestrische Physik, Postfach,1312, Giessenbachstr., 85741 Garching, Germany\vspace*{-2pt}\\
\scriptsize $^{12}$ Center for Cosmology and Particle Physics, New York University, New York, NY 10003, USA\vspace*{-2pt} \\ 
\scriptsize $^{13}$ School of Physics and Astronomy, University of St Andrews, St Andrews, KY16 9SS, UK\vspace*{-2pt} \\ 
\scriptsize $^{14}$ Instituto de F\'isica Te\'orica, (UAM/CSIC), Universidad Aut\'onoma de Madrid, Cantoblanco, E-28049 Madrid, Spain\vspace*{-2pt}\\
\scriptsize $^{15}$ Leibniz Institut f\"ur Astrophysik Potsdam (AIP), An der Sternwarte 16, D-14482 Potsdam, Germany\vspace*{-2pt} \\ 
\scriptsize $^{16}$ Campus of International Excellence UAM+CSIC, Cantoblanco, E-28049 Madrid, Spain\vspace*{-2pt}\\
\scriptsize $^{17}$ Departamento de F\'isica Te\'orica M8, Universidad Auton\'oma de Madrid (UAM), Cantoblanco, E-28049, Madrid, Spain\vspace*{-2pt}\\       
\scriptsize $^{18}$ Department of Physics and Astronomy, Sejong University, Seoul, 143-747, Korea.\vspace*{-2pt}\\
\scriptsize $^{19}$ Department of Physics and Astronomy, Ohio University, 251B Clippinger Labs, Athens, OH 45701, USA\vspace*{-2pt} \\ 
\scriptsize $^{20}$ Department of Physics and Astronomy, University of Utah, 115 S 1400 E, Salt Lake City, UT 84112, USA\vspace*{-2pt}\\
\scriptsize $^{21}$ Department of Chemistry and Physics, King's College, 133 North River St, Wilkes Barre, PA 18711, USA\vspace*{-2pt}\\
}
\date{\today}
\pagerange{\pageref{firstpage}--\pageref{lastpage}}   \pubyear{2016}
\maketitle
\label{firstpage}

\begin{abstract}

We investigate the potential sources of theoretical systematics in the anisotropic Baryon Acoustic Oscillation (BAO)  distance scale measurements  from the clustering of galaxies in configuration space using the final Data Release (DR12) of the Baryon Oscillation Spectroscopic Survey (BOSS). We perform a detailed study of the impact on BAO measurements from choices in the methodology such as fiducial cosmology, clustering estimators, random catalogues, fitting templates, and covariance matrices. 
 The theoretical systematic uncertainties in BAO parameters are found to be 0.002 in the isotropic dilation $\alpha$ and 0.003 in the quadrupolar dilation $\epsilon$. \change{The leading source of systematic uncertainty is related to the reconstruction techniques. Theoretical uncertainties  are sub-dominant compared with the statistical uncertainties  for BOSS survey,  accounting  $0.2\sigma_{stat}$ for $\alpha$ and $0.25\sigma_{stat}$ for $\epsilon$ 
 ($\sigma_{\alpha,stat} \sim$0.010 and $\sigma_{\epsilon,stat}\sim$ 0.012 respectively).}
We also present BAO-only distance scale constraints from the anisotropic analysis of the correlation function. Our constraints on the angular diameter distance $D_A(z)$ and the Hubble parameter $H(z)$, including both statistical and theoretical systematic uncertainties, are 1.5\% and 2.8\% at $z_{\rm eff}=0.38$, 1.4\% and 2.4\% at $z_{\rm eff}=0.51$, and 1.7\% and 2.6\% at $z_{\rm eff}=0.61$.  This paper is part of a set that analyzes the final galaxy clustering dataset from BOSS. The measurements and likelihoods presented here are cross-checked with other BAO analysis in  \citet{Acacia16}. The systematic error budget concerning the methodology on post-reconstruction BAO analysis presented here is used in  \citet{Acacia16} to produce the final cosmological constraints from BOSS.
 \\

\end{abstract}
\begin{keywords}
    reconstruction;
    galaxies;
    galaxies: statistics;
    cosmological parameters;
    large-scale structure of the Universe
\end{keywords}
\newpage
\newpage
\section{Introduction}
The Baryon Oscillation Spectroscopic Survey (BOSS, \citep{Daw12}) Data Release 12  (DR12, \citealt{DR12}) is the largest galaxy spectroscopic survey available to date, providing the most precise measurements on Baryonic Acoustic Oscillations (BAO).   
Since the previous Data Release, (i.e. DR11), the scientific goal of the experiment of 1 per cent measurement in the distance measurements was achieved. eBOSS \cite{eBOSS} is going to expand the redshift range of the distance measurements with precision similar to its predecessor BOSS.
Improvements on this precision are expected for the next generation of dark energy experiments as well, (Stage IV from the Dark Energy Task Force).  
Planned large galaxy surveys such as the Dark Energy Spectroscopic Instrument (DESI)  \citep{DESIOverview, DesiWhitePaper}, Euclid  \citep{EuclidAssesment, EuclidDefinition}, LSST \citep{LSST}, SKA \citep{SKA}, and WFIRST \citep{WFIRST, WFIRSTReport} are expected to measure $H(z_{\rm eff})$ and $D_A(z_{\rm eff})$ to sub-percent level. In order to approach this level of precision, a careful investigation of potential sources of systematic effects is required. 

In this work, we aim to examine the methodology of Baryon Acoustic Oscillation (BAO) analysis, focusing on the various potential sources of theoretical systematics. \change{We call ``theoretical systematics" to any uncertainty related to the methodology for extracting the BAO signal}. 
\change{Previous work related to the exploration of theorethical systematics have been performed partially in several references: concerning reconstruction-related systematics we have studies performed by \cite{PadWhi09, Seo10,bur14,Vargas14,bur15,Vargas15,Seo15}, concerning the modeling and anisotropic fitting methodology, also different works have explored different aspects \citep{Xeaip,And13,And14,Vargas14, Ross12}. This work distinguishes from previous works in that we perform a systematic exploration of the theoretical systematics with the aim to account for the error budget associated with the BAO methodology.}
We explore these systematic uncertainties by performing a BAO anisotropic analysis in configuration space over a large number of realistic mock catalogues generated for the final analysis of the BOSS galaxy clustering. The studies performed in this work provide a basis for the systematic error included in the final BAO measurements of the BOSS DR12 galaxy samples \citep{Acacia16}.  

The observational systematics in the BOSS combined galaxy catalogues are covered in \cite{Ross16}, \change{the observational systematics covers how the angular selection functions of the BOSS galaxy samples are defined and test for any systematic uncertainty that is imparted into BAO measurements based on this process.}
. This work is part of the final analysis of the BOSS DR12 completed samples for the BAO and Redshift-Space Distortions (RSD) analyses \citep[companion papers]{Ross16, Sanchez16a, Beutler16a, Beutler16b, Grieb16, Satpathy16}. 

In order to clarify the applicability of this analysis, we note to the reader that this work is focused on configuration space analysis and heavily focused on post-reconstruction results, whereas \cite{Acacia16} uses the average of Fourier space and configuration analysis, and also includes pre-reconstruction data for the full shape analysis. Thus, the systematics budget account derived from this study is relevant only for the BAO-only fits, and in configuration space and post-reconstruction in particular. We also provide pre-reconstruction results, however the larger part of the analysis will be centred in the post-reconstruction results.

The outline of the paper is as follows. In Section \ref{sec:method}, we describe the data and mock catalogues used for the studies and include a brief description of the anisotropic BAO methodology. In Section \ref{main_sys}, we define the fiducial (base)  methodology and we present the procedure we follow to analyze the variations of the methodology in order to account for an error budget.
In Sections \ref{sec:estimators}-\ref{sec:model}, we revisit the potential sources of systematics following step-by-step the anisotropic BAO analysis methodology. 
Thus, Section \ref{sec:estimators} is devoted to testing different 2-point estimators in configuration space (multipoles, wedges, and $\omega$-estimators);
Section \ref{sec:randoms} studies the effect of the random catalogues used on the 2-point estimator for the multipoles case; 
Section \ref{sec:covariance} explores the effects of the covariance matrix;  
Section \ref{sec:reconstruction} revisits the effect of smoothing scale in the reconstruction of the linear density field and its impact on BAO anisotropic analysis; 
Section \ref{sec:cosmology} tests the effect of the assumed fiducial cosmology in the analysis;
Section \ref{sec:model} studies the systematics related to the anisotropic fitting methodology and modelling.
In Section \ref{sec:bossdata}, we apply variations of the methodology to the BOSS DR12 measurements. 
In Section \ref{discussion}, we discuss the methodology used to establish a systematic error budget for the DR12 measurements. Finally, in Section \ref{BAO} we present the final constraints  on the angular diameter distance $D_A(z)$ and the Hubble parameter distance $H(z)$, including the theoretical systematic uncertainties. We conclude in Section \ref{conclusions}.

\section{Methodology}\label{sec:method}

\subsection{Galaxy Sample Creation:  BOSS Final Dataset}  \label{data}
The SDSS-III \citep{Eis11} BOSS \citep{Daw12} survey uses the Sloan Foundation 2.5-meter telescope at Apache Point Observatory \citep{Gun98,Gun06}. The targets use wide-field CCD photometry in five passbands u,g,r,i,z \citep{Fuk96}. 
The spectra are obtained using the double-armed BOSS spectrograph  \citep{Smee13}. The data is then reduced using the algorithms described in \cite{Bolton12}.
We use the BOSS DR12 \citep{DR12} galaxy sample and provide a brief description. For detailed information, please refer to \cite{Acacia16} and \cite{Reid2015}.
The BOSS galaxy samples have traditionally included the Constant Stellar Mass sample, or CMASS, covering redshifts in the range 0.43 $<$ z $<$ 0.70 and a fiducial redshift of 0.57, and the low-redshift sample, or LOWZ, covering redshifts of 0.15 $<$ z $<$ 0.43 with an effective redshift of 0.32 \citep{Reid2015}. Also included in the final analysis are the early LOWZ samples that were targeted with a different target selection algorithm; this set is called ``LOWZ Early."
For the final analysis we combine all these samples (hereafter ``BOSS Combined Sample") and define three redshift slices that overlap by $\sim$ 200 $h^{-1}$Mpc and are chosen to be approximately equal in effective volume. These are: $V_{\rm low}$, which includes galaxies in the redshift range $z= 0.2 - 0.5$ (denoted hereafter ``Bin 1"); $V_{\rm mid}$, which considers the redshift range $z= 0.4 - 0.6$ (denoted hereafter ``Bin 2"); and $V_{\rm high}$, which includes the redshift range $z= 0.5 - 0.75$ (denoted hereafter ``Bin 3"). $V_{\rm mid}$ completely overlaps with the other two, but also contains the peak number density. The effective redshifts $z_{\rm eff}$ for each bin are 0.38, 0.51, and 0.61, respectively. The three different redshift bins have 604001, 686370, and 594003 galaxies, respectively.

\subsection{ Mitigating Observational Systematics in the Galaxy Catalogues } \label{obssys}
Extensive discussions of observational systematics in BOSS galaxy catalogues are available in \cite{Ross11,Ross12}, and in the final BOSS data release. \cite{Ross16} describes the detailed observational systematics in BOSS combined galaxy catalogues. We will use all of the observational systematics weights that are determined in \cite{Ross16} throughout our entire work.

\subsection{Mock Catalogue Generation} \label{mocks} 
In this paper, we use SDSS III-BOSS mock catalogues extensively. They consist of  
PTHALOS \citep{Man12}, QPM \citep{QPM}, and MD-PATCHY \citep{PATCHY}. \change{The purpose of using different sets of approximative mock catalogs it to check if the BAO analysis is robust agains this choice. In particular in terms of the covariance matrix derived from the mocks.}

\subsubsection{PTHALOS Mocks}
PTHALOS mocks \citep{Man12} are based on Second Order Lagrangian Perturbation Theory (2LPT) for the density fields. The method for generating galaxy mocks follows the PTHALOS methodology described in \cite{ScoShe02}.
The mocks were generated at a fixed redshift $z=0.55$ and in cubic boxes (L = 2400 $h^{-1}$Mpc) using 1280$^3$ dark matter particles. The halos were found using a friends-of-friends algorithm and their masses were calibrated using $N$-body simulations. To populate the halos with galaxies, a Halo Occupation Distribution prescription was used, previously calibrated to match the observed clustering at small scales (30--80 $h^{-1}$Mpc).  The angular and radial masks from DR10/DR11 were applied on the original boxes to match the survey footprint. The mocks include redshift distortions but do not include other systematic corrections such as close-pair collisions, stellar density correlations, or redshift evolution.
A full description of the mocks can be found in \cite{Man12}.
The cosmology is $\Omega_M = 0.274,\Omega_\Lambda = 0.726,
\Omega_b = 0.045, \sigma_8 = 0.8$ and $h = 0.7.$

\subsubsection{QPM Mocks}

Quick Particle Mesh (QPM) mocks were generated for BOSS clustering analysis. These mock catalogues use low mass
and force resolution particle-mesh simulations employing $1280^3$ particles in a (2560 $h^{-1}$Mpc)$^3$ box run with large time steps. At selected
times, the particles and their local density smoothed on 2 $h^{-1}$Mpc scales were dumped; these particles were then sampled (with a
density-dependent probability) to form a set of mock halos that were then populated using a halo occupation distribution \citep{QPM}.
The mock catalogues match the observed number density of BOSS galaxies and follow the radial and angular selection functions of BOSS galaxy samples.
The cosmology is $\Omega_M = 0.29$, $\Omega_\Lambda$ = 0.71, $\Omega_b$ = 0.048, $\sigma_8$ = 0.8 and $h = 0.7.$

We use two different versions of the QPM mocks: 1) a version that matches the DR12 CMASS samples used in previous analyses \citep{Cuesta16, Vargas15, Gil15} and 2) a new set of mocks for the DR12 combined sample that match the samples used for the final BOSS clustering analysis. The mocks combine in an optimal way all BOSS galaxy data, including CMASS, LOWZ, and early chunks not used in previous BOSS analyses referred to as ``LOWZ Early" data. 
These samples are used in the final BAO/RSD analyses \citep[companion papers]{Ross16, Sanchez16a, Beutler16a, Beutler16b, Grieb16, Satpathy16}.\subsubsection{MULTIDARK-PATCHY Mocks}
MD-PATCHY mocks are based on Augmented Lagrangian Perturbation Theory and a 
non-linear bias stochastic scheme \citep{PATCHYtheory}, where the bias parameters are fitted to match the 
proper clustering of the BigMultiDark Planck simulation for each 
redshift snapshot \citep{Multidark}. The HADRON code was used for mass assignment to halos 
\citep{HADRON}. Light-cone mocks are built using the SUGAR code (Rodriguez-Torres et al. in preparation) and the Mock-Factory products (https://github.com/mockFactory).
These mocks were designed to reproduce the observed evolution of clustering, and its dependence with stellar mass.
Initial conditions for the MD-PATCHY code were created using the BigMultiDark simulation (also described in \citealt{Multidark}).
The cosmology is based on Planck \citep{Planck2015Cosmo}: $\Omega_M$= 0.307115, $\Omega_\Lambda $= 0.692885, 
$\Omega_b = 0.048$, $\sigma_8$ = 0.8288 and $h = 0.6777$.
We used Version 6C of the mocks (where  ``C" stands for ``Covariance matrix"
of the clustering measurements). These mocks are tuned to reproduce the clustering of observed data in terms of one-point,
two-point, and three-point clustering statistics in redshift space.

\begin{table*}
\begin{center}
\caption{Fiducial cosmologies used in this work. }
\label{tab:cosmofid}
\begin{tabular}{@{}lccccccc}
\hline
Cosmology&$\Omega_{CDM}$&$\Omega_M$& $\Omega_B$&$\Omega_\Lambda$&h&Samples\\
\hline
\\[-1.5ex]
PTHALOS&0.228286&0.274&0.0457143&0.726&0.7&CMASS DR11\\
QPM&0.244143&0.29&0.0458571&0.71&0.7 & CMASS DR11/DR12\\
\citet{Acacia16} & 0.261857&0.31& 0.0481426&0.69&0.676& COMBINED DR12\\
\hline
MD-PATCHY&0.258909&0.307115&0.048206&0.692885&0.676&-\\
\hline
\end{tabular}
\end{center}
\end{table*}

\subsection{Fiducial Cosmology}
For the analysis of our DR12 combined sample defined in Section \ref{data}, we assume the following fiducial cosmology:
 $\Omega_m=0.31$, $\Omega_{\Lambda}=0.69$, $\Omega_k=0$, $\Omega_bh^2=0.022$, $\Omega_{\nu}h^2=0.00064$, $w=-1$, $w_a=0$, $h=0.676$, $n_s=0.97$, and $\sigma_8=0.8$. Section~\ref{sec:cosmology} tests the effect of using different fiducial cosmologies in the analysis. The cosmologies explored in this case are presented in Table~\ref{tab:cosmofidtest}.

\subsection{Choosing Estimators} 
Anisotropic BAO analysis requires the computation of the 2D correlation function $\xi(r,\mu)$ (or power spectrum $P(k,\mu)$).
Using the full 2D correlation function is impractical, since it requires a covariance matrix that is precise for a relatively large number of parameters. 
There are, however, several ways of  compressing the information that are usually implemented in the different BAO analyses. In configuration space, these include the multipole method \citep{Xeaip,Xu12b}, the $\omega_l$ statistic \citep{Xu10}, and the wedges method \citep{Kaz13,Sanchez13}. 
We will define a few parameters to make our discussion of estimators more transparent. 
We first compute the two-point correlation function in 2D, 
decomposing the separation $r$ between two galaxies into two components: parallel to line of sight $r_{||}$ and perpendicular to line of sight, $r_{\perp}$, where $r$ is defined as follows:
\begin{equation}
r^2=r^2_{||}+r^2_{\perp}.
\end{equation}
We denote $\theta$ the angle  between the galaxy pair separation and the LOS direction, and we define $\mu=\cos \theta$ so that:
\begin{equation}
\mu^2=\cos^2 \theta = \frac{r_{||}^2}{r^2}.
\end{equation}

The 2D-correlation function $\xi(r,\mu)$ (for the pre-reconstructed case) 
is then computed using the Landy-Szalay estimator \citep{LanSza93} that reads as follows:
\begin{equation}
\xi(r, \mu)=\frac{DD(r, \mu)-2DR(r, \mu) +RR(r, \mu)}{RR(r, \mu)},
\end{equation}
where $DD(r,\mu), RR(r,\mu)$, and $DR(r, \mu)$ are the number of pairs of galaxies which are separated by a radial separation $r$ and angular separation $\mu$ from data-data, random-random, and data-random pairs, respectively.  
\commentH{you should define the post-recon version here, to reference it in sec 3.2.2. for example}

After computing the two-point correlation function in 2D, we can then compress it into multipoles, wedges, and $\omega_l$. 
\subsubsection{Multipoles}
The multipoles are Legendre moments of the 2D correlation function $\xi(r, \mu)$. They can be computed through the following equation: 
\begin{equation} 
\xi_{\ell}(r) = \frac{2\ell +1}{2} \int_{-1}^{+1} d\mu \; \xi(r,\mu) \; L_{\ell}(\mu),
\end{equation} 
where $L_{\ell}(\mu)$ is the $\ell$-th order Legendre polynomial.  We focus primarily on the monopole and the quadrupole ($\ell=0$ and $\ell=2$), although we will have a discussion on hexadecapole ($\ell=4$) in this work. 
\commentH{maybe comment that in linear theory these are all contributing multipoles}
\subsubsection{Clustering Wedges}
The clustering wedges are an integral of the correlation function over a range of $\mu$: 
\begin{equation}  
\xi_{\Delta \mu}(r) = \frac{1}{\Delta \mu} \int_{\mu_{min}}^{\mu_{min}+\Delta \mu} d\mu \; \xi(r,\mu).
\end{equation} 
We choose $\Delta \mu=0.5$ in our work, following \cite{Kaz12,Kaz13} and \cite{Sanchez13}. We define, in particular, the wedge parallel to line of sight to be $\xi_{||}(r)$ for $0.5<\mu<1$ and the wedge perpendicular to line of sight to be $\xi_{\perp}(r)$ for $0<\mu<0.5$.

The clustering wedges and multipoles  are complementary bases as shown in \cite{Kaz12}; they can be related (discarding multipoles with $\ell > 4$) by:
\begin{eqnarray}  \label{eqn:wedgesmultip}
\xi_\perp(r)=\xi_0(r)-\frac{3}{8} \xi_2(r) +\frac{15}{128}\xi_4(r),\\
\xi_\parallel(r)=\xi_0(r)+\frac{3}{8}\xi_2(r) -\frac{15}{128}\xi_4(r),
\end{eqnarray}  

\subsubsection{$\omega$-Estimator}
As in \cite{Xu10}, we define $\omega_l$ as the redshift space correlation function, $\xi_s\left(r, \mu\right)$, convolved with a compact and compensated filter $W_{\ell}\left(r,\mu, r_{c}\right)$ as a function of characteristic scale $r_{c}$:
\begin{equation}
\label{eqn:wLdef}
\begin{aligned}
\omega_l & = i^{\ell}\int d^3r \ \xi_s\left(r,\mu\right)W_{\ell}\left(r,\mu, r_{c}\right),\\
\end{aligned}
\end{equation}
where we have taken advantage of the orthogonality of the Legendre polynomials and set $W_{\ell} \left(r,\mu, r_c \right) = W_{\ell}\left(r, r_c \right) L_{\ell}\left(\mu\right)$.
Following \cite{Pad07} and \cite{Xu10}, we define a smooth, low order, compensated filter independent of $\ell$:
\begin{equation}
\label{eqn:filter}
W_\ell\left(x\right) = \left(2x\right)^2\left(1-x\right)^2\left(\frac{1}{2}-x\right)\frac{1}{r_c^3},
\end{equation}
where $x = \left(r/r_c\right)^3$.
This choice of filter gives the $\omega_l$ statistic several advantages.
By design, $\omega_l$ probes a narrow range of scales near the BAO feature and is not sensitive to large scale fluctuations or to poorly measured or modelled large scale modes \citep{Xu10}.

\subsection{Anisotropic Parametrization: $\alpha$ and $\epsilon$}

To analyze the anisotropic BAO signal, we need a model with a parametrization of the anisotropic clustering signal. 
There are two sources of anisotropies: the RSD and the anisotropies generated from assuming a wrong cosmology, also known as the Alcock-Paczynski effect (AP) \citep{AP}. As $D_A(z)$ and H(z) depend on the cosmological parameters differently, 
if one assumes a fiducial cosmology different from the one that matches the sample,
$H(z)$ and $D_A(z)$ will generate artificial anisotropies in the clustering along Line-of-Sight (LOS) and perpendicular directions \citep{Mat00, Oku08, PadWhi09, Seo08, Xu12b}. 

For the extraction of the cosmological information, we follow the methodology described in \cite{Xu12b} and \cite{And13}, which derives measurements
of the isotropic dilation of the coordinates parametrized by $\alpha$ and the anisotropic warping of the coordinates parametrized by $\epsilon$. 
Here $\alpha$ and $\epsilon$ parametrize the geometrical distortion derived from assuming a wrong cosmology when calculating the galaxy correlation function.
The parameters $\alpha$ and $\epsilon$ are defined as in \cite{PadWhi09}: 

\begin{eqnarray}\label{eq:ae1}
\alpha = \alpha_{\perp}^{2/3} \alpha_{||}^{1/3} \nonumber \,, \\
1+ \epsilon = \left( \frac{\alpha_{||}}{\alpha_{\perp}} \right)^{1/3} \,.
\end{eqnarray}
where $\alpha_\perp$ and $\alpha_{||}$ are defined  in terms of dilation in the transverse and line-of-sight directions, given by a difference between the fiducial cosmology and the ``true cosmology"\footnote{Note that $\alpha =  1$ and $\epsilon = 0$ for the mocks, if we use their natural cosmology as the fiducial cosmology for the analysis.}.

The parameters $\alpha$ and $\epsilon$ are related to $D_A(z)/r_s$ and $cz/(H(z)r_s)$ where $D_A(z)$ is the angular diameter distance to redshift $z$, $r_s$  is the sound horizon at radiation drag, and $H(z)$ is the Hubble parameter. 
Finally, $\alpha_{\perp}$ and $\alpha_{||}$ are related to $D_A(z)/r_s$ and $cz/(H(z)r_s)$ in the following way: 
\begin{equation}\label{eq:aper}
\alpha_{\perp} = \frac{D_{A} (z) r^{\rm fid}_{s}}{D^{\rm fid}_{A} r_{s}},
\end{equation}
and
\begin{equation}\label{eq:apar}
\alpha_{||} = \frac{H^{\rm fid}(z) r^{\rm fid}_{s}}{H(z) r_{s}}.
\end{equation}

\subsection{BAO Fitting, From Correlation Functions to Distance Estimates}

\subsubsection{Nonlinear Model for the Correlation Function}\label{sec:NLM}

For modelling the nonlinear correlation function we follow \citet{Xeaip,Xu12b}.
We start from a template for the 2D nonlinear power spectrum \citep{Fisher94} considered as follows:
\begin{equation}\label{fisher}
P(k, \mu) =(1+\beta \mu^2 )^2 F(k, \mu, \Sigma_s)P_{\rm NL} (k, \mu).
\end{equation}
The term $(1+\beta \mu^2)^2$  corresponds to the Kaiser model for large scale redshift space distortions. 
$F(k, \mu, \Sigma_s)$ is the streaming model for Fingers-of-God (FoG) given by: 
\begin{equation}\label{eq:streamming}
F(k, \mu, \Sigma_s)=\frac{1}{(1+k^2\mu^2 \Sigma_s^2)}
\end{equation}
where $\Sigma_s$ is the streaming scale. $P_{\rm NL}(k)$ is the nonlinear power spectrum. \change{In this work, we 
consider the  de-wiggled power spectrum $P_{\rm dw}=P_{\rm NL}(k)$ for the non linear power spectrum}, defined as: 
\begin{equation}\label{nonlinearpk}
\begin{array}{ll}
P_{\rm dw}(k, \mu)=&[P_{\rm lin}(k) -P_{\rm nw}(k)] \\
&\times \exp\left[ -\frac{k^2 \mu^2 \Sigma_{||}^2+k^2(1-\mu^2)\Sigma_{\perp}^2}{2}\right ] +P_{\rm nw} 
\end{array}
\end{equation}
where $P_{\rm lin}(k)$ is the linear theory power spectrum and $P_{\rm nw}(k)$ is a power spectrum without the acoustic oscillations \citep{Eis98}. $\Sigma_{||}$ and $\Sigma_\perp$ are the radial and transverse components of the standard Gaussian damping of BAO $\Sigma_{NL}$.

 In order to get the templates for the multipoles, we then decompose the 2D power spectrum into its Legendre moments:
\begin{equation}
P_{l,t}(k)=\frac{2l+1}{2}\int_{-1}^{1}P_t(k, \mu)L_l(\mu)d\mu,
\end{equation}
which can then be transformed to configuration space using
\begin{equation}
\xi_{l,t}(r) = i^l \int \frac{k^3 d \log(k)}{2 \pi^2}P_{l,t} j_l(kr),
\end{equation}
where $j_l(kr)$ is the $l$-th order spherical Bessel function and $L_l(\mu)$ is the $l$-th order Legendre polynomial.

The model we fit to our observed multipoles $\xi_0(r)$ and
$\xi_2(r)$ is: 
\begin{align}
\xi_{0}(r) &= B_0^2 \xi_{0, \rm t}(r) + A_0(r) \nonumber, \\ 
\xi_{2}(r) &= \xi_{2, \rm t}(r) + A_2(r),
\label{eqn:monoquadt} 
\end{align}
\commentH{why xi2 does not have an amplitude parameter?}
where 
\begin{equation}
A_{\ell}(r) = \frac{a_{\ell,1}}{r^2} + \frac{a_{\ell,2}}{r} + a_{\ell,3}; \,
\ell=0,2,\perp, \parallel \,
\label{eqn:fida}
\end{equation}
These $A_\ell(r)$ terms are used to marginalise out broadband (shape)
information
through the $a_{\ell, 1}\ldots a_{\ell, 3}$ nuisance parameters. 

Our model for wedges is derived from the templates  for $\xi_{\ell}$ by applying Eq.~\ref{eqn:wedgesmultip} (derived in \cite{Kaz12}), discarding contributions from multipoles with $\ell > 4$  to the expected observations for $\xi_{\ell}$ given in Eq.~\ref{eqn:monoquadt}. 

Our models for $\omega_\ell$ are derived from the templates for $\xi_{\ell}$ by applying Eq.~\ref{eqn:wLdef} with the filter defined in Eq.~\ref{eqn:filter} to the expected observations for $\xi_{\ell}$ given in Eq.~\ref{eqn:monoquadt}. 
In short, $\omega_\ell$ is a filtered integral of $\xi$. 

As in the multipoles fitting, when fitting $\omega_\ell$, we fit $\omega_0\left(r_c\right)$ and $\omega_2\left(r_c\right)$ simultaneously with the same nonlinear parameters.\commentH{which ones?}

In order to find the best-fit values of $\alpha$ and $\epsilon$, we 
minimise the $\chi^2$ function, 
\begin{equation}
\chi^2 = (\vec{m} - \vec{d})^T C^{-1} (\vec{m}-\vec{d}),
\end{equation}
where $\vec{m}$ is the model and $\vec{d}$ is the data correlation function (binned in radial bins). We scale the
inverse sample covariance matrix estimated from the mocks, $C^{-1}_s$, using the equation:
\begin{equation}\label{eq:cov}
C^{-1} = C^{-1}_s \frac{N_{\rm mocks} - N_{\rm bins} - 2}{N_{\rm mocks} - 1},
\end{equation}
correcting for the fact that it is a biased estimate of the true inverse
covariance matrix $C^{-1}$ \citep{Har07,Per14}.
Error estimates for $\alpha$ and $\epsilon$ are obtained by evaluating $\chi^2$ on a grid 
in these two parameters to map out the
likelihood surface. Assuming the likelihood surface is Gaussian allows
us to estimate $\sigma_\alpha$ and $\sigma_\epsilon$ as the standard
deviations of the marginalised likelihoods of $\alpha$ and $\epsilon$, respectively.

\subsection{Reconstruction}
Reconstruction has become part of the essential toolkit for BAO analysis  \citep{EisSeoWhi07,Eis07a,Pad12,Vargas14,bur14,bur15,Vargas15}. 
The reconstruction algorithm has proved to be effective in partially correcting the effects of nonlinear evolution, increasing the statistical precision of the measurements. The main idea of reconstruction is to use information encoded in the density field to estimate the displacement field and use this displacement field to move back the particles and partially remove the effect of the nonlinear growth of structure. This is possible since nonlinear evolution of the density field at the BAO scale is dominated by the infall velocities; these bulk flows are generated by the same structures observed in the density field. The algorithm used here is similar in spirit to \cite{Pad12}; our particular implementation is the same as in \cite{Vargas15}, where a detailed description can be found.

\section{Theoretical Systematics of the BAO analysis} \label{main_sys}

In Sections \ref{sec:estimators}-\ref{sec:model}, we will go through the following steps of the BAO analysis and discuss the potential systematics associated with each step. 
\begin{enumerate}
\item Two-point statistics estimators
\item Random Set
\item Covariance
\item Reconstruction
\item Fiducial Cosmology
\item Modelling of the 2-point statistics
\end{enumerate}

 Our goal is to establish a theoretical systematic error budget by considering the effect on the BAO measurements with regards to variations on the methodology. 
  In order to give a reference frame to the reader, we start defining our fiducial methodology in Table~\ref{tab:fidmetho}.
We describe in detail the results of our fiducial methodology in Section \ref{sec:estimators}. The results of the fiducial methodology are characterized by the mean $\bar{x}$ and standard deviation $S_x$ of  the distributions of the best parameters ($\alpha$, $\epsilon$) and their uncertainties ($\sigma_\alpha$ and $\sigma_\epsilon$). For the variants of the methodology, we will present only the variations of the distributions compared with the fiducial methodology. \change{The detailed results of the distributions can be found in the Appendice \ref{sec:tables}.}

\begin{table}
\begin{center}
\caption{ Fiducial Methodology }
\label{tab:fidmetho}
\begin{tabular}{@{}ll} 
\hline
Analysis & Fiducial Methodology\\
\hline
Estimator& Multipoles up to $\ell=2$\\
Randoms & Post-reconstruction: $7x$ SS, $50x$ SR (covariance mocks)\\
Randoms & Post-reconstruction: $50x$ (data)\\
Covariance& Sample Covariance from 1000 MD-Patchy mocks\\
Fid. Cosmo.& $\Omega_m=0.31$, $\Omega_{\Lambda}=0.69$, $\Omega_bh^2=0.022$\\
Alam et al. (2016).
&  $h=0.676$, $n_s=0.97$, and $\sigma_8=0.8$ \\ 
Reconstruction& Smoothing scale, $15 h^{-1}$Mpc\\
Modelling&  Gaussian Damping Model\\
 \change{Modelling}& \change{$\Sigma_{||}=5, \Sigma_\perp=5$}\\
Modelling&  Fitting Range: [55, 160]  $h^{-1}$Mpc\\
Modelling&  Binning: 5 $h^{-1}$Mpc\\
Modelling& De-Wiggled P(k) template\\
Modelling& $\alpha$-$\epsilon$ parametrization\\
Modelling& Nuisance terms: 3-term $A_l(r)$\\

\hline
\end{tabular}
\end{center}
\end{table}

In Sections \ref{sec:estimators}-\ref{sec:model}, for the convenience of the reader,  we follow the same structure: we start reminding  the reader of the fiducial choice; we then present which variations of the methodology we will explore in the section; and we finish by presenting the results. 
To explore the systematic errors associated to each step, we perform the anisotropic fits to the mocks catalogues following the fiducial methodology 
as well as 
the fits for each variant of the methodology. 
When analyzing the results, we determine which variations produce statistically significant biases on the measurement; the bias is defined as:
\begin{equation}\label{eq:bias}
 b_{x}=\bar{x}-x_{\rm exp},
 \end{equation}
  where $\bar{x}$ is the mean of the variable $x$ and  the variable $x={\alpha, \epsilon, \sigma_\alpha, \sigma_\epsilon}$; $x_{\rm exp}$ is the value expected for the variable. 
To determine the systematic error associated with one step of the analysis we compared each variation to the methodology with the fiducial case and we determine the variations:
\begin{equation}\label{eq:delta}
 \Delta x=\bar{x}_{\rm var}-\bar{x}_{\rm fid}.
 \end{equation}

The systematic errors associated to each step will be used in Section~\ref{discussion} to establish the final systematic error budget. \change{For the final error budget, we will take into account only the choices that do not generate significant shifts (bias) in the best fits parameters (i.e we consider a bias is significant if it is greater that $1-\sigma$).}
We expect all of them to make a small contribution to the final error budget for the BOSS samples; however, an accurate account of the potential sources of systematics is necessary in the current era of per cent precision in distance measurements.

As an additional result in cases where the uncertainties are affected by the variation in the methodology, we also quote the differences in the error distributions using the same notation definition \ref{eq:delta}.

\section{Estimators}  \label{sec:estimators}
In this section, we quantify the systematic error related to the estimator choice.  We followed two approaches: in the first part of the section, we explored how using different estimators in configuration space generates variations in the anisotropic fits. In the second part, we examine 
the analysis performed with the DR12 combined sample by comparing it with other results.
We compare it first with an analog analysis performed in Configuration Space  using multipoles \citep{Ross16}; we include results from an analysis using multipoles in Fourier Space (\cite{Beutler16a}) and also compare it with the consensus results presented in \cite{Acacia16}.

For the first part, we implement the different estimators in the same pipeline so that the variations will be dependent only on the estimator and not on other details of the fits that could vary between different clustering analyses. The estimators used in our fiducial methodology are the multipoles. In addition to the multipoles analysis, we analyzed the wedges ($\xi_{\parallel, \perp}$) and $\omega_\ell$ estimators (all of them explained in the methodology section).

\begin{figure*}
   \centering     
    \includegraphics[width=2.6in]{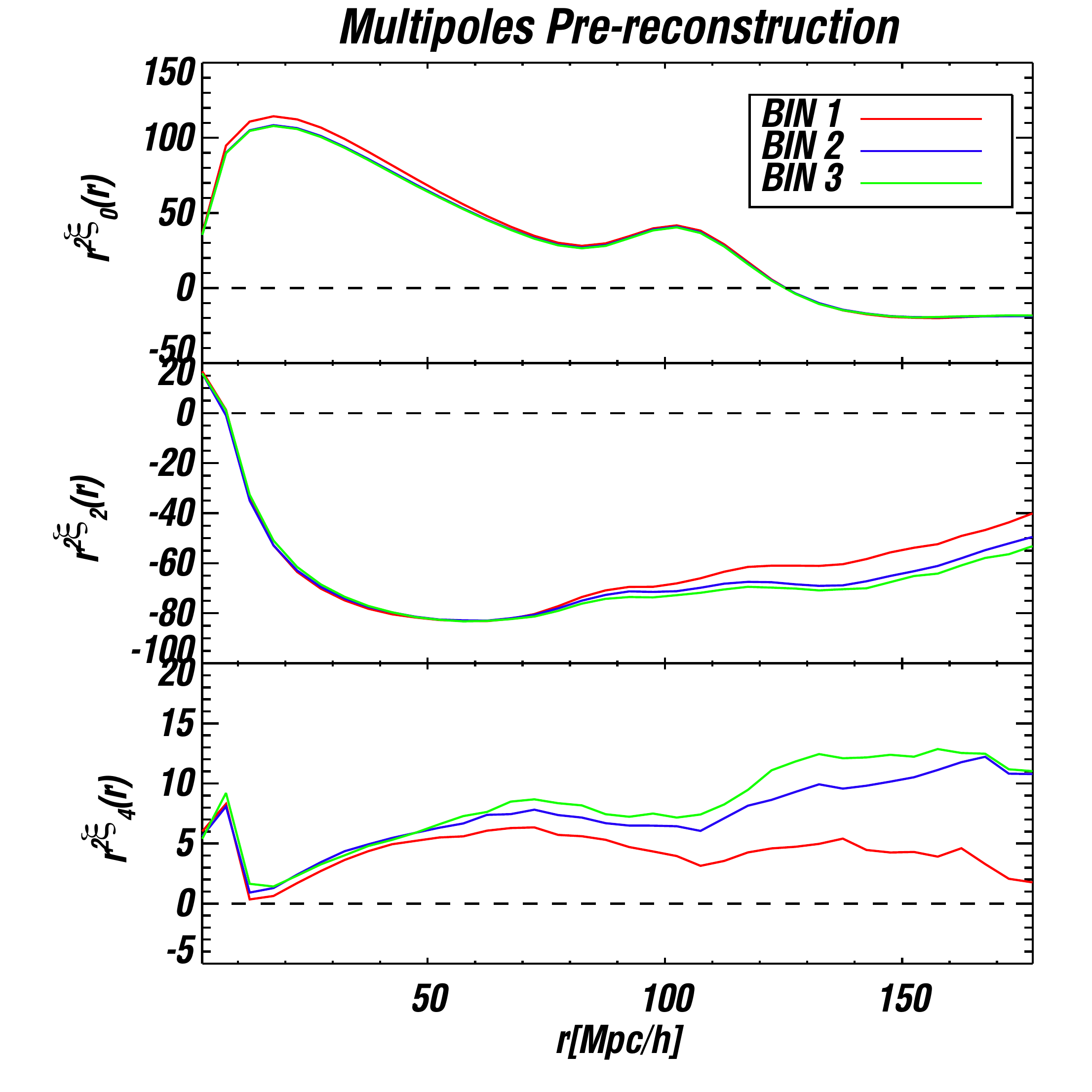}
    \includegraphics[width=2.6in]{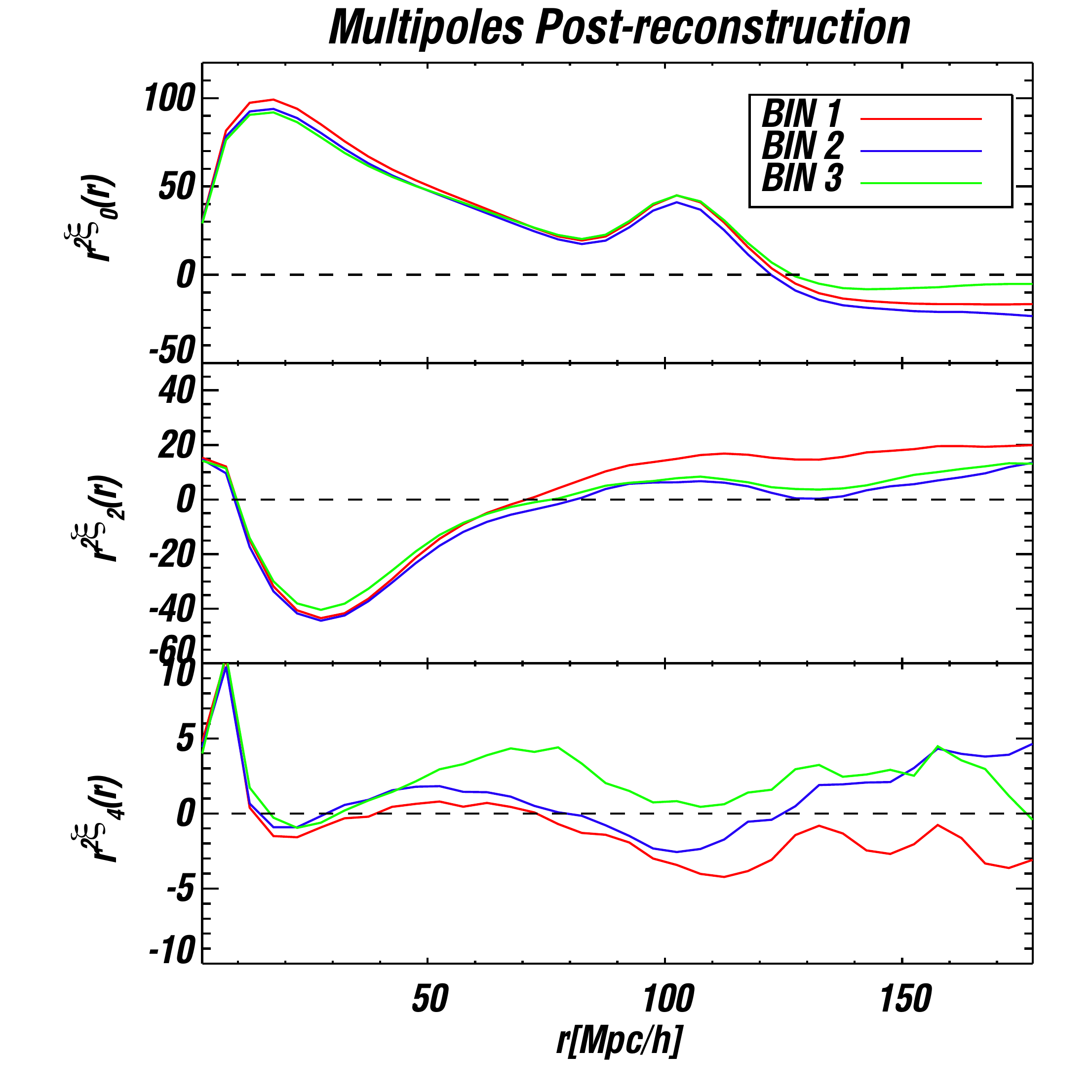}    
    \includegraphics[width=2.6in]{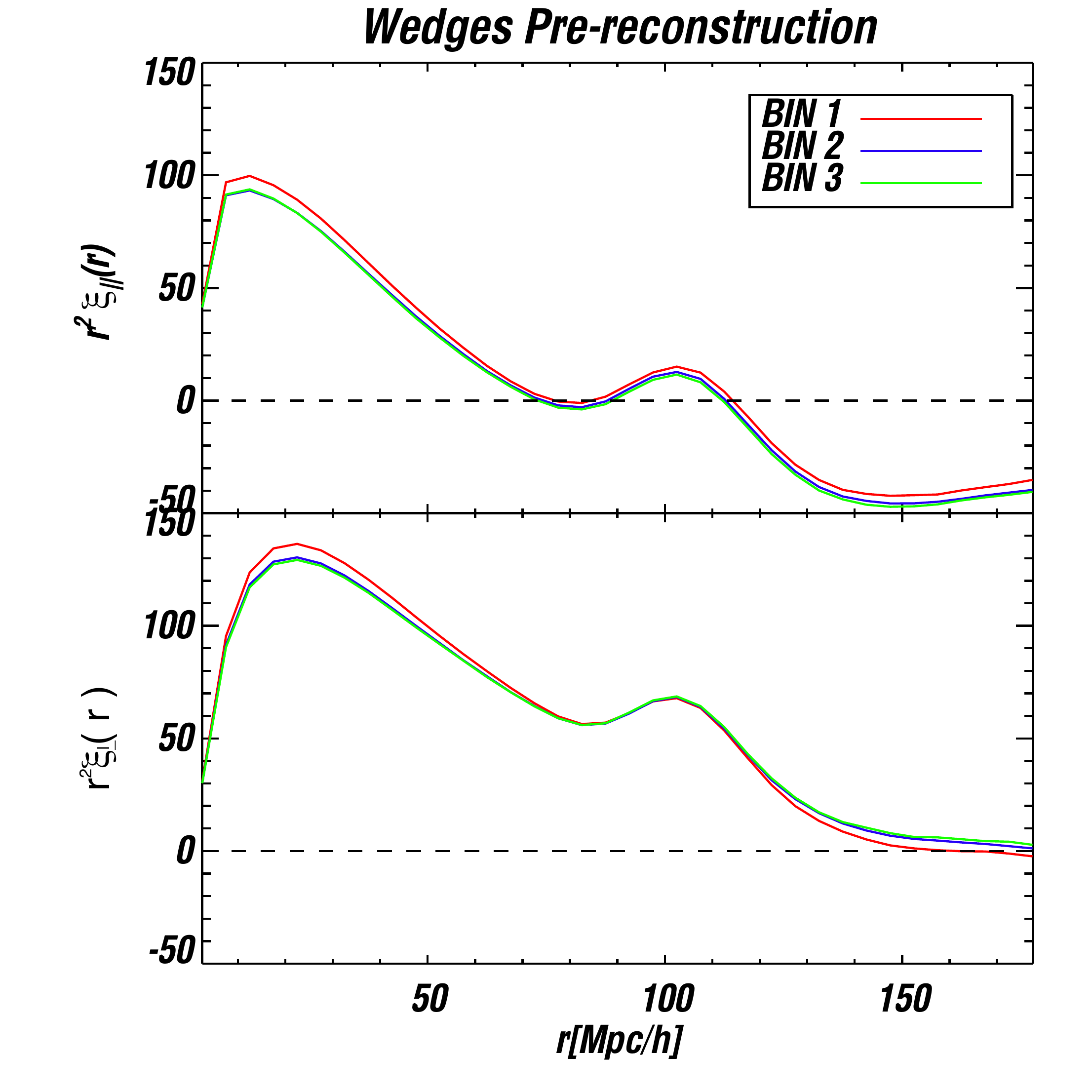}
    \includegraphics[width=2.6in]{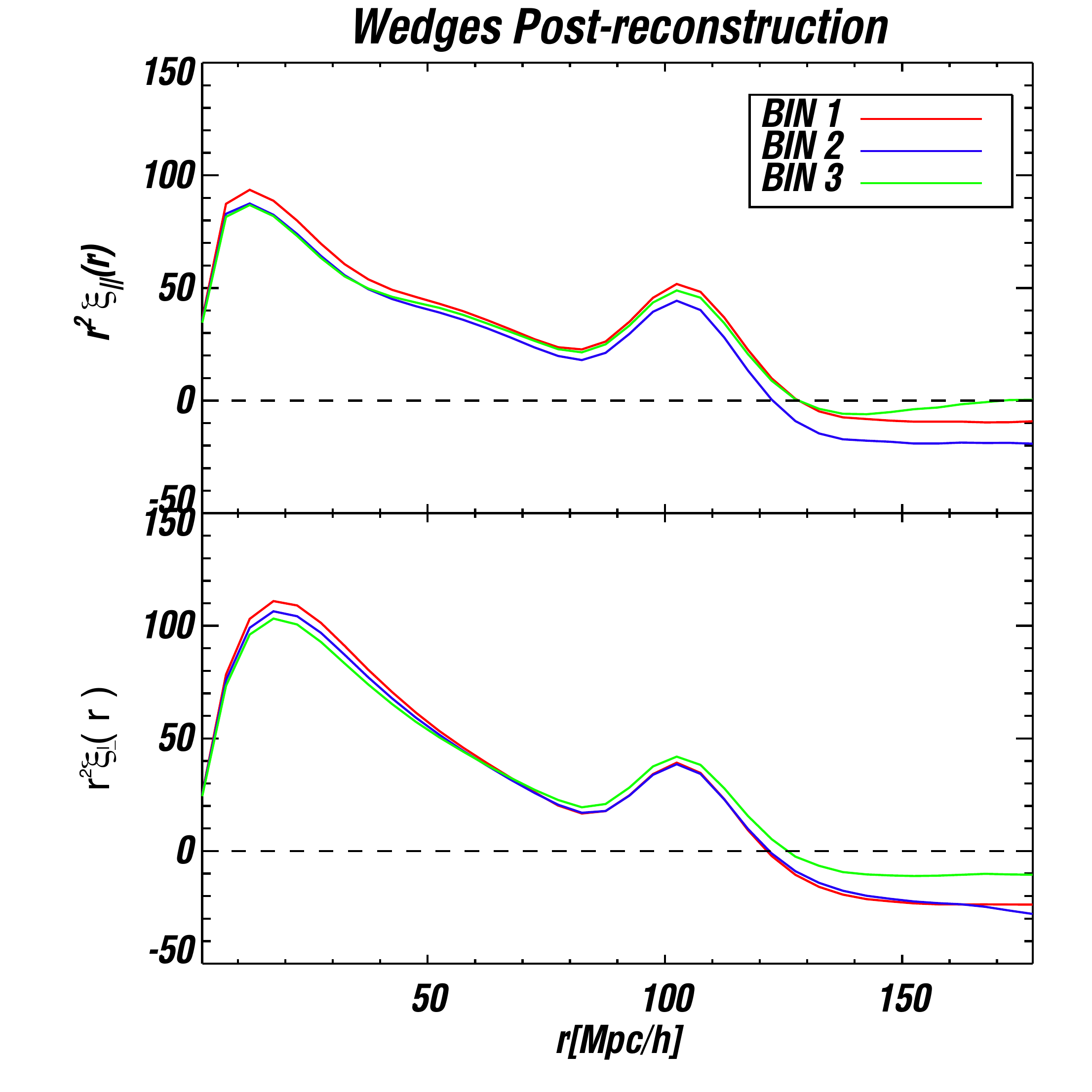}   
    \includegraphics[width=2.6in]{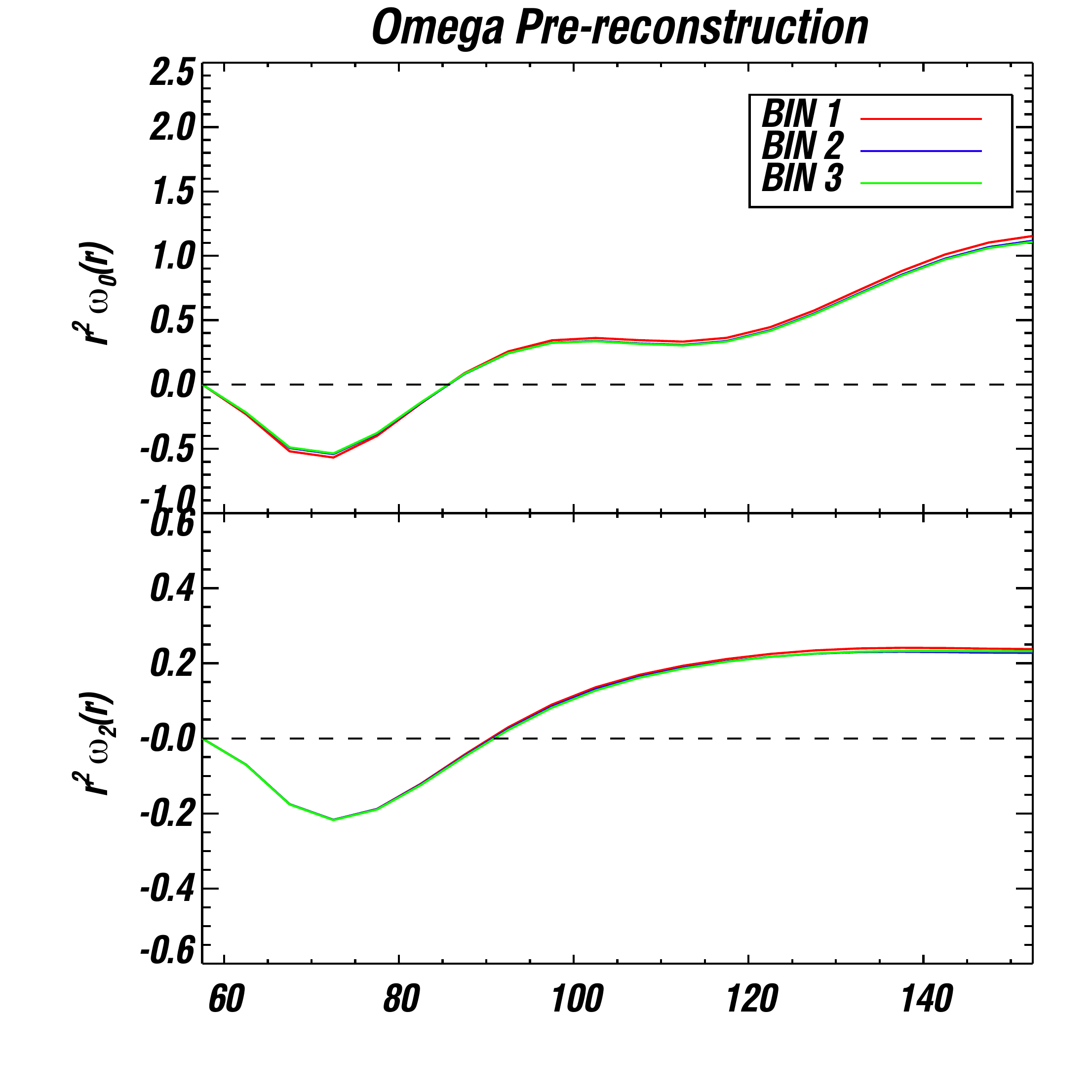}
    \includegraphics[width=2.6in]{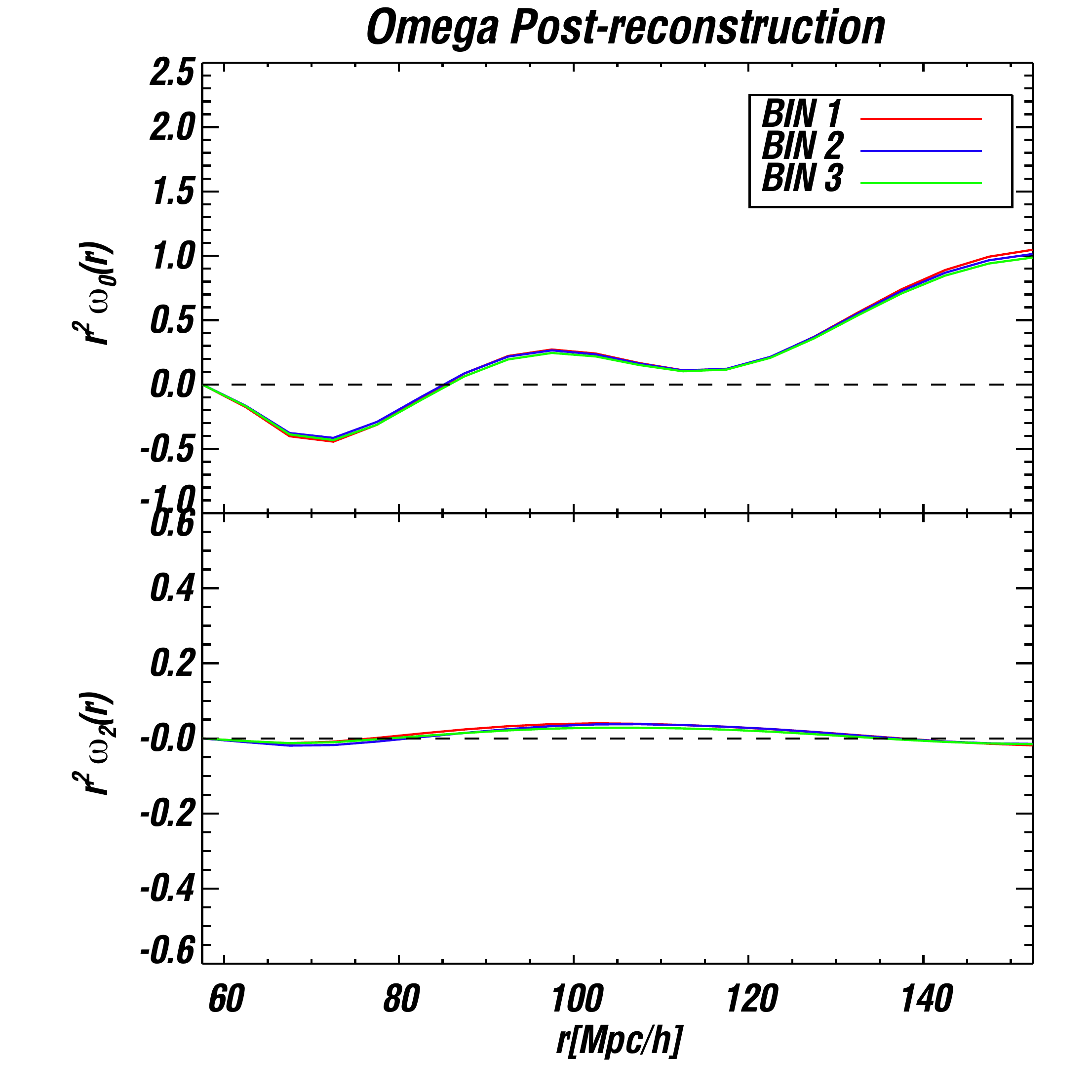}
 \caption{ [Top panels Multipoles]. Mean monopole, quadrupole, and hexadecapole from 1000 MD-PATCHY mocks of BOSS Combined Samples pre- (left) and post-reconstruction (right).\commentH{Add 'Multipoles' in the figure title'}
   [Intermediate Panels] Wedges clustering estimator: Mean of 1000 MD-PATCHY pre-reconstruction (left) and post-reconstruction (right) mocks for the 3 redshift bins. \commentH{Labels are inversed, parallel and perp.}   [Bottom Panels] $\omega_\ell$ clustering estimator: Mean of 1000 MD-PATCHY mocks pre-reconstruction (left) and post-reconstruction (right) for the 3 redshift bins. ``Bin 1" refers to the lower redshift bin ($z= 0.2 - 0.5$);  ``Bin 2" considers the intermediate redshift range ($z= 0.4 - 0.6$), and ``Bin 3" refers to higher redshift range ($z= 0.5 - 0.75$).\commentH{x-axis label missing}
   }
   \label{fig:estimators}
\end{figure*}

The multipoles measurements are shown in Figure \ref{fig:estimators}; the top panels show the mean monopole, the intermediate panels display the mean quadrupole, and the bottom panels show the mean hexadecapole from 1000 MD-PATCHY Combined mocks pre- (left) and post-reconstruction (right) for the three redshift bins. The hexadecapole contribution will be discussed in \ref{sec:hexa}; here we concentrate on the multipoles expansion up to the quadrupole, as that represents the standard analysis. 
The intermediate panels of Figure~\ref{fig:estimators} show the mean of the mocks  using a wedges clustering estimator pre- (left) and post-reconstruction (right) for 1000 MD-PATCHY for the 3 redshift bins. 
The analogue plot is shown in the bottom panels of Figure~\ref{fig:estimators}  for the $\omega_\ell$ clustering estimator pre/post-reconstruction (left/right panels) for the three redshift bins. 

We perform the anisotropic fits following the methodology described in Section~\ref{sec:method}.  We use the sample covariance pre-/post-reconstruction for performing the fits computed from the 1000 MD-PATCHY mocks for the different estimators. We show in Figure~\ref{fig:covs} the respective correlation matrices pre-/post-reconstruction for the different estimators: the top panels for the multipoles estimator; the intermediate panels for the wedges estimator and the bottom panels for the $\omega$-estimator (each column represents different redshift bins). The post-reconstruction correlation matrices of multipoles and wedges are clearly more diagonal compared with the pre-reconstructed ones\footnote{We will explore in Section \ref{sec:covariance} the effect of reconstruction in the covariance matrices itself and the fits for the multipoles case.}. 

In Table~\ref{tab:fitfiducial}, we present the results for the fiducial methodology using the multipoles. We focus first on the distributions of $\alpha$ and $\epsilon$.  The bias column of the table indicates that pre-reconstruction we have $b_\alpha^{\rm pre}=0.0022-0.0042$ and only 0.001 post-reconstruction. 
For $\epsilon$, we find a bias $b_\epsilon^{\rm pre} \sim 0.001$ 
remains at the same level, except for bin 2 that increases slightly $b_\epsilon^{\rm post}=0.002 $. The standard deviations columns ($S_{\alpha, \epsilon}$) indicates that the dispersion of $\alpha$ and $\epsilon$ are reduced post-reconstruction ($S_\alpha^{\rm pre}=0.0181-0.0227$ is reduced to $S_\alpha^{\rm post}=0.0118-0.0136$, $S_\epsilon^{\rm pre}=0.0227-0.0265$ becomes $S_\epsilon^{\rm post}=0.0132-0.0163$). 
  Concerning the error distributions of the fitting parameters, we describe the distributions with the mean and standard deviation, and we find that both quantities reduce post-reconstruction. In particular we would like to highlight the changes to the mean: the mean of the error distribution in $\alpha$ parameter pre-reconstruction $\sigma_\alpha^{\rm pre}=0.0215-0.0262$  reduces to $\sigma_\alpha^{\rm post}=0.0130-0.0147$ post-reconstruction, and these results of the mean error are consistent with the dispersions measurements from the $\alpha$ and $\epsilon$ distributions.   
For the rest of the paper, when we study variations of the distributions of the best fitting parameters related to the variations in the methodology, instead of referring to the individual distributions of the best fitting results (i.e., $b_\alpha, b_\epsilon, S_\alpha, S_\epsilon, \bar{\sigma_\alpha}, \bar{\sigma_\epsilon}, ...$) we will focus on differences of these variables compared to our fiducial case (i.e., $\Delta \alpha, \Delta \epsilon, \Delta \sigma_\alpha, \Delta \sigma_\epsilon$).

\begin{table*}
\caption{Fitting results from MD-PATCHY  mocks pre-/post-reconstruction for the fiducial methodology using multipoles $\xi_\ell$ (Section~\ref{sec:estimators}). The different columns are the mean of the distributions of the best fits parameters denoted by $\bar{x}$, the mean of distributions of the uncertainties denoted by $\bar{\sigma_x}$, the standard deviation of the distributions denoted by $S_x$, the bias defined as the difference of the mean value compared to the expected value for the variable, $b_{x}=\bar{x}-x_{\rm exp}$, where $x_{\rm exp}$ is the expected value and the mean $\chi^2/d.o.f$. 
$\alpha_{\rm exp}=[0.9993, 0.9996, 0.9999]$ and $\epsilon_{\rm exp}=[0.0002,0.0003, 0.0004]$.
}
\label{tab:fitfiducial}
\begin{tabular}{@{}lccccccccccc}
\hline

\multicolumn{12}{c}{Fiducial Methodology}\\
\hline
\multicolumn{12}{c}{DR12 Combined MD-PATCHY mocks, Pre-Reconstruction}\\
\hline
Estimator&
$\bar{\alpha}$& $S_\alpha$&
$b_\alpha$&
$\bar{\epsilon}$& $S_\epsilon$&
$b_\epsilon$&$<\chi^2/d.o.f.>$&$\bar{\sigma_\alpha}$&$S_{\sigma_\alpha}$& $\bar{\sigma_\epsilon}$&$S_{\sigma_\epsilon}$\\
\hline

Bin 1 ($0.2 < z < 0.5$)&
$1.0015$&$0.0227$&
$0.0022$&
$-0.0003$&$0.0265$&
$-0.0005$&$29.9/30$&
$0.0262$&$0.0103$&
$0.0334$&$0.0135$
\\
\hline

Bin 2 ($0.4 < z < 0.6$)&
$1.0038$&$0.0181$&
$0.0042$&
$0.0014$&$0.0231$&
$0.0012$&$29.8/30$&
$0.0222$&$0.0085$&
$0.0291$&$0.0119$\\
\hline

Bin 3 ($0.5< z < 0.75$)&
$1.0039$&$0.0181$&
$0.0040$&
$0.0003$&$0.0227$&
$0.0001$&$29.8/30$&
$0.0215$&$0.0082$&
$0.0282$&$0.0119$\\
\hline
\multicolumn{12}{c}{DR12 Combined MD-PATCHY mocks, Post-Reconstruction}\\
\hline
Bin 1 ($0.2 < z < 0.5$)&
$0.9986$&$0.0136$&
$-0.0007$&
$0.0009$&$0.0163$&
$0.0007$&$30.6/30$&
$0.0147$&$0.0067$&
$0.0188$&$0.0104$\\
\hline
Bin 2 ($0.4 < z < 0.6$)&
$1.0006$&$0.0118$&
$0.0010$&
$0.0023$&$0.0132$&
$0.0021$&$30.9/30$&
$0.0130$&$0.0058$&
$0.0163$&$0.0088$
\\
\hline

Bin 3 ($0.5 < z < 0.75$)&
$1.0007$&$0.0121$&
$0.0008$&
$0.0011$&$0.0138$&
$0.0009$&$30.6/30$&
$0.0133$&$0.0052$&
$0.0166$&$0.0082$\\
\hline
\end{tabular}
\end{table*}

We now move on to discuss the results from using different estimators.   As we are interested in measuring the systematic error associated with the estimators, we look at the variations on the mean ($\Delta \alpha, \Delta \epsilon$ defined by Equations \ref{eq:delta})\footnote{The detailed results of the fits are presented in Appendix~\ref{sec:tables}.}. 
We show the results in Table~\ref{tab:deltaest} for quantities pre-/post-reconstruction. Pre-reconstruction, we found large differences when using different estimators; further exploration should be done to investigate whether tuning the parameters of the model or using different templates results in a reduction of the differences observed in the distributions of $\alpha,\epsilon$, and their uncertainties. Pre-reconstruction correlation function fits are particularly sensitive to the model used in the fits. For the post-reconstruction cases, we find differences $\Delta \alpha < 0.002$ and $\Delta \epsilon < 0.001$ between $\omega$-multipoles estimators for the three redshift bins.  For wedges-multipoles, the difference in the mean of the best fitting values is $\Delta \alpha, \Delta \epsilon \le 0.001$ for all redshift bins. We report the RMS of the different cases as the error associated with the estimator choice in configuration space.

\begin{figure}
   \centering     
    \includegraphics[width=3.5in]{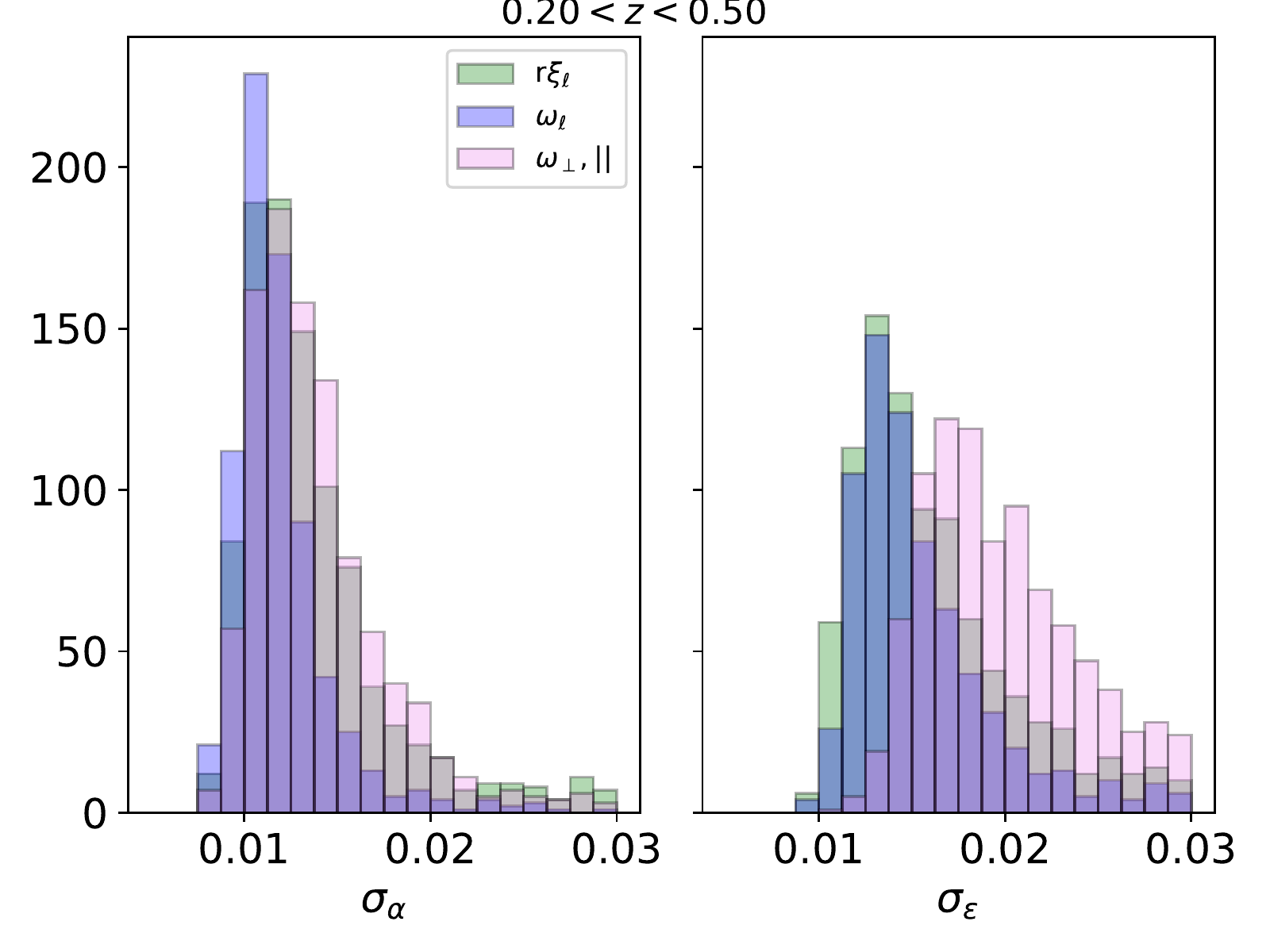}
   \caption{Error Histograms from different clustering estimators $\xi_{0, 2},\xi_{\parallel, \perp}, \omega_l$ for 1000 MD-PATCHY post-reconstruction mocks for the lowest redshift bin.
   Left panel shows distribution for $\sigma_\alpha$ and right panel for $\sigma_\epsilon$.
   Similar plots are obtained for the intermediate and higher redshift bins.}
   \label{fig:errordist}
\end{figure}

\begin{table}
\begin{center}
\caption{Estimator Systematic Error. We summarize the variations, $\Delta \alpha$, $\Delta \epsilon$ (defined by eq.\ref{eq:delta}) observed from the different estimators and their biases, $b_\alpha$, $b_\epsilon$ (defined by eq.\ref{eq:bias}).  As well as the variations in the uncertainties distributions, $\Delta \sigma_\alpha$, $\Delta \sigma_\epsilon$.
The RMS indicates the root mean square of the three redshift bins considering only the wedges and multipoles estimators.
}
\label{tab:deltaest}
\begin{tabular}{@{}lcccccc}
\hline
\multicolumn{7}{c}{\change{DR12 Combined MD-PATCHY mocks, Pre-Reconstruction}}\\
\hline
Est&$b_\alpha$&$b_\epsilon$&$\Delta \alpha$ &$\Delta \epsilon$  &$\Delta \sigma_\alpha$&$\Delta \sigma_\epsilon$ \\ 
\hline
\multicolumn{7}{c}{Bin 1 ($0.2 < z < 0.5$)}\\
\hline

$\xi_{\parallel, \perp}$&0.0091&0.0079	
&-0.0069&-0.0084&-0.0014&-0.0078\\

$\omega_\ell$	&-0.0020&-0.0044&
0.0042&	  	0.0039&0.0056&0.0061\\
\hline
\multicolumn{7}{c}{Bin 2 ($0.4 < z < 0.6$)}\\
\hline

$\xi_{\parallel, \perp}$&0.0084&0.0065&
-0.0042&-0.0053&-0.0015&-0.0084\\

$\omega_\ell$	&-0.0003&-0.0030&
0.0045&0.0041&0.0003&0.0004\\
\hline
\multicolumn{7}{c}{Bin 3 ($0.5 < z < 0.75$)}\\
\hline

$\xi_{\parallel, \perp}$	&0.0072&0.0077&
-0.0032&		-0.0076&-0.0006&-0.0081\\

$\omega_\ell$	&-0.0003&-0.0031&
0.0043&		0.003&0.0039&0.0038\\
\hline
RMS &-&-&0.0047&0.0057 & 0.0029 & 0.0064\\
\hline
\multicolumn{7}{c}{\change{DR12 Combined MD-PATCHY mocks, Post-Reconstruction}}\\
\hline
\multicolumn{7}{c}{Bin 1 ($0.2 < z < 0.5$)}\\
\hline

$\xi_{\parallel, \perp}$	&-0.0005&0.0004&
-0.0002&		0.0003&0.0002&0.0029\\

$\omega_\ell$	&-0.0026&0.0010&
0.0019&	  	-0.0003&0.0025&0.002\\
\hline
\multicolumn{7}{c}{Bin 2 ($0.4 < z < 0.6$)}\\
\hline

$\xi_{\parallel, \perp}$	&0.0013&0.0031&
-0.0003&		-0.001&0.0001&-0.0034\\

$\omega_\ell$	&-0.0004&0.0027&
0.0014&		-0.0007&$<$0.0001&-0.0033\\
\hline
\multicolumn{7}{c}{Bin 3 ($0.5 < z < 0.75$)}\\
\hline

$\xi_{\parallel, \perp}$	&0.0010&0.0015&
-0.0002&		-0.0006&0.0003&-0.0037\\

$\omega_\ell$	&-0.0010&0.0007&
0.0018&		$<$0.0001&0.0021&0.0014\\
\hline

RMS&-&-&0.0002 &0.0007&0.0002&0.0033\\
\hline
\end{tabular}
\end{center}
\end{table}

In addition to our results about two-point estimators in configuration space, we include the results from comparing similar BAO analyses performed with the DR12 combined sample \citep{Beutler16a,Ross16,Acacia16} to our Configuration Space fiducial methodology denoted $\xi_\ell$(V). The results are shown in Table~\ref{tab:estimatorAca}. From comparing the multipoles results from \cite{Ross16} to our fiducial case, we get variations $\Delta \alpha, \Delta \epsilon \le 0.001$ (for more precise figures check numbers in the Table~\ref{tab:estimatorAca}). From comparing the multipoles in Fourier Space \citep{Beutler16a} (FS) with our fiducial model, we also find  variations $\Delta \alpha < 0.001$, but the variations in $\epsilon$ are slightly larger ($\Delta \epsilon \le 0.002$)\footnote{Similar results are found when comparing FS with \cite{Ross16} results.}. Finally, we quote the BAO consensus results from \cite{Acacia16} (generated from \cite{Sanchez16b} methodology that optimally combines the multipoles in CS and FS) with our fiducial methodology. For the error budget for \cite{Acacia16}, we quoted the last one as the systematic error associated to the estimator step, i.e $\Delta \alpha=$0.0004  and $\Delta \epsilon=$0.0012  taking the RMS of the three redshift bins. 

\begin{table}
\begin{center}
\caption{Systematic errors from the choice of post-reconstruction estimator, from Alam et al. (2016).
We quote $\Delta \alpha$ and $\Delta \epsilon$ (defined by eq.\ref{eq:delta}) over every estimator considered for Alam et. al. (2016)  BAO-only, as well as their biases,  $b_\alpha$ and $b_\epsilon$ (defined by eq.\ref{eq:bias}). We include the results from using different multipoles analysis in configuration space (this work and  
 Ross et al. 2016 hereafter denoted by $\xi_\ell(R)$). We present also how the multipoles in configuration space (our fiducial methodology) compare to the multipoles in Fourier space using 
Beutler et al. (2016), hereafter denoted by $P_\ell$. Finally, we note the differences when using the consensus values from BAO-only (Alam et. al. 2016) methods as compared to our fiducial methodology. The RMS indicates the root mean square of the three redshift bins over the consensus lines.}
\label{tab:estimatorAca}
\begin{tabular}{@{}lllcccccc} 
\hline
Estimators&$b_\alpha$&$b_\epsilon$&$\Delta \alpha$& $\Delta \epsilon $\\
\hline
\multicolumn{5}{c}{Bin 1 ($0.2 < z < 0.5$)}\\
\hline
$\xi_\ell$(V) &-0.0007&0.0007&-&-\\
$\xi_\ell(R)$&0.0002&0.0009&0.0009&0.0002\\
$P_\ell$ &-0.0012&0.0030&-0.0005&0.0023\\
Consensus*&-0.0008&0.0025&$<$0.0001&0.0018\\
\hline
\multicolumn{5}{c}{Bin 2 ($0.4 < z < 0.6$)}\\
\hline
$\xi_\ell$(V) &0.0010&0.0021&-&-\\
$\xi_\ell(R)$&0.0014&0.0011&0.0004&-0.0009\\
$P_\ell$ &0.0005&0.0021&-0.0005&0.0012\\
Consensus*&0.0008&0.0025&-0.0002&0.0005\\
\hline
\multicolumn{5}{c}{Bin 3 ($0.5 < z < 0.75$)}\\
\hline
$\xi_\ell$(V) &0.0008&0.0009&-&-\\
$\xi_\ell(R)$&0.0015&-0.0007&0.0007&-0.0015\\
$P_\ell$ &-0.0004&0.0013&-0.0012&0.0006\\
Consensus*&0.0002&0.0016&-0.0006&0.0008\\
\hline
RMS & -&-&0.0004&0.0012\\
\hline
\end{tabular}
\end{center}
\end{table}

In addition to the variations in the best fits for $\alpha$ and $\epsilon$, for establishing the theoretical systematic error budget, we also analyse the variations in the uncertainties derived from using different estimators and we quote the differences of the mean of the error distributions $\Delta \sigma_\alpha, \Delta \sigma_\epsilon$. The mean differences in the uncertainty, $\Delta \sigma_\alpha,$ varies 0.0001--0.0025 between estimators for the post-reconstruction mocks in  the three redshift bins. As an illustration of this difference, we show in Figure~\ref{fig:errordist} the error distributions for the first redshift bin using the three different estimators. We observe that the distributions of $\sigma_\alpha$ for the $\omega$-estimator  is narrower with a slightly lower mean error compared with the multipoles errors distribution. The wedges $\sigma_\alpha$ distribution follows a similar trend to the multipoles. For $\epsilon$, we observe again a slightly narrower distribution for the $\omega$-estimator compared with the multipoles; for wedges, error distribution is centred in a larger value compared with the other two distributions.  We verified the likelihoods surfaces and 1-D probability distributions on a mock-by-mock basis as a sanity check to detect problems in the fits, and we found  that the larger uncertainties are coming from the wider $\chi^2$ surfaces that generate wider $p(\epsilon)$ distributions when marginalising over $\alpha$. 
The fact that the multipoles method should do slightly better than two wedges was shown in \cite{Ross15}. Also it is not surprising that the $\omega$-statistic  provides smaller errors, as it might represent a better compression of the data given that it is less sensitive to nuisance terms in the fitting.

To summarize our systematic errors from BAO distance measurements associated with using different estimators, we conclude that $\Delta \alpha < 0.0012$ and $\Delta \epsilon= 0.0006$.
 However, as in \cite{Acacia16} we are only using the multipoles approach (in configuration and Fourier spaces) for getting the consensus values, and as we combined the results from the multipoles in CS and FS optimally using \cite{Sanchez16b} methodology, we decided to set the systematic error as the difference between the consensus value and the Fiducial methodology in this paper for consistency with the rest of the methodology followed in the paper. Thus the systematic error associated with the estimators to be used in the final error budget is $\Delta \alpha < 0.0004$ and $\Delta \epsilon= 0.0012$.

\section{Randoms} \label{sec:randoms}
The computation of two-point estimators requires the generation of a random set matching the survey geometry and completeness. The error on the correlation function due to the  number of randoms is related to the error in the determination of the correlation function for a given sample. In the limit of an infinite random set, the noise contribution from the randoms  is zero. However, in practise, using a large number of randoms significantly increases  the cost of computing the estimator in the analysis.  In this section, we test the effect of using different sizes of the random catalogue in anisotropic fits.

For the pre-reconstruction case, the RR and RD pair-counts used in the Landy-Szalay estimator are computed once for all the mocks. The finite size of the randoms coupled with the fact that we use the same random file for all of the mocks is a potential source of scatter that is not in the covariance matrix. We test using two different sets of randoms. If we use a single random catalogue, we might get some small shift that is purely from that random catalogue not being infinite. The difference between that shift and the shift found from a much larger random catalogue should indeed tell us something about the size of the systematic uncertainty.

For the post-reconstruction case, the approach is slightly different.  In the post-reconstruction case, there are two sets of random catalogues: the denominator randoms used to determine the geometry of the survey, and the numerator randoms that we shift according to the displacement field inferred from data; we call this shifted random catalogue. The shifted random catalogue is  different for every single mock, thus the DS and SS counts need to be recomputed for each mock.  From the definition of the Landy-Szalay estimator, we know that the variance in the correlation functions is dominated by the DR pair-counts, which goes to $4/x$ (where $x$ is the ratio of randoms-to-data), with the RR paircounts being subdominant, scaling as $1/x^2$. Thus, if we use $x=4$, the variance is 10 times bigger than if we use $x=40$. Thus, we decided to test the effect of the the size of the random catalogues in this case for the $SS$ term for balancing the errors in the correlation function estimate. \footnote{In particular, computing the post-reconstruction correlation function requires the computation of very expensive random-random pair-counts for each one of our thousands of mock catalogues. Thus, this test is critical for minimizing the computational cost of  doing so for 1000 post-reconstruction mocks for the 3 redshift bins.}

\subsection{Random size Pre-reconstruction}

 We test the effect of the randoms pre-reconstruction with 100 QPM CMASS mocks; we use as fiducial cosmology the natural cosmology of the QPM mocks (see Table~\ref{tab:cosmofid}).
We test two random sets: the first case is a random set 20 times the size of the data sample, denoted by ``$20\times$", and a second set 50 times the size of the data sample, denoted by ``$50\times$". We computed the multipoles using these two set of randoms and then 
we performed the anisotropic analysis described in Section~\ref{sec:method}.  
The results are shown in Table~\ref{tab:randtest_delta}; we report the mean difference of fits performed on the correlation functions using $20\times$ compared to the correlation functions computed with $50\times$ randoms.

\begin{table}
\begin{center}
\caption{Fitting results from QPM CMASS mocks Pre/Post-Reconstruction: Random Test (Section~\ref{sec:randoms}). The variations in the best fits are defined as: $\Delta \alpha=\alpha_{50\times}-\alpha_{20\times}$ and $\Delta \epsilon=\epsilon_{50\times}-\epsilon_{20\times}$ for pre-reconstruction test.
$\Delta \alpha=\alpha_{50\times}-\alpha_{4\times}$ and $\Delta \epsilon=\epsilon_{50\times}-\epsilon_{4\times}$ for post-reconstruction test. The RMS indicates the root mean square of the three redshift bins.  ``Bin 1" refers to the lower redshift bin ($z= 0.2 - 0.5$);  ``Bin 2" considers the intermediate redshift range ($z= 0.4 - 0.6$), and ``Bin 3" refers to higher redshift range ($z= 0.5 - 0.75$).}
\label{tab:randtest_delta}
\begin{tabular}{@{}lcc} 
\hline
Set&$|<\Delta \alpha>| \pm$ RMS & $|<\Delta \epsilon>| \pm$ RMS \\
\hline
CMASS Pre-Rec&0.0008 $\pm$ 0.0038&0.0006 $\pm$ 0.0066 \\
\hline
Bin 1 Post-Rec&0.0002 $\pm$ 0.0017&0.0003 $\pm$ 0.0035\\
\\[-1.5ex]
Bin 2 Post-Rec&0.0002 $\pm$ 0.0011&$<$0.0001 $\pm$ 0.0015\\
\\[-1.5ex]
Bin 3 Post-Rec& 0.0001 $\pm$ 0.0015&$<$0.0001 $\pm$ 0.0017\\
\hline
RMS &0.0002&0.0002\\
\hline
\end{tabular}
\end{center}
\end{table}

The main conclusion is that randoms make very small differences in the mean of isotropic/anisotropic fits for both of these two cases. The differences observed are $\le 0.0002$. 

\subsection{Different number of randoms for DR and RR terms post-reconstruction}
 We test the effect of the randoms post-reconstruction with 100 QPM CMASS mocks; we use as fiducial cosmology the natural cosmology of the QPM mocks (see Table~\ref{tab:cosmofid}). We test the effect of randoms differently than in pre-reconstruction. We reduce the number of randoms from $50\times$ to $4\times$  for post-reconstruction correlation functions in the numerator  SS (not for the SR term)  and we test the impact on the fitting parameters. 
We show the results in Table \ref{tab:randtest_delta} of the best fits distributions obtained with 100 mock catalogues.
 The differences between the $50\times$ and the $4\times$ anisotropic fits are very small: $0.0002$ for $\alpha$ and $0.0002$ for $\epsilon$. 
The variations in the anisotropic fits when using different numbers of randoms must be random (i.e., not systematic); these variations must be statistical fluctuations and they should be in our covariance matrices if we use the same numbers of randoms. However, the finite size of the randoms is a potential source of scatter that is not in the covariance matrix, and what is demonstrated is that it is not contributing significantly to the scatter.

As an additional output of this test, the results validate the use of $50\times$ pair-counts for DS pair-counts and $7\times$ for SS for the mock covariance  for \cite{Acacia16}. With the current two-point correlation code and the number of mocks required for doing BAO/RSD analysis using the correlation function, a total of 132,000 CPU-hours were required for estimating the post-reconstruction pair-counts using EDISON at NERSC. DR12 correlation functions were computed with $50\times$ randoms pre- and post-reconstruction.
  
We conclude that there is no systematic error in BAO distance measurements associated with the number of randoms. 
The variations in the scatter are $<0.0001$ in $\alpha$ and $\epsilon$.

\section{Covariances} \label{sec:covariance}
The covariance matrix, in particular its inverse, is crucial for obtaining the cosmological parameters from the galaxy surveys \citep{Per14}.
The usual way to get the covariance matrix is to use the brute force approach, which consists of using a large number of mock catalogues to estimate the sample covariance matrix.
For the analysis of BOSS, two sets of mock catalogues were considered, QPM and MD-PATCHY. Different methods were used to generate these sets of mocks, and so we test the systematic consequences of using the corresponding covariance matrices for BAO fitting.

New approaches have been proposed in the existing literature to avoid using mock catalogues given their computational cost \citep{Pope08,delaTorre13,Paz15,Schneider11,Oco15}.\commentH{use citation}
In addition to the testing different sets of mocks, we tested the theoretical approach developed by \cite{Oco15} based on a Gaussian model where few parameters are calibrated with the simulations. 

In \ref{sub:cov_mocks}, we discuss general differences between the (inverse) covariance matrices provided by the QPM and MD-PATCHY mocks. In \ref{sub:cov_model}, we discuss the model covariance matrix and compare its general features to the mock covariance matrices. In \ref{sub:cov_summary}, we present the results of using each matrix for non-linear BAO fitting. Throughout this section, the fiducial choice for the covariance matrix is the sample covariance obtained from the MD-PATCHY mocks. 

\subsection{\label{sub:cov_mocks}Comparison of Sample Covariance Matrix in DR12 }

We analyze the sample covariances generated from two sets of DR12 mock catalogues: QPM and MD-PATCHY\footnote{We test in Appendix~\ref{sec:covarDR11} how the covariance generated with the previous generation of mock catalogues (PTHALOS) compares with one generated with the new sets of mock catalogues available (QPM).}. We separate the analysis into two different aspects:  1) Sample Covariance differences related to reconstruction, 2) Sample Covariance differences related to different mock catalogues. 
We study how the structure of the covariance matrix changes between different cases (pre- vs post-reconstruction mocks and QPM vs MD-PATCHY mocks) and how these differences \change{affect} the anisotropic fits. 
For the comparisons, we chose to work with  the inverse of the covariance matrix, i.e., \change{the precision matrix denoted by $\Psi$}. The main reason is that it is the precision matrix which is used directly in the fitting process, and the mapping between the covariance matrix and its inverse is highly non-linear.
An additional advantage of working with the precision matrices is that the structure of the precision matrix is simpler, thus it is easier to determine the similarities/discrepancies in the matrices.

\begin{figure*}
\begin{centering}
\hspace*{-6.em}
\includegraphics[width=8.5in]{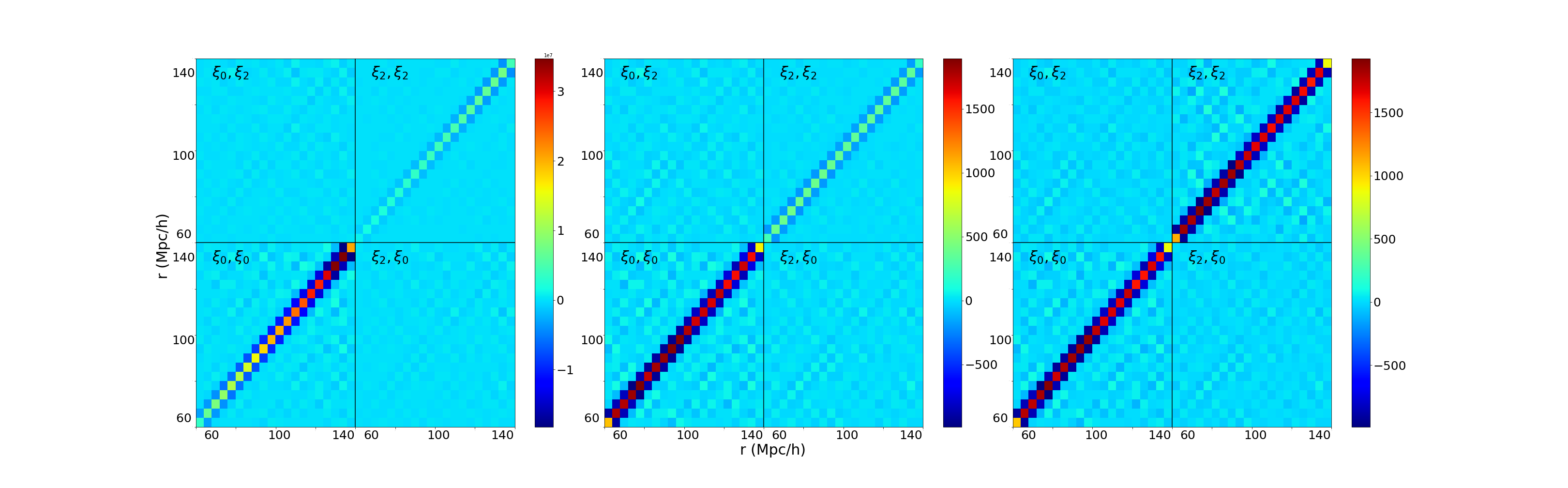}
\par\end{centering}
\caption{Precision matrix for the post-reconstruction MD-PATCHY mocks (for the intermediate redshift bin, $z= 0.4 - 0.6$) with different scalings. On the left, the precision matrix  with no rescaling applied. It illustrates that the bulk of the structure in the precision matrix is on the diagonal and first off-diagonal. In the middle, we divide out the naive scaling with $r$ and plot $\Psi_{ab}/r_{a}r_{b}$.This illustrates that the naive scaling largely captures the $r$-dependence of the precision matrix. On the right, we apply the radial scaling and scale up the quadrupole by a factor chosen to put the monopole and quadrupole on equal footing, which helps assess the amplitude of the main monopole-quadrupole entries.}\label{fig:covscaling}
\end{figure*}

In order to make plots of the precision matrices easier to read, we provide a discussion of the general features of the precision matrices considered in this paper.
As is readily seen in Figure \ref{fig:covscaling}, the structure of the precision matrix is primarily (but not exclusively) concentrated on the diagonal and first off-diagonal of the matrix.
The naive $r$-dependence of these diagonal contributions is $\Psi_{ab}\sim r_a r_b$, and so throughout this section we will plot $\Psi_{ab}/r_a r_b$. This rescaling is applied to the middle plot in Figure \ref{fig:covscaling}, which illustrates that this removes most (but not all) of the $r$-dependence from the plot.
The middle plot also indicates that our measurements of the monopole (lower left) are significantly more precise than our measurements of the quadrupole (upper right). We can compensate for this by dividing the quadrupole by a compensating factor of $\sqrt{4.7}$, then propagating this rescaling through to the precision matrix. The result, shown in the rightmost plot in Figure \ref{fig:covscaling}, is that we can see that the precision for the quadrupole has essentially the same structure as the precision for the monopole, and that after these rescalings the noise on the precision matrix is approximately homogeneous. We can also assess the significance of the monopole-quadrupole terms, relative to the monopole-monopole and quadrupole-quadrupole terms.
In all of these plots, a sharp decline in precision is observed for the bins with the smallest and largest $r$. This arises because these are the smallest/largest bins that we consider, and does not indicate anything else unusual about those bins.


Before assessing the detailed structure of the various precision matrices, we consider whether any of these matrices are, in total, ``more precise" than any of the others.
We do this comparison by evaluating $\log\left(\det\left(\Psi\right)\right)$, which captures the overall amplitude, on each matrix. 
In order to facilitate comparison between each matrix and the fiducial case (post-reconstruction MD-PATCHY
mocks), we also calculate the overall rescaling that would have to be applied to the \emph{fiducial} matrix in order to make its amplitude match that of the other cases. We report these as percent changes, with negative values indicating that the matrix is \emph{less} precise than the fiducial case, and positive values indicating that the matrix is \emph{more} precise than the fiducial case.
We discuss this table in the following subsections.

\begin{table}
\caption{Evaluation of $\log\left(\det\left(\Psi\right)\right)$ for each matrix. This quantity captures the overall amplitude of each matrix, i.e. facilitate the comparison between each matrix and the fiducial case (post-reconstruction MD-PATCHY 
mocks). We also calculate the overall rescaling that would have to be applied to the \emph{fiducial} matrix in order to make its amplitude match that of the other cases.
We report these as percent changes, with negative values indicating that the matrix is \emph{less} precise than the fiducial case, and positive values indicating that the matrix is \emph{more} precise than the fiducial case.} \label{tab:detcov}
\begin{tabular}{@{}lcc}
\hline
Values of $\log\left(\det\left(\Psi\right)\right)$\\
\hline 
Mocks & $\log\left(\det\left(\Psi\right)\right)$ & Percent change\\
\hline 
\multicolumn{3}{c}{Bin 1 ($0.20 < z < 0.50$)}\\
\hline 
QPM, pre-reconstruction & 594.2 & -7.6\%\\
MD-PATCHY, pre-reconstruction & 588.2 & -19.0\%\\
QPM, post-reconstruction & 600.5 & +8.1\%\\
MD-PATCHY, post-reconstruction & 597.1 & -\\
Model & 597.1 & -0.7\%\\
\hline
 \multicolumn{3}{c}{Bin 2 ($0.40 < z < 0.60$)}\\
\hline 
QPM, pre-reconstruction & 601.9 & -10.1\%\\
MD-PATCHY, pre-reconstruction & 598.0 & -16.8\%\\
QPM, post-reconstruction & 607.8 & +4.1\%\\
MD-PATCHY, post-reconstruction & 606.1 &-\\
Model & 605.8 & -0.9\%\\
\hline 
\multicolumn{3}{c}{Bin 3 ($0.50 < z < 0.75$)}\\
\hline 
QPM, pre-reconstruction & 586.2 & -11.6\%\\
MD-PATCHY, pre-reconstruction & 585.0 & -12.8\%\\
QPM, post-reconstruction & 591.5 & +0.9\%\\
MD-PATCHY, post-reconstruction & 591.1 &-\\
Model & 591.2 & +0.3\%\\
\hline 
\end{tabular}
\end{table}

\subsubsection{Pre-/Post-reconstruction Covariances}
\begin{figure*}
   \centering     
\hspace*{-5.em}	
     \includegraphics[width=8.5in]{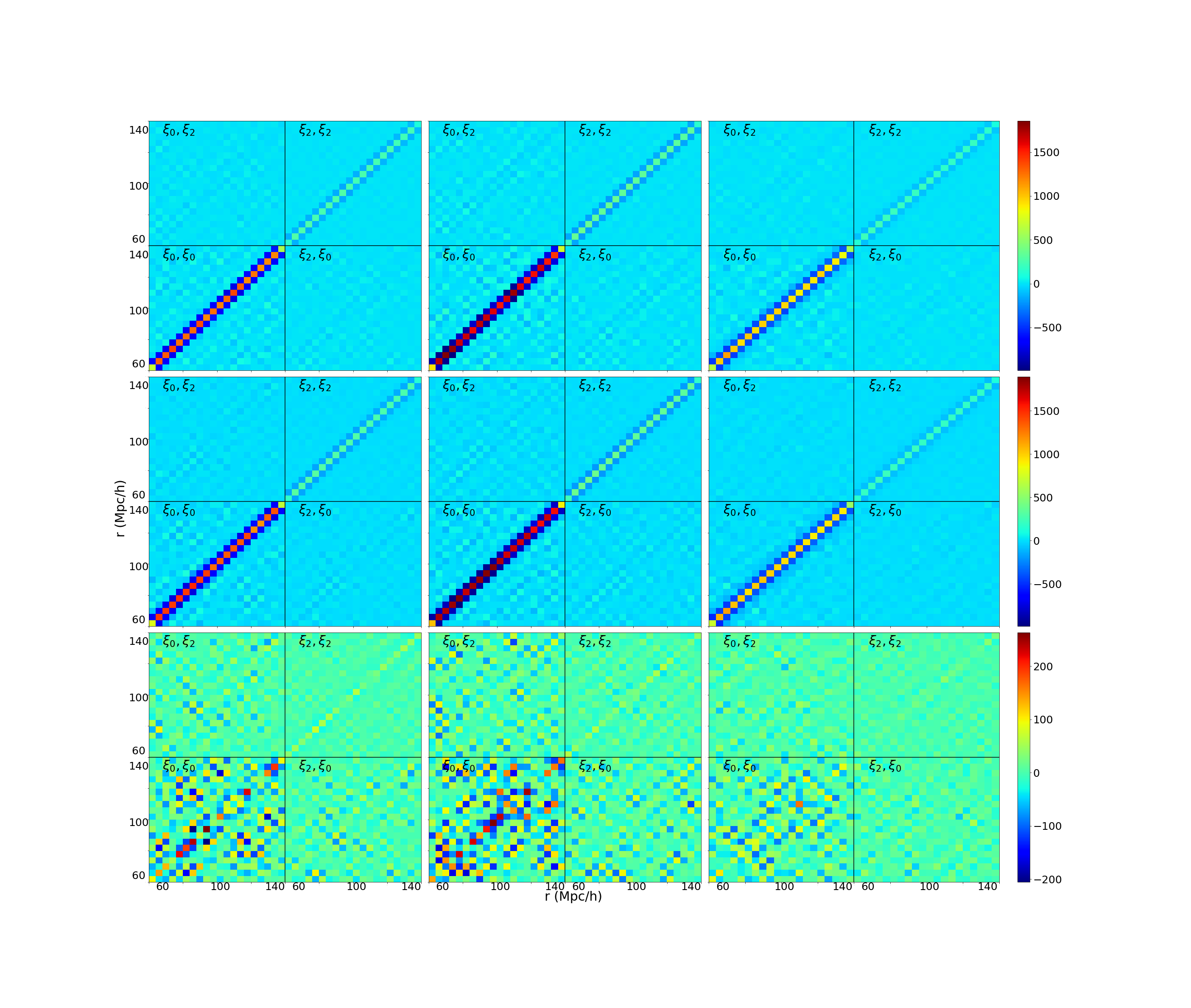}
   \caption{ 
   Precision matrix for 1000 MD-PATCHY mocks. In the first row, the pre-reconstruction precision matrix obtained  for the 3 redshift bins: low, intermediate, and high redshift bin (from left to right). In the second row, the post-reconstruction precision matrices for the corresponding redshift bins. In the third row, the difference between post- and pre-reconstruction precision matrices.}
   \label{fig:covQPMPATpre_post}
\end{figure*}

The first question we address is what  are the effects of reconstruction on the precision matrix. 
In Figure~\ref{fig:covQPMPATpre_post}, we show the precision matrix for the MD-PATCHY mocks. In the first row are the pre-reconstruction precision matrices obtained  for the three redshift bins: low, intermediate, and high redshift bin (from left to right). In the second row are the post-reconstruction precision matrices for the corresponding redshift bins. In the third row are the difference between the post- and pre-reconstruction precision matrices.
From the third row we notice a different trend for the three redshift bins, the higher redshift bin shows smaller differences compared to the lower and intermediate bins. We observe that the differences in the precision matrices pre-reconstruction are concentrated in the diagonal terms of the monopole-monopole, quadrupole-quadrupole, monopole-quadrupole; everywhere else we find a noise pattern indicating the consistency between precision matrices.
The noise level seems equivalent for pre- and post-reconstruction precision matrices. 

In Table \ref{tab:detcov}, we quantify the overall rescaling of the precision matrix required to make the values of $\log(\det(\Psi))$ match. We observe a $\sim$16\% increase in the precision post-reconstruction for the MD-PATCHY mocks (taking the RMS of the three redshift bins), and a $\sim14\%$ in MD-PATCHY mocks likely results from the effects of reconstruction on the higher-point correlation functions, since reconstruction has little impact on the overall amplitude of the two-point correlation function or the survey volume, and those together determine the Gaussian contributions to the covariance matrix. MD-PATCHY improves more than QPM, although post-reconstruction they have similar values.

\begin{figure*}
   \centering     
   \includegraphics[width=3.5in]{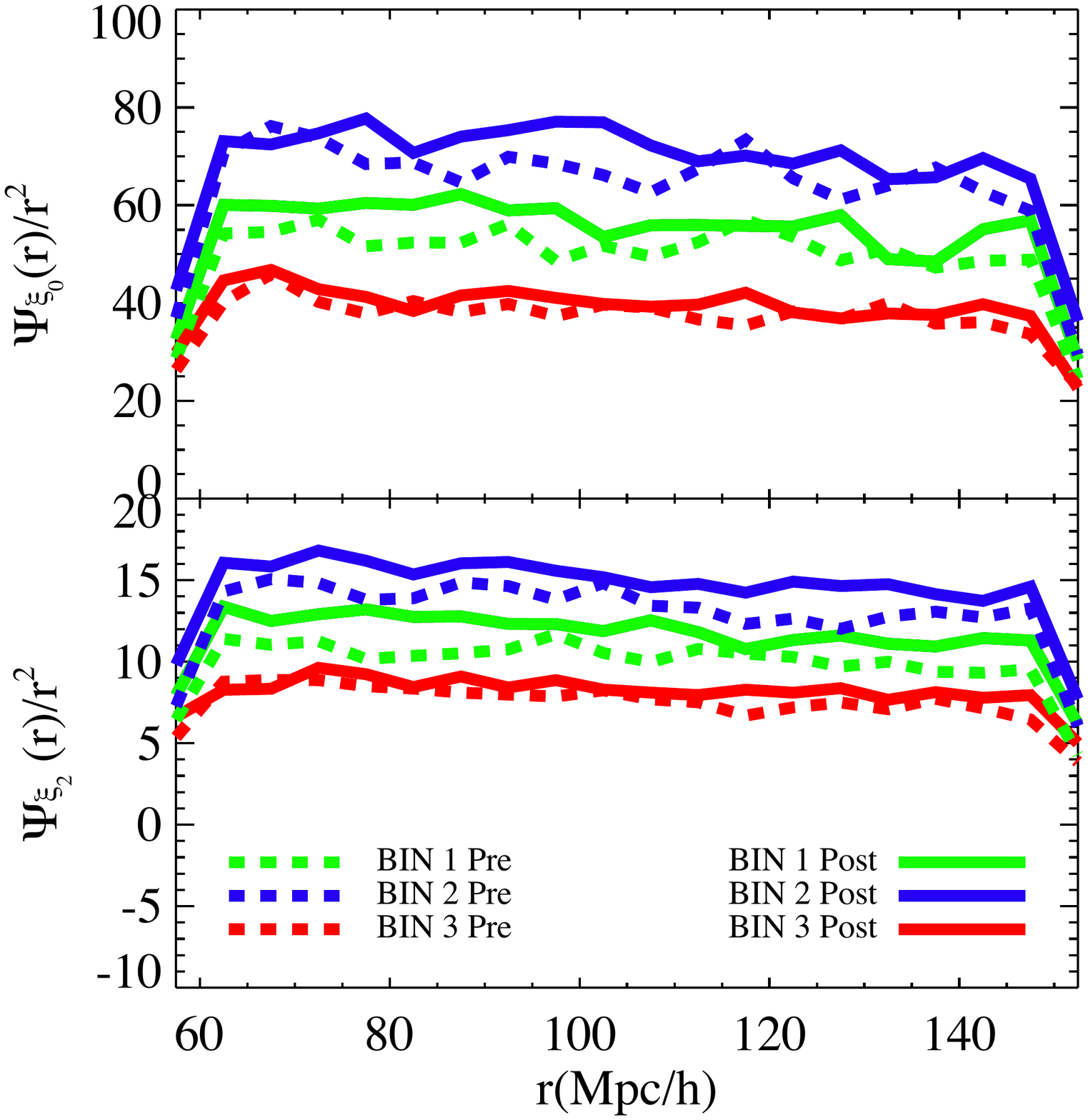} \hspace*{-3.em}
    \includegraphics[width=3.5in]{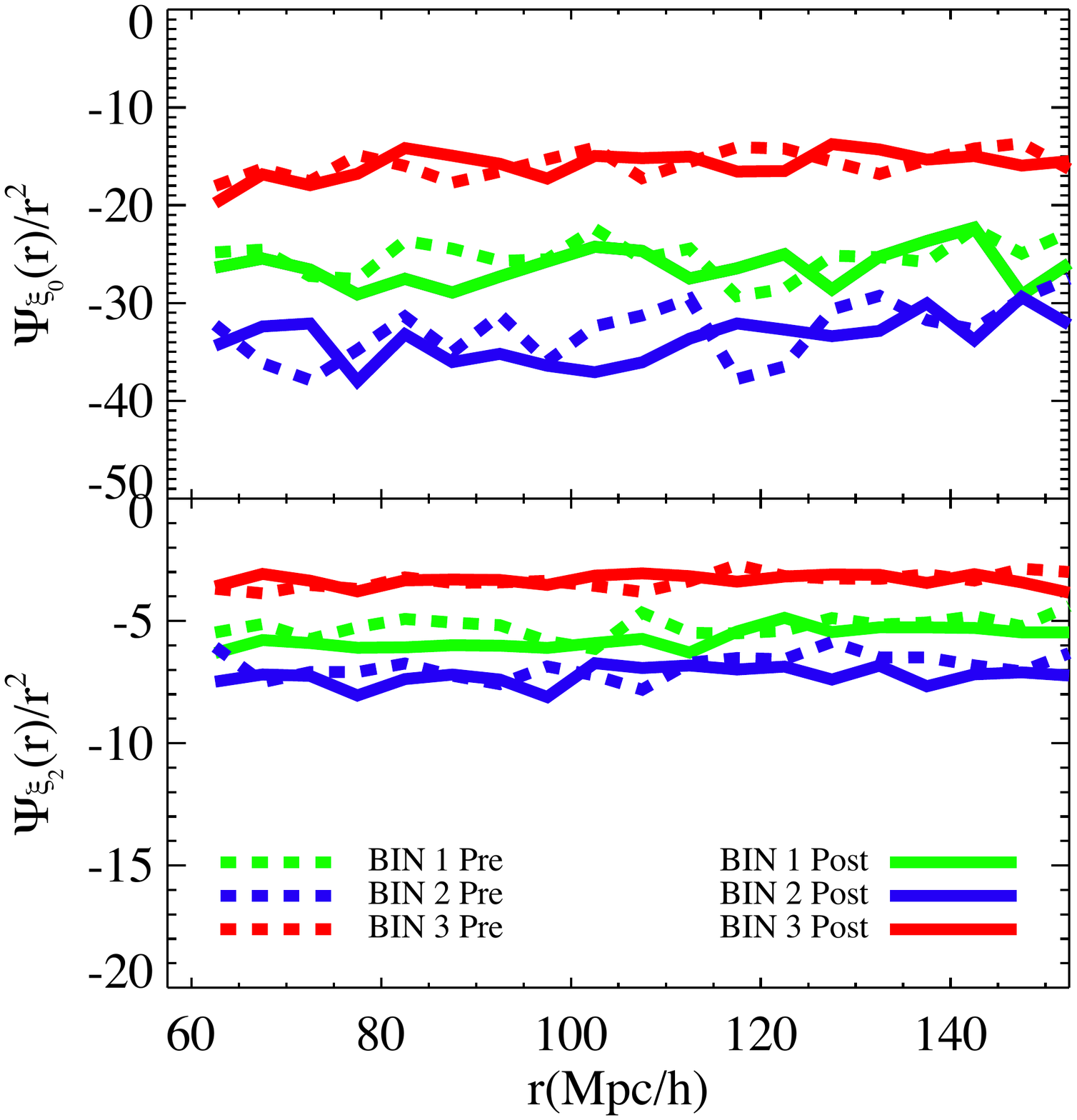}
   \caption{ Diagonal [left] and \change{first} off-diagonal [right] terms of precision matrices of MD-PATCHY for 1000 mocks for the 3 redshift bins, pre-(dashed lines) /post-reconstruction (solid lines). ``Bin 1" refers to the lower redshift bin ($z= 0.2 - 0.5$);  ``Bin 2" considers the intermediate redshift range ($z= 0.4 - 0.6$ ), and ``Bin 3" refers to higher redshift range ($z= 0.5 - 0.75$). Top panels are monopole terms, bottom panels are quadrupole terms.}
   \label{fig:diagmocksPREandPOST}
\end{figure*}

We show in the left panel of Figure~\ref{fig:diagmocksPREandPOST} the diagonal terms of the precision matrices for the three redshift bins for MD-PATCHY. In the right panel, we show the \change{first} off-diagonal terms of the precision matrices for the same three redshift bins. The top panels correspond to the monopole terms and the bottom ones to the quadrupole terms. There is a re-scaling by $r^2$ to eliminate radial trends. 
As stated above, the precision matrices are approximately uniform across $r$, once the $r_a\times r_b$ scaling is taken into account. In these plots we can see by eye the slight residual $r$-dependence. 
It also appears that the improvement in precision from reconstruction is localized to the diagonal bins, as the off-diagonal bins are largely unchanged.

For the fitting results, we quantify how much of the  improvement in the best fitting parameters and their uncertainties comes from the differences in covariance and how much comes from the reconstruction of the sample.  
In Table~\ref{tab:fitresdr12reca}, we show the results of performing the BAO anisotropic analysis \change{on the} 1000 MD-PATCHY/QPM mocks using the different covariances. We study first the bias associated with the $\alpha$ measurements for MD-PATCHY: $b_\alpha$ reduces post-reconstruction as expected from 0.002 to $< $0.001 for the lower redshift bin, and from 0.004 to $\sim$0.001 for the intermediate and higher redshift. QPM bias shows no reduction of the shift post-reconstruction for the lower bin (0.003 for both cases), while the intermediate bin decreases from 0.005 to 0.001 and the higher redshift bin decreases from 0.003 to 0.001. The results post reconstruction for the bias  are in agreement for both mock catalogues except for the lowest redshift bin.
  Comparing the post-reconstruction results using post-reconstruction covariances against the post-reconstruction results using pre-reconstruction covariances, we observe, as expected, that the decrease in the error bars also is mostly related to the reconstruction of the catalogue and not to the covariance changes post-reconstruction. 
 The decrease in the mean  error distributions obtained from analyzing post reconstruction catalogues compared to their value pre-reconstruction is 0.011 for $\alpha$ and 0.015 for $\epsilon$ for the first redshift bin.
  The difference when we use the pre-reconstruction covariance instead of the post-reconstruction covariance, is only 0.004 for both $\alpha$
   and 0.007 for both $\epsilon$; similar numbers result from analyzing the intermediate and higher redshift bins. 
   The differences in the standard deviation $\Delta S_\alpha$ between pre- and post-reconstruction distributions are 0.006-0.009, and the difference in the dispersion when we use the covariance pre-reconstruction for fitting the post-reconstruction correlation functions is 0.001-0.005, indicating again that the post-reconstruction covariance is also contributing to the reduction of the dispersion on the fits.
   Finally, we move on to the uncertainties: the reduction of the mean uncertainties when comparing pre- and post-reconstruction results is 38-44\% for $\alpha$ and $\epsilon$, whereas we compared the error distributions using the covariance pre-reconstruction (instead of the post-reconstruction) in the post-reconstruction fits the improvement is only 24-31\%, so the reduction of the errors observed is coming exclusively from the covariance post-reconstruction.
   However, the impact is mostly in the dispersion of the uncertainties distributions; using the post-reconstruction covariance, these are narrower compared to the pre-reconstruction ones. The dispersion reduces from pre-reconstruction to post-reconstruction in 31-36\% and when we use pre-reconstruction covariance instead for the fits, the reduction  is only 20-28\%; again the difference in the dispersion is related only to the use of the post-reconstruction covariance matrix.

\subsubsection{Differences in Sample Covariances from different mock catalogues}

In this section, we check whether there are differences in the structure of the sample covariance matrices generated from different sets of mock catalogues: QPM and MD-PATCHY.  We then quantify if these differences have a significant effect on the fitting results and the uncertainties. We proceed to analyze only the post-reconstruction results, as for the BAO analysis, the covariance used is always post-reconstruction. 

\begin{figure*}
   \centering
    \hspace*{-5.em}	
     \includegraphics[width=8.5in]{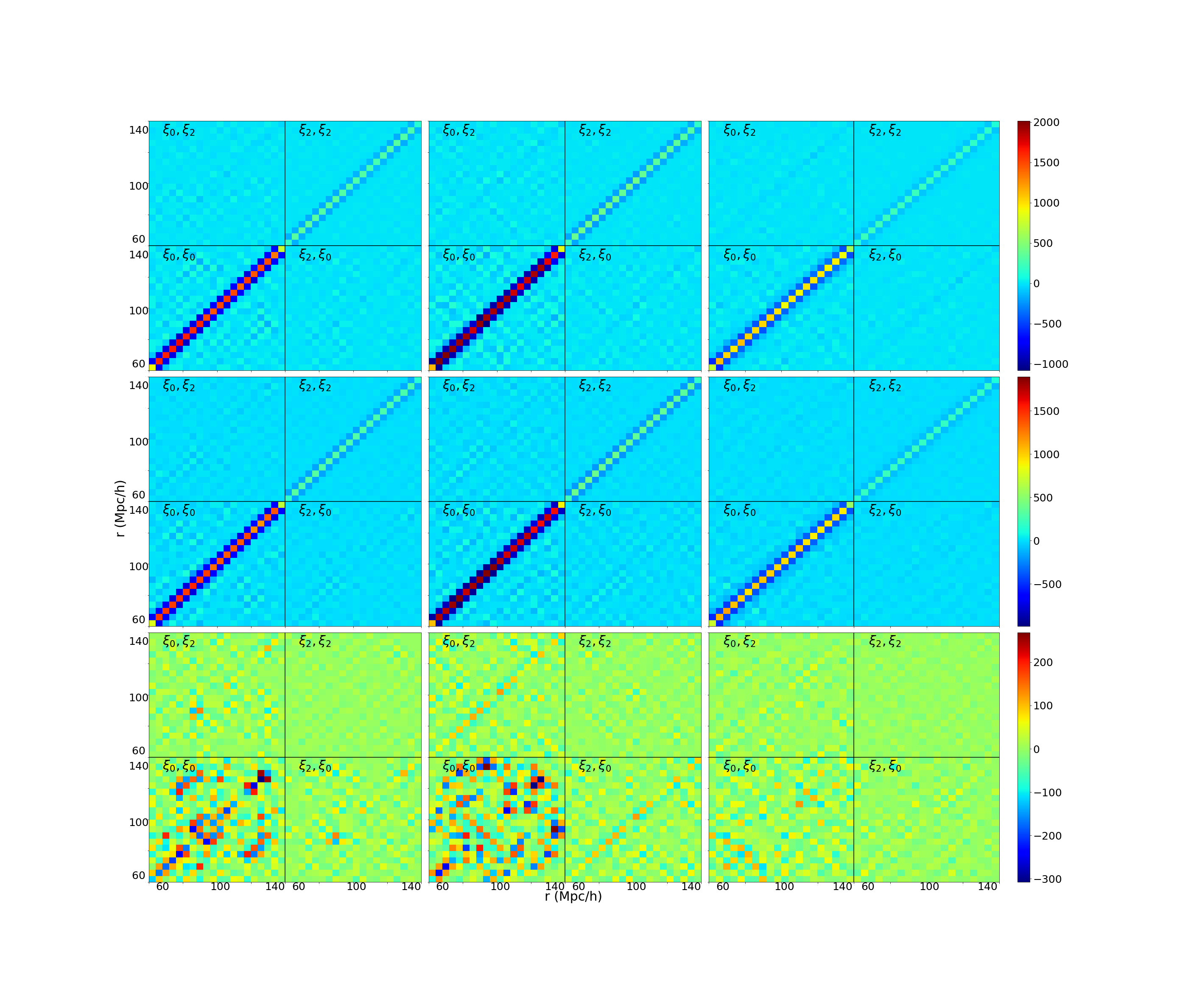}  
    \caption{ Precision matrix for different mock catalogues post-reconstruction, Top panel shows the sample covariance from 1000 from MD-PATCHY, intermediate panel show the sample covariance from 1000 QPM mocks  and the botton panel show the difference between precision matrix  of both sets of mocks. The three columns corresponds to the 3 redshift bins, from left to right, low, intermediate and hight redshift bins. The differences are re-scaled by $r^2$. Each precision matrix has four blocks bottom left is the monopole-monopole term, top-right the quadrupole-quadrupole and bottom-right and top-left the cross-terms.}
   \label{fig:covQPMPATpre_postdiff}
\end{figure*}

In Figure~\ref{fig:covQPMPATpre_postdiff}, we show the precision matrix for the different MD-PATCHY sets of mocks post-reconstruction in the top [MD-PATCHY] and intermediate panels [QPM]; in the bottom panels, we show the difference between the precision matrix generated from MD-PATCHY and the precision matrix generated from QPM for the three redshift bins, low, intermediate and higher redshift, respectively. 

We notice different trends for the three redshift bins; the higher redshift bin  (earliest time) shows that QPM and MD-PATCHY are completely consistent, while the lower and intermediate bins show differences. We observe a similar trend in the $\log(\det(\Psi))$, where the values are pretty similar for the high redshift bin (earliest time) between QPM and MD-PATCHY, then
 grow in bin 2 and grow further in bin 1. We speculate that this behaviour may result from QPM  developing non-Gaussian structure less rapidly than MD-PATCHY, but  further exploration would be required to verify this hypothesis.

\begin{figure*}
   \centering     
    \includegraphics[width=3.5in]{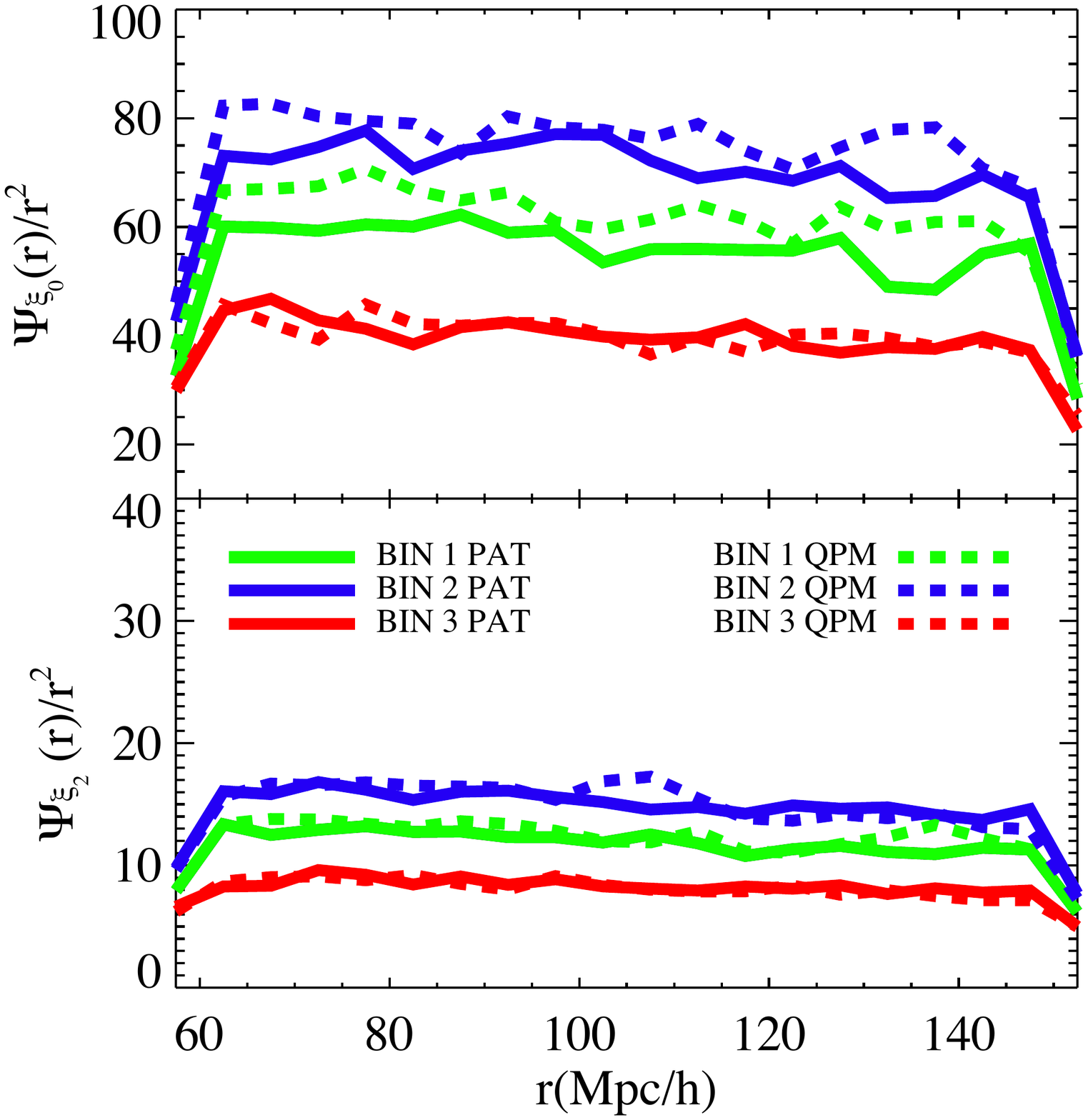} \hspace*{-3.em}
     \includegraphics[width=3.5in]{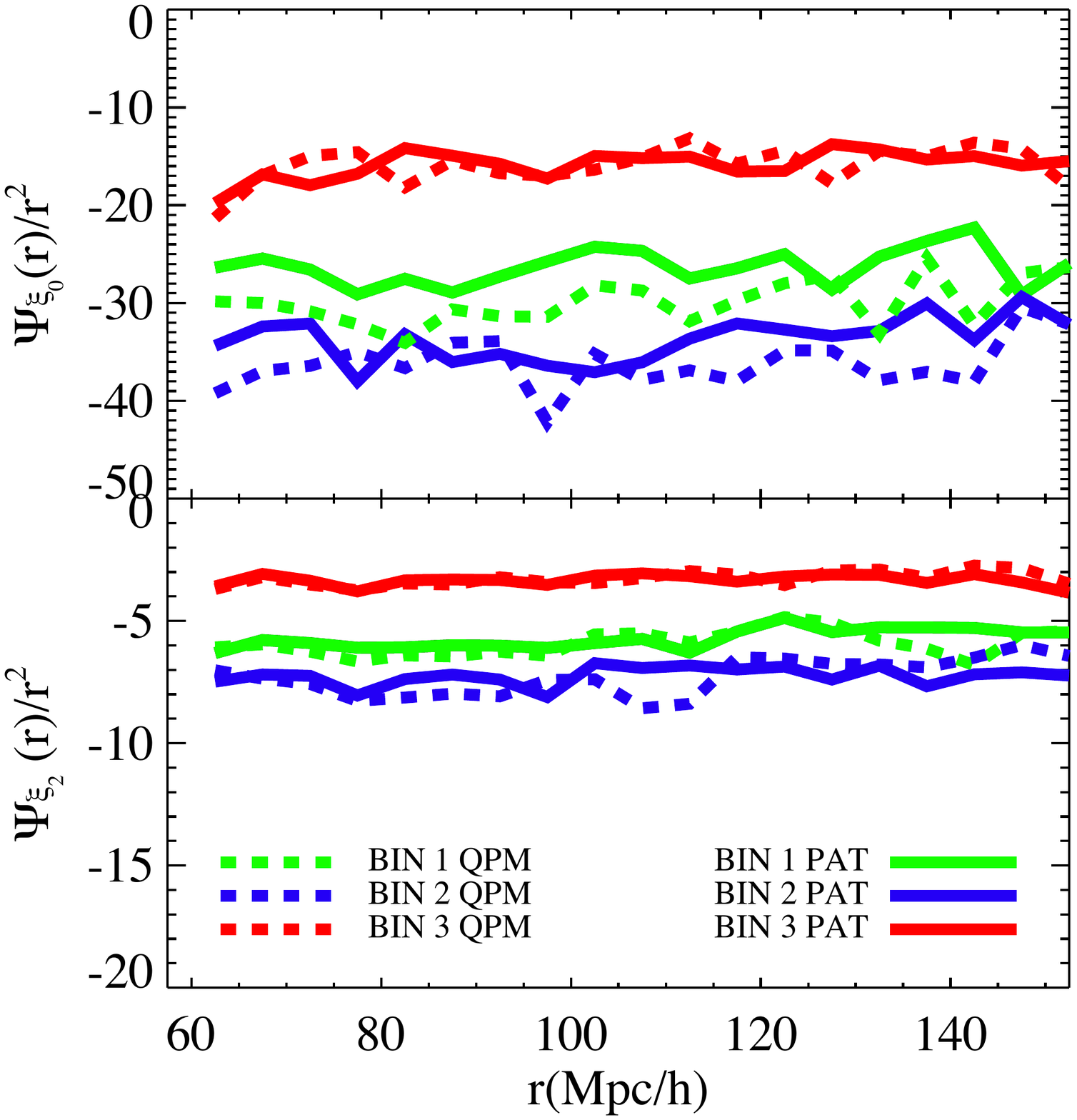} 
   \caption{ Diagonal [left] and \change{first} off-diagonal [right] terms of precision matrices of QPM (dashed lines) and MD-PATCHY (solid lines) for 1000 mocks for the 3 redshift bins, post-reconstruction. ``Bin 1" refers to the lower redshift bin ($z= 0.2 - 0.5$);  ``Bin 2" considers the intermediate redshift range ($z= 0.4 - 0.6$), and ``Bin 3" refers to higher redshift range ($z= 0.5 - 0.75$). Top panels monopole terms, bottom panels quadrupole terms.}
   \label{fig:diagmocks}
\end{figure*}

 In the left panel of Figure~\ref{fig:diagmocks},  we see the diagonal terms of the precision matrices for the three redshift bins in different colours for both sets of mock catalogues: QPM [dashed lines] and MD-PATCHY [solid lines]. In the right panel, we see the first off-diagonal terms of the precision matrices for the same three redshift bins for both sets of mocks catalogues. The top panels again refer to the monopole terms and the bottom ones to the quadrupole terms. The values of the precision are re-scaled to remove the naive $r$-dependence.
 In contrast to the pre-/post-reconstruction comparison, we see clear differences in both the diagonal and \change{first} off-diagonal plots.
 The diagonal terms of the precision matrix for QPM mocks show slightly larger values (10-20\%) in  the monopole terms, while for the quadrupole the variations are smaller ($\sim10$\%). For the first off-diagonal terms, the QPM monopole terms are $\sim20$\% smaller, and for the quadruple terms we observe 20\% variations, showing QPM smaller values than MD-PATCHY.

In Table~\ref{tab:fitresdr12rec}, we show the results of performing the BAO anisotropic analysis on 1000 mocks using the different covariances. Regarding the best fitting parameters, first we verify the bias column, in which the different covariances provides unbiased estimates of the parameters $\alpha$ and $\epsilon$, and the mean bias for the most of the mocks is $b_\alpha\sim 0.001$ and $b_\epsilon= 0.001-0.002$ for MD-PATCHY and $b_\epsilon= 0.003-0.004$ for QPM.
The dispersion is very similar for both mock catalogues. We also find differences in the dispersion of $\alpha$ and $\epsilon$ distributions ($\Delta S_\alpha =0.001-0.002$, $\Delta S_\epsilon < 0.001$)

The error distributions are also very similar; we find differences in the mean values between 0.001-0.002 for $\sigma_\alpha$ and $\sigma_\epsilon$. Additionally, the dispersion of the error distributions shows variations of the same order. 

We also tested the BAO fitting methodology on one set of mocks using the covariance given for the second set of mocks to quantify how the best fitting parameters  could be affected by  using a covariance matrix derived from a different set of mocks. The variations in the best fitting parameters are  around 0.001 for $\alpha$ and $\epsilon$ for the three redshift bins as well as the differences on the mean values of the uncertainties distributions. The dispersion also shows variations of the same order for $\alpha$. 
Finally, we proceed to quantify the differences in the mean of $\alpha$ and $\epsilon$ distributions. For $\alpha$, we found the low redshift bin differs between mock sets by 0.0003-0.0005, 
while the intermediate and high redshift bins shows 0.0013-0.0017 and 0.0002-0.0005.

\subsection{\label{sub:cov_model}Comparison between Sample and Theoretical covariance}

The second kind of covariances we want to test against the fiducial choice is using semi-analytic models for the covariance matrix. We tested the theoretical approach proposed by \cite{Oco15}  based on a Gaussian model were two parameters  (shot noise and volume) 
  are calibrated with the mock catalogues. Testing hybrid methods for generating covariance matrices  will be important in the context of future surveys as we will need to apply hybrid methods to the large scale structure analysis given the increasing quantity of data and volumes that increase the computational requirements for the estimation of the covariances in the large scale structure analysis.

\subsubsection{Model Covariance Matrix for DR12}

The approach in \cite{Oco15} makes a few simplifying assumptions in order to produce a tractable, realistic model for the covariance matrix. Most importantly, the galaxy field is assumed to be Gaussian. For a Gaussian random field, the covariance of the two-point correlation function can be written in terms of integrals of the four-point correlation function. Gaussianity implies that the four point correlation function has no connected piece, and can be written entirely in terms of the two-point correlation function.

We then assume that the galaxy survey is a Poisson-sampled version of the underlying galaxy field. The covariance of the two-point function in the galaxy survey is then a sum of integrals over configurations of two, three, and four points. Since the underlying field is still assumed to be Gaussian, the integrands are composed only of the underlying two-point correlation function and the number density. 
 A key innovation of \cite{Oco15} 
 is to allow the number density to vary, incorporating both the position-dependent survey mask and redshift-dependent number density of galaxies. The resulting integrals must be performed numerically, but are found to converge quite efficiently. The estimates below used $\approx 600$ CPU hours for each redshift bin.

The galaxy field is known to be non-Gaussian, and \cite{Oco15}  saw clear evidence of the effects of non-Gaussianity on the covariance matrix. The technique introduced there to model the effects of non-Gaussianity is simply to increase the level of shot noise in the survey. The intuition behind this is simple -- the three- and four-point functions primarily affect the galaxy field at relatively short scales, which are also where the effects of shot noise are most relevant. The shot noise re-scaling is implemented as an overall re-scaling of the two-, three-, and four-point integrals. 
The fitting is performed using a likelihood function based on the Kullback-Leibler divergence introduced in \cite{Oco15}, where it was shown that the shot-noise rescaling could be determined using a small number of mocks. Here, since our goal is simply to construct a usable model, the shot noise was determined using the full suite of 1000 MD-PATCHY mocks.

The implementation of the described method for this study differs slightly from the implementation in \citep{Oco15}. First, the present analysis utilises a monopole-quadrupole decomposition of the correlation function rather than evaluating the correlation function in $r-\mu$ bins. To accommodate this, the covariance matrix was determined using 20 $\mu$ bins, then projected down to the monopole+quadrupole covariance matrix. This projection was performed separately for the two-, three-, and four-point contributions to the covariance matrix, before fitting the shot-noise re-scaling.

Second, the present analysis combines counts of galaxies in the North and South galactic caps (NGC and SGC), since the survey masks and redshift-dependent number densities used to compute the model covariance matrix are different in the NGC and SGC. The two-, three-, and four-point contributions to the covariance matrix from the NGC and SGC were computed separately and then combined, again prior to fitting of the shot-noise re-scaling. The estimates of the random-random pair counts in each galactic cap were used to determine the relative weight given to each cap. Specifically, the combination was performed as \commentH{is it mean or sum?}

\begin{align*}
\begin{split}
C_\mathrm{comb} = \frac{\mathrm{mean}(RR_\mathrm{NGC})}{\mathrm{mean}(RR_\mathrm{NGC}+RR_\mathrm{SGC})} C_\mathrm{NGC} + \\
+\frac{\mathrm{mean}(RR_\mathrm{SGC})}{\mathrm{mean}(RR_\mathrm{NGC}+RR_\mathrm{SGC})} C_\mathrm{SGC}\,.
\end{split}
\end{align*}

Third, in addition to allowing the shot-noise re-scaling to vary in the fitting, the overall survey volume is also a free parameter. This was done to accommodate minor differences between the volume and number density described by the survey mask and observed redshift-dependent galaxy number density, and the volume and number density implemented in the mocks. Minor discrepancies between the two can arise, e.g. from survey defects, such as bright stars, that are accounted for in the mocks but that are too small to be accurately reflected in the survey mask. 

\subsubsection{Results from Model Covariance}

\begin{figure*}
   \centering
   \hspace*{-5.em}	
   \includegraphics[width=8.0in]{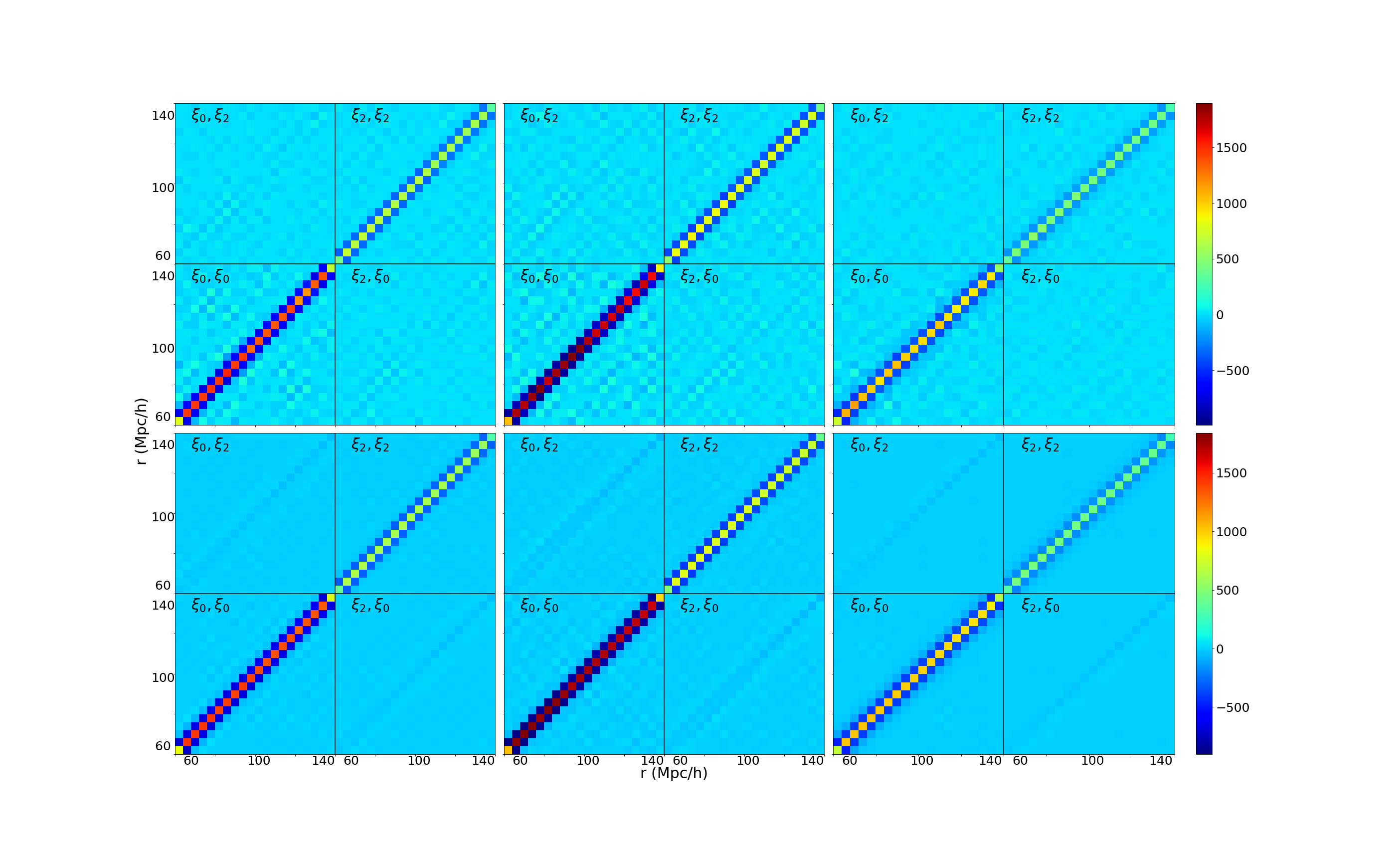}
   \caption{ Precision matrices of MD-PATCHY for 1000 mocks  [top panels] and model precision matrices [bottom panels] for the 3 redshift bins, post-reconstruction.}
   \label{fig:precmocks}
\end{figure*}

First, we compare the structure of the sample precision matrix estimated from mock catalogues to the model precision matrix. The comparison between precision matrices is shown in Figure~\ref{fig:precmocks}. The top panels are the precision matrix for the three redshift bins from MD-PATCHY mock catalogues; the lower panels are from model covariance. The most noticeable aspect is the clear noise reduction of the model covariance compared with the sample covariance from a finite number of mocks.
 
\begin{figure*}
   \centering     
     \includegraphics[width=3.5in]{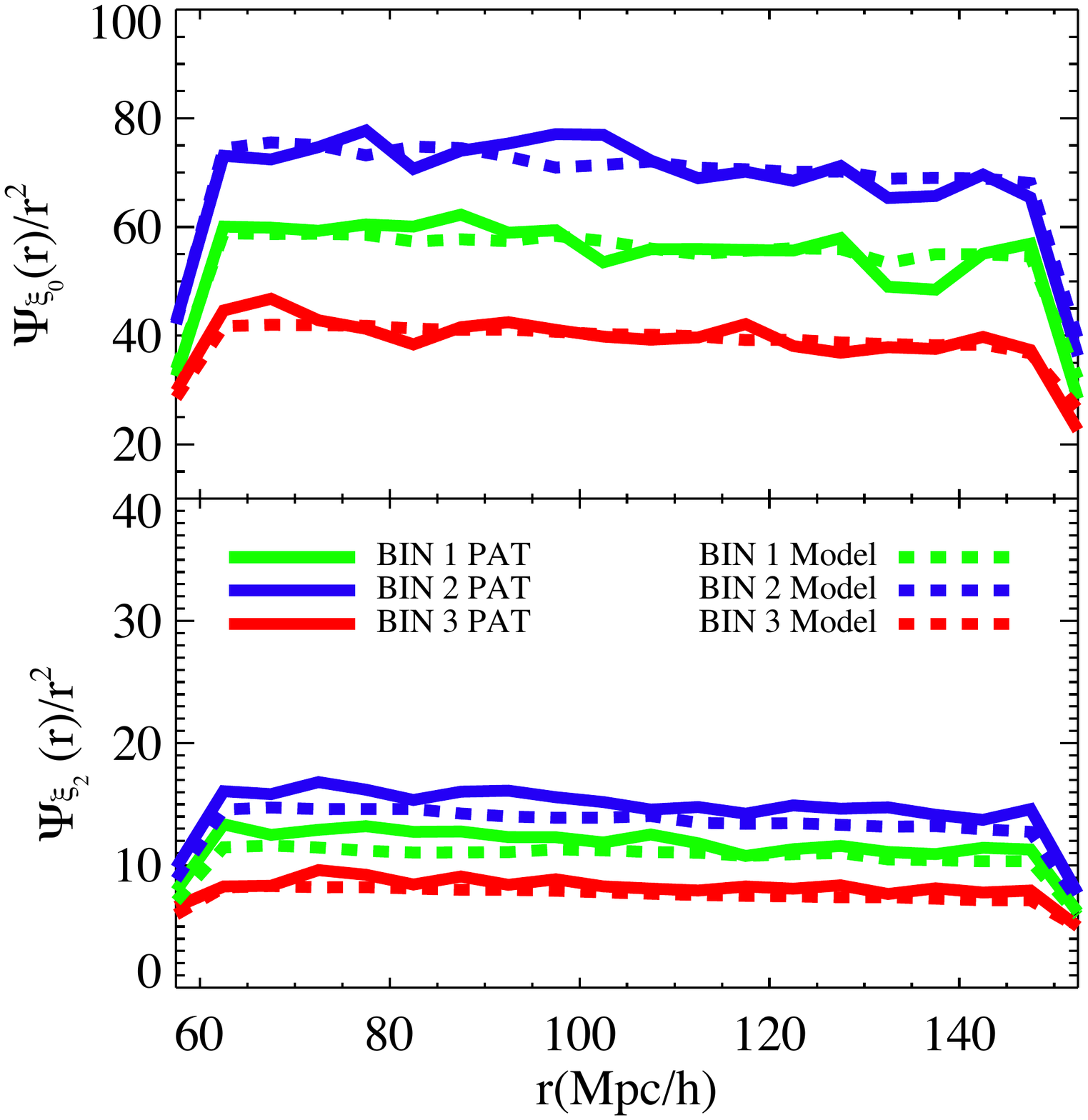} \hspace*{-3.em}
     \includegraphics[width=3.5in]{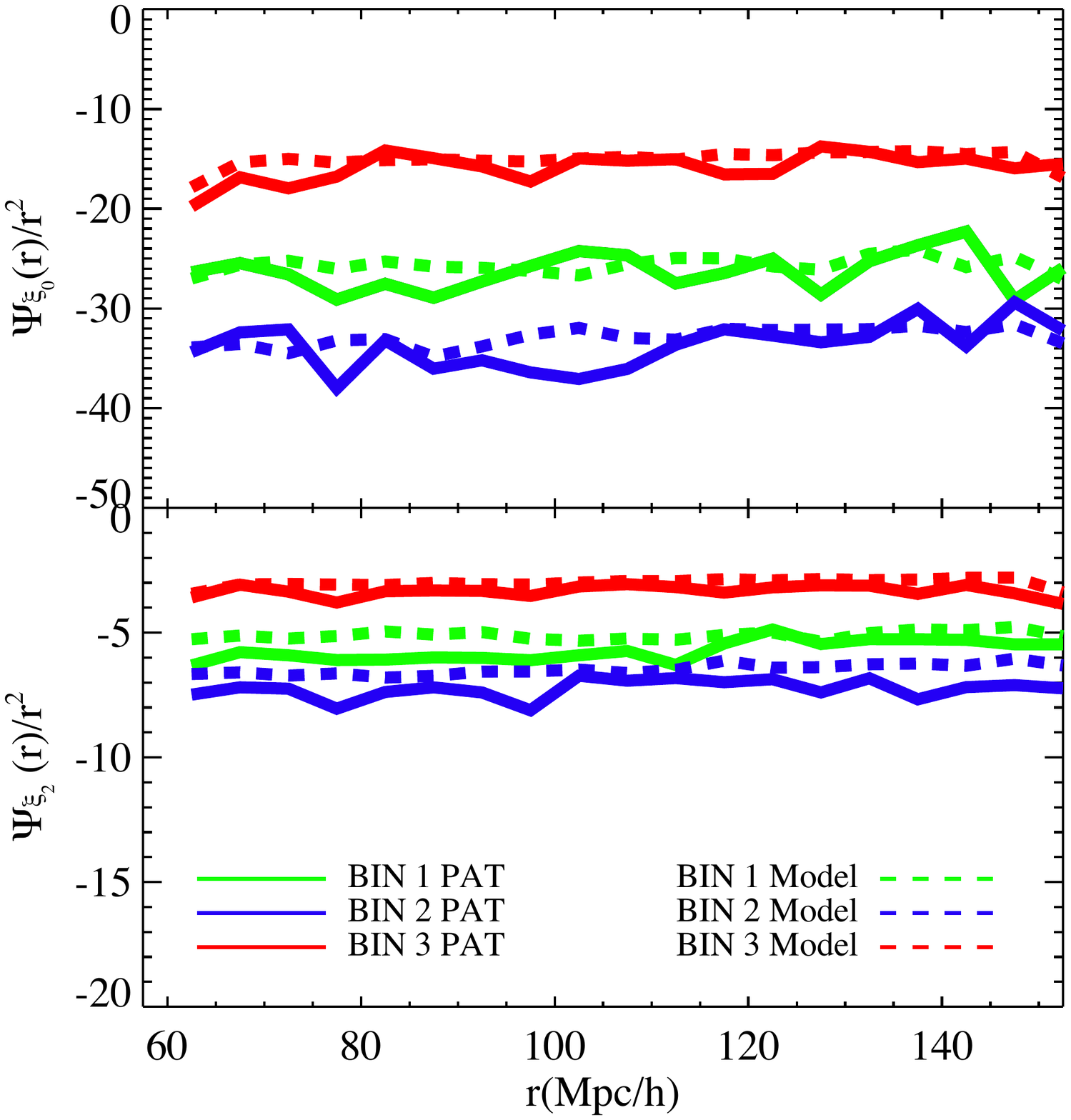}
   \caption{ Diagonal [left] and \change{first} off-diagonal [right] terms of model (dashed lines) and sample precision (solid lines) matrices of MD-PATCHY for 1000 mocks for the 3 redshift bins, post-reconstruction.``Bin 1" refers to the lower redshift bin ($z= 0.2 - 0.5$);  ``Bin 2" considers the intermediate redshift range ($z= 0.4 - 0.6$ ), and ``Bin 3" refers to higher redshift range ($z= 0.5 - 0.75$).}
   \label{fig:diagmocksMOD}
\end{figure*}

 We show in the left panel of Figure~\ref{fig:diagmocksMOD} the diagonal terms of the precision matrices for the three redshift bins for the model covariance and the sample covariance computed from MD-PATCHY mock catalogues. In the right panel are the \change{first} off-diagonal terms of the covariance matrices for the same three redshift bins. The top panels correspond to the monopole terms and the bottom to the quadrupole terms. There is a re-scaling by the $r_a\times r_b$ to remove the naive scaling with $r$. We observe again this clear reduction in noise comparing the noisy estimate from sample covariance with the smooth line coming from model covariance.  The diagonal terms for the monopole  are almost identical in the sample and the model precision matrices. The quadrupole diagonal terms for the sample covariance are slightly larger than the model covariance. For the off-diagonal terms we have a similar trend, in that the monopole terms are very similar, and the quadrupole terms show slightly larger variations, while the sample covariance has more negative values. Comparing the values of $\log(\det(\psi))$ in Table \ref{tab:detcov}, we observe the re-scaling between the sample covariance from MD-PATCHY and model covariance are very small  compared with the other cases, showing the model  is in exceedingly good agreement with post-reconstruction MD-PATCHY, as expected as it has been fit to post-reconstruction MD-PATCHY.

The excellent agreement between the mock and model precision matrices indicates that, for future surveys, better-understood and more realistic mocks should be a higher priority than generating large numbers of mocks. This follows from the simple observation that the differences between the QPM and MD-PATCHY mock covariance matrices are larger than the differences between the MD-PATCHY mock covariance and the model covariance, produced as in \cite{Oco15} and fit to the MD-PATCHY mock covariance. 

Secondly, we analyze the performance of the model covariance compared with the sample covariance in term of anisotropic fits. Table~\ref{tab:fitresdr12rec} summarizes the results of performing the BAO anisotropic analysis on the 1000 MD-PATCHY mocks using the model and sample covariances. The distributions of best fitting parameters as well as the error distributions are extremely similar: the bias for all cases are $b_\alpha,b_\epsilon \sim1$. The dispersion of $\alpha$ and $\epsilon$ distributions differs by less than $0.001$

The differences in the mean uncertainties and the differences in their dispersion for all the redshift bins are also $\sim$0.001. Finally, the differences between the means of the fits using the model covariance matrix compared to the fiducial covariance matrix are $\Delta \alpha=(0.0002, 0.0005, 0.0003)$ and $\Delta \epsilon=(<0.0001, 0.0007, 0.0008)$.

\subsection{\label{sub:cov_summary}Summary for Covariances-Systematic Uncertainty}
To finish this section, we want to give an estimate of the systematic error associated with the covariance step in the BAO analysis. From Table~\ref{tab:fitresdr12reca}, we have nine estimates for the variation in $\alpha, \epsilon$ parameters for different choices of covariance matrices; we compute those differences and present them in Table~\ref{tab:deltacov}. We will generate a final estimate of the systematic uncertainty from the RMS of the nine values. We decided to not include the differences coming from the model covariance cases as the variations are very small and they will artificially decrease the error estimate.

 Summarizing the covariance results, we found that using different mock catalogues produces variations in the best fits 0.0002-0.0017 in $\alpha$ and 0.0002-0.0013 in $\epsilon$. The variations between model and sample found are lower than 0.0010. We conclude that the systematic error in BAO distance measurements associated with using different techniques for estimating the covariance matrix is 0.0009 for $\alpha$  and 0.0009 for $\epsilon$, from taking the RMS of all the combinations analyzed in this section. The differences in the uncertainties distributions are $\Delta \sigma_\alpha=0.0005$ and $\Delta \sigma_\epsilon=0.0010$. 

\begin{table}
\begin{center}
\caption{Covariance Matrix Systematics. We summarize the variations $\Delta \alpha, \Delta \epsilon, \Delta \sigma_\alpha, \Delta \sigma_\epsilon$ observed from the different combinations from Table~\ref{tab:fitresdr12reca}. PP denotes the results from fitting MD-PATCHY mocks using the MD-PATCHY sample covariance, QQ denotes the results from fitting QPM mocks using QPM Sample Covariance, QP denotes the results from fitting QPM mocks using MD-PATCHY Sample Covariance, PQ denotes the results from fitting  MD-PATCHY mocks using QPM Sample Covariance and PP-model denotes the results from fitting  MD-PATCHY mocks using model covariance. The PP-model cases are not take into account in the RMS calculation because they are too small.} \label{tab:deltacov}
\begin{tabular}{@{}lcccc}
\hline
\multicolumn{5}{c}{DR12 Sample Covariance Pre-Reconstruction}\\
\hline
Cov&$\Delta \alpha$ &$\Delta \epsilon$ &$\Delta \sigma_\alpha$ &$\Delta \sigma_\epsilon$ \\ \\[-1.5ex]
\hline
\multicolumn{5}{c}{Bin 1 ($0.20 < z < 0.50$)}\\
\hline
PP-QQ&-0.0012&		-0.0031&0.0038&0.0014\\
\hline
\multicolumn{5}{c}{Bin 2 ($0.40 < z < 0.60$)}\\
\hline
PP-QQ&-0.0002&		-0.0021&0.0024&0.0011\\
\hline
\multicolumn{5}{c}{Bin 3 ($0.50 < z < 0.75$)}\\
\hline
PP-QQ&0.0008&		-0.0035&0.0017&0.0002\\
\hline
RMS &0.0008&0.0030 & 0.0028 & 0.0010 \\
\hline
\multicolumn{5}{c}{DR12 Sample Covariance Post-Reconstruction}\\
\hline
Cov&$\Delta \alpha$ &$\Delta \epsilon$ &$\Delta \sigma_\alpha$ &$\Delta \sigma_\epsilon$ \\[-1.5ex]
\\
\hline
\multicolumn{5}{c}{Bin 1 ($0.20 < z < 0.50$)}\\
\hline
PP-PQ&-0.0003&		0.0002&0.0009&0.0014\\
QQ-QP&-0.0005&	  	0.0004&-0.0008&-0.0019\\
PP-model$^{*}$&-0.0002&	  	$<$0.0001&0.0003&-0.0002\\
\hline
\multicolumn{5}{c}{Bin 2 ($0.40 < z < 0.60$)}\\
\hline
PP-PQ&0.0017&		0.0013&0.0002&0.0007\\
QQ-QP&-0.0013&		-0.0006&0.0001&-0.0005\\
PP-model$^{*}$&0.0005&	  	0.0007&$<$0.0001&-0.0008\\
\hline
\multicolumn{5}{c}{Bin 3 ($0.50 < z < 0.75$)}\\
\hline
PP-PQ&0.0002&		0.0011&-0.0004&0.0003\\
QQ-QP&0.0005&		-0.0013&0.0002&-0.0002\\
PP-model$^{*}$&-0.0005&	  	0.0008&$<$0.0001&$<$0.0001\\
\hline
RMS &0.0009 &0.0009&0.0005&0.0010\\
\hline
\end{tabular}
\end{center}
\end{table}

\section{Reconstruction}\label{sec:reconstruction}
There are numerous works which have looked into different sources of systematics error related to the reconstruction algorithm, especially within the BOSS collaboration.  We refer the reader to these reconstruction-related systematics studies \citep{PadWhi09, Seo10,bur14,Vargas14,bur15,Vargas15,Seo15}. 
In this work, we devote two sections (this section and Section \ref{sec:cosmology}) to the study of systematic errors related to reconstruction. This section is devoted to revisiting the effect of the smoothing scale used on the reconstruction of the density field in the anisotropic BAO parameters. Section \ref{sec:cosmology} tests the effect of using a different fiducial cosmology, including the effects related to reconstruction. New potential systematics related to reconstruction are explored  in Vargas-Magana et al. in prep, where we study the effects on anisotropic BAO fits from redshift distortions corrections in Fourier space reconstruction. \change{Previous work from \cite{Xeaip}  tested the effect of varying bias and growth rate on the anisotropic results and showed the effects are smaller than 0.3\%; these effects are not revisited for the current exploration and are not included in the final error budget.}

Previous results are summarized  in Table \ref{tab:rec}.  \cite{bur14} have studied the effect on smoothing scale when performing isotropic fits to the power spectrum monopole. They found variations in the $\alpha \le 0.001$ when testing the smoothing scale using PTHALOS mocks; the precise numbers for the differences in best fits parameters between the smoothing scales tested are  enumerated in Table \ref{tab:rec}. \cite{Vargas15} extended the study to anisotropic BAO analysis and found the variations are between 0.004-0.005. The tests were performed with different simulations, QPM mocks and also using different reconstruction implementations \citep{Vargas15,Pad12}. The specific numbers for the different smoothing scales and reconstruction implementations are enumerated in the Table \ref{tab:rec}. Given the differences in the systematic bias found in previous studies,  in this section we revisit previous results related to the effect of smoothing scale in the anisotropic BAO parameters in the context of the combined sample and using a different set of MD-PATCHY mocks.  

\begin{table*}
\begin{center}
\caption{Reconstruction-related systematics: variations in anisotropic BAO parameters for different smoothing scales (R). We present the results from previous works. We quote the $\Delta \alpha$ and $\Delta \epsilon$ (eq.\ref{eq:delta}) given by the differences between the fiducial smoothing scale 15 $h^{-1}$Mpc and the variant of the smoothing scale being tested.}
\label{tab:rec}
\begin{tabular}{@{}lllcccc} 
\hline
Mocks&Sample&R\change{(Mpc/h)}&Reconstruction Implementation&Reference &$\Delta \alpha$& $\Delta \epsilon $\\
\hline
QPM(200)&CMASS&5&Vargas FS&\cite{Vargas15}& 0.005&0.001\\
QPM(200)&CMASS& 10&Vargas FS&\cite{Vargas15}&0.004&$<$0.001\\
\hline
PTHALOS(600)&CMASS&5&Burden FS&\cite{bur14}\footnote{This work analysis the isotropic BAO signal.}&$<$0.001&-\\
PTHALOS(600)&CMASS&10&Burden FS&\cite{bur14}&0.001&-\\
\hline
\end{tabular}
\end{center}
\end{table*}

We performed reconstruction over 100 MD-PATCHY mocks for bin 2 of combined sample, for which we used a bias $b=2.1$ and a growth factor $f=\Omega_m^\gamma$ for the fiducial values and $\gamma$ value for General Relativity. We use three different smoothing scales 5,10 and 15 $h^{-1}$Mpc. We compute the multipoles 
and perform the fits using fiducial methodology (Table~\ref{tab:fidmetho}). We present the results in Table~\ref{tab:rectest}. We use the covariance matrix from the 1000 mocks to prevent that the noise in the covariance generates large fluctuations in the best fitting values. This approximation neglects the impact of the smoothing scale on the covariance matrix properties. However, we experimented with fitting the mocks using the noisy covariance matrix generated from the 100 mocks and found very large bias in the distributions. Thus we decided to fix the covariance for all the cases to the one generated from 1000 mocks and the 15 $h^{-1}$Mpc smoothing scale. 

\begin{figure}
   \centering     
   \hspace*{-6.em}
        \includegraphics[width=2in]{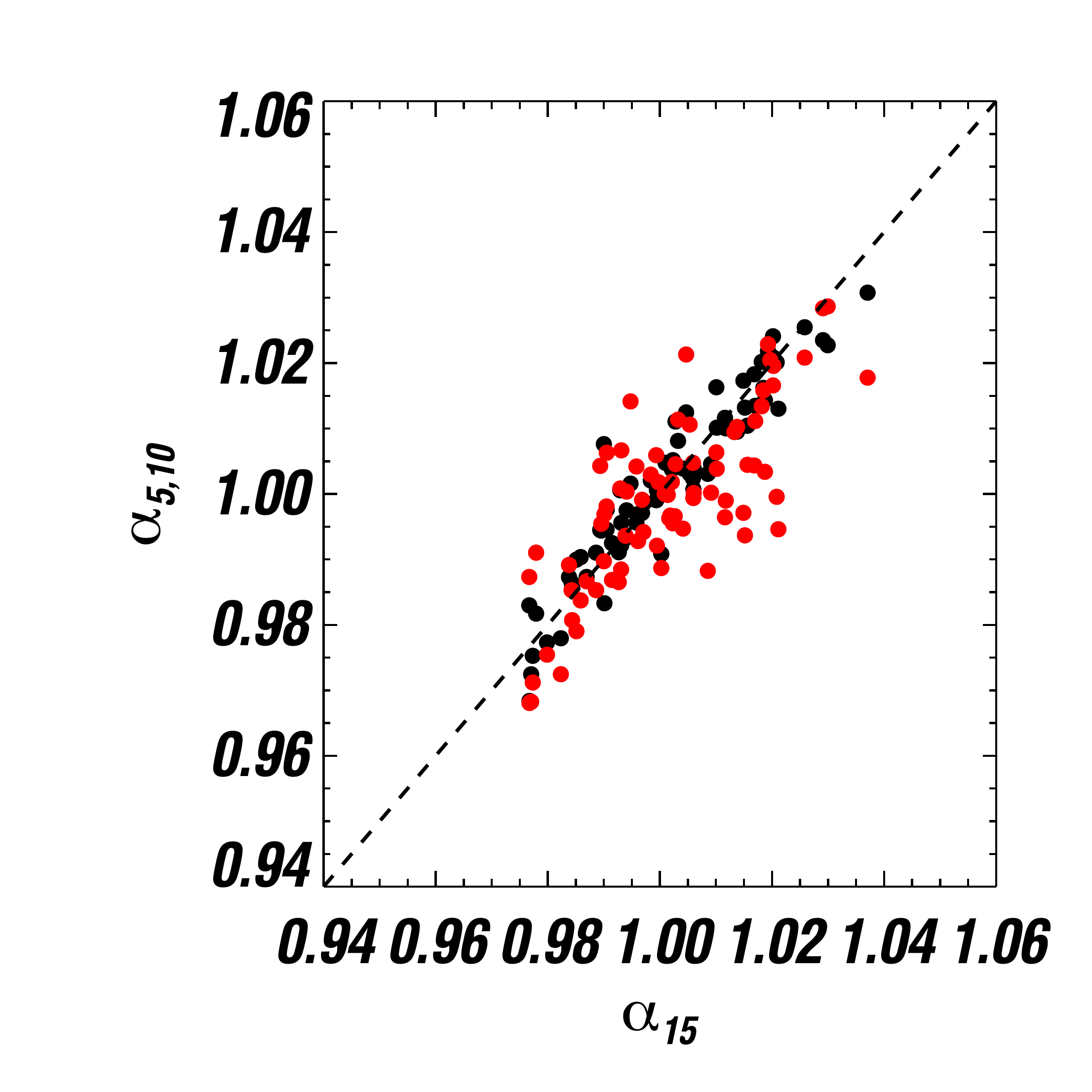}\hspace*{-3.em}
    \includegraphics[width=2in]{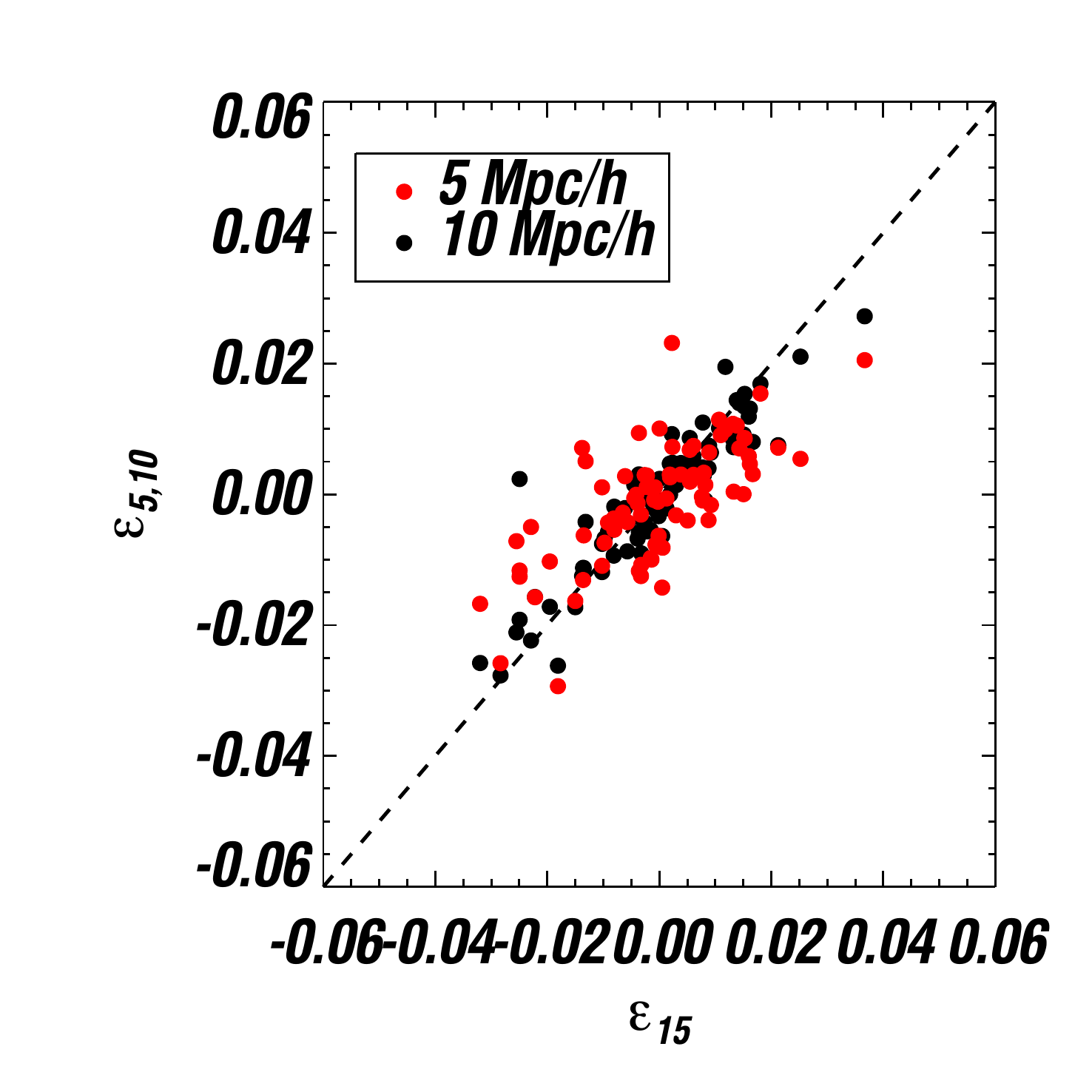}
   \caption{ Dispersion plots of best fits from different smoothing scales 5,10 $h^{-1}$Mpc for 100 MD-PATCHY  post-reconstruction mocks for the intermediate redshift bin ($z= 0.4 - 0.6$). Left panel dispersion plots for $\alpha$ and right panel for $\epsilon$.}
   \label{fig:dispsmoothing}
\end{figure}

We present in Figure \ref{fig:dispsmoothing} the dispersion plots from comparing the fiducial case (15 $h^{-1}$Mpc) with the two variants (5,10 $h^{-1}$Mpc), we observed the values are well correlated for the three cases, indicating there are no issues with the fits. The biases measured in the best fits values for the three cases are very small $<0.002$ for $\alpha$ and $<0.001$ for $\epsilon$. 
Finally, to obtain the systematic error associated with reconstruction, in particular related to the smoothing scale, we compute the variations in the best fit parameter related to this step (the values are shown in Table~\ref{tab:deltarec}.) We  quoted the RMS of the two combinations that will serve as our systematic error associated with the reconstruction variations tested in this work.


\begin{table}
\begin{center}
\caption{ Reconstruction-related systematics: variations in anisotropic BAO parameters for different smoothing scales. We quote the $\Delta \alpha$ and $\Delta \epsilon$ (eq.\ref{eq:delta}) given by the differences between the fiducial smoothing scale 15 $h^{-1}$Mpc and the variant of the smoothing scale being tested. We use 100 PATCHY mocks for the intermediate redshift bin ($z= 0.4 - 0.6$). We also include the RMS of the  $\Delta \alpha$ and $\Delta \epsilon$ considering the two smoothing scales tested.}
\label{tab:deltarec}
\begin{tabular}{@{}lllcccc} 
\hline
Mocks&R &$b_\alpha$&$b_\epsilon$&$\Delta \alpha$& $\Delta \epsilon $\\
\hline
PATCHY(100) &5&0.0033&-0.0006&0.0022 & 0.0009\\
PATCHY(100)&10&0.0024&-0.0004&0.0009& 0.0002\\
\hline
RMS &&&&0.0017&0.0006\\
\hline
\end{tabular}
\end{center}
\end{table}

 We conclude that the systematic error in BAO distance measurements associated with the smoothing scale used in  reconstruction of the density field are $0.0017$ for $\alpha$ and $0.0006 $ for $\epsilon$.

\section{Fiducial Cosmology assumed in the analysis} \label{sec:cosmology}
The cosmology affects three stages of the analysis: first, the comoving coordinate calculation; second, the fitting template; and third, the cosmological parameters assumed in reconstruction. We test in this section the effect of the three stages together instead of studying the individual contributions at each stage, as we wanted to study the overall effect on the best fits. Previous works have explored the dependency on cosmology in the fitting template and reported variations of 
$\Delta \alpha < 0.001$\footnote{The fits were performed with 15\% larger $\Omega_m$ than the fiducial value.} (\cite{Xu12b}), tests over the distance-redshift relation and tests varying the cosmological parameters are not reported in previous works. 
In order to study the fiducial cosmology dependence of BAO anisotropic analysis, we analyze QPM  mocks assuming a fiducial cosmology different from the one used to generate the mocks. 
We perform the anisotropic BAO analysis and compare to the results when we assume the ``true" cosmology. The fiducial cosmologies used for analyzing the QPM mocks are summarized in Table \ref{tab:cosmofidtest}. 
Anderson is a flat cosmology that is shifted in $\Omega_m$ by 0.5\% compared to QPM cosmology, but \change{has} exactly the same $\Omega_b, h$.

We estimate the mean shift that we should observe in the fitted parameters due to the fact that we are using the wrong cosmology. The shifts in $\alpha$ and $\epsilon$ are given by:
\begin{eqnarray}\label{eq:ae}
\alpha=\frac{D_{V,\rm{fid}}}{D_{V,\rm{true}}}\frac{r_{s}^{\rm true}}{r_s^{\rm fid}}=  \left [ \frac { (D^{\rm fid}_{A}(z))^2 H^{\rm true}(z)}{(D_{A}^{\rm true}(z))^2 H^{\rm fid}(z) } \right ]^{1/3} \frac{r_{s}^{\rm true}}{r_s^{\rm fid}},\\\
\epsilon=\left [ \frac {H^{\rm true}(z) D_{A}^{\rm true}(z)}{H^{\rm fid}(z) D^{\rm fid}_{A}(z) }\right ]^{1/3} -1. 
\end{eqnarray}
where
\begin{equation}\label{eq:dv}
D_V=[cz(1+z)D_A^2(z)/H(z)]^{1/3}.
\end{equation}

The bias $b_\alpha$ and $b_\epsilon$ are defined as follows:
\begin{eqnarray}
b_\alpha=<\alpha> -\alpha_{\rm exp} \quad,\\
b_\epsilon=<\epsilon> -\epsilon_{\rm exp}.
\end{eqnarray}
The expected values for each cosmology are:  $\alpha_{\rm exp}^{\rm QPM}=1.0, \epsilon_{\rm exp}^{\rm QPM}=0.0$,$\alpha_{\rm exp}^{\rm And}=1.0064, \epsilon_{\rm exp}^{\rm And}=-0.0021$.
Substituting the cosmological parameters in equations \ref{eq:ae} and \ref{eq:dv}, and evaluating at $z=0.57$, we find that the expected shifts are $b_\alpha=0.6\%$ and $b_\epsilon=-0.2\%$ for Anderson Cosmology (AND) (Table \ref{tab:cosmo2} ).  

\begin{table}
\begin{center}
\caption{Fiducial cosmologies tested in Section \ref{sec:cosmology} on QPM mocks.}
\label{tab:cosmofidtest}
\begin{tabular}{@{}lcccccc}
\hline
Cosmology&$\Omega_{CDM}$&$\Omega_M$& $\Omega_B$&$\Omega_\Lambda$&h\\
\hline
\\[-1.5ex]
AND&0.228286&0.274&0.0457143&0.726&0.7\\
QPM&0.244143&0.29&0.0458571&0.71&0.7\\
\hline
\end{tabular}
\end{center}
\end{table}

\begin{table}
\begin{center}
\caption{ Derived Distances at z=0.57. $\alpha_{\rm exp}$ and $\epsilon_{\rm exp}$ are the expected shifts in the fitted parameters due to using the wrong cosmology. The sound horizon is evaluated using \textsc{CAMB} (Lewis et al. 2000)}
\label{tab:cosmo2}
\begin{tabular}{@{}lcccccc}
\hline
Cosmo&H(z)&$D_A(z)$&$r_s$&$D_V(z)$&$\alpha_{\rm exp}$&$\epsilon_{\rm exp}$\\
         & km /s/Mpc & Mpc & Mpc & Mpc & & \\ 
\hline
\\[-1.5ex]
AND&93.6&1359.6&149.28&2027.0&1.0064&-0.0021\\
QPM&94.7 &1351.1&147.21&2009.5&1.0&0.0\\
\hline
\end{tabular}
\end{center}
\end{table}

\begin{figure}
   \centering     
   \hspace*{-5.em}
       \includegraphics[width=2in]{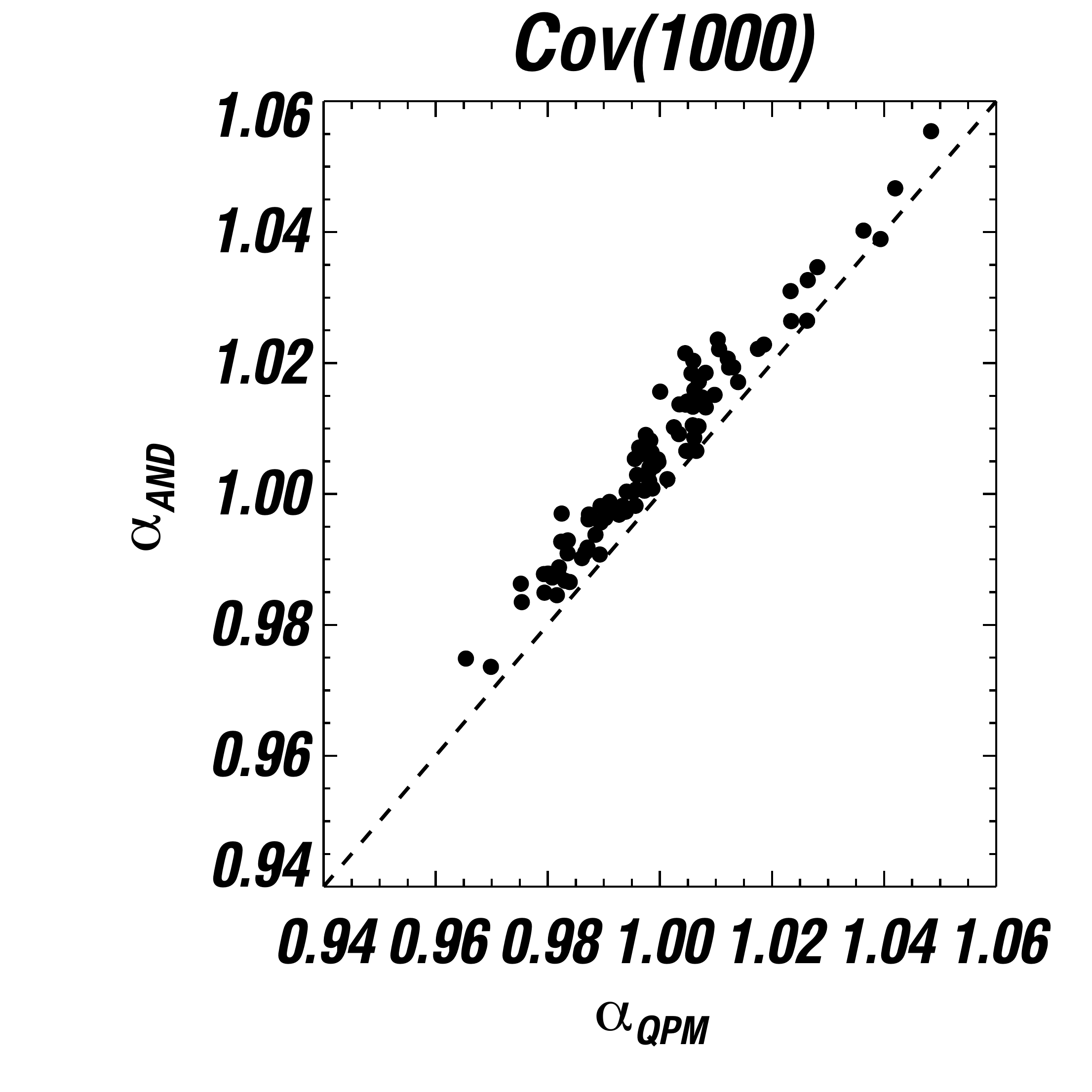}\hspace*{-3.em}
   \includegraphics[width=2in]{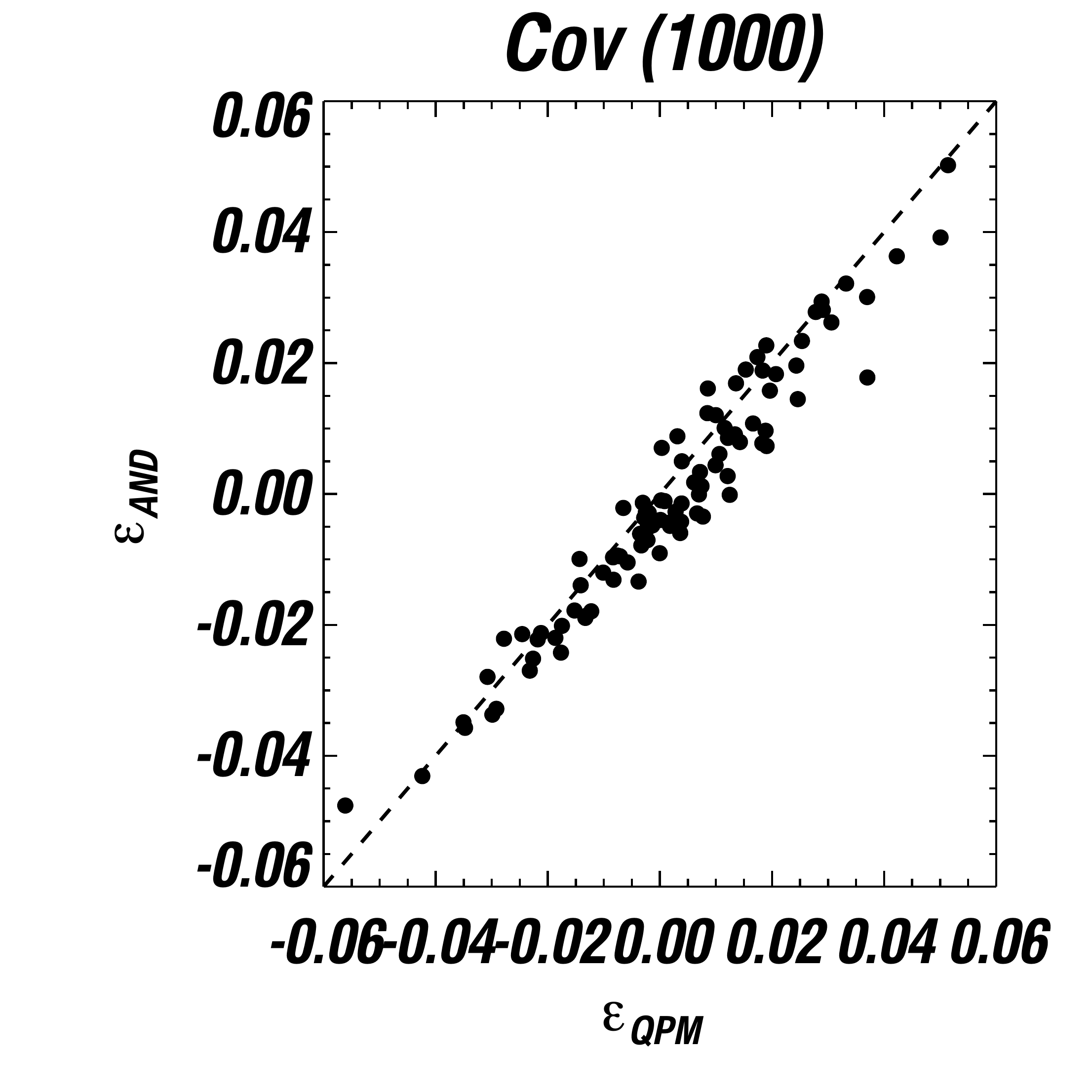}
   \caption{ Dispersion plots of $\alpha, \epsilon$ from performing BAO anisotropic analysis using two different fiducial cosmologies. \change{We observe that the best fitting values are well correlated but need to be re-scaled accordingly to the expected shift. When we apply the re-scaling (red dots) we recover the one-to-one relation as expected.}}
   \label{fig:fiddisp}
\end{figure}

In Table~\ref{tab:fidcosmotest}, we show the results of fitting 100 mocks for this test, pre- and post reconstruction  using a different cosmology in the analysis from the natural cosmology of the mocks. The covariance matrix used was generated from 1000 NGC mocks in QPM cosmology. Figure \ref{fig:fiddisp} shows the dispersion plots of $\alpha, \epsilon$ from performing BAO anisotropic analysis using two different fiducial cosmologies; we observe that the best fitting values are well correlated but need to be re-scaled accordingly to the expected shift.
The bias observed in the two cosmologies pre-reconstruction are very similar $\sim0.2$ \%; the bias is reduced post-reconstruction, the mocks in the QPM cosmology show a bias $b_\alpha=-0.0006$. The mocks in the Anderson cosmology shows a bias of 0.0003\%. We summarize the differences in  the bias observed in Table~\ref{tab:fidcosmodelta}. These differences in bias will serve to estimate the systematic shift associated to the fiducial cosmology.
When analyzing the mocks with the ``true'' cosmology, we measure the intrinsic  bias coming from: 1) Nonlinear evolution of the density perturbations and 2) bias of the fitting methodology. We do expect to reduce the first contribution in post-reconstruction. When considering a ``wrong cosmology", any extra contribution to the bias can be associated with the cosmology mismatch.
We have a shift of  0.0024  on $\alpha$ for the true cosmology, which becomes 0.0006  post-reconstruction. The reduction observed is related just to the removal of the non linear evolution of the density field. The extra bias of 0.0009  in $\alpha$ and 0.0010 in $\epsilon$ ($\Delta b_\alpha, \Delta b_\epsilon$ in Table~\ref{tab:fidcosmodelta}) observed when using Anderson cosmology is related to the ``wrong cosmology" assumed.

\begin{table}
\begin{center}
\caption{Fiducial cosmology-related systematics. Fitting results from 100 QPM NGC mocks pre-/post-reconstruction using a different cosmology in the analysis from the natural cosmology of the mocks. 
The different columns are the mean difference between bias, i.e. $\Delta b_\alpha=b_\alpha^{\rm true}-b_\alpha^{\rm wrong}$, the analogue for $\epsilon$. 
 The covariance matrix used was generated from 1000 NGC mocks in QPM cosmology.}
\label{tab:fidcosmodelta}
\begin{tabular}{@{}lcc}
\hline
\multicolumn{3}{c}{Fiducial Cosmology related Systematics}\\
\hline

\multicolumn{3}{c}{DR11 CMASS  QPM mocks, Pre/Post-Reconstruction}\\
\hline
Sample &$\Delta b_\alpha $& 
$\Delta b_\epsilon$\\
\hline
Pre-recon&0.0002&$<$0.0001\\
Post-recon&0.0009&0.0010\\
\hline
\end{tabular}
\end{center}
\end{table}

 We conclude that the systematic error in BAO distance measurements associated with fiducial methodology is $0.0009$ for $\alpha$ and $0.0010$ for $\epsilon$.

\section{Systematics uncertainties related to anisotropic fitting methodology}\label{sec:model}

The anisotropic fitting methodology has been extensively tested \citep{Xeaip,And13,And14,Vargas14, Ross12} and the methodology is well established. However, we can still explore the sub-percent uncertainties coming from the fitting methodology. In this section, we explore some of the uncertainties that have not been previously explored, and need to be revisited in the context of the BOSS final analysis.

\subsection{Damping Model}
\begin{figure}
   \centering     
    \includegraphics[width=3.6in]{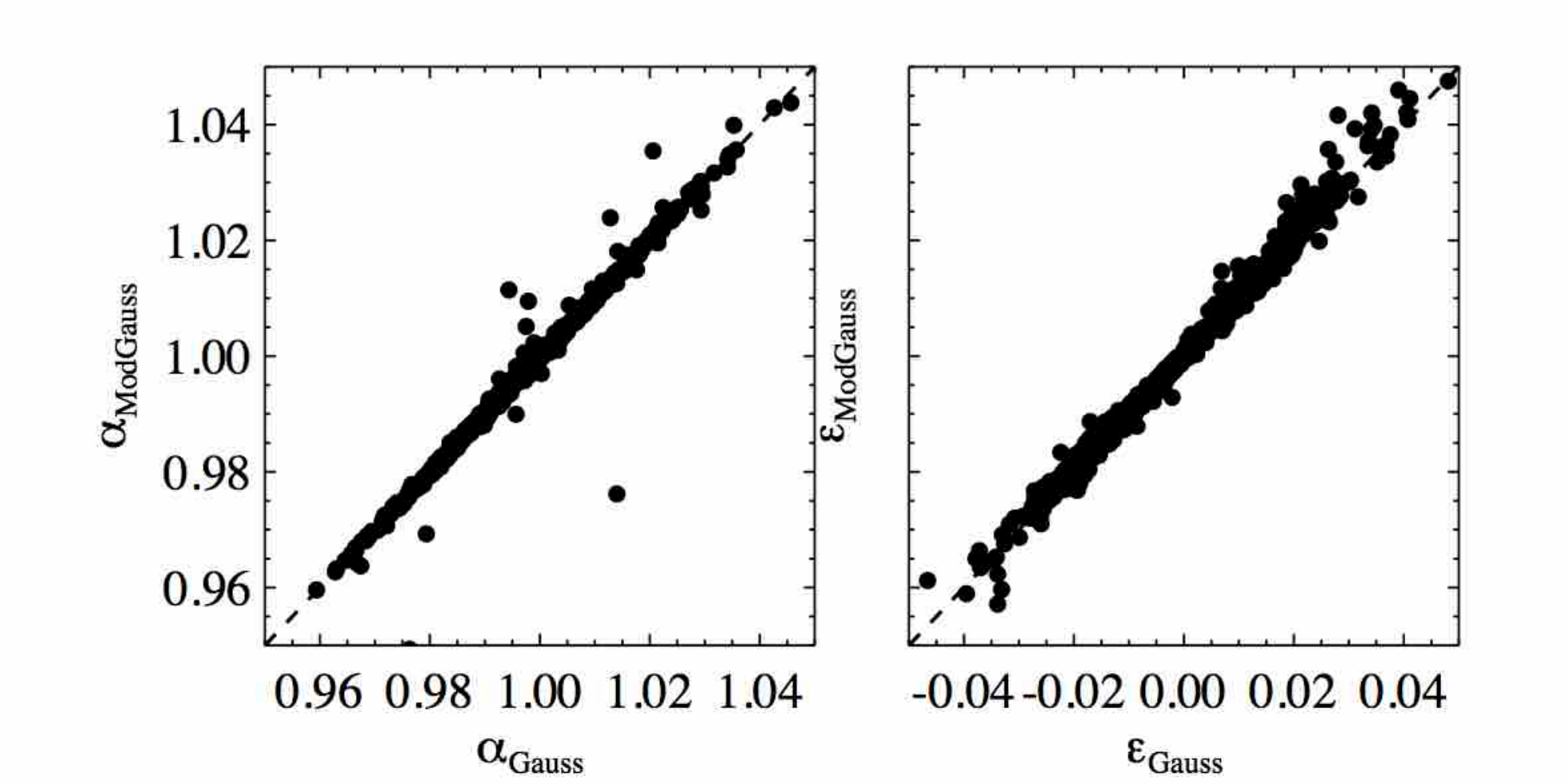}
   \caption{ Dispersion plots of $\alpha$ and $\epsilon$ from MD-PATCHY for 1000 mocks lower redshift bin ($z= 0.2 - 0.5$) post-reconstruction using two variants of the model for the damping: Gaussian and Modified Gaussian.}
   \label{fig:gauss}
\end{figure}

Recent developments in the modelling of the damping have been proposed and tested \citep{Seo15,Ross16,Beutler16a,Acacia16}.
We test the so-called ``Gaussian Damping Model," our fiducial damping model, \change{$C_G$, given by the Kayser term, $(1+\beta \mu^2)^2$, presented in Equation \ref{fisher} and the non-linear gaussian damping model introduced in Equation \ref{nonlinearpk}}:
\begin{equation}
\label{gaussdamp}
C_G(k,\mu,z)=(1+\beta \mu^2)\exp \left [ -\frac{k^2(1-\mu^2)\Sigma_{\perp}-k^2\mu^2 \Sigma_{||}}{4} \right ],
\end{equation}
against the modified Gaussian model, given by the following  equation:
\begin{equation}
C_{MG}(k,\mu,z)=(1+\beta\mu^2(1-S(k))
\exp \left [ -\frac{k^2(1-\mu^2)\Sigma_{\perp}-k^2\mu^2 \Sigma_{||}}{4} \right ].
\end{equation}
\change{where $S(k)=\exp^{-k^2 \Sigma_r^2/2}$ accounts for the smoothing scale applied to the density field during reconstruction $\Sigma_r$}. 
This modified Gaussian damping model better suits the reconstruction algorithm applied in this paper \citep{Seo15}. The Gaussian model is used in DR12 CMASS/LOWZ analysis \citep{Cuesta16,Vargas15,Gil15} and the Modified Gaussian model is used in the final BOSS analysis \citep{Ross16,Beutler16a,Acacia16}. \change{For this test we fixed the values of the $\Sigma_{||,\perp}$ to the fiducial values. The systematic error associated to the variations of those parameters were explored in previous work \citep{Vargas15} and are considered in the final error budget in Section \ref{discussion}.}
 In Figure \ref{fig:gauss}, we show the dispersion plots of the best fits using both variants of the damping model  applied to the lower redshift bin. The best fits are not affected by this variant of the model; there is just a small dispersion on $\epsilon$.

\begin{table}
\begin{center}
\caption{Fitting Systematic Error: Damping Model. 
We summarize the variations, $\Delta \alpha$, $\Delta \epsilon,$ observed from variations of the non-linear damping, the numbers are obtained from different combinations from Table~ \ref{tab:fitsys}. As well as the variations in the uncertainties distributions, $\Delta \sigma_\alpha$ and $\Delta \sigma_\epsilon$,  and the respective biases ($b_\alpha$ and $b_\epsilon$). }
\label{tab:deltabincenter}
\begin{tabular}{@{}lcccccc}
\hline
\multicolumn{7}{c}{DR12  Pre-Reconstruction}\\
\hline
Model&$b_\alpha$&$b_\epsilon$&$\Delta \alpha$ &$\Delta \epsilon$  &$\Delta \sigma_\alpha$&$\Delta \sigma_\epsilon$ \\ 
\hline
\multicolumn{7}{c}{Bin 1 ($0.2 < z < 0.5$)}\\
\hline
$C_{G}$&-0.0007&0.0007&-&-&-&-\\
$C_{MG}$&-0.0007&0.0011&$<$0.0001 &0.0004 &0.0003 &0.0019\\\\[-1.5ex]
\hline
\multicolumn{7}{c}{Bin 1 ($0.2 < z < 0.5$)}\\\\[-1.5ex]
\hline
$C_{G}$&0.0010&0.0021&-&-&-&-\\
$C_{MG}$&0.0010&0.0023&$<$0.0001 &0.0005 &0.0004 &0.0022\\\\[-1.5ex]
\hline
\multicolumn{7}{c}{Bin 1 ($0.2 < z < 0.5$)}\\
\hline 
$C_{G}$&0.0008&0.0009&-&-&-&-\\
$C_{MG}$&0.0007&0.0008&0.0001&0.0010& 0.0002 &0.0019\\\\[-1.5ex]
\hline
\end{tabular}
\end{center}
\end{table}

Table \ref{tab:fitsys} shows the results of fitting using these two versions of the damping model in the combined sample for the three redshift bins. For determining the systematical error, we compute as before the variation in the mean values. We get $\Delta \alpha=[<0.0001, <0.0001, 0.0001]$ and $\Delta \epsilon=[0.0004, 0.0005, 0.0010]$ for the three redshift bins, and taking the RMS of the three cases we get $\Delta \alpha<0.0001$, $\Delta \epsilon=0.0007$.

\begin{figure}
   \centering     
     \includegraphics[width=3.4in]{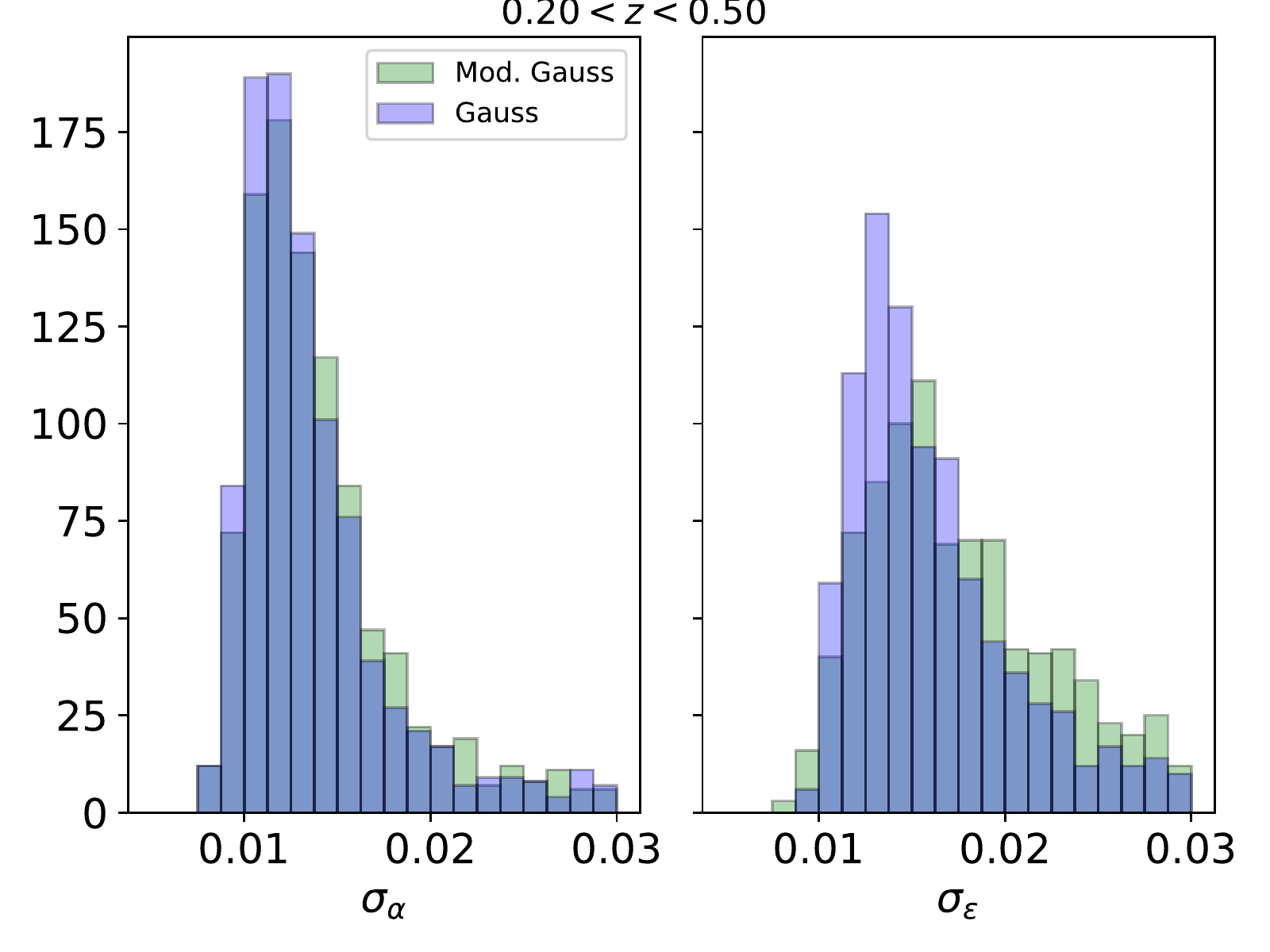}
   \caption{ Error distributions from MD-PATCHY for 1000 mocks lower redshift bin ($z= 0.2 - 0.5$) post-reconstruction using two variants of the model for the damping: Gaussian and Modified Gaussian.}
   \label{fig:errgauss}
\end{figure}

In addition to the determining the variation in the mean values, we would like to check the effects on the error distributions. In the case of the Gaussian damping, the results indicate that the variation of the damping model affects $\epsilon$ uncertainties. Figure~\ref{fig:errgauss} shows the error distributions in both cases: Gaussian and Modified Gaussian damping.  
 The Modified Gaussian model is giving smaller errors compared to the Gaussian model. According to Table \ref{tab:fitsys}, the differences in the error distributions are $\Delta \sigma_\alpha=[0.0003, 0.0004, 0.0002]$ and  $\Delta \sigma_\epsilon=[0.0019, 0.0022, 0.0019]$, and taking the RMS we get $\Delta \sigma_\alpha=0.0003$ and $\Delta \sigma_\epsilon=0.0020$.


We conclude that the systematic error in BAO distance measurements associated with the Damping Model is $\Delta \alpha<0.0001$, $\Delta \epsilon=0.0007$. The variations in the error distributions are $\Delta \sigma_\alpha=0.0003$ and $\Delta \sigma_\epsilon=0.0020$.

\subsection{Hexadecapole contribution}\label{sec:hexa}
Usually the hexadecapole is not used on BAO fits because of its small signal-to-noise ratio. Furthermore, after reconstruction its amplitude is reduced even more, decreasing its importance. In this subsection, we compare our fiducial methodology considering only monopole+quadrupole against the fits using monopole+quadrupole+hexadecapole.  In the top panels of Figure~\ref{fig:estimators}, we show the mean monopole, quadrupole, and hexadecapole from 1000 MD-PATCHY Combined mocks pre- and post-reconstruction for the three redshift bins. The figure demonstrates that, before reconstruction, the hexadecapole is significantly smaller compared with the monopole and quadrupole. Post-reconstruction reduces even more the hexadecapole contribution. Also the mean hexadecapole of 1000 mocks is significantly noisier compared to the means of the multipoles $\ell=0,2$. 

We perform anisotropic BAO fits following the methodology described in Section~\ref{sec:method}, but this time we consider also the hexadecapole and compare it with the fits using only the monopole+quadrupole (denoted by $l=0,2$ hereafter). The sample covariance used on the fits is shown in the top panels of Figure~\ref{fig:covs} for the multipoles estimator. The correlation reveals that post-reconstruction results in a more diagonal hexadecapole contribution, and the covariance is reduced for the off-diagonal terms.

\begin{table}
\begin{center}
\caption{Fitting Systematic Error: Hexadecapole. We summarize the variations, $\Delta \alpha$, $\Delta \epsilon$ (defined by eq.\ref{eq:delta}) observed considering monopole+quadrupole fits (denoted by $\ell =2$) with monopole+quadrupole+hexadecapole fits (denoted by $\ell=4$) for 1000 MD-PATCHY  post-reconstruction.
We also show the biases, $b_\alpha$, $b_\epsilon$ (defined by eq.\ref{eq:bias}). The detailed results of the fits are presented in Table~\ref{tab:fitsyshexa}}
\label{tab:deltahexa}
\begin{tabular}{@{}lcccccc}
\hline
\multicolumn{7}{c}{DR12  Pre-Reconstruction}\\
\hline
$\ell$&$b_\alpha$&$b_\epsilon$&$\Delta \alpha$ &$\Delta \epsilon$  &$\Delta \sigma_\alpha$&$\Delta \sigma_\epsilon$ \\ 
\hline
\multicolumn{7}{c}{Bin 1 ($0.2 < z < 0.5$)}\\
\hline
$\xi_{\ell=2}$ &-0.0007&0.0007&-&-&-&-\\
$\xi_{\ell=4}$ &-0.0009&0.0004&0.0002 &0.0003 &0.0016 &0.0034\\\\[-1.5ex]
\hline
\multicolumn{7}{c}{Bin 1 ($0.2 < z < 0.5$)}\\\\[-1.5ex]
\hline
$\xi_{l=2}$ &0.0010&0.0021&-&-&-&-\\
$\xi_{l=4}$&0.0007&0.0018&0.0003 &0.0003 &0.0014 &0.0030\\\\[-1.5ex]
\hline
\multicolumn{7}{c}{Bin 1 ($0.2 < z < 0.5$)}\\
\hline 
$\xi_{l=2}$&0.0008&0.0009&-&-&-&-\\
$\xi_{l=4}$&0.0008&0.0009&$<$0.0001&$<$0.0001& 0.0014 &0.0030\\\\[-1.5ex]
\hline
\end{tabular}
\end{center}
\end{table}

 In Table~\ref{tab:deltahexa}, we show the results of performing the BAO anisotropic analysis to the 1000 PATCHY mocks, pre- and post-reconstruction. The best fit results indicate that differences of using $\ell=0,2$ compared with $\ell=0,2,4$ are not significant. In the $\alpha-\epsilon$ fits, the variation in the mean values are, taking the RMS of the three cases, $\Delta \alpha, \Delta \epsilon$=0.0002 post-reconstruction. 
The small effect on the best fitting values confirms previous results from \cite{Ross15} that demonstrate the monopole and quadrupole represent a complete and optimal set of estimators in the case where information is spherically distributed.

\begin{figure}
   \centering     
     \includegraphics[width=3.5in]{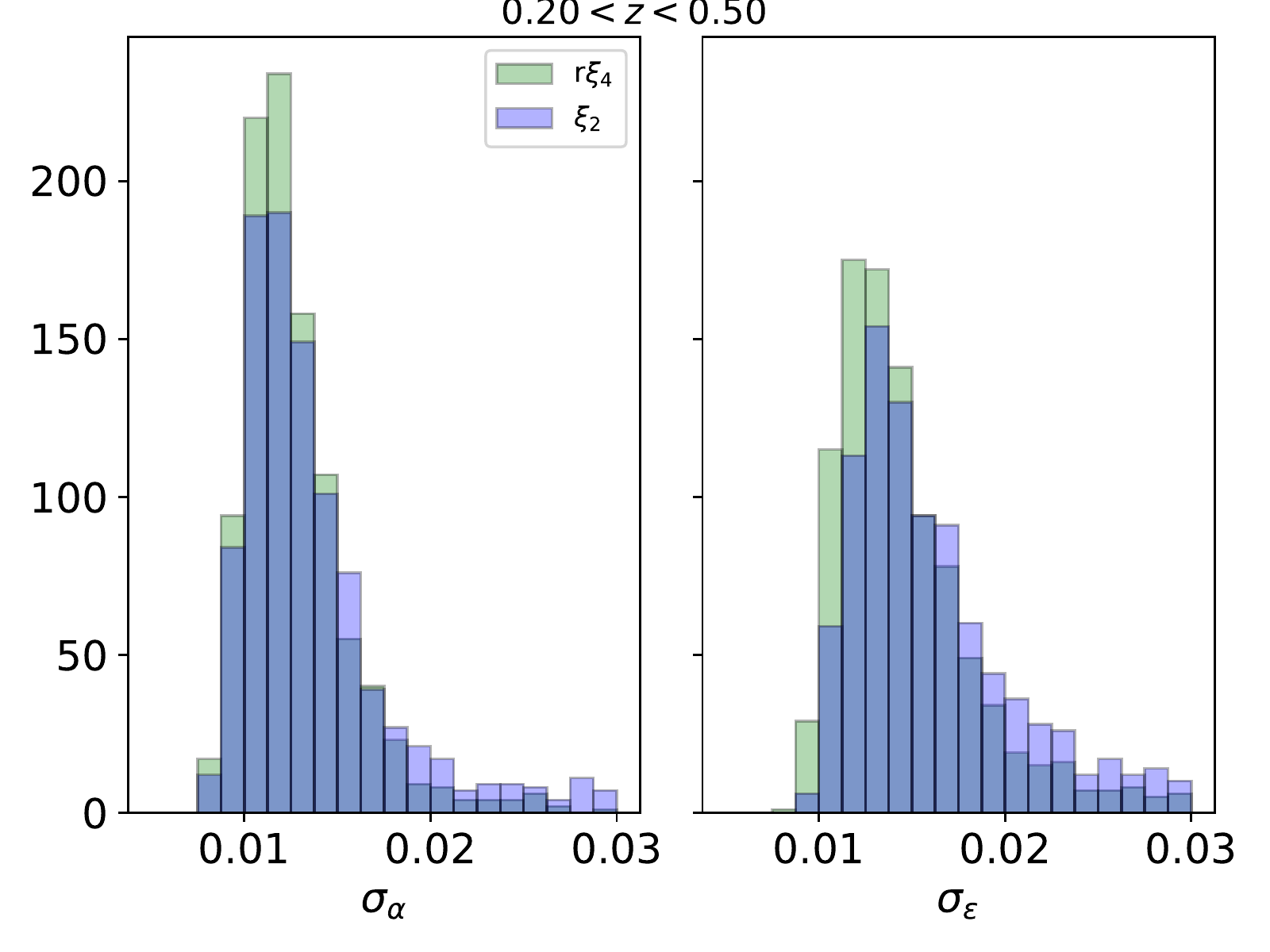}
   \caption{ Error Distribution  for 1000 MD-PATCHY  post-reconstruction for the intermediate redshift bin comparing fits using monopole+quadrupole fits (denoted by $\ell =2$) with monopole+quadrupole+hexadecapole fits (denoted by $\ell=4$) for 1000 MD-PATCHY  post-reconstruction for the lower redshift bin ($z= 0.2 - 0.5$). }
   \label{fig:disthexa}
\end{figure}

In addition to quantifying systematic errors we also test the effect on the uncertainties.   Figure \ref{fig:disthexa} shows the error distributions of the post-reconstruction fits in the three redshift bins. 
The figure illustrates for the lower redshift bin that when including hexadecapole information (i.e., monopole+quadrupole+hexadecapole), the error distributions show differences with respect to only the monopole-quadrupole fits. Results indicate that using the hexadecapole reduces the mean uncertainties by taking the RMS, $\Delta \sigma_\alpha=0.0015$ and  $\Delta \sigma_\epsilon=0.0032$.

  \cite{Beutler16b} (companion paper), analyses the same mocks in Fourier Space, comparing also the monopole+quadrupole+hexadecapole fits with the monopole+ quadrupole case for performing RSD analysis. Their results indicate the dispersion is decreased when considering the hexadecapole; in our results, the reduction in the dispersion is smaller in the best fit values.

 We conclude that the systematic error in BAO distance measurements associated with using monopole+ quadrupole against the fits using monopole+quadrupole+hexadecapole is $ 0.0002$ for $\alpha$ and  $\epsilon$. The variations in the mean error distributions are $\Delta \sigma_\alpha=0.0015$ and  $\Delta \sigma_\epsilon=0.0032$.

\begin{figure}
   \centering     
   \hspace*{-5.em}
    \includegraphics[width=2in]{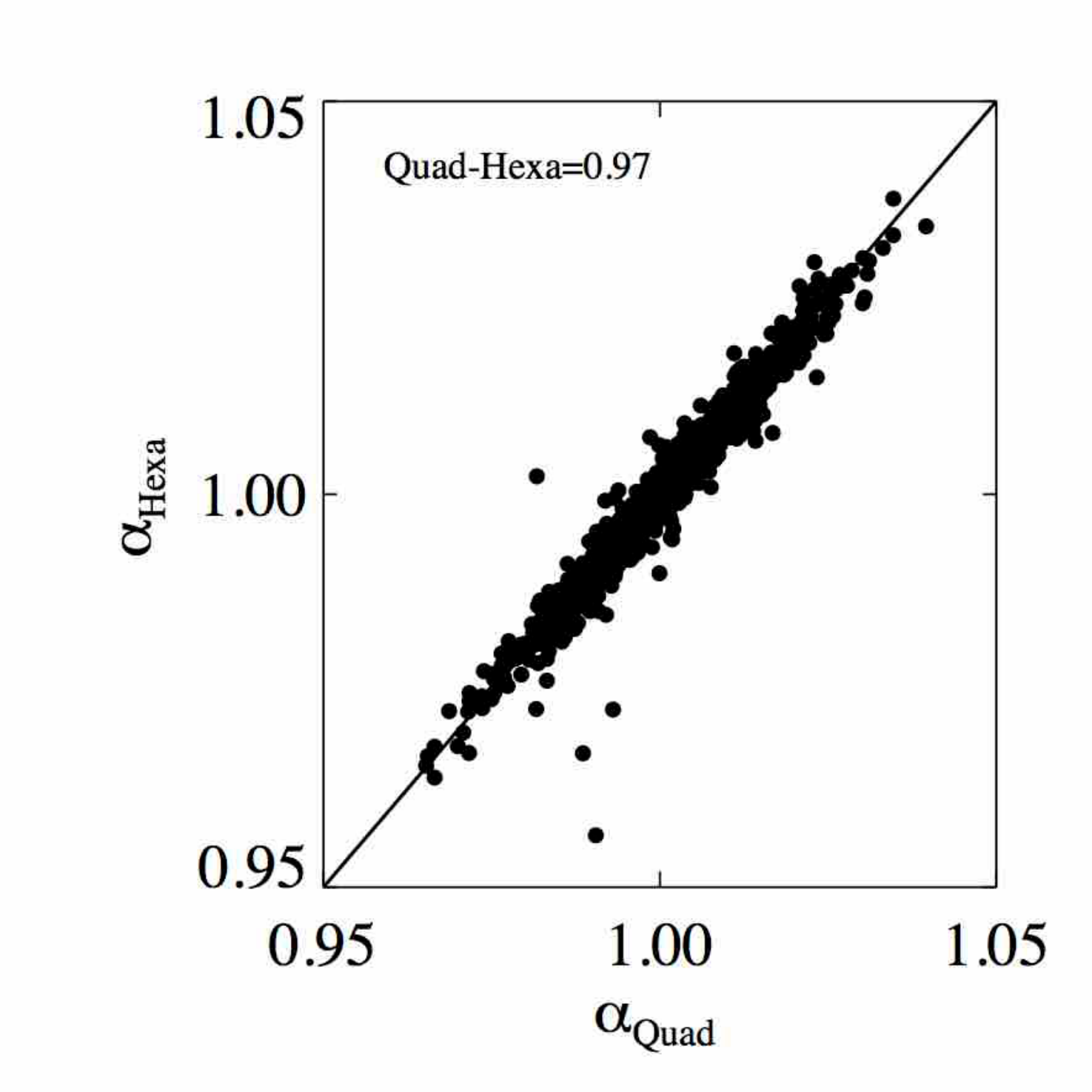}\hspace*{-3.em}
    \includegraphics[width=2in]{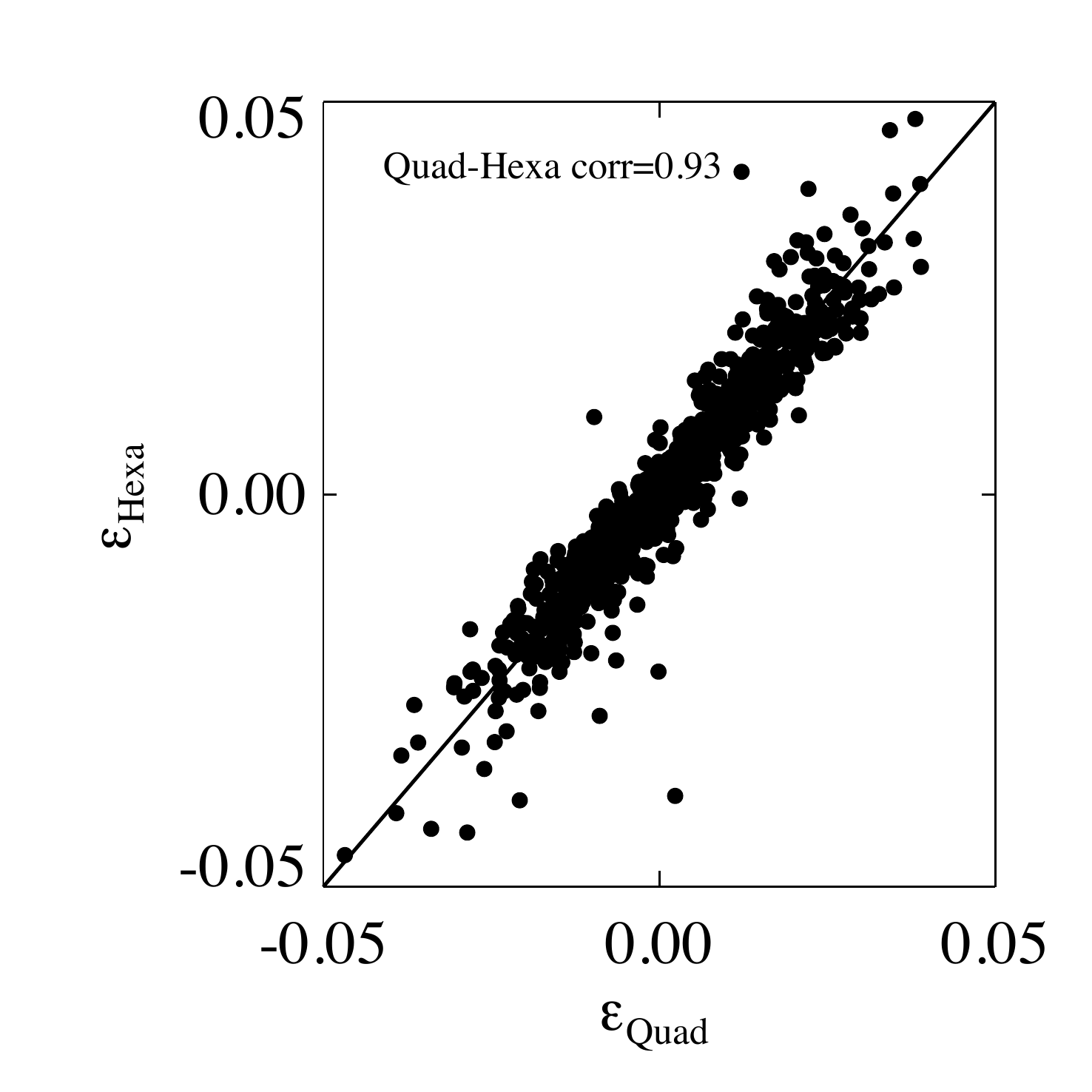}
   \caption{ Dispersion plots comparing fits using monopole+quadrupole fits (denoted by $\ell =2$) with monopole+quadrupole+hexadecapole fits (denoted by $\ell=4$) for 1000 MD-PATCHY  post-reconstruction for the lower redshift bin ($z= 0.2 - 0.5$). }
      \label{fig:errhexa}
\end{figure}

\subsection{Fitting range}
We revisit the impact of the fitting range on the anisotropic BAO results. The fitting range depends on the template and the noise. Increasing the maximum scale included in the fitting implies a trade-off between increasing the number of bins, and increasing the noise contribution to the fit. On the other hand, the minimum scale is related directly to the accuracy of the template and its ability to fit the small scales accurately, or of the broadband terms to absorb the mismatch between the template and the measurements. We examine the results in terms of the $\chi^2/d.o.f$ values.  Our fiducial choice is a range of (55,155 $h^{-1}$Mpc), using 5 $h^{-1}$Mpc bin results in 20 bins.

\begin{table}
\begin{center}
\caption{Fitting Systematic Error: Range. 
We summarize the variations, $\Delta \alpha$, $\Delta \epsilon$ (defined by eq.\ref{eq:delta}) observed considering variations of the fitting range, and their biases, $b_\alpha$, $b_\epsilon$ (defined by eq.\ref{eq:bias}). The detailed results of the fits are presented in Table~\ref{tab:fitsyspost1}. 
First block refers to variations of the minimum scale fixing the upper limit to 160 $h^{-1}$Mpc. Second Block, refers to variations of the maximum scale fixing the lower limit to 55 $h^{-1}$Mpc.}
\label{tab:deltarange}
\begin{tabular}{@{}lcccccc}
\hline
\multicolumn{7}{c}{DR12  Pre-Reconstruction}\\
\hline
$r_{min}$&$b_\alpha$&$b_\epsilon$&$\Delta \alpha$ &$\Delta \epsilon$  &$\Delta \sigma_\alpha$&$\Delta \sigma_\epsilon$ \\ 
\hline
\multicolumn{7}{c}{Bin 1 ($0.2 < z < 0.5$)}\\\\[-1.5ex]
\hline
30&$0.0041$&$0.0034$&0.0031&0.0013&0.0031&0.0013\\\\[-1.5ex]
40&$0.0011$&$0.0023$&0.0001&0.0002&0.0002&0.0006\\\\[-1.5ex]
50&$0.0010$&$0.0021$&0.0001&0.0002&0.0003&0.0005\\\\[-1.5ex]
60&$0.0011$&$0.0023$&-&-&-&-\\\\[-1.5ex]
70&$-0.0007$&$0.0010$&-0.0017&-0.0011&-0.0004&-0.0020\\\\[-1.5ex]
80&$0.0208$&$0.0248$&0.0195&0.0229&0.0061&0.0174\\
\hline
$r_{max}$&$b_\alpha$&$b_\epsilon$&$\Delta \alpha$ &$\Delta \epsilon$  &$\Delta \sigma_\alpha$&$\Delta \sigma_\epsilon$ \\ 
\hline
120&$ 0.0017$&$0.0010$&0.0007&-0.0011&0.0033&0.0053\\\\[-1.5ex]
130&$-0.0001$&$0.0011$&-0.0011&-0.0010&0.0023&0.0039\\\\[-1.5ex]
140&$0.0005$&$0.0017$&0.0004&-0.0002&-0.0013&-0.0024\\\\[-1.5ex]
155&$0.0010$&$0.0021$&-&-&-&-\\\\[-1.5ex]
160&$0.0009$&$0.0019$&-0.0001&-0.0002&-0.0001&-0.0003\\\\[-1.5ex]
170&$0.0011$&$0.0020$&-0.0002&0.0001&0.0010&0.0016\\\\[-1.5ex]
180&$0.0011$&$0.0021$&-0.0002&0.0002&0.0011&0.0019\\

\hline
\end{tabular}
\end{center}
\end{table}

Table~\ref{tab:deltarange} shows the results of fitting 1000 PATCHY mocks for the intermediate redshift bin.  The results show that the fits are robust against fitting range variations except when the lower bound is close to the BAO scale (70-80 $h^{-1}$Mpc), or going to very small scales (30 $h^{-1}$Mpc), which gives significantly biased results.
In the case of the upper bounds, the results indicate that the fits are very stable against variations of the upper bound of the fits for the values not too close to the BAO (i.e larger than 130 $h^{-1}$Mpc).
For our error budget account, we consider the results that show the smaller bias (i.e only lower bounds of 40-60 and upper bounds larger than 140) and we found the variations in $\Delta \alpha, \Delta \epsilon$
quoting the RMS of the different cases : $\Delta \alpha =0.0002$ and $\Delta \epsilon=0.0002$.

Concerning the variations in the mean error of the distributions, they are small (again only considering the cases that shows small biases i.e only lower bounds of 40-60 $h^{-1}$Mpc and upper bounds larger than 140 $h^{-1}$Mpc) and we find and RMS variation of  $\Delta \sigma_\alpha=0.0011$ and $\Delta \sigma_\epsilon=0.0018$. The performance of the fits depends on the details in the modelling and anisotropic methodology. A better model allows us to include smaller scales in the fitting. Using large scales can provide more information but also increases the noise. In \cite{Ross16} (companion paper), the fitting range is also tested. They found an optimal range of 60-160 $h^{-1}$Mpc, which is the fiducial range we use in this paper. %
 However, according to our results, in our methodology increasing the upper bound to 180 $h^{-1}$Mpc will decrease the error by 0.0011 (8\%) in $\alpha$ and  0.0019 (12\%) in $\epsilon$.

  We conclude that the systematic error in BAO distance measurements associated with the range of the fits considering only non-biased results (i.e lower limit between 40-60 $h^{-1}$Mpc and upper limit larger than 140 $h^{-1}$Mpc) is $0.0002$ for $\alpha$ and 0.0002 for $\epsilon$. The variations in the error distributions  could be as large as 0.0011 in $\sigma_\alpha$ and 0.0019 in $\sigma_\epsilon$.

\subsection{Bin centres }
We vary the bin centre, maintaining the bin width to test whether this change affects the anisotropic fitting results. Fixing the bin size to 5 $h^{-1}$Mpc, we shifted the bin centre by 1, 2 ,3, and 4 $h^{-1}$Mpc compared with the fiducial choice of 0 $h^{-1}$Mpc shift. \footnote{This test also allows us to better compare the power spectrum with the configuration space fits. This is because we expect that the average of the measurements performed over different centre bins will be strongly correlated with the power spectrum measurements.}

Table \ref{tab:deltabincenter} shows the fitting results with the different bin centres shifted compared with the fiducial choice. The bias of all measurements is similar for all cases ($<0.0006$). We measure variations on the mean values of $\alpha$ and $\epsilon$, 
taking the RMS of the different cases: $\Delta \alpha=0.0002$ and  $\Delta \epsilon=0.0004$. We conclude that the systematic error in BAO distance measurements associated to bin centres is $0.0002$ for $\alpha$ and 0.0004 for $\epsilon$ for the three redshift bins.

\begin{table}
\begin{center}
\caption{Fitting Systematic Error: Bin Centre. We summarize the variations, $\Delta \alpha$, $\Delta \epsilon$ (defined by eq.\ref{eq:delta}) observed from varying the bin center, and their biases, $b_\alpha$, $b_\epsilon$ (defined by eq.\ref{eq:bias}). The detailed results of the fits are presented in Table~\ref{tab:fitsys}. }

\label{tab:deltagauss}
\begin{tabular}{@{}lcccc}
\hline
\multicolumn{5}{c}{DR12  Pre-Reconstruction}\\
\hline
Bin Centre&$b_\alpha$&$b_\epsilon$&$\Delta \alpha$ &$\Delta \epsilon$  \\
\hline
\multicolumn{5}{c}{Bin 1 ($0.2 < z < 0.5$)}\\
\hline
0 $h^{-1}$Mpc&$0.0008$&$0.0021$&-&-\\
1 $h^{-1}$Mpc&$0.0010$&$0.0021$&0.0002&$<$0.0001\\
2 $h^{-1}$Mpc&$0.0010$&$0.0017$&$<$0.0001&0.0004\\
3 $h^{-1}$Mpc&$0.0007$&$0.0015$&0.0003&0.0006\\
4 $h^{-1}$Mpc&$0.0009$&$0.0018$&0.0001&0.0003\\
\hline
\end{tabular}
\end{center}
\end{table}

\section{Application to BOSS Data} \label{sec:bossdata}

\begin{figure*}
   \centering
         \includegraphics[width=2.9in]{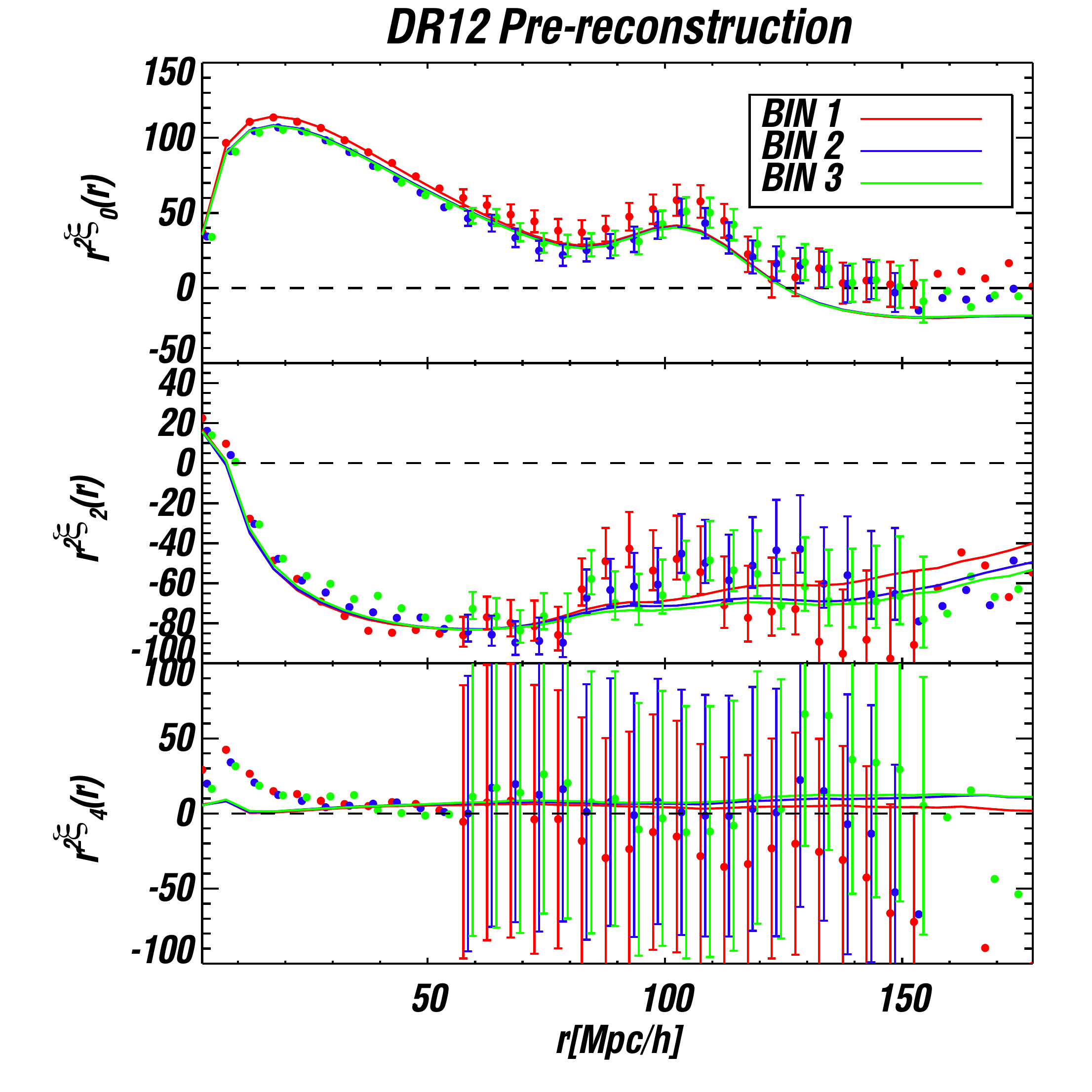}
     \includegraphics[width=2.9in]{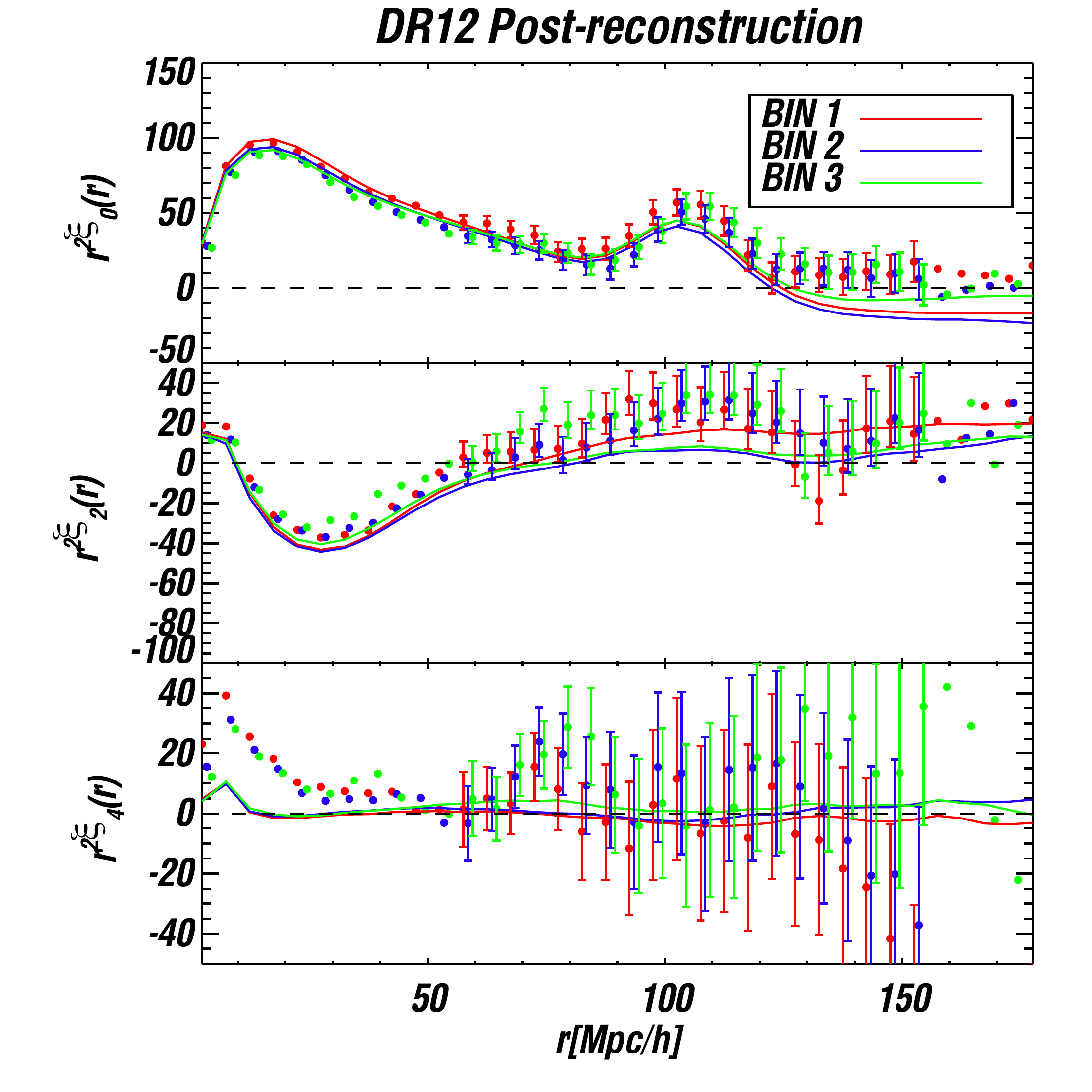}
    \includegraphics[width=2.9in]{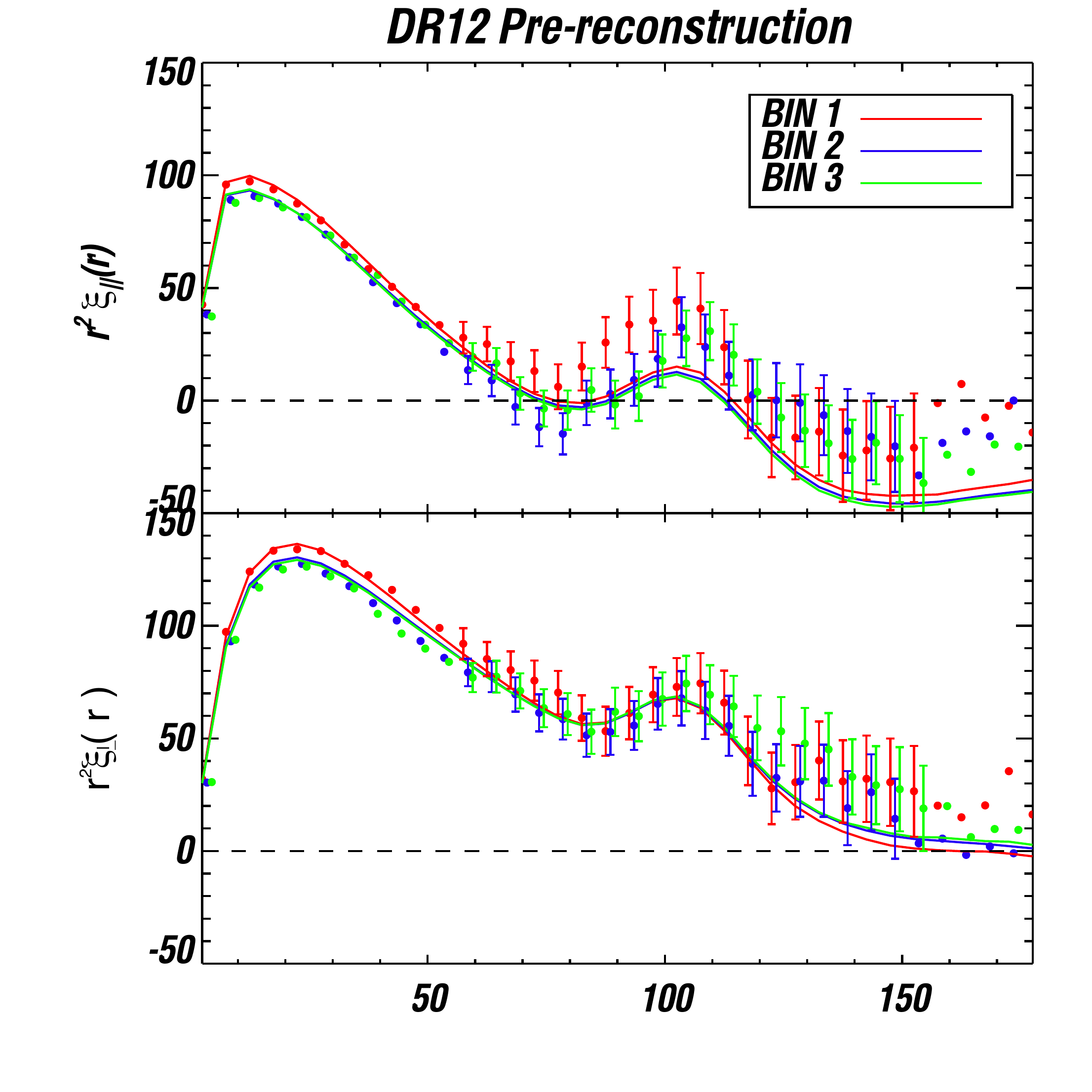}  
    \includegraphics[width=2.9in]{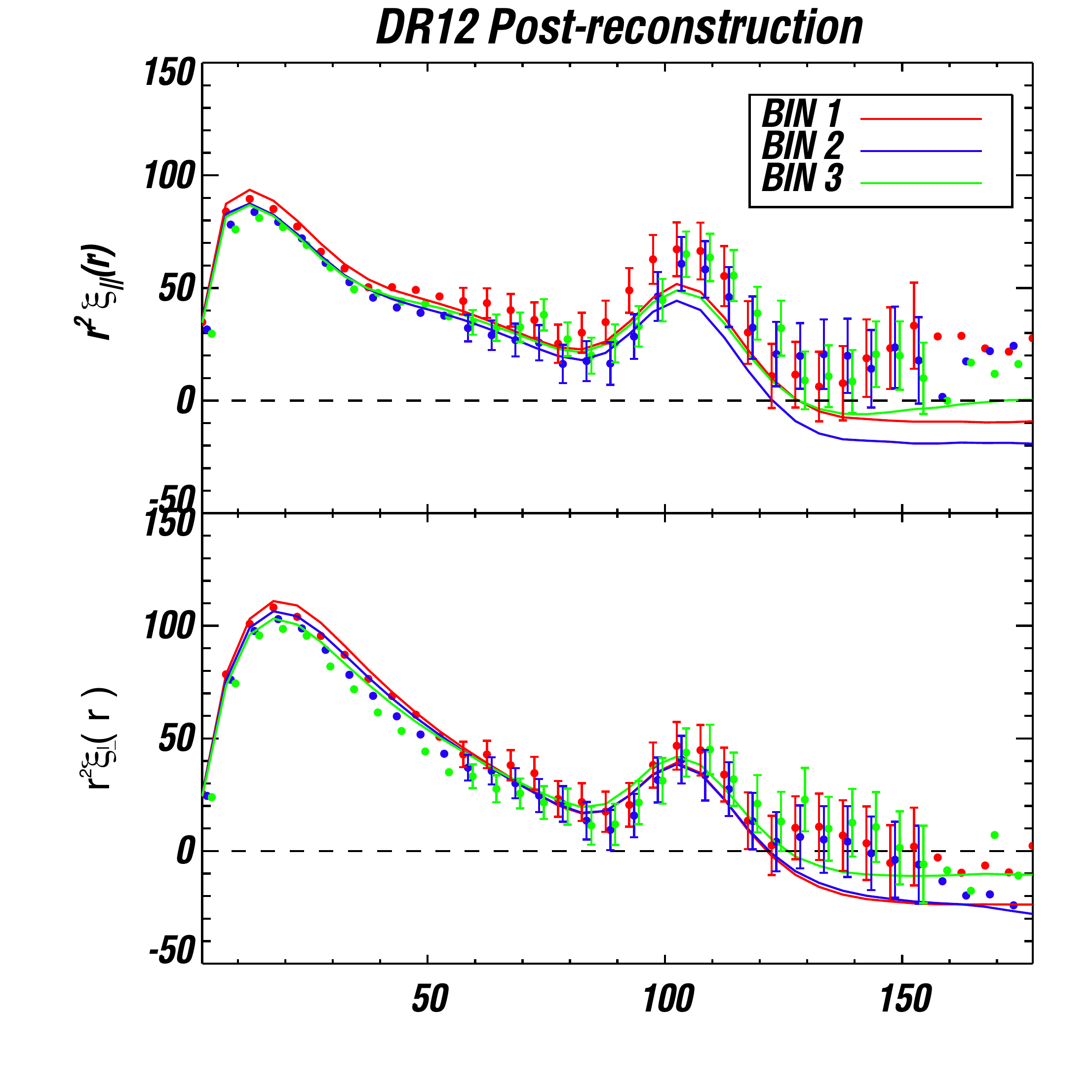}
   \includegraphics[width=2.9in]{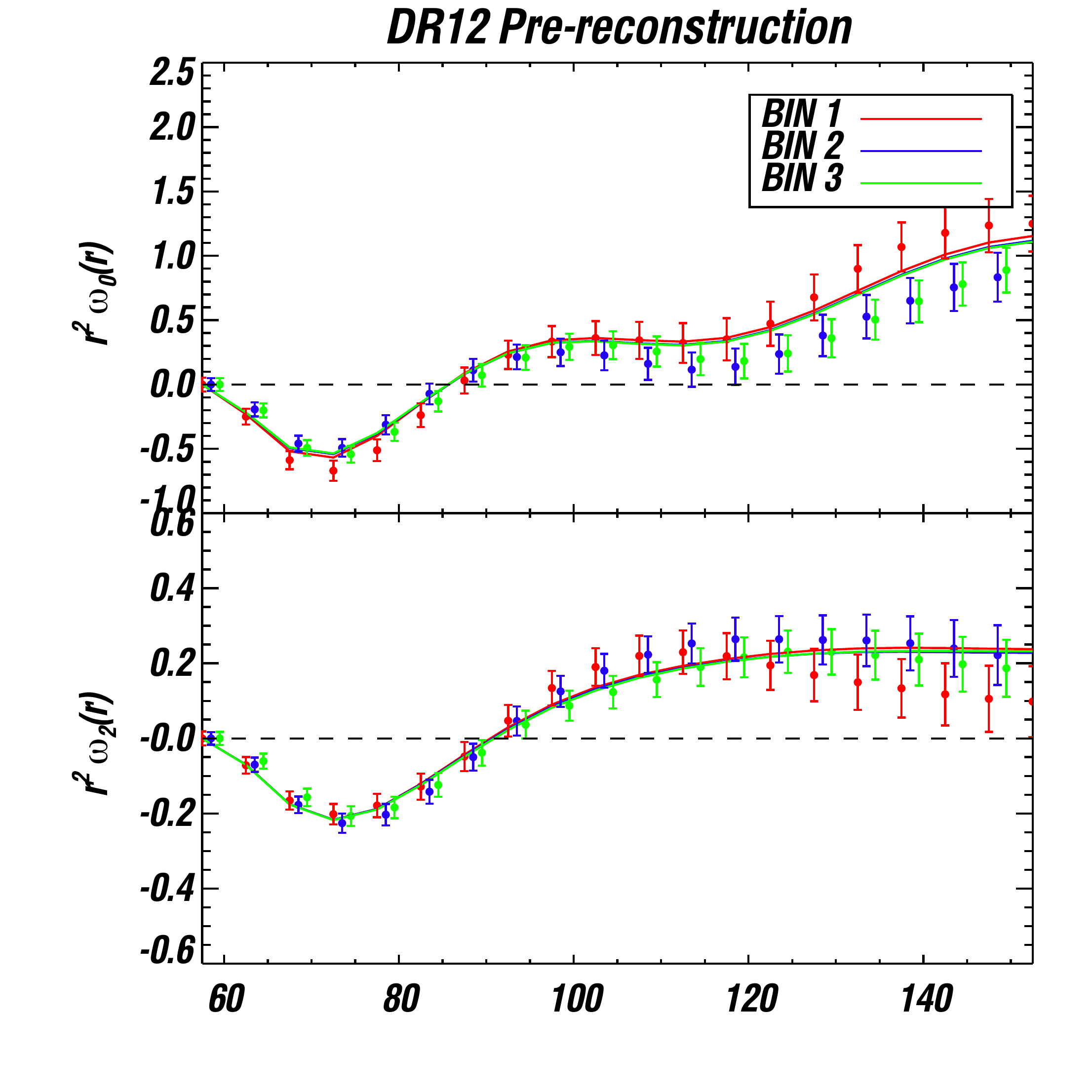}   
    \includegraphics[width=2.9in]{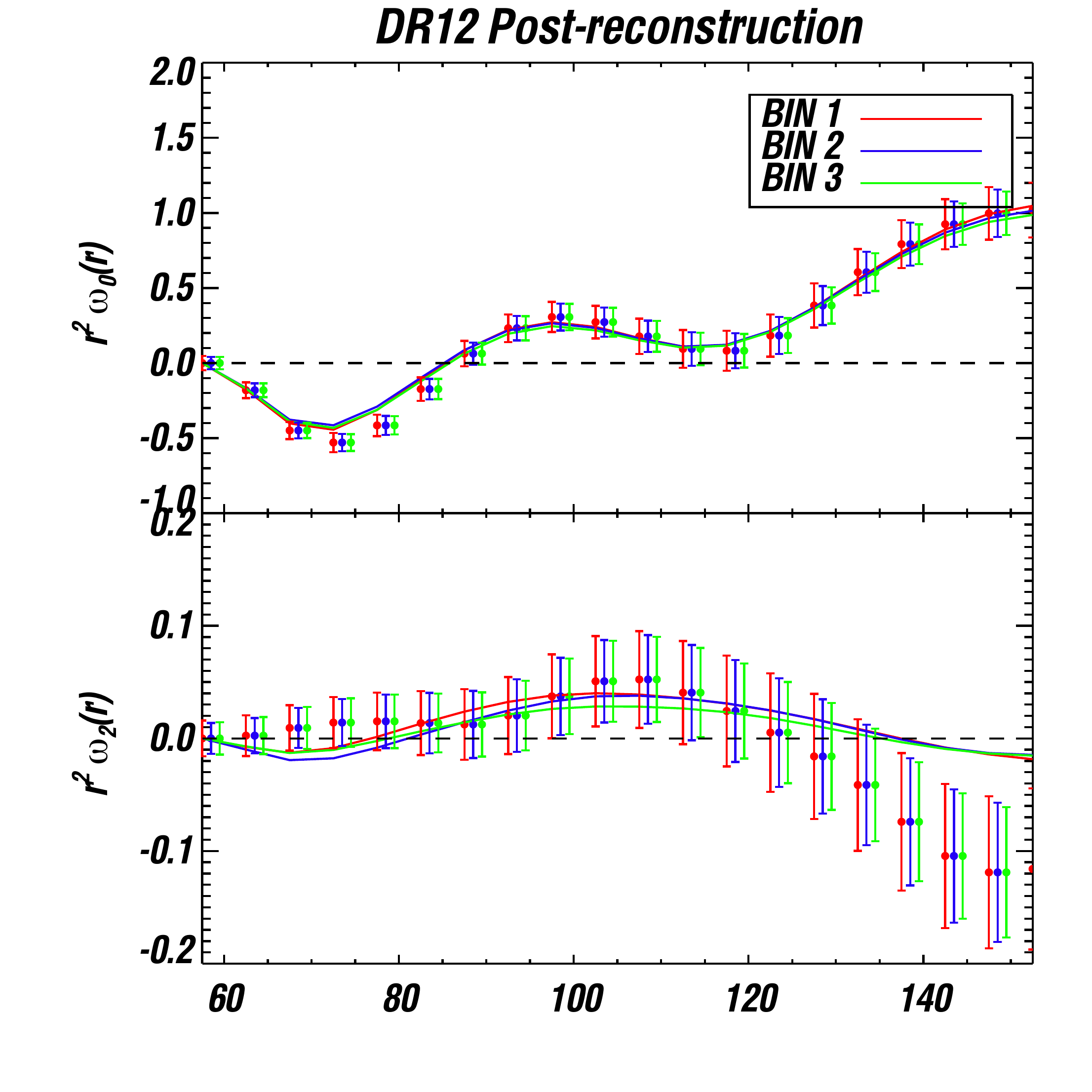}
    \caption{ 
 Top panels: Multipoles of the final BOSS combined sample pre-reconstruction (left panel) and post-reconstruction (right panel) in three redshift bins. Intermediate panels: Same as before for clustering wedges. Bottom panels: $\omega_l$ clustering statistics. Error bars represent the data, and solid line represent the mean of the mocks. ``Bin 1" refers to the lower redshift bin ($z= 0.2 - 0.5$);  ``Bin 2" considers the intermediate redshift range ($z= 0.4 - 0.6$ ), and ``Bin 3" refers to higher redshift range ($z= 0.5 - 0.75$). \commentH{ Add the best fit}}
      \label{fig:data}
\end{figure*}

We now apply our methodology as described in Section~\ref{sec:method} to our BOSS dataset. We test if the offsets between measured data parameters from variants of the methodology and the fiducial best-fit values are consistent with those from the mocks. We use the final BOSS combined catalogue from SDSS-DR12, as described in Sec.~\ref{data}. 
We  apply all weights described in \cite{Ross16} which included weights related to seeing, stellar densities, fibre collisions, redshift failures and the FKP weights to reduce the variance in clustering measurements. 
We first calculate the pair-counts from the catalogues, and then compute the various clustering estimators before reconstruction as seen in the left panels of Figure~\ref{fig:data} top panels for multipoles, intermediate panels for wedges and bottom panels for the $\omega_l$ clustering estimator. 
 The data points are the error bars, and the solid lines are the mean from the mocks. We then apply our reconstruction algorithms to our data, and show the resulting clustering estimators for the data in the right panels of Figure~\ref{fig:data} for all three redshift bins in the catalogue. The error bars are given by the diagonal terms of the covariance matrix computed from the 1000 MD-PATCHY mock catalogues. 
We notice the data measurements: monopole, quadrupole, and hexadecapole are consistent with the behaviour \change{observed in the mocks}.
 The new feature in this plot is the hexadecapole contribution, which is very small even pre-reconstruction and noisy. 
 The hexadecapole contribution post-reconstruction presents a reduction of the amplitude and of the associated error bars. 
The clustering wedges  ($\xi_{||,\perp}$) and the $\omega$ data measurements are also consistent with the behaviour observed on the mocks. 

The results of applying the anisotropic fitting methodology to the combined samples of BOSS are presented in Table~\ref{tab:fitsdr12a} 
for the three redshift bins, considering the different variations of the methodology explored in the paper.
One aspect to notice is that at least for the DR12 data, the uncertainty found from the $\omega$-estimator is almost the same as for the standard multipoles (some bins are a little better, some are a little worse), so arguably it would make very little difference which one we choose for the consensus result, but we might expect the $\omega$-estimator to perform the best in the future.

 We analyze the differences in the values of $\alpha$ and $\epsilon$ generated for the variations of the methodology ($\Delta\alpha$ and $\Delta\epsilon$) in Table \ref{tab:fitsdr12a}, and their consistency with the mocks results. We start analyzing the lower redshift bin. The results are shown in Figure \ref{fig:deltasdata}; we show how the $[\Delta \alpha^{\rm DR12}, \Delta \epsilon^{\rm DR12}]$ observed in DR12 compares with the $\Delta \alpha^{\rm MOCKS} $, $\Delta \epsilon^{\rm MOCKS} $ observed in the mocks for the three redshift bins. 
 From the figures we conclude the variations observed on the data are completely consistent with the results obtained with the mocks for the lower redshift bin, as most of them lie within the $1-\sigma$ contour. 
Only the hexadecapole case lies outside the $1-\sigma$ contours for the lower redshift bin, \change{we should notice the hexadecapole is particularly noise domitted (Figure \ref{fig:data})}.

\begin{figure*}
   \centering
   \includegraphics[width=7in]{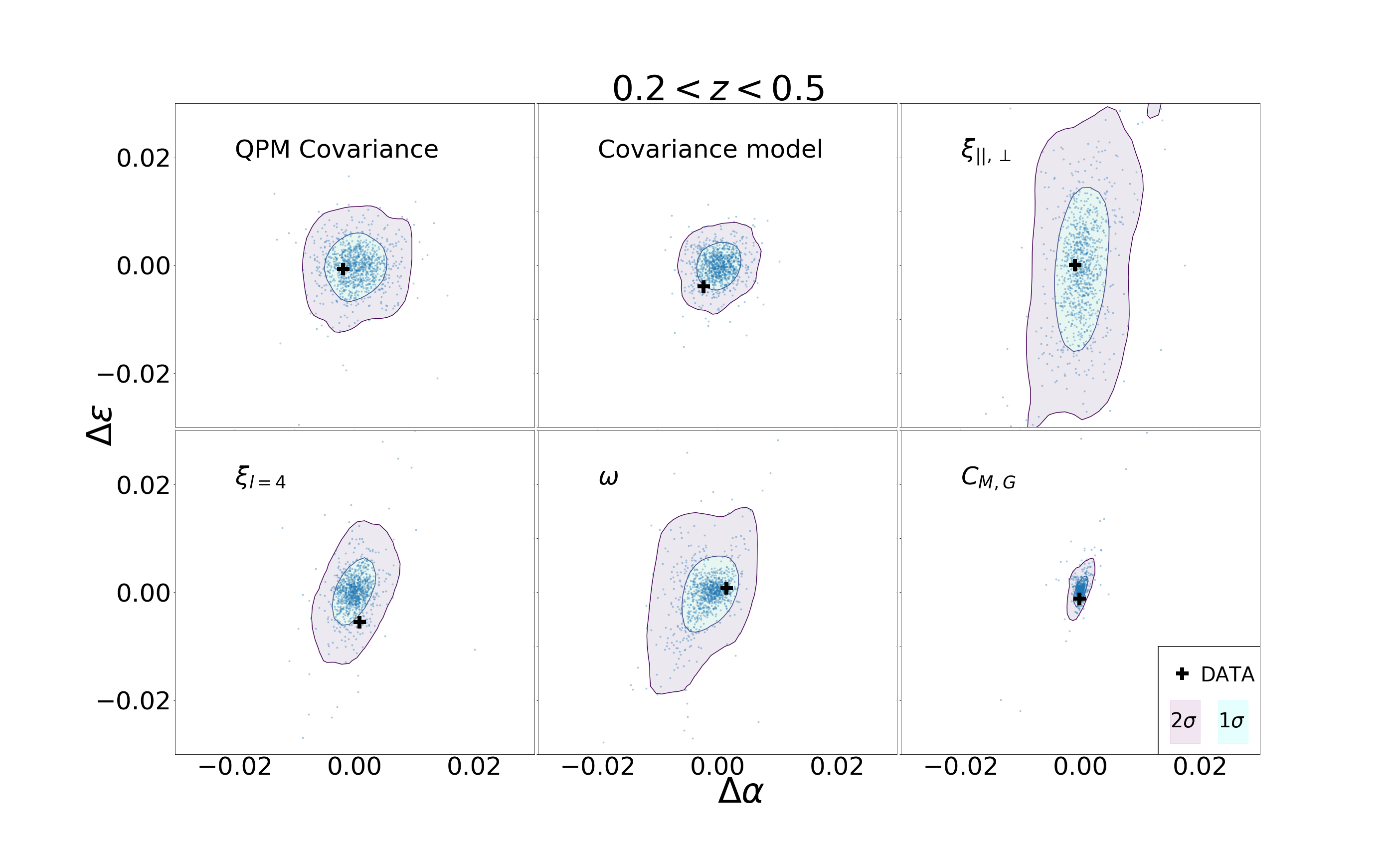}
          \caption{ We show how the $[\Delta \alpha^{\rm DR12}, \Delta \epsilon^{\rm DR12}]$ observed in DR12 [black cross] compares with the $\Delta \alpha^{\rm MOCKS} $, $\Delta \epsilon^{\rm MOCKS} $ observed in the mocks [blue dots] for the following six variants of the methodology (from left to right, and top to bottom): QPM covariance, model covariance, wedges estimator, hexadecapole contribution, $\omega_\ell$-estimator and modified Gaussian damping model, respectively. The six panels corresponds to the lower redshift bin.}
      \label{fig:deltasdata}
\end{figure*} 
\begin{figure*}
   \centering 
    \includegraphics[width=7in]{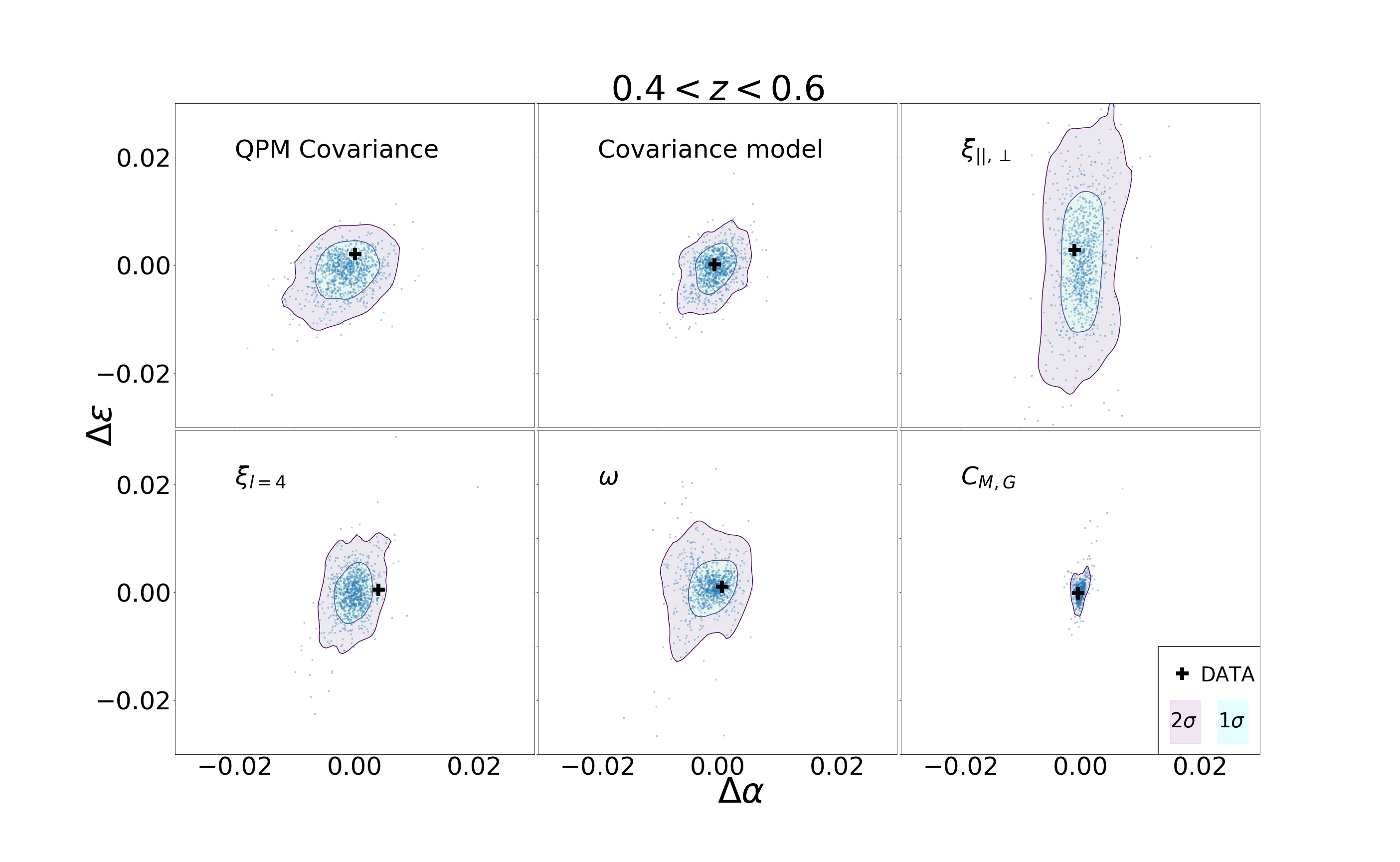}   
       \caption{ We show how the $[\Delta \alpha^{\rm DR12}, \Delta \epsilon^{\rm DR12}]$ observed in DR12 [black cross] compares with the $\Delta \alpha^{\rm MOCKS} $, $\Delta \epsilon^{\rm MOCKS} $ observed in the mocks [blue dots] for the following six variants of the methodology (from left to right, and top to bottom): QPM covariance, model covariance, wedges estimator, hexadecapole contribution, $\omega_\ell$-estimator and modified Gaussian damping model, respectively. The six panels corresponds the intermediate redshift bin.}
      \label{fig:deltasdata}
\end{figure*}
\begin{figure*}
   \centering
 \includegraphics[width=7in]{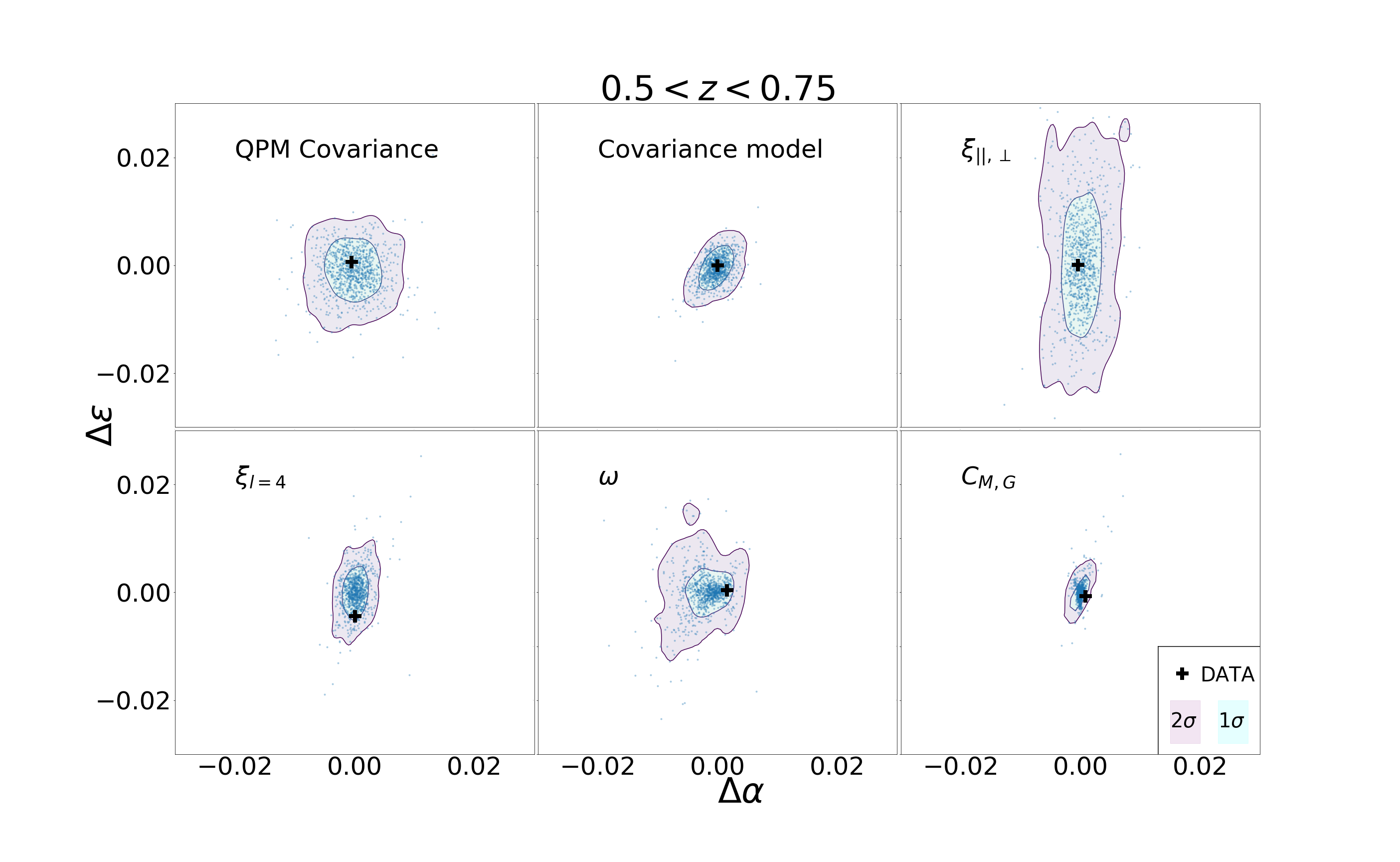}	

    \caption{ We show how the $[\Delta \alpha^{\rm DR12}, \Delta \epsilon^{\rm DR12}]$ observed in DR12 [black cross] compares with the $\Delta \alpha^{\rm MOCKS} $, $\Delta \epsilon^{\rm MOCKS} $ observed in the mocks [blue dots] for the following six variants of the methodology (from left to right, and top to bottom): QPM covariance, model covariance, wedges estimator, hexadecapole contribution, $\omega_\ell$-estimator and modified Gaussian damping model, respectively. The six panels corresponds to the higher redshift bin.}
      \label{fig:deltasdata}
\end{figure*}

\begin{figure*}
   \centering
     \includegraphics[width=7.1in]{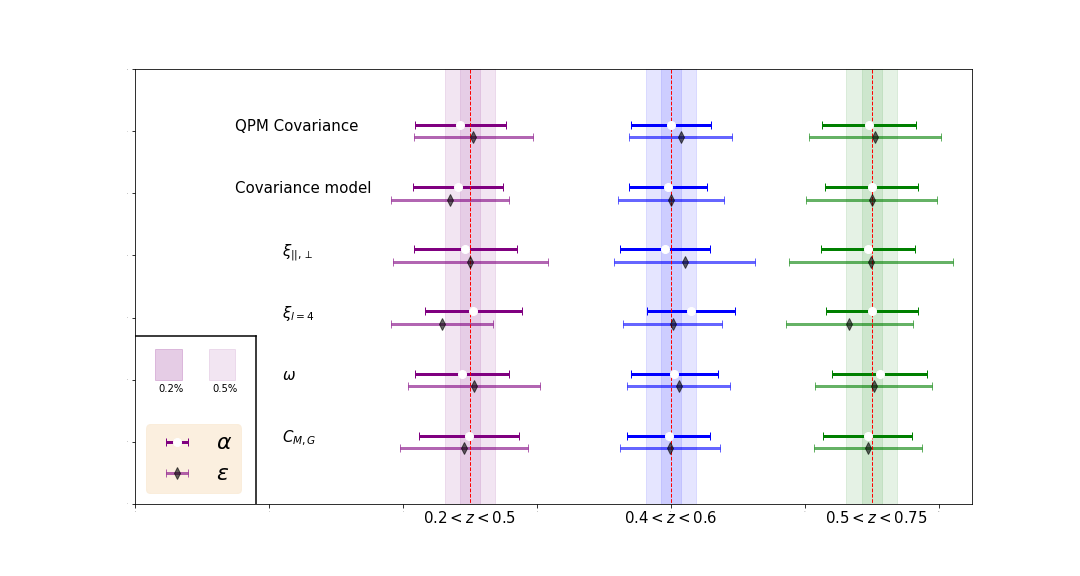}   
   \caption{ Best fits and errors from fitting the three redshift bin of BOSS data. The white circles show the $\alpha$ and the black diamond the $\epsilon$ obtained from the variant of the methodology with their respective error bars.
   The red dashed lines are the fiducial case of the different measurements, the light shadow region correspond to a difference from the mean of 0.2\% (light shadow), and  0.5\%(dark shadow) respectively.\commentH{remove values in the x axis}}
   \label{fig:dataComp}
\end{figure*}

Figure \ref{fig:dataComp} shows the best fits obtained from applying the variants  of the methodology studied in the  paper to DR12 final data samples. The white dot shows results for $\alpha$ best fits values for the three redshift bins. 
 The dashed line indicates the measurement for the fiducial methodology and the dark shadow region are the 0.2\% variations with respect to the fiducial case and the dark shadow regions are the $\Delta x = \pm0.005$, for $x=\alpha, \epsilon$.  As we can observe, the different measurements from all three redshift bins lie within $\Delta x = \pm0.002$ except one measurement in the intermediate redshift bin for the hexadecapole. 
The black diamond shows $\epsilon$ measurements. For $\epsilon$, the results indicate that, for the intermediate and high redshift bins, almost all  measurements are within the dark shadow region. 

\begin{table*}  
\caption{Fitting results from Combined DR12 samples in the low, intermediate and hight redshift bins for different variants of the methodology. $\Delta x$ is the difference with respect to the fiducial case (first row)}
\label{tab:fitsdr12a}

\begin{tabular}{@{}lcccccccc}
\hline
\multicolumn{8}{c}{Testing Variants of Methodology on DR12}\\
\hline
\multicolumn{8}{c}{DR12 Combined Samples Post-reconstruction}\\
\hline
\multicolumn{8}{c}{BIN 1 ($0.2 < z < 0.5$)}\\
\hline

Variable&$\alpha \pm \sigma_\alpha$ & $\Delta \alpha$&$\Delta \sigma_\alpha$&$\epsilon \pm \sigma_\epsilon$ &$\Delta \epsilon$&$\Delta \sigma_\epsilon$&$ \chi^2$ \\
\hline

PAT cov &
$0.9995\pm 0.0098$&-&-&$0.0152 \pm 0.0125$&&&$1.41$\\
\hline
QPM cov &
$0.9976\pm 0.0091$& -0.0019& -0.0007&$0.0159\pm 0.0119$&0.0007&-0.0006&1.47\\
\\[-1.5ex]

Model Cov&
$0.9971\pm 0.0090$& -0.0023& -0.0008&$0.0113\pm 0.0118$&-0.0039&-0.0006&1.37\\
\hline
$\xi_{\parallel, \perp}$ &
$0.9986\pm 0.0103$& -0.0009& 0.0005&$0.0153\pm 0.0154$&0.0001&0.0029&1.16 \\
\\[-1.5ex]

$\xi_{l=4}$&
$1.0002\pm 0.0097$& 0.0008& -0.0001&$0.0097\pm 0.0102$&-0.0055&-0.0022&1.32\\
\\[-1.5ex]

$\omega_l$ &
$0.9980 \pm 0.0094$&0.0015&-0.0004&$0.0160 \pm 0.0132$&0.0008&-0.0007&1.41\\
\hline

$C_{MG,60}$&
$0.9993\pm 0.0099$& -0.0002& 0.0000&$0.0140\pm 0.0127$&-0.0012&0.0002&1.36\\
%
\hline


\multicolumn{8}{c}{BIN 2 ($0.4 < z < 0.6$)}\\
\hline
Variable&$\alpha \pm \sigma_\alpha$ & $\Delta \alpha$&$\Delta \sigma_\alpha$&$\epsilon \pm \sigma_\epsilon$ &$\Delta \epsilon$&$\Delta \sigma_\epsilon$&$ \chi^2$ \\
\hline
PAT&
$0.9928\pm 0.0084$&-&-&$-0.0036 \pm 0.0107$&-&-&0.63\\
\hline
QPM cov&
$0.9929\pm 0.0080$& 0.0001& -0.0004&$-0.0016\pm 0.0103$&0.0021&-0.0004&0.69 \\
\\[-1.5ex]
Model cov&
$0.9923\pm 0.0078$& -0.0005& -0.0006&$-0.0035\pm 0.0106$&0.0002&-0.0001&0.68 \\
\hline
$\xi_{\parallel \perp}$&
$0.9917\pm 0.0089$& -0.0010& 0.0005&$-0.0008\pm 0.0140$&0.0029&0.0034&0.66\\
\\[-1.5ex]
$\xi_{l=4}$&
$0.9968\pm 0.0088$& 0.0040& 0.0004&$-0.0032\pm 0.0099$&0.0005&-0.0008&2.22\\
\\[-1.5ex]
$\omega_l$&

$0.9935\pm0.0087$&0.0007&0.0003&$-0.0020\pm0.0102$&0.0011&-0.0005&0.70 \\
\hline
$C_{MG 60}$&
$0.9924\pm 0.0083$& -0.0004& -0.0001&$-0.0037\pm 0.0100$&-0.0001&-0.0007&0.62\\
\hline
%
\multicolumn{8}{c}{BIN 3 ($0.5 < z < 0.75$)}\\
\hline
Variable&$\alpha \pm \sigma_\alpha$ & $\Delta \alpha$&$\Delta \sigma_\alpha$&$\epsilon \pm \sigma_\epsilon$ &$\Delta \epsilon$&$\Delta \sigma_\epsilon$&$ \chi^2/d.o.f$ \\
\hline
PAT cov&
$0.9820\pm 0.0091$&-&-&$-0.0109 \pm 0.0125$&-&-&0.97 \\
\hline
QPM cov&
$0.9815\pm 0.0094$& -0.0005& 0.0003&$-0.0102\pm 0.0131$&0.0006&0.0006&1.04\\
\\[-1.5ex]
Model cov&
$0.9820\pm 0.0092$& 0.0000& 0.0001&$-0.0109\pm 0.0130$&0.0000&0.0005&0.97 \\
\hline
$\xi_{\parallel \perp}$&
$0.9813\pm 0.0094$& -0.0004& 0.0003&$-0.0110\pm 0.0163$&0.0001&0.0038&1.08\\
\\[-1.5ex]
$\xi_{l=4}$&
$0.9821\pm 0.0091$& 0.0001& 0.0000&$-0.0153\pm 0.0127$&-0.0044&0.0001&1.59 \\
\\[-1.5ex]
$\omega_l$&
$0.9836\pm 0.0094$&0.0016&0.0010&$-0.0105\pm 0.0116$&0.0004&0.0009&1.23\\
\hline
$C_{MG 60}$&
$0.9812\pm 0.0088$& -0.0008& -0.0003&$-0.0116\pm 0.0108$&-0.0007&-0.0017&1.00\\
\hline
\end{tabular}
\end{table*}

\section{Systematic Error budget} \label{discussion}

 The aim of this paper is to provide final numbers for the distance measurements  $D_V(z)$, $D_A(z)$, $H(z)$  for the three redshift bins which include the systematic error coming from the theoretical systematics explored in this paper.
In previous sections, we have tested several variations of the fiducial methodology. We conclude that all of them give statistically unbiased measurements with comparable error bars, so there is no strong reason to prefer one method over the rest. 
In previous sections, we picked a fiducial pipeline, and we tested how variations of this pipeline change the results using the mock catalogues measuring the difference in the mean values of the best fit parameters  between fiducial and the variant; this provides us the contribution to the error budget for a particular variant of the methodology. Every change was made one at a time, i.e in this first estimate of the error budget, we are neglecting possible correlations between effects.

In Tables \ref{tab:sysonebyonepre}, \ref{tab:sysonebyonepost}, \ref{tab:sysonebyonepostb}, we summarize our findings for pre-reconstruction and post-reconstruction results.  For these three tables, we quote the $\Delta \alpha$ and $\Delta \epsilon$ (eq.\ref{eq:delta}) over every method variation computed using the main table of the corresponding section (RMS of differences between the means obtained of each kind of variations and the fiducial case). The numbers post-reconstruction are also quoted in the concluding statement its respective section. We consider only the cases that produce unbiased measurements, i.e we eliminate the cases where the modification is significantly biasing the measurements (ex. extending the lower bound of the fits produces a bias of 0.3\%). 

We start by briefly discussing pre-reconstruction results. These results are presented in Table \ref{tab:sysonebyonepre}. We present two blocks: the first block is devoted to the results obtained from this work and a second block includes previous results on theoretical systematics uncertainties in BAO analysis. The results indicate that the main contribution for the error comes from the estimator choice, which shows large variations in the best fits of $\Delta \alpha=0.004$ and $\epsilon=0.006$). From the second block, the larger contribution is coming from the template in the case of $\alpha$ (0.006); all other contributions are contributing less than 0.001. The total error budget pre-reconstruction adding in quadrature all the terms is 0.007 in $\alpha$ and $\epsilon$.

Post-reconstruction results are split into Tables \ref{tab:sysonebyonepost} and \ref{tab:sysonebyonepostb}, the first of which summarizes the error budget for \cite{Acacia16} and the second of which shows the results of the tests that do not contribute to the error budget, but still provide insights about variations in the best fits parameters.   
The first block in Table\ref{tab:sysonebyonepost} presents the results on the combined sample from this work.  We include separately the cosmology test (in a second block), as this test was performed with fewer mocks and with a different sample (CMASS). The third block includes results from \cite{Acacia16} related to the estimator, and the variations on the estimator relates to the consensus values\footnote{The consensus values result from combining optimally Fourier space multipoles \citep{Beutler16a} and configuration space multipoles.} compared to the results in configuration space from this work. We also include a fourth block with the findings of previous research on theoretical systematics uncertainties in BAO analysis.  
 \cite{Vargas14} analysed the potential systematics in BAO fitting methodology using mocks and data from BOSS DR10 and DR11 for the CMASS sample.The methodological changes tested are: (i) Model Templates ; (ii) Fitting Range and Bin Sizes; (iii) Nuisance Terms Model; iv) Priors; v) Non Linear Damping parameters  $\Sigma_{||,\perp}$  and Streaming model $\Sigma_{s}$. The variations in $\alpha$ and $\epsilon$ observed in the first block are all below $0.0017$ for $\alpha$ and  $0.0019$ for $\epsilon$. The variations in $\alpha$ and $\epsilon$ observed post-reconstruction are in all cases $\Delta \alpha,\Delta \epsilon < 0.001$. The dominant term in $\alpha$ error comes from the covariances (first block), fiducial cosmology (both contributing similarly) and the reconstruction step that is the largest error contribution to the systematic error budget. If we just consider the smoothing scale of 10 $h^{-1}$Mpc and not smaller smoothing scales this contribution decrease to a similar level as the covariance and the fiducial tests. For $\epsilon$, the dominant terms are the estimator, covariance, fiducial cosmology, broadband terms and non linear damping, all of them contributing in similar proportions, followed by the damping model.
 
\begin{table*}
\begin{center}
\caption{ Summary of Theoretical Systematic Errors Pre-reconstruction. We quote the maximal $\Delta \alpha$ and $\Delta \epsilon$ (eq.\ref{eq:delta}) over every method variation computed using the main table of the corresponding section (maximal  difference between the means obtained of each kind of variations and the fiducial case). The first block corresponds to the findings in this work. The case of Fiducial Cosmology is treated separately as it was performed with a small number of simulations and with the CMASS sample. The third block corresponds to the findings of previous works. We also include a total.}
\label{tab:sysonebyonepre}
\begin{tabular}{@{}lllcc} 
\hline
\multicolumn{4}{c}{DR12  Pre-Reconstruction}\\
\hline
Mocks&Sample&Source of Uncertainty&$\Delta \alpha$& $\Delta \epsilon $\\
\hline
PATCHY(1000)&COMBINED&Estimator (Section~\ref{sec:estimators})&0.0047&0.0057\\
QPM(1000)&COMBINED&Randoms (Section~\ref{sec:randoms})&0.0008&0.0006\\
QPM/PATCHY(1000)&COMBINED&Covariance (Section~\ref{sec:covariance})&0.0008&0.0030\\
*PATCHY(1000)&COMBINED&Hexadecapole (Section~\ref{sec:model})&0.0002 &0.0003\\
\hline
QPM(100)&CMASS&Fiducial Cosmology (Section~\ref{sec:cosmology})&0.0002&$<0.0001$\\
\hline
PTHALOS(600)&CMASS&Fitting Templates \citep{Vargas14}&0.0055&0.0004\\
PTHALOS(600)&CMASS&Fitting Bin Size \citep{Vargas14}&0.0003&0.0002\\
PTHALOS(600)&CMASS&Fitting Priors \citep{Vargas14}&0.0002&0.0006\\
PTHALOS(600)&CMASS&Fitting Nuisance T. \citep{Vargas14}&0.0005&0.0006\\
PTHALOS(600)&CMASS&Fitting Streaming  \citep{Vargas14}&0.0003&0.0007\\
PTHALOS(600)&CMASS&Fitting NL damping \citep{Vargas14}&0.0005&0.0033\\
\hline
Total & &&0.0071&0.0074\\
\hline
\end{tabular}
\end{center}
\end{table*}

\begin{table*}
\begin{center}
\caption{ Summary of Theoretical Systematic Errors Post-reconstruction. We quote the  $\Delta \alpha$ and $\Delta \epsilon$ (eq.\ref{eq:delta}) over every method variation considered for  \citep{Acacia16} BAO-only error budget. We computed using the main table of the corresponding section (RMS of the differences between the means obtained of each kind of variation and the fiducial case).The first block corresponds to the findings in this work. The cases of Reconstruction/Fiducial Cosmology are presented separately, as they were performed with an small number of simulations. We also include in the table as a reference the result from comparing our fiducial methodology to the consensus values from Alam et al. (2016).  We consider this number as the final account of the systematic error budget instead of the estimators Multp-Wed, as the consensus value includes the Fourier Space estimator contribution and also the optimal  combination method following Sanchez et al 2016. 
The second block includes the results from previous works on the  systematics related to the modelling. We present partial sums of the items as well as a global sum that corresponds to the final error budget for Alam et al. (2016). }
\label{tab:sysonebyonepost}
\begin{tabular}{@{}lllcccccc} 
\hline
\multicolumn{4}{c}{DR12  Post-Reconstruction}\\
\hline
Mocks&Sample&Source of Uncertainty&$\Delta \alpha$& $\Delta \epsilon $\\
\hline
QPM/PATCHY(1000)&COMBINED&Covariance (Section~\ref{sec:covariance})&0.0009&0.0009\\
PATCHY(1000)&COMBINED&Fitting:Damping (Section~\ref{sec:model})&$<$0.0001&0.0007\\
PATCHY(1000)&COMBINED&Fitting:Range (Section~\ref{sec:model})&0.0002&0.0002\\
PATCHY(1000)&COMBINED&Fitting:Bin Centre (Section~\ref{sec:model})&0.0002&0.0004\\

\hline
PATCHY(100)&COMBINED&Reconstruction Smoothing Scale (Section~\ref{sec:reconstruction})&0.0017&0.0006 \\
QPM(100)&CMASS&Fiducial Cosmology (Section~\ref{sec:cosmology})&0.0009&0.0010\\
\hline
PATCHY(1000)&COMBINED&Estimator Consensus-Fid \citep{Acacia16}& $0.0004$&0.0012\\
\hline
Total 1& &&0.0022&0.0021\\
\hline
Mocks&Sample&Source of Uncertainty&$\Delta \alpha$& $\Delta \epsilon $\\
\hline
PTHALOS(600)&CMASS&Fitting Templates \citep{Vargas14}&0.0003&0.0003\\
PTHALOS(600)&CMASS&Fitting Bin Size \citep{Vargas14}&0.0003&0.0003\\
PTHALOS(600)&CMASS&Fitting Priors \citep{Vargas14}&0.0001&0.0004\\
PTHALOS(600)&CMASS&Fitting Nuisance T. \citep{Vargas14}&0.0005&0.0015\\
PTHALOS(600)&CMASS&Fitting Streaming  \citep{Vargas14}&$<$0.0001&0.0005\\
PTHALOS(600)&CMASS&Fitting NL damping \citep{Vargas14}&0.0002&0.0009\\
\hline
Total 2 & &&0.0007&0.0018\\ 
\hline
Total 1+Total 2& &&0.0023&0.0028\\
\hline
\end{tabular}
\end{center}
\end{table*}

\begin{table*}
\begin{center}
\caption{ Summary of Test Post-reconstruction: Best fitting values. We quote the maximal $\Delta \alpha$ and $\Delta \epsilon$ (eq.\ref{eq:delta}) over the method variations not considered in the error budget of Alam et al (2016). We computed using the main table of the corresponding section (RMS of differences between the means obtained of each kind of variation and the fiducial case).}
\label{tab:sysonebyonepostb}
\begin{tabular}{@{}lllcc} 
\hline
Mocks&Sample&Source of Uncertainty&$\Delta \alpha$& $\Delta \epsilon $\\
\hline
PATCHY(1000)&COMBINED&Estimator (all) (Section~\ref{sec:estimators})&0.0012&0.0006\\
QPM(1000)&COMBINED&Randoms (Section~\ref{sec:randoms})& 0.0002&0.0002\\ 
PATCHY(1000)&COMBINED&Hexadecapole (Section~\ref{sec:model})&0.0002&0.0002\\
\hline
\end{tabular}
\end{center}
\end{table*}

Our estimate of the error budget is obtained  by adding  in quadrature all the sources of the systematic errors.   
One caveat on this estimate are the tests performed with a small number of mocks, for while testing with $\sim$100 mocks is enough to determine if a methodology is biased, \change{in} order to get the bias to the precision required for our measurements, we need more mocks. Nevertheless, we decided to consider the bias obtained from the test with few mocks in the error budget. 
Further work should be done to determine with more precision these biases in future surveys as eBOSS and DESI. The numbers in parenthesis for $\epsilon$ indicate the total when, instead of using our accounting of the systematic error related to the estimators, we use the systematic error coming from comparing our fiducial results with the consensus values. The main difference is that the systematic error in $\alpha$ decreases to be negligible, while on the other side, the systematic error in $\epsilon$ increases and becomes the most important contributor to the error budget. The difference in $\epsilon$ values in this case is driven by the differences between FS and CS estimators. 

Combining all  sources of theoretical systematic uncertainty, we find $\Delta \alpha\approx0.002$ and $\Delta \epsilon\approx0.003$ for all three redshift bins (see the ``Total 1+ Total 2" row in Table \ref{tab:sysonebyonepostb} for more precise figures). Simply combining these results in quadrature implicitly assumes that each source of uncertainty is independent, and so we regard these estimates as upper bounds rather than the best possible estimates. 
Crucially, we find that non-reconstruction-related sources of theoretical systematic uncertainty (everything in Table~\ref{tab:sysonebyonepost} except the second block) in $\alpha$ all combine to make a contribution that is dominated by the uncertainty related to reconstruction techniques (second block in Table ~\ref{tab:sysonebyonepost}).

\section{Final Distance Constraints from DR12}\label{BAO}

In this section, we present our final BAO measurements for the combined galaxy samples from BOSS DR12 in the three redshift bins. We translate our measurements on $\alpha$ and $\epsilon$ estimated with our fiducial cosmology into distance measurements following equations \ref{eq:ae1}, \ref{eq:aper}, and \ref{eq:apar}.
We include the systematic error in $\alpha$ and $\epsilon$ as estimated in previous section.  We quote in Table~\ref{tab:finalresults} our final distance constraints from the analysis of the BAO in the correlation function of BOSS combined sample. We quote our results in the angle-averaged distance $D_V(z)$, the Hubble parameter $H(z)$, and the angular diameter $D_A(z)$. We quote the two contributions to the error as ($\sigma_{\rm stat}, \sigma_{\rm sys}$), so that the total error is the sum in quadrature of both quantities ($\sigma_{\rm tot}^2=\sigma_{\rm stat}^2+\sigma_{\rm sys}^2$).

\begin{table*}
\begin{center}
\caption{ Distance constraints from the analysis of the BAO in the correlation functions of combined samples using the fiducial methodology. We quoted our results in the angle average distance $D_V(z)$, the Hubble parameter $H(z)$, the angular diameter $D_A(z)$, and the correlation $\rho_{D_A,H}$. We also quote $\alpha$ and $\epsilon$ with the statistical error\commentH{is the error in alpha and epsilon the total error as in the distances? or is it just the statistical error?}. The sound horizon is evaluated using \textsc{CAMB} (Lewis et al. 2000). A fiducial $r_s^{\rm fid}=147.78$ Mpc is assumed. \commentH{I think the fiducial cosmology uses rs=147.78Mpc} The error bars in the constraints include the contribution from the systematic error budget. We quoted the two contributions to the error as ($x\pm\sigma_{\rm stat}\pm\sigma_{\rm sys}$). The total error is the sum in quadrature of both quantities $\sigma_{tot}^2=\sigma_{\rm stat}^2+\sigma_{\rm sys}^2$. For the systematic error we consider $\Delta \alpha=0.002, \Delta \epsilon=0.003$. 
 The $\sigma_{sys}$ considers $\rho_{\alpha,\epsilon}=0.$}
\label{tab:finalresults}
\begin{tabular}{@{}lccccccc}
\hline
z&$D_V(z) r_s^{\rm fid}/r_s$&$H(z)r_s/r_s^{\rm fid}$&$D_A(z) r_s^{\rm fid}/r_s$&$D_M=(1+z)D_A$&$\rho_{D_A, H}$&$\alpha$&$\epsilon$\\
      &   Mpc & km s$^{-1}$ Mpc$^{-1}$ & Mpc & Mpc & & \\
\hline
\\[-1.5ex]
0.38& 1475 $\pm  14 \pm 3$ & 80.5$\pm 2.2 \pm0.5$&1092 $\pm 16\pm4$&$1507\pm22\pm6$&0.47&$0.9995\pm 0.0098$&$0.0152 \pm 0.0125$\\
0.51&  1872 $\pm 16 \pm 4$& 90.9 $\pm 2.1 \pm 0.6$&1308 $\pm 18\pm5$&$1975\pm27\pm8$&0.50&$0.9928\pm 0.0084$&$-0.0036 \pm 0.0107$\\
0.61& 2131$\pm 20 \pm 4$&99.1$\pm 2.5 \pm 0.6$&1423$\pm 23\pm5$&$2291\pm37\pm8$&0.56&$0.9820\pm 0.0091$ &$-0.0109 \pm 0.0125$\\

\hline

\end{tabular}
\end{center}
\end{table*}

The final constraints  on the angular diameter distance $D_A(z)$ and the Hubble parameter $H(z)$, including both statistical and theoretical systematic uncertainty, are 1.5\% and 2.8\% for the low redshift bin ($z_{\rm eff}=0.38$), 1.4\% and 2.4\% for the intermediate redshift bin ($z_{\rm eff}=0.51$), and 1.7\% and 2.6\% for the high redshift  bin ($z_{\rm eff}=0.61$). The constraints on $D_V(z)$ are 1.0\%, 0.9\%, and 1.0\% for these three redshift bins.

\begin{figure*}
   \centering   
    \includegraphics[width=2.3in]{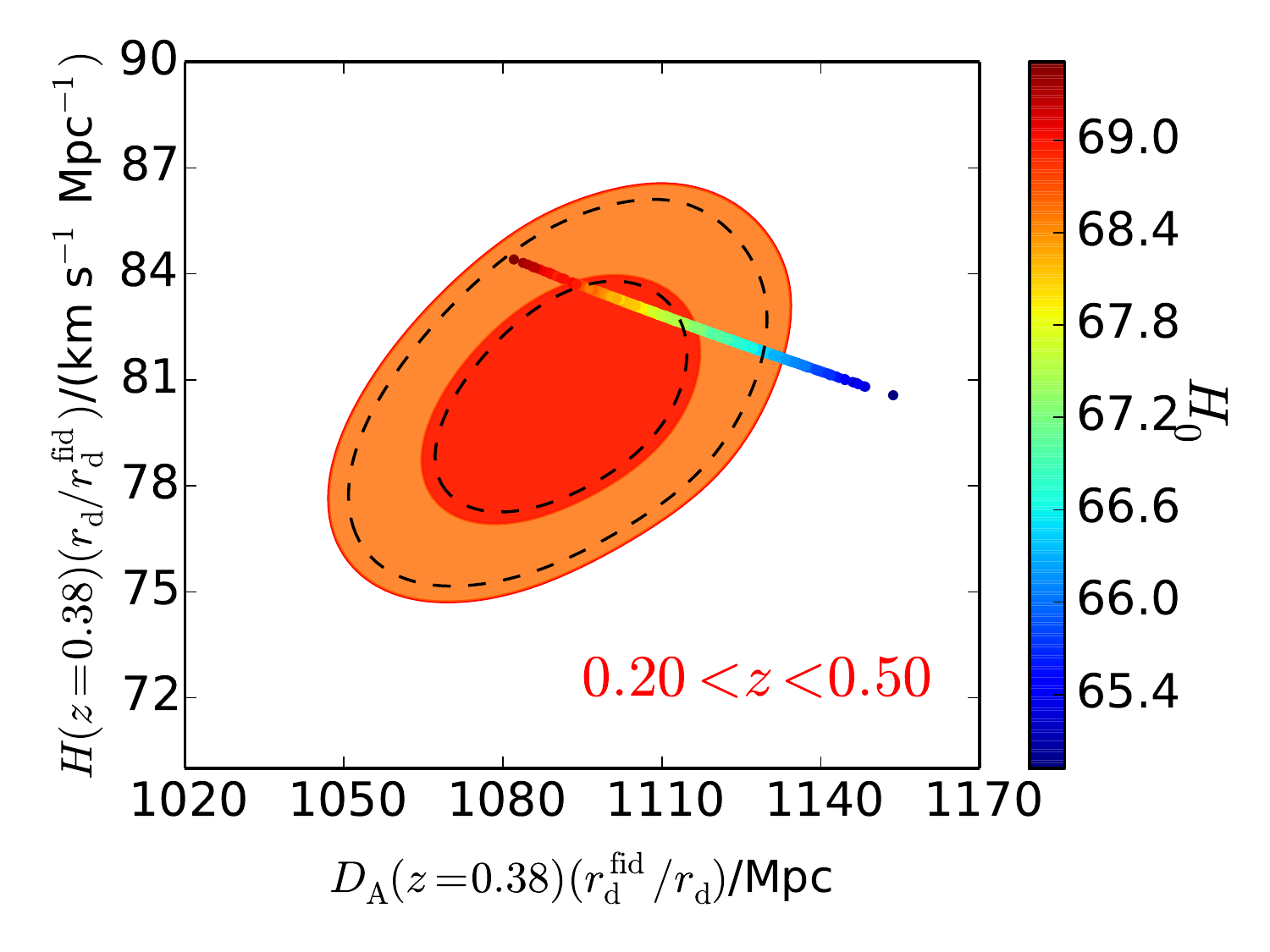}
    \includegraphics[width=2.3in]{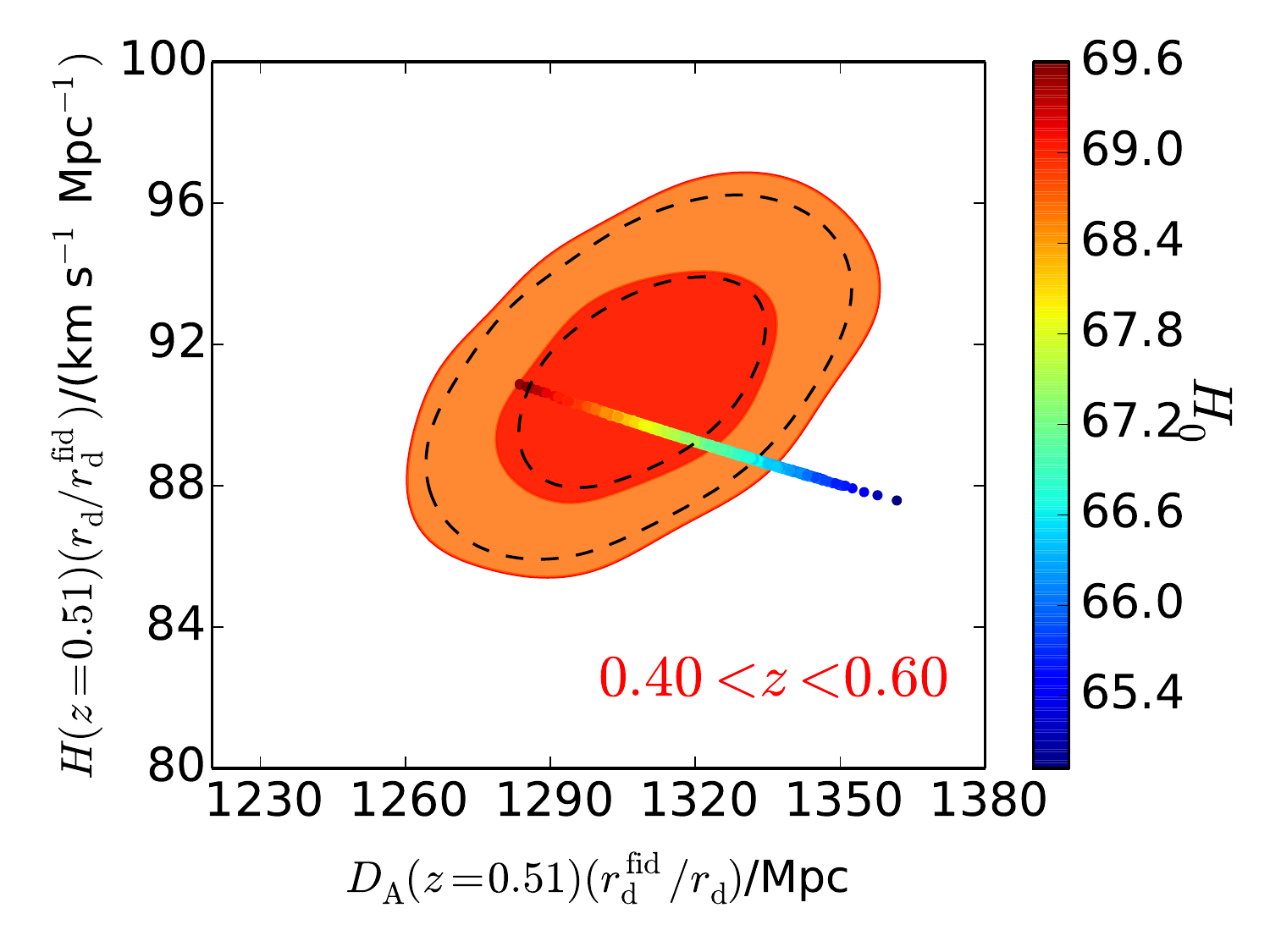}
    \includegraphics[width=2.3in]{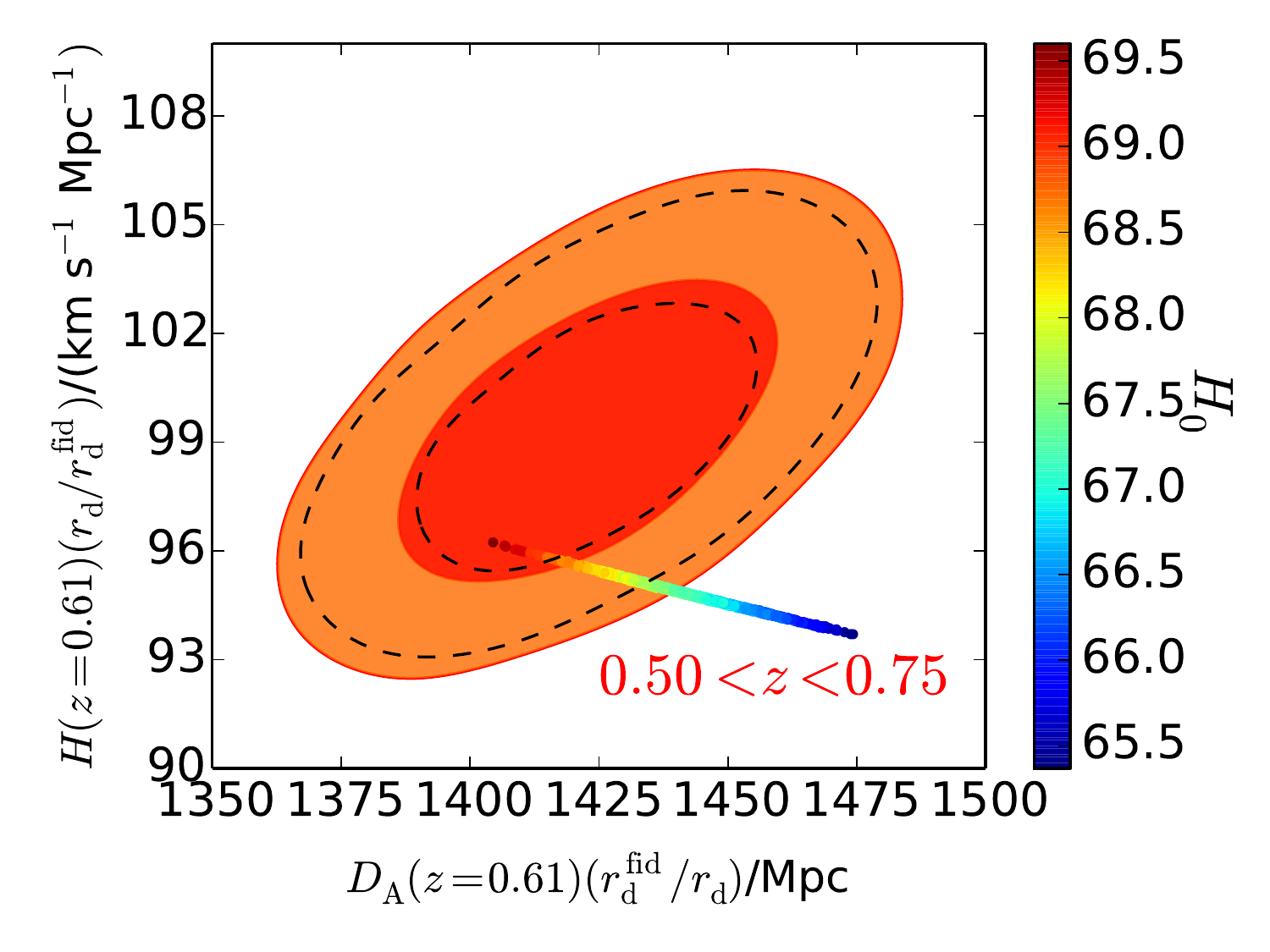}
     \caption{Constraints 1-2$\sigma$ for $D_A(z)$ and $H(z)$ for the three samples at $z_{\rm eff}=0.38,0.51,0.61$ from the fiducial methodology: multipoles analysis with MD-PATCHY covariance. Also included is the constraints from Planck 2015 temperature and polarisation power spectrum data assuming a $\Lambda CDM$ model. \commentM{mention the units of $H_0$} \commentM{Remake figure }}
     \label{fig:dahz}
\end{figure*}

In Figure \ref{fig:dahz}, we present the constraints 1-2$\sigma$ for $D_A(z)$ and $H(z)$ for the three samples at $z_{\rm eff}=0.38, 0.51, 0.61$ from the fiducial methodology: analysis of correlation function multipoles using the covariance matrix from MD-PATCHY mock catalogues. We include the contours without the contribution of the systematic errors in dashed lines just as a reference. 
We also include the constraints from the Planck 2015 \citep{Planck2015Cosmo} temperature and polarisation power spectrum data assuming a $\Lambda CDM$ model. BOSS results are in agreement with Planck results as shown in the figure. The cosmological implications are presented in the BOSS collaboration paper (\citealt{Acacia16}, companion paper), where the different measurements are cross-checked and combined.

\section{Conclusions} \label{conclusions}
In this paper, we have presented a detailed investigation of possible sources of theoretical systematics in anisotropic BAO measurements in configuration space. \change{We defined a fiducial methodology 
and described the results of our fiducial methodology. 
We, then examined the various steps  of the analysis and studied the potential systematics associated with each step presented in the same order as in the analysis. We have applied variations of the methodology to mock catalogues. To determine the systematic error associated with one step of the analysis we compared each variation to the methodology with the fiducial case and we determine the variations best fits distributions.  \footnote{For the variants of the methodology, we presented only the variations of the best fits distributions compared with the fiducial methodology in the main sections of the paper, however, the detailed results of the distributions can be found in the Appendice \ref{sec:tables}.}}

We can summarize our findings as follows:

\begin{itemize}
\item {\bf Estimators.} We analyzed different estimators in configuration space: multipoles, wedges, and $\omega_\ell$ estimators. We studied the systematic uncertainties derived from this step in the clustering analysis post-reconstruction. We found differences $\Delta_{\alpha} = 0.0012$, $\Delta_\epsilon = 0.006$  between different estimators for the three redshift bins. This systematic error could be reduced when combining optimally different estimators, as demonstrated in \cite{Sanchez16b}.
Concerning the uncertainties, we found  differences in the mean uncertainty between $\Delta\sigma_{\alpha} = 0.0013$ and $\Delta\sigma_\epsilon = 0.0029$ for the post-reconstruction mocks in  the three redshift bins. 

\item {\bf Randoms.} We tested the effect of using different sizes of the random catalogue in anisotropic fits. For the pre-reconstruction case, we tested two cases: $20\times$ and $50\times$. The main conclusion is that the randoms make very small differences in the isotropic/anisotropic fits. The differences observed are $< 0.0003$ for CMASS mocks. 
For the post-reconstruction case, we chose a different approach.  We reduced the number of randoms from $50\times$ to $4\times$  for post-reconstruction correlation functions in the numerator SS (not for the DS term) and tested the impact in the correlation function and fitting parameters. We found that differences between the $50\times$ and the $4\times$ anisotropic fits are very small, $\sim0.0002$, for $\alpha$ and  $\epsilon$ for combined mocks for all three bins. 

\item {\bf Covariances.} We explored the effect of using different kinds of mocks to estimate the sample covariance matrix. Also, we tested the recently proposed approach of modelling the covariance using \cite{Oco15}  methodology calibrated with DR12 combined sample catalogues. We documented the differences we found in the structure of the different covariances; also we showed the results of performing the BAO anisotropic fittings to the mocks using the different covariances. We also found differences $\le 0.0009$  for $\alpha$  and $\le 0.0009$ for $\epsilon$ for the three redshift bins. The uncertainties distributions are also very similar, we found differences in the mean values $\Delta\sigma_\alpha = 0.0005$ and $\Delta\sigma_\epsilon = 0.0010$. 

\item {\bf Reconstruction.} We revisited the effect of the smoothing scale on the Combined MD-PATCHY mocks. We tested the three smoothing scales of 5,10 and 15 $h^{-1}$Mpc with 100 mocks. 
We found variations in the best fit parameter in BAO distance measurements associated with the smoothing scale used in  reconstruction of the density field  $0.0017$ for $\alpha$ and $0.0006 $ for $\epsilon$ if we consider smoothing scales of  5 and 10  $h^{-1}$Mpc. If we only consider differences between 15 and 10 $h^{-1}$Mpc, the variations are only 0.0009 on $\alpha$ and 0.0002 on $\epsilon$.

\item {\bf Fiducial Cosmology.} We studied the fiducial cosmology dependence of BAO anisotropic analyses. We analyzed mocks assuming a different fiducial cosmology from the one used to generate the mocks and we compared with results obtained assuming their ``true cosmology." We tested flat cosmologies that are shifted in $\Omega_m$ by 0.5\% compared to true cosmology, but used exactly the same $\Omega_b, h$. We found that the variations in the cosmology generate up to  0.0010 variations in the $\alpha, \epsilon$ values.

\item {\bf Fitting/Modelling.} We  explored a few sub-percent uncertainties coming from the fitting methodology: 

1) We tested the ``Modified Gaussian  Damping Model" template against  the ``Gaussian  Damping Model" template used in a previous analysis.  The observed variation in the mean are: $\le 0.0001$ for $\alpha$ and  $\le 0.0007$ for $\epsilon$. The differences in the error distributions are  0.0003 in $\sigma_\alpha$ mean values, and $\le 0.0020$ in $\sigma_\epsilon$. The Modified Gaussian model gives smaller errors compared to the Gaussian model. 

2) We tested the effects of considering the hexadecapole in the fits. The best fit results indicate that differences of using monopole+quadrupole compared with monopole+quadrupole+hexadecapole are not significant. The differences in $\alpha$ and $\epsilon$ are $<0.0002$  post-reconstruction. The error distributions pre-reconstruction show a reduction of 0.0015 in all three redshift bins for $\sigma_\alpha$ and 0.0032 for $\sigma_\epsilon$ with respect to the monopole+quadrupole fits.  

3) We revisited the fitting range and derived the optimal fitting range. We found an RMS  variations of 0.0002 $\alpha$ and $\epsilon$ for different choices of the smallest/largest scale included in the fitting. We also found that by increasing  upper bounds we can reduce the error about 0.0011 in $\alpha$ and 0.0019 in $\epsilon$.

4) We revisited the dependence with the choice of bin centre. We found variations of $0.0002$  on the mean values of $\alpha$ and 0.0004 on the mean values of $\epsilon$.  The error distributions are unchanged across the different bin centres. In $\sigma_\alpha$ and $\sigma_\epsilon$, the dispersion is also approximately constant.
\end{itemize}

Based on this research we establish a systematic error budget for the new measurements. \change{For the systematic error budget, we consider only the choices that do not generate significant shifts (bias) in the best fits parameters. 
We set this requirement because the fiducial cosmology by definition should minimize the biases , thus any variation that biases significantly our analysis will be naturally discarted.}
 Our error budget do not consider correlations between variations of the methodology, i.e it only considers changes in the methodology performed one-by-one. We find that $\sigma^{sys}_\alpha=0.002$ and $\sigma^{sys}_\epsilon=0.003$. We notice that the non-reconstruction-related sources of theoretical systematic uncertainty in $\alpha$ all combine to make a contribution that is dominated by the uncertainty related to reconstruction techniques. 
 The theoretical systematic uncertainty due to reconstruction is smaller for $\epsilon$ and is sub-leading compared to our upper bound of $\Delta\epsilon \approx 0.0025$, so that upper bound is essentially unchanged by reconstruction-related systematics.

 We have applied all the variations to the final DR12 Combined Sample of BOSS. The different measurements (except for the monopole+quadrupole+hexadecapole fits that lie in some cases outside bounds) in all three redshifts are consistent with the mock catalogues results lying within the $1-\sigma$ contours. 
 
We have derived from the fiducial methodology fits the distance measurements for the three redshift bins of the BOSS Combined Sample. The final constraints  on the angular diameter distance $D_A(z)$ and the Hubble parameter $H(z)$, including statistical and theoretical systematic uncertainties, are 1.5\% and 2.8\% for the low redshift bin ($z_{\rm eff}=0.38$), 1.4\% and 2.4\% for the intermediate redshift bin ($z_{\rm eff}=0.51$), and 1.7\% and 2.6\% for the high redshift  bin ($z_{\rm eff}=0.61$). The constraints on $D_V(z)$ are 1.0\%, 0.9\%, and 1.0\% for the three redshift bins.
 
 Finally our work serves also to evaluate alternative methodologies for future applications, i.e. to see if there would be demonstrable benefits from adopting any of the other approaches. Between the variations we explored in the paper, we summarized the results of those that can be considered alternative choices to the usual clustering anisotropic BAO analysis:
\begin{itemize}
\item $\omega$-estimator. Though few BAO analyses have been performed using this clustering estimator, our results suggests this choice is preferred for BAO analysis as the errors are reduced. 
\item Model Covariance. We tested for the first time a model covariance approach from \cite{Oco15} applied to a BOSS sample. We found that the performance of the model covariance is competitive with the sample covariance approach, validating its use in future surveys.
\item Hexadecapole. We showed that the use of the hexadecapole decreases the mean error and the dispersion of the error distributions. \commentH{ do you think that this could reconcile the different error in $H(z)$ from configuration space and Fourier space?}
\end{itemize}

 In the era of precision cosmology, the next generation of dark energy experiments with galaxy surveys will push forward the precision of the measurements one order of magnitude, thus the comprehension of all sources of systematics errors will be critical for obtaining the sub-percent precision required in the distance measurements. This contribution serves as a starting point for future investigation in the theoretical systematics affecting the BAO-only distance measurements.

\section*{Acknowledgements}
We acknowledge to Sebastien Fromenteau for useful discussions in the revised version. MV is partially supported by Programa de Apoyo a Proyectos de Investigaci\'on e Innovaci\'on Tecnol\'ogica (PAPITT) No  IA102516 and No IA101518, Proyecto Conacyt Fronteras No 281 and from Proyecto LANCAD-UNAM-DGTIC-319. SH is supported by NSF AST1412966, NASA-EUCLID11-0004, and NSF AST1517593 for this work. AJC is supported by the European Research Council under the European Community's Seventh Framework Programme FP7-IDEAS-Phys.LSS 240117. Funding for this work was partially provided by the Spanish MINECO under projects AYA2014-58747-P and MDM-2014-0369 of ICCUB (Unidad de Excelencia ``Mar{\'\i}a de Maeztu"). GR is supported by the National Research Foundation of Korea (NRF) through NRF-SGER 2014055950 funded by the Korean Ministry of Education, Science and Technology (MoEST), and by the faculty research fund of Sejong University in 2016. AGS, JNG, and SSA acknowledge support from the Trans-regional Collaborative Research Centre TR33 The Dark Universe of the German Research Foundation (DFG).

Funding for SDSS-III has been provided by the Alfred P. Sloan
Foundation, the Participating Institutions, the National Science
Foundation, and the U.S. Department of Energy Office of Science. The
SDSS-III web site is http://www.sdss3.org/.

SDSS-III is managed by the Astrophysical Research Consortium for the
Participating Institutions of the SDSS-III Collaboration including the
University of Arizona,
the Brazilian Participation Group,
Brookhaven National Laboratory,
University of Cambridge,
Carnegie Mellon University,
University of Florida,
the French Participation Group,
the German Participation Group,
Harvard University,
the Instituto de Astrofisica de Canarias,
the Michigan State/Notre Dame/JINA Participation Group,
Johns Hopkins University,
Lawrence Berkeley National Laboratory,
Max Planck Institute for Astrophysics,
Max Planck Institute for Extraterrestrial Physics,
New Mexico State University,
New York University,
Ohio State University,
Pennsylvania State University,
University of Portsmouth,
Princeton University,
the Spanish Participation Group,
University of Tokyo,
University of Utah,
Vanderbilt University,
University of Virginia,
University of Washington,
and Yale University.

This research used resources of the National Energy Research Scientific
Computing Center, which is supported by the Office of Science of the
U.S. Department of Energy under Contract No. DE-AC02-05CH11231.

\appendix

\section{Estimators Fitting Results}

 In Table \ref{tab:fitest}, we show the results of performing the BAO anisotropic analysis to the 1000 MD-PATCHY mocks using different clustering estimators, pre- and post-reconstruction. For the analysis in this section, we concentrate on the post-reconstruction results because those are the ones used for the BAO analysis. We focus first on the distributions of $\alpha$ and $\epsilon$.  The bias column of the table indicates the different estimators are unbiased estimators, $b_\alpha,b_\epsilon \le $0.001 for $\alpha$ and $\epsilon$ for multipoles and wedges, except for bin 2 where the bias in $\epsilon$ is slightly larger but still consistent with previous works ($b_\epsilon = 0.001-0.003$). In the case of $\omega$-estimator, it shows a slightly larger bias in some bins compared to multipoles and wedges, its bias varies between 0.0004--0.003 in $\alpha$ and $\epsilon$. Further investigation needs to be done to characterize the bias when using this estimator, we suspect varying the range should affect the results significantly, i.e there is still room to reduce the bias observed. 

We now move on to discuss the dispersion of the distributions, the column of standard deviations indicates the distributions for $\alpha$ are very similar for all estimators($\Delta S_\alpha = 0.0001-0.0002$). However the $\epsilon$ distributions show larger scatter compared to the multipoles case, the differences in dispersion between $\omega$ and multipoles is $\Delta S_\epsilon^{\omega-M} = 0.0016-0.0025$, and between wedges and multipoles is $\Delta S_\epsilon^{\omega-M} = 0.0048-0.0057$. 

\begin{figure}
   \centering     
    \includegraphics[width=1.5in]{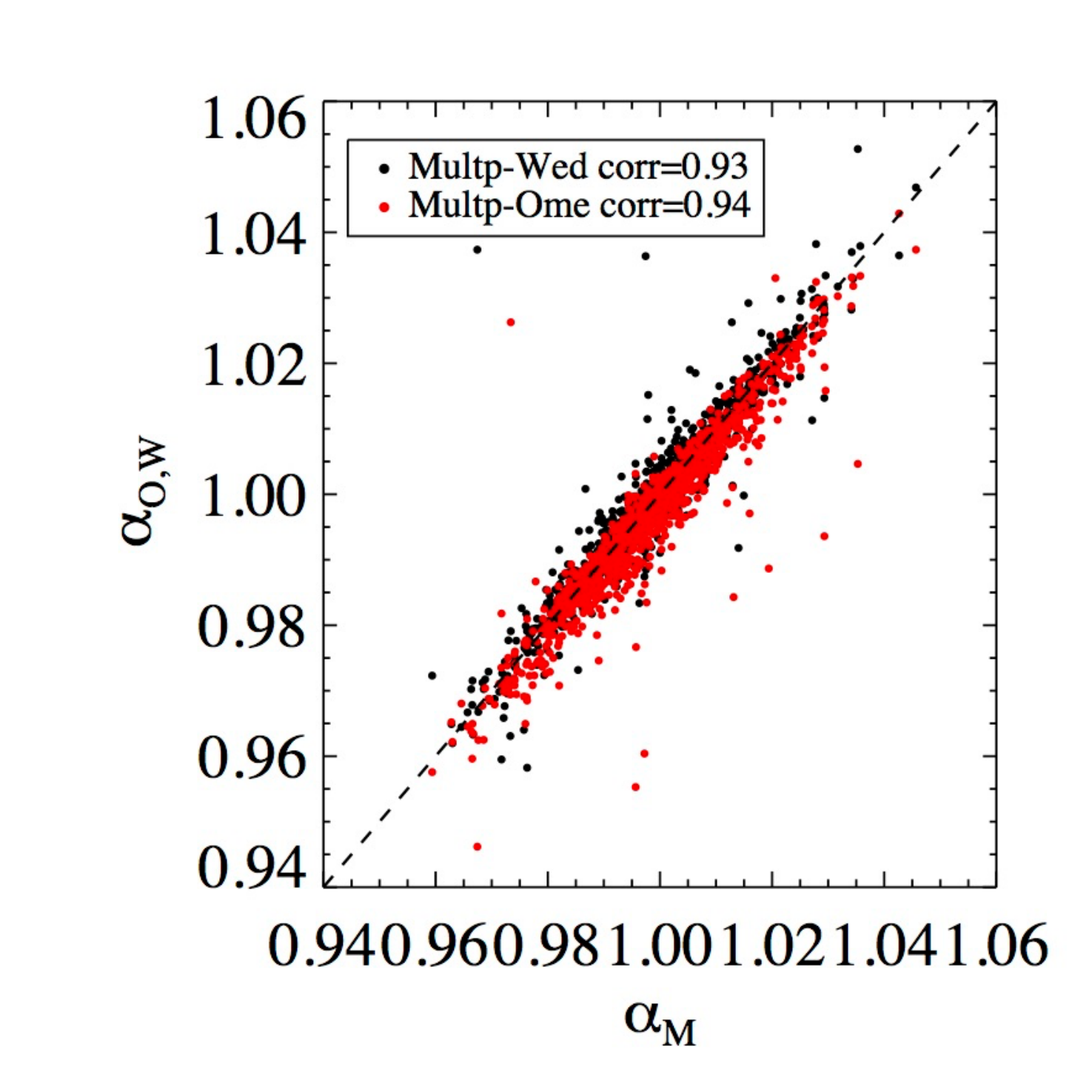} 
        \includegraphics[width=1.5in]{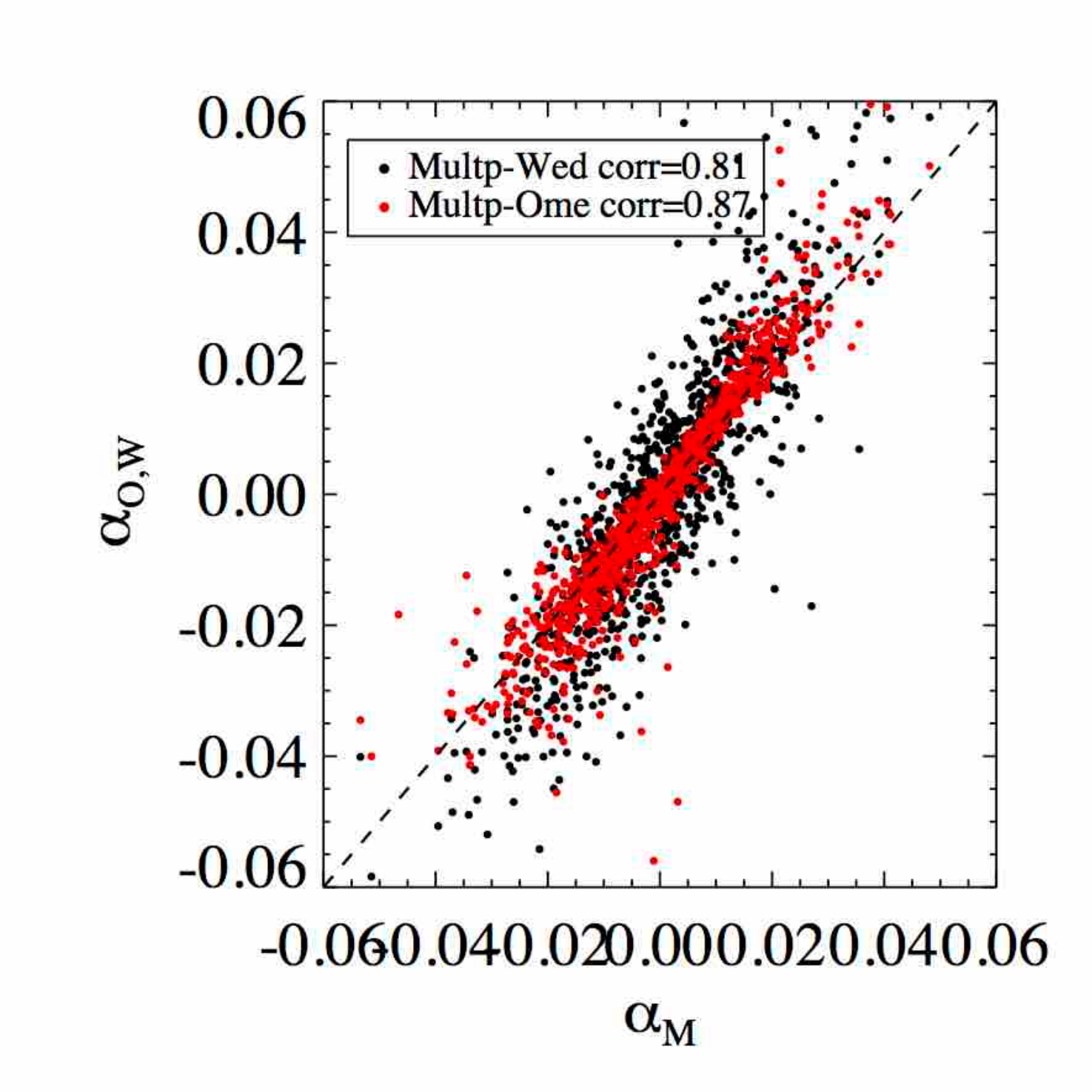} 
   \caption{ Dispersion plots of best fits from different clustering estimators $\xi_{0, 2},\xi_{\parallel, \perp}, \omega_l$  for 1000 MD-PATCHY  post-reconstruction mocks for the lowest redshift bin ($z= 0.2 - 0.5$). Left panel dispersion plots for $\alpha$ and right panel for $\epsilon$.
   Similar plots are obtained for the intermediate and higher redshift bins.}
      \label{fig:dispestimators}
\end{figure}

As a crosscheck, we look at the consistency of the results. Figure~\ref{fig:dispestimators} shows the scatter plots comparing the best fits obtained from using different estimators for the lower redshift bin (intermediate and higher bins show similar numbers and plots), on the left $\alpha$ and on the right $\epsilon$. The correlation between best fit parameters using different estimators is high, the correlation between $\alpha$ coming from Multipoles and those coming from $\omega$ analysis is  $C_\alpha^{M-\omega} = 0.94$, the correlation between $\epsilon$ coming from Multipoles and $\epsilon$ coming from wedges is $C_\alpha^{M-\omega} = 0.93$ and slightly lower for  $\epsilon$ coming from Multipoles and $\epsilon$ coming from $\omega$, $C_\epsilon^{M-\omega} = 0.87$ and finally for  $\epsilon$ coming from Multipoles and $\epsilon$ coming from wedges is $C_\epsilon^{M-\omega} = 0.81$. The results indicate the consistency between the results obtained from the different estimators.

\section{Comparison of Sample Covariance Matrix in DR11: PTHALOS and QPM}\label{sec:covarDR11}

We test the variations of BAO fitting results from the sample covariance in DR11 from two sets of mocks: PTHALOS \citep{Man12} used in previous data releases (DR9 , DR10, and DR11) and the new QPM mocks used in the DR12 analysis. We limit this test to pre-reconstruction mock catalogues.  The PTHALOS mocks are based on Second Order Lagrangian Perturbation Theory while the QPM mocks are generated from N-body simulations with big time steps. Although the QPM mocks have low resolution, they are expected to recover more of the non-linearities to an order higher than the second; thus, we expect them to be more realistic than PTHALOS. These differences could impact the shifts observed in the pre-reconstruction results.

 In Table \ref{tab:dr11}, we show the fitting results obtained for the different sets of mocks; the results from PTHALOS were taken from \cite{Vargas15}. 
When we start comparing the bias obtained from both mock catalogues, we find small differences in the bias values ($\sim0.001$). We find also small differences in the dispersion of the distributions of $\alpha$ and $\epsilon$, $\Delta S_\alpha \sim 0.001$ and $\Delta S_\alpha \sim 0.002$.

 The error distributions, when applying the methodology to both sets of mocks, are also in agreement. But QPM does result in larger error bars ($\Delta \sigma_\alpha=0.0013$ and $\Delta \sigma_\epsilon=0.0017$). The scatter differences are 0.0024 and 0.0042.
 
  The fitting results are in reasonable agreement  between both sets of mocks. The small differences we observe must be related to the differences in the methodology. 
The DR11 PTHALOS analysis was performed using a bin size of 8 $h^{-1}$Mpc compared to 5 $h^{-1}$Mpc used in the new QPM mocks; the fitting range has also changed, which improves the $\chi^2/d.o.f.$. However, it was shown in \cite{Vargas14} that bin size has a $<0.001 $ bias effect from the best fitting parameters and 0.001 in the errors. We used  the same values for the non-linear damping in both sets of mock catalogues: $(\Sigma_{\parallel},\Sigma_{\perp})=(6,12) h^{-1}$ Mpc. It was also shown in  \cite{Vargas14}, that variations in the non-linear damping $\Sigma_{\parallel, \perp}$ affect the results in ~0.001 or less in both best fitting values and the uncertainties. Because the cosmology of the mocks is slightly different, when we compare the covariance values, we are not able to disentangle the differences related to the mocks and the cosmology. 
\begin{table}
\caption{Covariance Matrix Systematics. Fitting results from BOSS mocks pre-reconstruction. The different columns are the mean of the distributions  of the best fits parameters and their respective uncertainties, denoted by $\bar{x}$ with $x=\alpha, \epsilon, \sigma_\alpha, \sigma_\epsilon$ and the bias defined as the difference of the mean value compared to the expected value for the variable, $b_{x}=\bar{x}-x_{\rm exp}$ where $x_{\rm exp}$ is the expected value. There are few differences between the two analyses: the DR11 PTHALOS analysis was performed using a bin size of 8 $h^{-1}$Mpc  compared to 5 $h^{-1}$Mpc used in the new QPM mocks; the fitting range has also changed, which improves the $\chi^2/d.o.f.$. }

\label{tab:dr11}
\begin{tabular}{@{}lcccccc}
\hline
\multicolumn{7}{c}{Covariance Matrix Systematics}\\
\hline
\multicolumn{7}{c}{DR11 CMASS MD-PATCHY-QPM NGC mocks, Pre-Reconstruction}\\
\hline
Cov&$\bar{\alpha}$&
 $b_\alpha$&%
$\bar{\epsilon}$&
$ b_\epsilon$&
$\bar{\sigma_\alpha}$&
$\bar{\sigma_\epsilon}$
\\
\hline
PTHALOS &
$1.0044$&
$0.0044$&
$0.0021$&
$0.0019$&
$0.0153$&
$0.0186$
\\
\\[-1.5ex]
QPM &
$1.0027$&
$0.0027$&
 $0.0011$&
 $0.0011$& 
 0.0191& 
 0.0250
 \\
\\[-1.5ex]
\hline
\end{tabular}
\end{table}
\section{Tables}\label{sec:tables}

\begin{table}
\caption{Two-Point Statistics Estimator Systematics. Fitting results from MD-PATCHY mocks pre- and post-reconstruction using different estimators (Section~\ref{sec:estimators}). The different columns are the mean of the distributions of the best fits parameters and their respective uncertainties denoted by $\bar{x}$ with $x=\alpha, \epsilon, \sigma_\alpha, \sigma_\epsilon$, the bias defined as the difference of the mean value compared to the expected value for the variable, $b_{x}=\bar{x}-x_{\rm exp}$, where $x_{\rm exp}$ is the expected value. \commentH{The d.o.f. is better stated next to the chi2 value e.g. 29.8/30. This is useful for homogeneity with other tables in this paper where each row will have a different number of d.o.f} 
}
\label{tab:fitest}
\begin{tabular}{@{}lcccccc}
\hline

\multicolumn{7}{c}{Two-Point Statistics Estimator Systematics}\\
\hline
\multicolumn{7}{c}{DR12 Combined MD-PATCHY mocks, Pre-Reconstruction}\\
\hline
Est&
$\bar{\alpha}$&
$b_\alpha$&
$\bar{\epsilon}$&
$b_\epsilon$&
$\bar{\sigma}_\alpha$&
$\bar{\sigma}_\epsilon$
\\
\hline
\multicolumn{7}{c}{Bin 1 ($0.2 < z < 0.5$)}\\
\hline
$\xi_\ell$&
$1.0015$&
$0.0022$&
$-0.0003$&
$-0.0005$&
$0.0262$&
$0.0334$
\\\\[-1.5ex]
$\xi_{\parallel, \perp}$&
$1.0084$&
$0.0091$&
$0.0081$&
$0.0079$&
$0.0276$&
$0.0412$
\\\\[-1.5ex]
$\omega_\ell$&
$0.9973$&
$-0.0020$&
$-0.0042$&
$-0.0044$&
$0.0206$&
$0.0273$
\\\\[-1.5ex]

\hline
\multicolumn{7}{c}{Bin 2 ($0.4 < z < 0.6$)}\\
\hline
$\xi_\ell$&
$1.0038$&
$0.0042$&
$0.0014$&
$0.0012$&
$0.0222$&
$0.0291$
\\\\[-1.5ex]
$\xi_{\parallel, \perp}$&
$1.0080$&
$0.0084$&
$0.0067$&
$0.0065$&
$0.0237$&
$0.0375$
\\\\[-1.5ex]
$\omega_\ell$&
$0.9993$&
$-0.0003$&
$-0.0027$&
$-0.0030$&
$0.0219$&
$0.0287$
\\\\[-1.5ex]

\hline
\multicolumn{7}{c}{Bin 3 ($0.5< z < 0.75$)}\\
\hline
$\xi_\ell$&
$1.0039$&
$0.0040$&
$0.0003$&
$0.0001$&
$0.0215$&
$0.0282$
\\\\[-1.5ex]
$\xi_{\parallel, \perp}$&
$1.0071$&
$0.0072$&
$0.0079$&
$0.0077$&
$0.0221$&
$0.0363$
\\\\[-1.5ex]
$\omega_\ell$&
$0.9996$&
$-0.0003$&
$-0.0027$&
$-0.0031$&
$0.0176$&
$0.0244$
\\\hline
\multicolumn{7}{c}{DR12 Combined MD-PATCHY mocks, Post-Reconstruction}\\
\hline
Est&
$\bar{\alpha}$&
$b_\alpha$&
$\bar{\epsilon}$& 
$b_\epsilon$&
$\bar{\sigma_\alpha}$&
$\bar{\sigma_\epsilon}$
\\
\hline
\multicolumn{7}{c}{Bin 1 ($0.2 < z < 0.5$)}\\
\hline
$\xi_\ell$&
$0.9986$&
$-0.0007$&
$0.0009$&
$0.0007$&
$0.0147$&
$0.0188$
\\\\[-1.5ex]
$\xi_{\parallel, \perp}$ &
$0.9988$&
$-0.0005$&
$0.0006$&
$0.0004$&
$0.0145$&
$0.0217$
\\\\[-1.5ex]
$\omega_\ell$&
$0.9967$&
$-0.0026$&
$0.0012$&
$0.0010$&
$0.0122$&
$0.0168$
\\
\hline
\multicolumn{7}{c}{Bin 2 ($0.4 < z < 0.6$)}\\
\hline
$\xi_\ell$&
$1.0006$&
$0.0010$&
$0.0023$&
$0.0021$&
$0.0130$&
$0.0163$
\\\\[-1.5ex]
$\xi_{\parallel, \perp}$&
$1.0009$&
$0.0013$&
$0.0033$&
$0.0031$&
$0.0129$&
$0.0197$
\\\\[-1.5ex]
$\omega_\ell$&
$0.9992$&
$-0.0004$&
$0.0030$&
$0.0027$&
$0.0130$&
$0.0196$
\\
\hline
\multicolumn{7}{c}{Bin 3 ($0.5 < z < 0.75$)}\\
\hline
$\xi_\ell$&
$1.0007$&
$0.0008$&
$0.0011$&
$0.0009$&
$0.0133$&
$0.0166$
\\\\[-1.5ex]
$\xi_{\parallel, \perp}$&
$1.0009$&
$0.0010$&
$0.0017$&
$0.0015$&
$0.0130$&
$0.0203$
\\\\[-1.5ex]
$\omega_\ell$&
$0.9989$&
$-0.0010$&
$0.0011$&
$0.0007$&
$0.0112$&
$0.0152$
\\
\hline
\end{tabular}
\end{table}
\begin{table}
\caption{Covariance Matrix Systematics. Fitting results from BOSS mocks post-reconstruction using different mocks and covariances. The first column indicates the mocks used for the fits (first letter) and covariance (second letter). For example, P-Q indicates the fits of MD-PATCHY mocks using the sample covariance from QPM. The different columns are the mean of the distributions of the best fits parameters and the mean of their respective uncertainties denoted by $\bar{x}$ with $x=\alpha, \epsilon, \sigma_\alpha, \sigma_\epsilon$,  the bias defined as the difference of the mean value compared to the expected value for the variable, $b_{x}=\bar{x}-x_{\rm exp}$, where $x_{\rm exp}$ is the expected value. }
\label{tab:fitresdr12reca}
\begin{tabular}{@{}lccccccc}
\hline
\multicolumn{7}{c}{Covariance Matrix Systematics: Sample Covariance from Different simulations.}\\
\hline
\multicolumn{7}{c}{DR12 Combined Sample  MD-PATCHY mocks, Pre-Reconstruction}\\
\hline
S-C&$\bar{\alpha}$&
 $b_\alpha$&
$\bar{\epsilon}$&
$b_\epsilon$& 
$\bar{\sigma_\alpha}$&
$\bar{\sigma_\epsilon}$
\\
\\[-1.5ex]
\hline
\multicolumn{7}{c}{BIN 1 ($0.2 < z < 0.5$)}\\
\hline

P-P&
$1.0015$&
$0.0022$&
$-0.0003$&
$-0.0005$&
$0.0262$&
$0.0334$
\\
\\[-1.5ex]

Q-Q&
$0.9801$&
$0.0034$&
$0.0028$&
$0.0026$&
$0.0224$&
$0.0320$
\\
\hline
\multicolumn{7}{c}{BIN 2 ($0.4 < z < 0.6$)}\\
\hline

P-P&
$1.0038$&
$0.0042$&
$0.0014$&
$0.0012$&
$0.0222$&
$0.0291$
\\
\\[-1.5ex]

Q-Q&
$0.9838$&
$0.0046$&
$0.0035$&
$0.0033$&
$0.0198$&
$0.0280$
\\
\hline
\multicolumn{7}{c}{BIN 3 ($0.5 < z < 0.75$)}\\
\hline

P-P&
$1.0039$&
$0.0040$&
$0.0003$&
$0.0001$&
$0.0215$&
$0.0282$
\\
\\[-1.5ex]

Q-Q&
$0.9842$&
$0.0032$&
$0.0038$&
$0.0036$&
$0.0198$&
$0.0280$
\\
\hline
\multicolumn{7}{c}{DR12 Combined Sample  MD-PATCHY/QPM mocks, Post-Reconstruction}\\
\hline
\multicolumn{7}{c}{BIN 1 ($0.2 < z < 0.5$)}\\
\hline
P-P&
$0.9986$&
$-0.0007$&
$0.0009$&
$0.0007$&
$0.0147$&
$0.0188$
\\
\\[-1.5ex]
P-P*&
$0.9989$&
$-0.0004$&
$0.0011$&
$0.0009$&
$0.0188$&
$0.0253$
\\
\\[-1.5ex]
P-Q&
$0.9989$&
$-0.0004$&
$0.0007$&
$0.0005$&
$0.0138$&
$0.0174$
\\\\[-1.5ex]

Q-Q&
$0.9797$&
$0.0030$&
$0.0032$&
$0.0030$&
$0.0136$&
$0.0190$
\\\\[-1.5ex]

Q-P&
$0.9802$&
$0.0035$&
$0.0028$&
$0.0026$&
$0.0144$&
$0.0209$
\\
\\[-1.6ex]
\hline
\multicolumn{7}{c}{BIN 2 ($0.4 < z < 0.6$)}\\
\hline
P-P&

$1.0006$&
$0.0010$&
$0.0023$&
$0.0021$&
$0.0130$&
$0.0163$
\\
\\[-1.5ex]
P-P*&
$1.0003$&
$0.0007$&
$0.0025$&
$0.0023$&
$0.0153$&
$0.0207$

\\\\[-1.5ex]
P-Q&

$0.9989$&
$-0.0007$&
$0.0010$&
$0.0008$&
$0.0128$&
$0.0156$
\\
\\[-1.5ex]

Q-Q&
$0.9805$&
$0.0013$&
$0.0043$&
$0.0041$&
$0.0125$&
$0.0172$
\\
\\[-1.5ex]
Q-P&
$0.9818$&
$0.0026$&
$0.0049$&
$0.0047$&
$0.0124$&
$0.0177$
\\
\hline
\multicolumn{7}{c}{BIN 3 ($0.5 < z < 0.75$)}\\
\hline
P-P&

$1.0007$&
$0.0008$&
$0.0011$&
$0.0009$&
$0.0133$&
$0.0166$
\\
\\[-1.5ex]

P-P*&
$1.0004$&
$0.0005$&
$0.0001$&
$-0.0001$&
$0.0156$&
$0.0204$
\\
\\[-1.5ex]
P-Q&
$1.0005$&
$0.0006$&
$-0.0000$&
$-0.0002$&
$0.0137$&
$0.0163$
\\
\\[-1.5ex]
Q-Q&
$0.9822$&
$0.0012$&
$0.0046$&
$0.0044$&
$0.0130$&
$0.0179$
\\
\\[-1.5ex]

Q-P&
$0.9817$&
$0.0007$&
$0.0059$&
$0.0057$&
$0.0128$&
$0.0181$
\\

\hline
\end{tabular}
\end{table}

\begin{table}
\caption{Covariance Matrix Systematics. Fitting results from BOSS mocks post-reconstruction using model covariance (denoted by M) and the sample covariance from MD-PATCHY (denoted by S). The different columns are the mean of the distributions of the best fits parameters and the mean of their respective uncertainties denoted by $\bar{x}$ with $x=\alpha, \epsilon, \sigma_\alpha, \sigma_\epsilon$,  the bias defined as the difference of the mean value compared to the expected value for the variable, $b_{x}=\bar{x}-x_{\rm exp}$, where $x_{\rm exp}$ is the expected value. }
\label{tab:fitresdr12rec}
\begin{tabular}{@{}lcccccc}
\hline
\multicolumn{7}{c}{Covariance Matrix Systematics: Model Covariance.}\\
\hline
\multicolumn{7}{c}{DR12 Combined Sample  MD-PATCHY mocks, Pre-Reconstruction}\\
\hline

Cov&$\bar{\alpha}$&
 $b_\alpha$&
$\bar{\epsilon}$&
$b_\epsilon$&
$\bar{\sigma_\alpha}$&
 $\bar{\sigma_\epsilon}$
 \\\\[-1.5ex]
\hline
\multicolumn{7}{c}{Bin 1 ($0.20 < z < 0.50$)}\\
\hline
M&
$1.0011$&
$0.0018$&
$-0.0007$&
$-0.0009$&
$0.0260$&
$0.0343$
\\\\[-1.5ex]
S&
$1.0015$&
$0.0022$&
$-0.0003$&
$-0.0005$&
$0.0262$&
$0.0334$
\\
\hline
\multicolumn{7}{c}{Bin 2 ($0.40 < z < 0.60$)}\\
\hline
M&
$1.0035$&
$0.0039$&
$0.0009$&
$0.0007$&
$0.0225$&
$0.0293$
\\\\[-1.5ex]
S&
$1.0038$&
$0.0042$&
$0.0014$&
$0.0012$&
$0.0222$&
$0.0291$
\\
\hline
\multicolumn{7}{c}{Bin 3 ($0.50 < z < 0.75$)}\\
\hline
M&
$1.0038$&
$0.0039$&
$0.0001$&
$-0.0001$&
$0.0219$&
$0.0287$
\\\\[-1.5ex]
S&
$1.0039$&
$0.0040$&
$0.0003$&
$0.0001$&
$0.0215$&
$0.0282$
\\
\hline
\multicolumn{7}{c}{DR12 Model Covariance, Post-Reconstruction}\\
\hline
\multicolumn{7}{c}{Bin 1 ($0.20 < z < 0.50$)}\\
\hline
M&
$0.9988$&
$-0.0005$&
$0.0009$&
$0.0007$&
$0.0144$&
$0.0190$
\\\\[-1.5ex]
S&
$0.9986$&
$-0.0007$&
$0.0009$&
$0.0007$&
$0.0147$&
$0.0188$
\\
\hline
\multicolumn{7}{c}{Bin 2 ($0.40 < z < 0.60$)}\\
\hline
M&
$1.0001$&
$0.0005$&
$0.0016$&
$0.0014$&
$0.0130$&
$0.0171$
\\\\[-1.5ex]
S&
$1.0006$&
$0.0010$&
$0.0023$&
$0.0021$&
$0.0130$&
$0.0163$
\\
\hline
\multicolumn{7}{c}{Bin 3 ($0.50 < z < 0.75$)}\\
\hline
M&
$1.0004$&
$0.0005$&
$0.0003$&
$0.0001$&
$0.0133$&
$0.0166$
\\\\[-1.5ex]
S&
$1.0007$&
$0.0008$&
$0.0011$&
$0.0009$&
$0.0133$&
$0.0166$
\\
\hline
\end{tabular}
\end{table}

\begin{table}
\begin{center}
\caption{Random catalogue test. Results of the BAO anisotropic fitting of QPM mocks post-reconstruction when the size of the random catalogue is varied. The different columns are the mean of the distributions of the best fits parameters denoted by $\bar{x}$ with $x=\alpha, \epsilon$, the bias defined as the difference of the mean value compared to the expected value for the variable, $b_{x}=\bar{x}-x_{\rm exp}$, where  $x_{\rm exp}$. Since the fiducial cosmology is not the natural cosmology of QPM mocks, the expected values for $\alpha$ and $\epsilon$ are: $\alpha_{\rm exp}=$ (0.9767,0.9792,0.9810), $1+\epsilon_{\rm exp}$=(1.0017,1.0023,1.0027). We use the covariance from 1000 QPM mocks\commentM{Check results changing the covariance matrix for 1000}. The case label as 4x means 4x for SS pair-counts and 50x for SR pair-counts.  ``Bin 1" refers to the lower redshift bin ($z= 0.2 - 0.5$);  ``Bin 2" considers the intermediate redshift range ($z= 0.4 - 0.6$), and ``Bin 3" refers to higher redshift range ($z= 0.5 - 0.75$)}
\label{tab:randtest}
\begin{tabular}{@{}lcccc} 
\hline
\multicolumn{5}{c}{Correlation Function Systematics: Random Systematics}\\
\hline
\multicolumn{5}{c}{DR12 Combined Sample  QPM mocks, Post-Reconstruction}\\\\[-1.5ex]
\hline
Imp&
$\bar{\alpha}$ & 
 $b_\alpha$&
$\bar{\epsilon}$& 
$b_\epsilon$\\

\hline

\\[-1.5ex]
Bin 1 50x&
$0.9782$&
$0.0015$&
$0.0033$&
$0.0016$\\
\\[-1.5ex]

Bin 2 50x&
$0.9804$&
$0.0012$&
$0.0038$&
$0.0015$\\
\\[-1.5ex]

Bin 3 50x&
$0.9826$&
$0.0016$&
$0.0051$&
$0.0024$\\
\\[-1.5ex]
\hline

Bin 1 4x &
$0.9780$&
$0.0013$&
$0.0031$&
$0.0014$\\
\\[-1.5ex]

Bin 2 4x &
$0.9803$&
$0.0011$&
$0.0038$&
$0.0015$\\
\\[-1.5ex]

Bin 3 4x &
$0.9827$&
$0.0017$&
$0.0051$&
$0.0024$\\
\\[-1.5ex]
\hline

\end{tabular}
\end{center}
\end{table}


\begin{table}
\begin{center}
\caption{Fiducial cosmology related systematics. Fitting results from QPM NGC mocks pre-/post- reconstruction using a different cosmology in the analysis from the natural cosmology of the mocks. Anderson and Cosmology 3 are flat cosmologies that are shifted in $\Omega_m$ by 0.5\% compared to QPM cosmology, but exactly the same $\Omega_b, h$. 
The different columns are the mean of the distributions of the best fits parameters denoted by $\bar{x}$, the bias defined as the difference of the mean value compared to the expected value for the variable, $b_{x}=\bar{x}-x_{\rm exp}$, where $x_{\rm exp}$ is the expected value. 
The expected shifts are:  $\alpha_{\rm exp}^{\rm QPM}=1.0, \epsilon_{\rm exp}^{\rm QPM}=0.0$,$\alpha_{\rm exp}^{\rm And}=1.0064, \epsilon_{\rm exp}^{\rm And}=-0.0021$. 
Tests post-reconstruction were performed with $N_{sim}=$96 mocks, $\sqrt{N_{sim}}\sim9.8$}
\label{tab:fidcosmotest}
\begin{tabular}{@{}lcccccccc}
\hline
\multicolumn{5}{c}{Fiducial Cosmology related Systematics}\\
\hline

\multicolumn{5}{c}{DR11 CMASS  QPM mocks, Pre-Reconstruction}\\
\hline
Cosmo&$\bar{\alpha}$&
$b_\alpha$ &
$\bar{\epsilon}$&
$b_\epsilon$\\
\hline
AND &
1.0083&
0.0019&
-0.0021&
$<$0.0001\\
\\[-1.5ex]
QPM &
1.0024 &
0.0024&
-0.0001&
0.0001\\
\\[-1.5ex]
\hline
\multicolumn{5}{c}{DR11 CMASS  QPM mocks, Post-Reconstruction}\\
\hline
AND & 
1.0061&
-0.0003&
-0.0014&
0.0005\\

\\[-1.5ex]
QPM&
0.9994&
0.0006&
0.0015&
0.0015\\
\\[-1.5ex]
\hline
\end{tabular}
\end{center}
\end{table}

\begin{table}
\begin{center}
\caption{Reconstruction related systematics. Fitting results from MD-PATCHY NGC mocks post-reconstruction using different smoothing scale in the reconstruction of the density field. 
The different columns are the mean of the distributions of the best fits parameters denoted by $\bar{x}$ with $x=\alpha, \epsilon$, the bias defined as the difference of the mean value compared to the expected value for the variable, $b_{x}=\bar{x}-x_{\rm exp}$ where $x_{\rm exp}$. The expected shifts are:  $\alpha_{exp}^{\rm PATCHY}=0.9993, \epsilon_{exp}^{\rm PATCHY}=0.0002$. This test was performed only for the intermediate redshift bin.}
\label{tab:rectest}
\begin{tabular}{@{}lcccc}
\hline
\multicolumn{5}{c}{Reconstruction related Systematics}\\
\hline
\multicolumn{5}{c}{DR12 MD-PATCHY Combined mocks, Post-Reconstruction }\\
\hline
\multicolumn{5}{c}{ Bin 2 ($z= 0.4 - 0.6$)}\\
\hline
Smoothing&$\bar{\alpha}$&
$b_\alpha$ &$\bar{\epsilon}$&
$b_\epsilon$\\
\\[-1.5ex]
\hline
5&0.9995& 
0.0002& -0.0011&
-0.0013\\
10&1.0026&
 0.0033&-0.0004&
 -0.0006\\
15&1.0016&
0.0024&-0.0002&
-0.0004\\
\hline
\end{tabular}
\end{center}
\end{table}

\begin{table}
\caption{Fitting Related Systematics. Fitting results from MD-PATCHY DR12 Combined Sample mocks post-reconstruction using different Non-linear damping models. The different columns are the mean of the distributions of the best fits parameters and the respective uncertainties denoted by $\bar{x}$ with $x=\alpha, \epsilon, \sigma_\alpha, \sigma_\epsilon$,  the bias defined as the difference of the mean value compared to the expected value for the variable, $b_{x}=\bar{x}-x_{\rm exp}$, where $x_{\rm exp}$ is the expected value. }
\label{tab:fitsys}

\begin{tabular}{@{}lccccccccccc}
\hline
\multicolumn{7}{c}{Fitting Methodology Systematics}\\
\hline
\multicolumn{7}{c}{DR12 Combined Sample MD-PATCHY mocks, Post-Reconstruction}\\
\hline
\\[-1.5ex]

$C_{G,MG}$&
$\bar{\alpha}$&
$b_\alpha$&
$\bar{\epsilon}$&
$b_\epsilon$&
$\bar{\sigma_\alpha}$&
$\bar{\sigma_\epsilon}$
\\
\hline
\multicolumn{7}{c}{Bin 3 ($0.2 < z < 0.5$)}\\
\hline
$C_{G}$&

$0.9986$&
$-0.0007$&
$0.0009$&
$0.0007$&
$0.0147$&
$0.0188$
\\
\\[-1.5ex]
$C_{MG}$&
$0.9986$&
$-0.0007$&
$0.0013$&
$0.0011$&
$0.0150$&
$0.0207$
\\
\hline
\multicolumn{7}{c}{Bin 3 ($0.4 < z < 0.6$)}\\
\hline
$C_{G}$&
$1.0006$&
$0.0010$&
$0.0023$&
$0.0021$&
$0.0130$&
$0.0163$
\\
\\[-1.5ex]
$C_{MG}$&
$1.0006$&
$0.0010$&
$0.0025$&
$0.0023$&
$0.0137$&
$0.0185$
\\
\hline
\multicolumn{7}{c}{Bin 3 ($0.5 < z < 0.75$)}\\
\hline
$C_{G}$&
$1.0007$&
$0.0008$&
$0.0011$&
$0.0009$&
$0.0133$&
$0.0166$
\\
\\[-1.5ex]
$C_{MG}$&
$1.0006$&
$0.0007$&
$0.0010$&
$0.0008$&
$0.0137$&
$0.0185$
\\
\hline
\end{tabular}
\end{table}

\begin{table}
\caption{Fitting Related Systematics. Fitting results from MD-PATCHY DR12 Combined Sample mocks pre-/post-reconstruction using different $\ell$ in multipoles estimator, $\ell=2$ denotes monopole+quadrupole fits and $\ell=4$ indicates monopole+quadrupole+hexadecapole fits. The different columns are the mean of the distributions of the best fits parameters and the respective uncertainties denoted by $\bar{x}$ with $x=\alpha, \epsilon, \sigma_\alpha, \sigma_\epsilon$, the bias defined as the difference of the mean value compared to the expected value for the variable, $b_{x}=\bar{x}-x_{\rm exp}$, where $x_{\rm exp}$ is the expected value. }
\label{tab:fitsyshexa}
\begin{tabular}{@{}lccccccccccc}
\hline
\multicolumn{7}{c}{Fitting Methodology Systematics: hexadecapole contribution}\\
\hline
\multicolumn{7}{c}{DR12 Combined Sample MD-PATCHY mocks, Pre-Reconstruction}\\
\hline
Multp&
$\bar{\alpha}$&
$b_\alpha$&
$\bar{\epsilon}$&
$b_\epsilon$&
$\bar{\sigma_\alpha}$&
 $\bar{\sigma_\epsilon}$
 \\
\hline
\multicolumn{7}{c}{Bin 1 ($0.2 < z < 0.5$)}\\
\hline
$\xi_{l=2}$ &
$1.0015$&
$0.0022$&
$-0.0003$&
$-0.0005$&
$0.0262$&
$0.0334$
\\\\[-1.5ex]
$\xi_{l=4}$ &
$1.0017$&
$0.0024$&
$0.0001$&
$-0.0001$&
$0.0204$&
$0.0242$
\\
\hline
\multicolumn{7}{c}{Bin 2 ($0.4 < z < 0.6$)}\\
\hline
$\xi_{l=2}$ &
$1.0038$&
$0.0042$&
$0.0014$&
$0.0012$&
$0.0222$&
$0.0291$
\\\\[-1.5ex]
$\xi_{l=4}$ &
$1.0036$&
$0.0040$&
$0.0011$&
$0.0009$&
$0.0196$&
$0.0227$
\\
\hline
\multicolumn{7}{c}{Bin 3 ($0.5 < z < 0.75$)}\\
\hline
 $\xi_{l=2}$ &
$1.0039$&
$0.0040$&
$0.0003$&
$0.0001$&
$0.0215$&
$0.0282$
\\
 $\xi_{l=4}$ &
 $1.0039$&
 $0.0040$&
$0.0003$&
$0.0001$&
$0.0112$&
$0.0137$
\\
\hline
\multicolumn{7}{c}{DR12 Combined Sample MD-PATCHY mocks, Post-Reconstruction}\\
\hline
\multicolumn{7}{c}{Bin 1 ($0.2 < z < 0.5$)}\\
\hline
$\xi_{l=2}$ &
$0.9986$&
$-0.0007$&
$0.0009$&
$0.0007$&
$0.0147$&
$0.0188$
\\\\[-1.5ex]
 $\xi_{l=4}$ &
 $0.9984$&
 $-0.0009$&
$0.0006$&
$0.0004$&
$0.0154$
\\
\hline
\multicolumn{7}{c}{Bin 2 ($0.4 < z < 0.6$)}\\
\hline
$\xi_{l=2}$ &
$1.0006$&
$0.0010$&
$0.0023$&
$0.0021$&
$0.0130$&
$0.0163$
\\\\[-1.5ex]
$\xi_{l=4}$&
$1.0003$&
$0.0007$&
$0.0020$&
$0.0018$&
$0.0116$&
$0.0133$
\\

\hline
\multicolumn{7}{c}{Bin 3 ($0.5 < z < 0.75$)}\\
\hline
$\xi_{l=2}$ &
$1.0007$&
$0.0008$&
$0.0011$&
$0.0009$&
$0.0133$&
$0.0166$
\\
$\xi_{l=4}$ &
$1.0007$&
$0.0008$&
$0.0011$&
$0.0009$&
$0.0119$&
$0.0136$
\\
\hline
\end{tabular}
\end{table}

\begin{table}
\begin{center}
\caption{ Fitting Related Systematics: Fitting results from MD-PATCHY DR12 Combined Sample mocks post-reconstruction varying the range of the fit. The first block shows the variation of $r_{min}$; the second block the variation of $r_{max}$. This test was performed for the intermediate redshift bin. The different columns are the mean of the distributions of the best fits parameters and the respective uncertainties denoted by $\bar{x}$ with $x=\alpha, \epsilon, \sigma_\alpha, \sigma_\epsilon$, the bias defined as the difference of the mean value compared to the expected value for the variable, $b_{x}=\bar{x}-x_{\rm exp}$, where $x_{\rm exp}$ is the expected value.}
\label{tab:fitsyspost1}
\begin{tabular}{@{}lcccccc}
\hline
\multicolumn{7}{c}{Fitting Methodology Systematics: Range }\\
\hline
\multicolumn{7}{c}{DR12 Combined Sample MD-PATCHY mocks, Post-Reconstruction}\\
\hline
\multicolumn{7}{c}{ Bin 2 ($z= 0.4 - 0.6$)}\\
\hline
$R_{min}$&
$\bar{\alpha}$&
$b_\alpha$&
$\bar{\epsilon}$&
$b_\epsilon$&
$\bar{\sigma_\alpha}$&
 $\bar{\sigma_\epsilon}$
 \\
\hline
30&
$1.0037$&
$0.0041$&
$0.0036$&
$0.0034$&
$0.0112$&
$0.0137$
\\
\\[-1.5ex]
40&
$1.0007$&
$0.0011$&
$0.0025$&
$0.0023$&
$0.0127$&
$0.0158$
\\
\\[-1.5ex]
50&
$1.0007$&
$0.0011$&
$0.0025$&
$0.0023$&
$0.0128$&
$0.0157$
\\
\\[-1.5ex]
60&
$1.0006$&
$0.0010$&
$0.0023$&
$0.0021$&
$0.0130$&
$0.0163$
\\
\\[-1.5ex]

70&
$0.9989$&
$-0.0007$&
$0.0012$&
$0.0010$&
$0.0126$&
$0.0143$
\\
\\[-1.5ex]

80&
$1.0201$&
$0.0208$&
$0.0252$&
$0.0248$&
$0.0191$&
$0.0337$
\\
\\[-1.5ex]

\hline
\multicolumn{7}{c}{DR12 Combined Sample MD-PATCHY mocks, Post-Reconstruction}\\
\hline
\multicolumn{7}{c}{ Bin 2 ($z= 0.4 - 0.6$)}\\
\hline
120&
$1.0013$&
$0.0017$&
$0.0012$&
$0.0010$&
$0.0163$&
$0.0216$
\\
\\[-1.5ex]
130&
$0.9995$&
$-0.0001$&
$0.0013$&
$0.0011$&
$0.0153$&
$0.0202$
\\
\\[-1.5ex]
140&
$1.0001$&
$0.0005$&
$0.0019$&
$0.0017$&
$0.0142$&
$0.0185$
\\
\\[-1.5ex]
155&
$1.0006$&
$0.0010$&
$0.0023$&
$0.0021$&
$0.0130$&
$0.0163$
\\
\\[-1.5ex]
160&
$1.0005$&
$0.0009$&
$0.0021$&
$0.0019$&
$0.0129$&
$0.0161$
\\

\\[-1.5ex]
170&
$1.0007$&
$0.0011$&
$0.0022$&
$0.0020$&
$0.0119$&
$0.0144$
\\
\\[-1.5ex]
180&
$1.0007$&
$0.0011$&
$0.0023$&
$0.0021$&
$0.0118$&
$0.0142$
\\
\\[-1.5ex]
\hline
\\[-1.5ex]
\end{tabular}
\end{center}
\end{table}

\begin{table}
\begin{center}
\caption{Fitting Related Systematics. Fitting results from MD-PATCHY DR12 Combined Sample mocks post-reconstruction varying the bin centre. This test was performed only for the intermediate redshift bin ($z= 0.4 - 0.6$). The different columns are the mean of the distributions of the best fits parameters and the respective uncertainties denoted by $\bar{x}$ with $x=\alpha, \epsilon$, the bias defined as the difference of the mean value compared to the expected value for the variable, $b_{x}=\bar{x}-x_{\rm exp}$, where $x_{\rm exp}$ is the expected value.}
\label{tab:fitsyspost2}
\begin{tabular}{@{}ccccc}
\hline
\multicolumn{5}{c}{Fitting Methodology Systematics:  Bin Centre}\\
\hline
\multicolumn{5}{c}{DR12 Combined Sample  MD-PATCHY mocks, Post-Reconstruction}\\
\hline
\multicolumn{5}{c}{ Bin 2 ($z= 0.4 - 0.6$)}\\
\hline
Centre&
$\bar{\alpha}$&
$b_\alpha$&
$\bar{\epsilon}$&
$b_\epsilon$\\
\\[-1.5ex]
\hline
0 $h^{-1}$Mpc&
$1.0006$&
$0.0010$&
$0.0023$&
$0.0021$
\\\\[-1.5ex]
1 $h^{-1}$Mpc&
$1.0004$&
$0.0008$&
$0.0023$&
$0.0021$
\\\\[-1.5ex]
2 $h^{-1}$Mpc&
$1.0006$&
$0.0010$&
$0.0019$&
$0.0017$
\\\\[-1.5ex]
3 $h^{-1}$Mpc&
$1.0003$&
$0.0007$&
$0.0017$&
$0.0015$

\\\\[-1.5ex]
4 $h^{-1}$Mpc&
$1.0005$&
$0.0009$&
$0.0020$&
$0.0018$
\\\\[-1.5ex]
\hline
\\[-1.5ex]

\end{tabular}
\end{center}
\end{table}

\section{Correlation Matrices}

\begin{figure*}
   \centering     
    \includegraphics[width=2.0in]{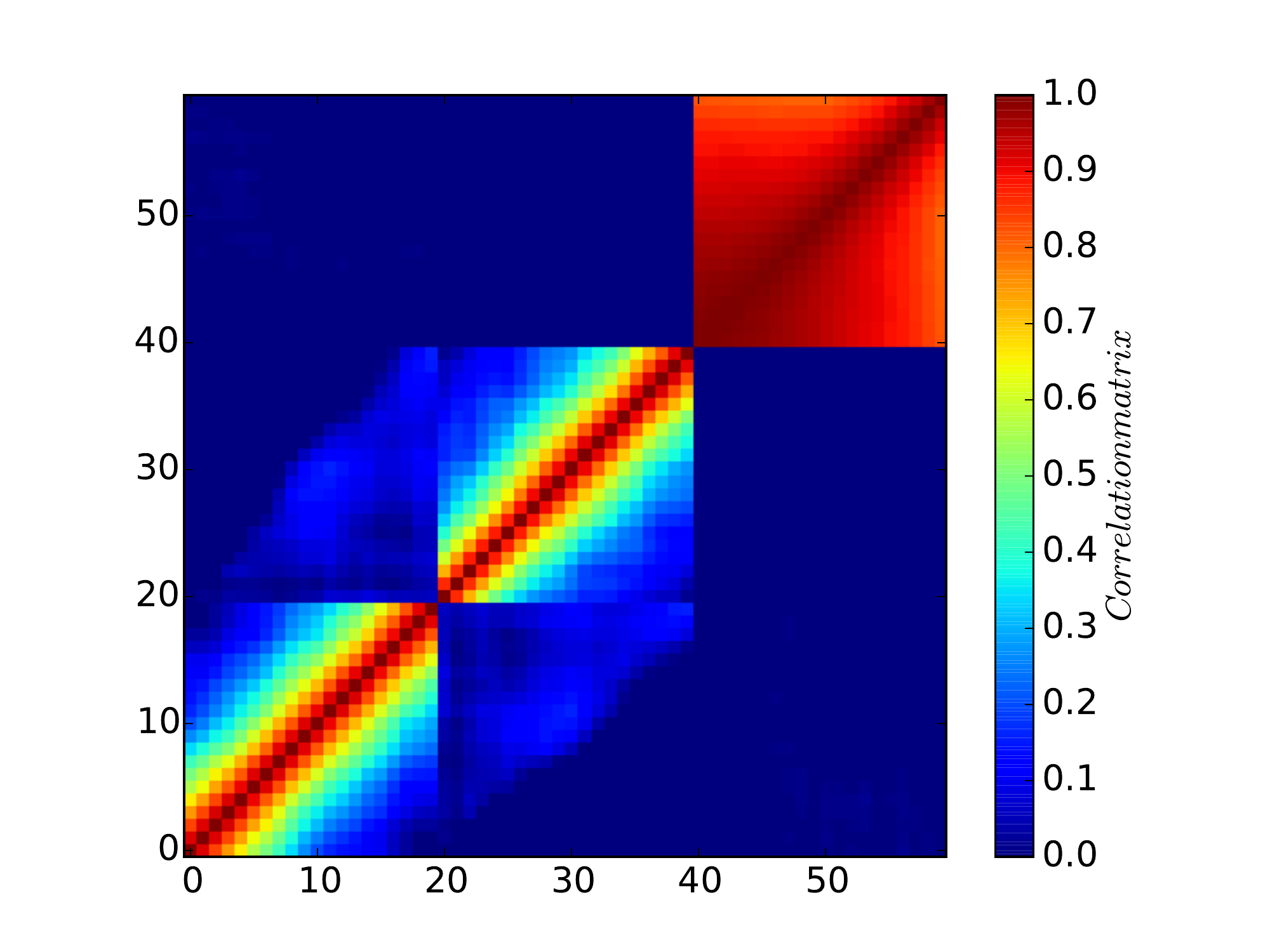}
    \includegraphics[width=2.0in]{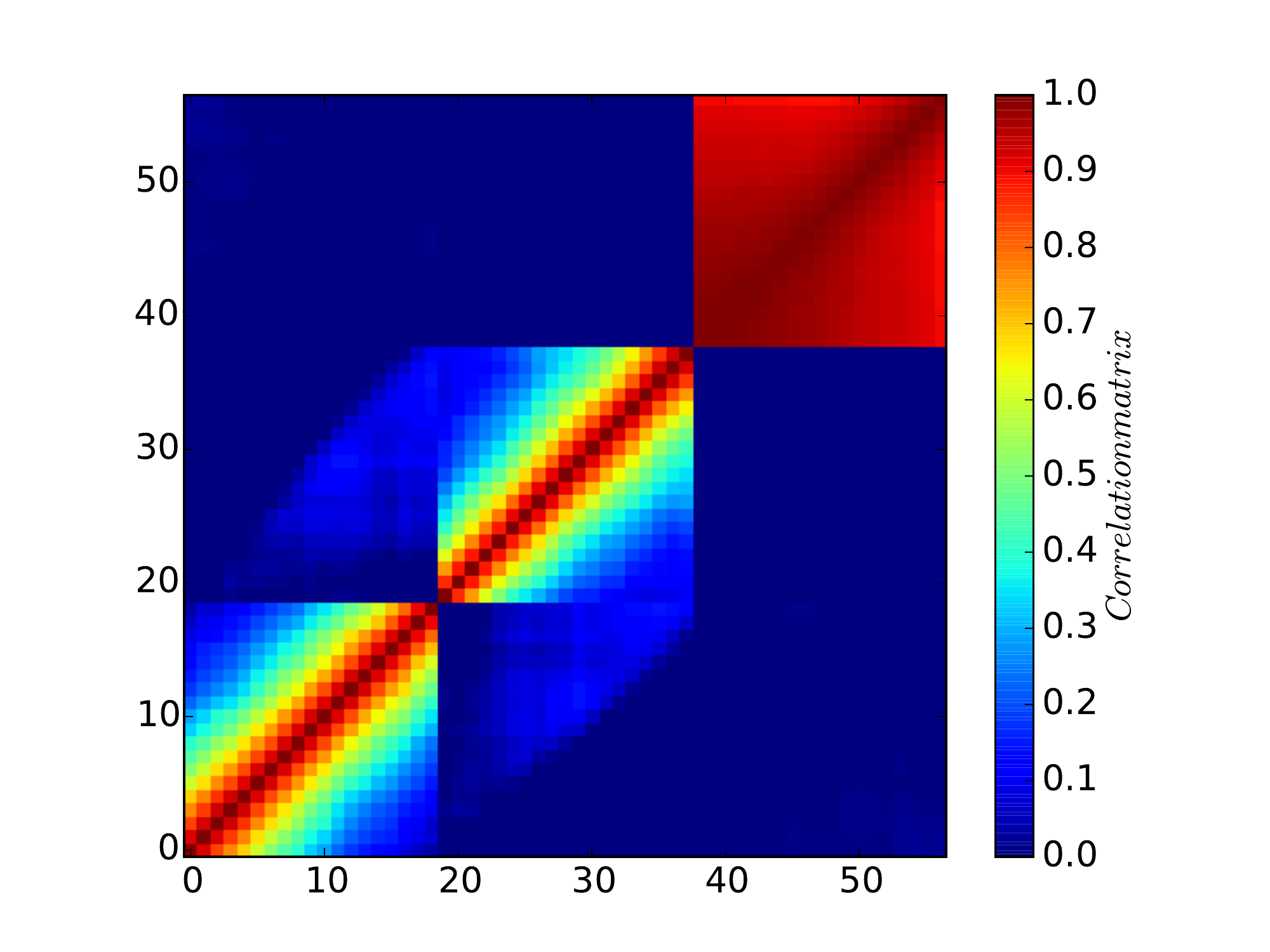}
    \includegraphics[width=2.0in]{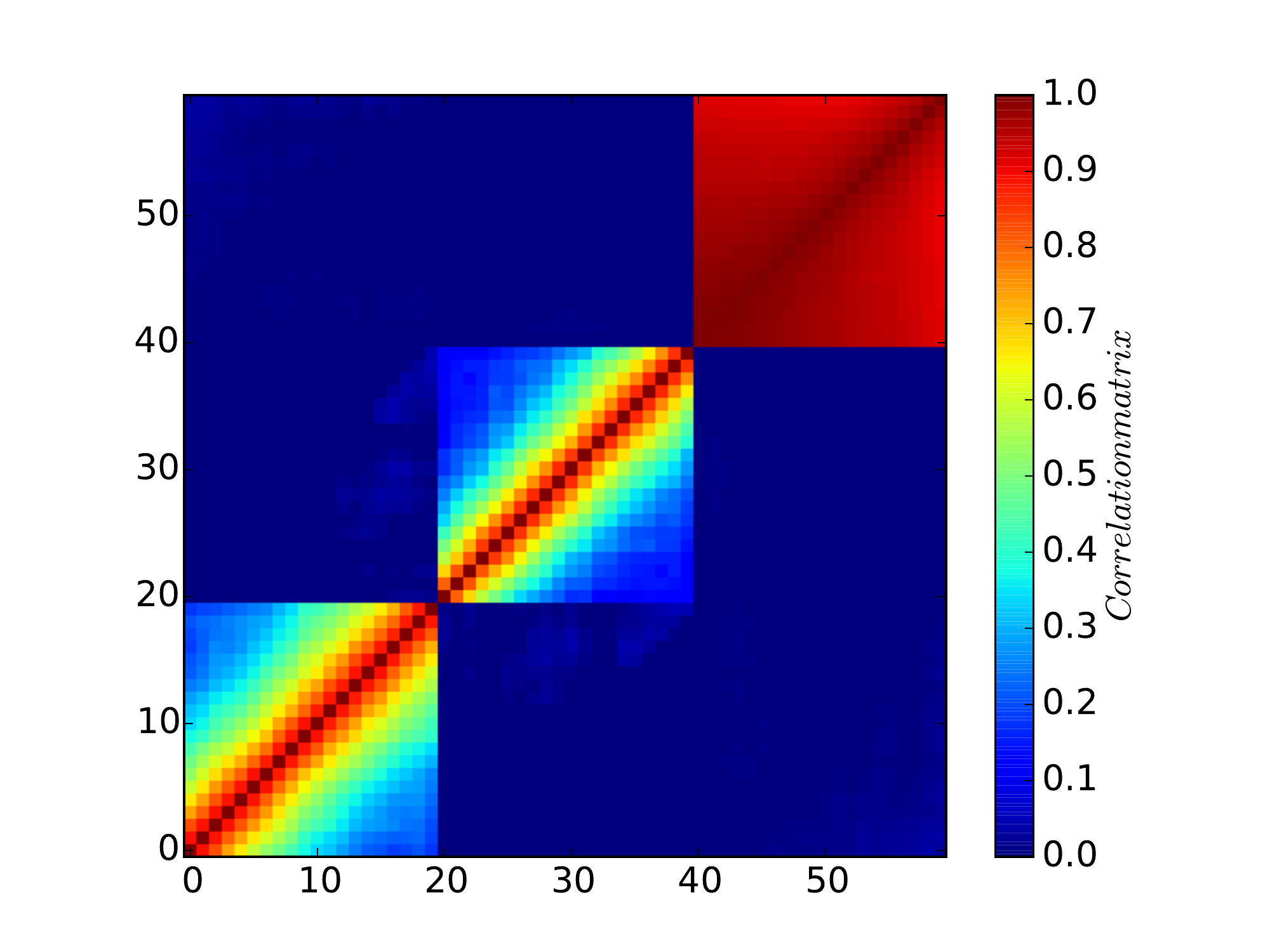}
    \includegraphics[width=2.0in]{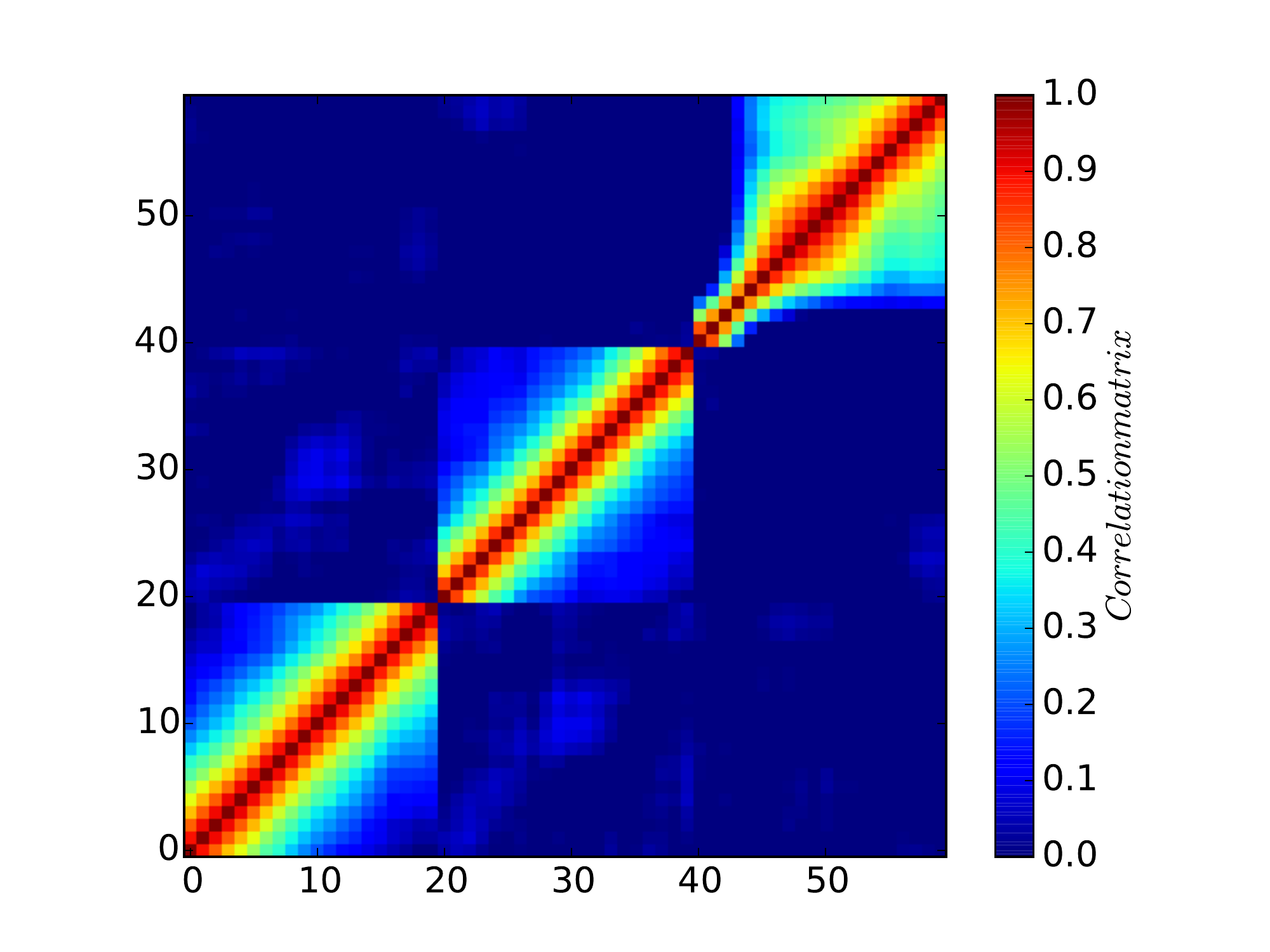}
    \includegraphics[width=2.0in]{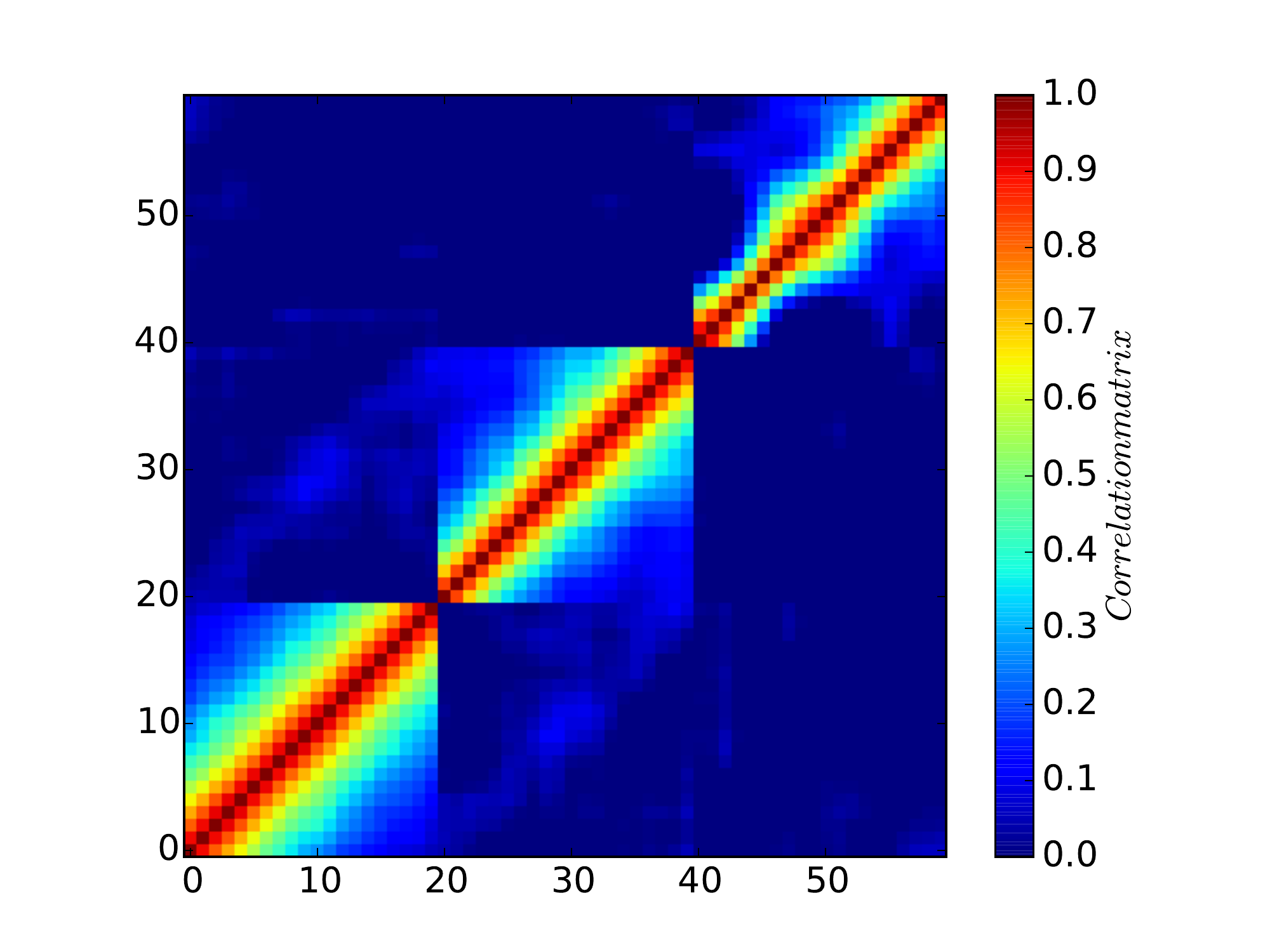}
    \includegraphics[width=2.0in]{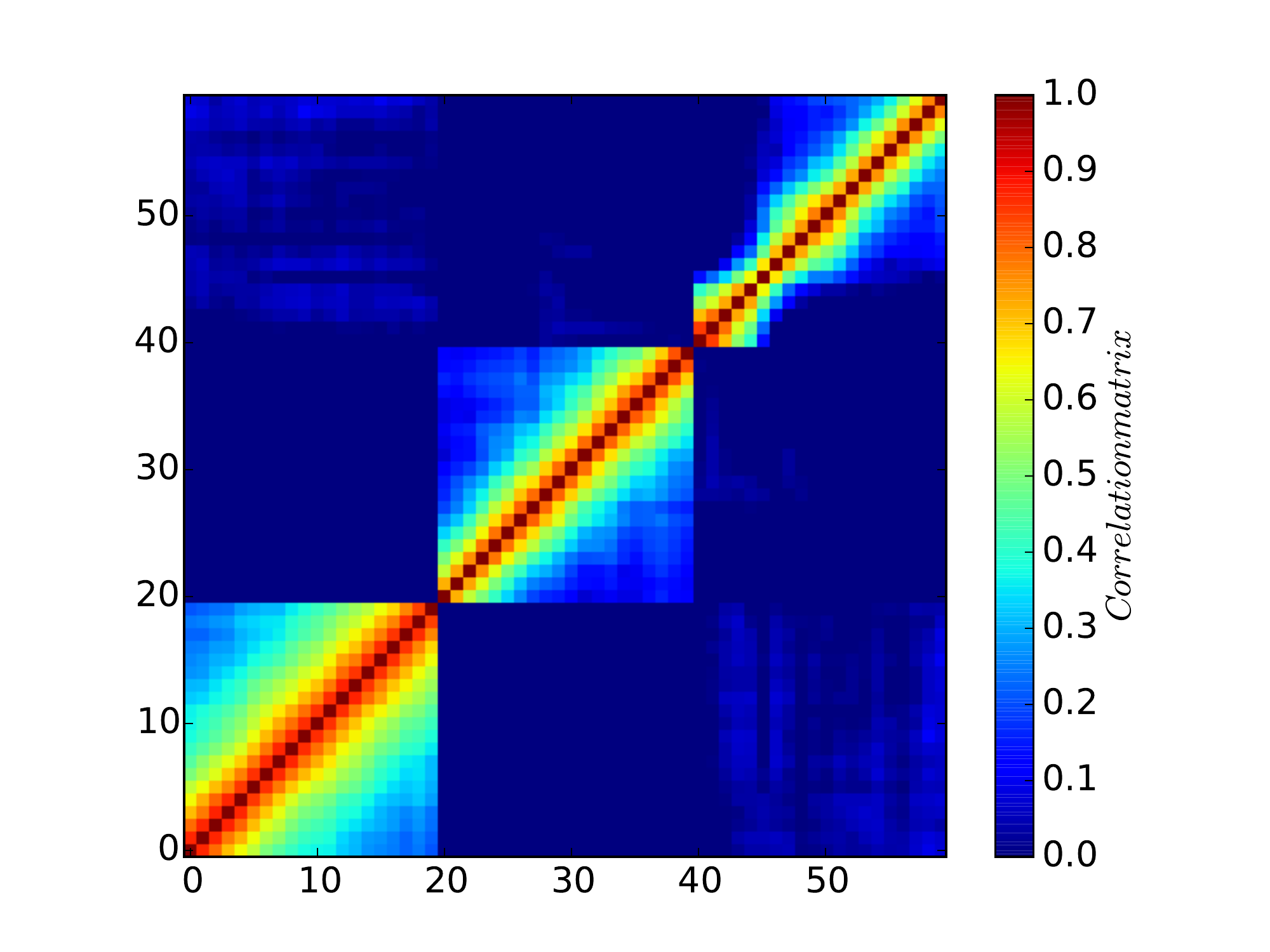}
    \includegraphics[width=2.0in]{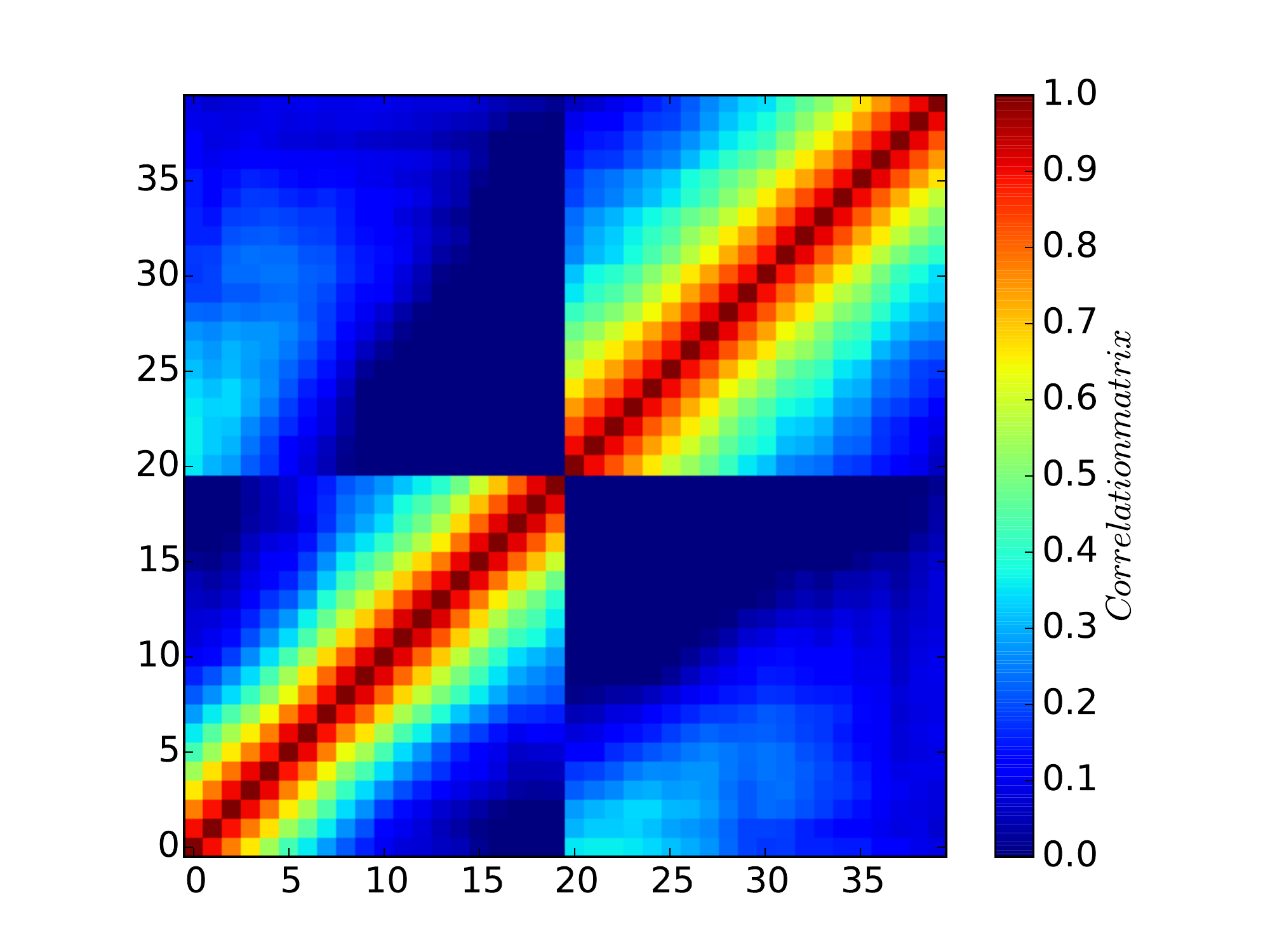}
    \includegraphics[width=2.0in]{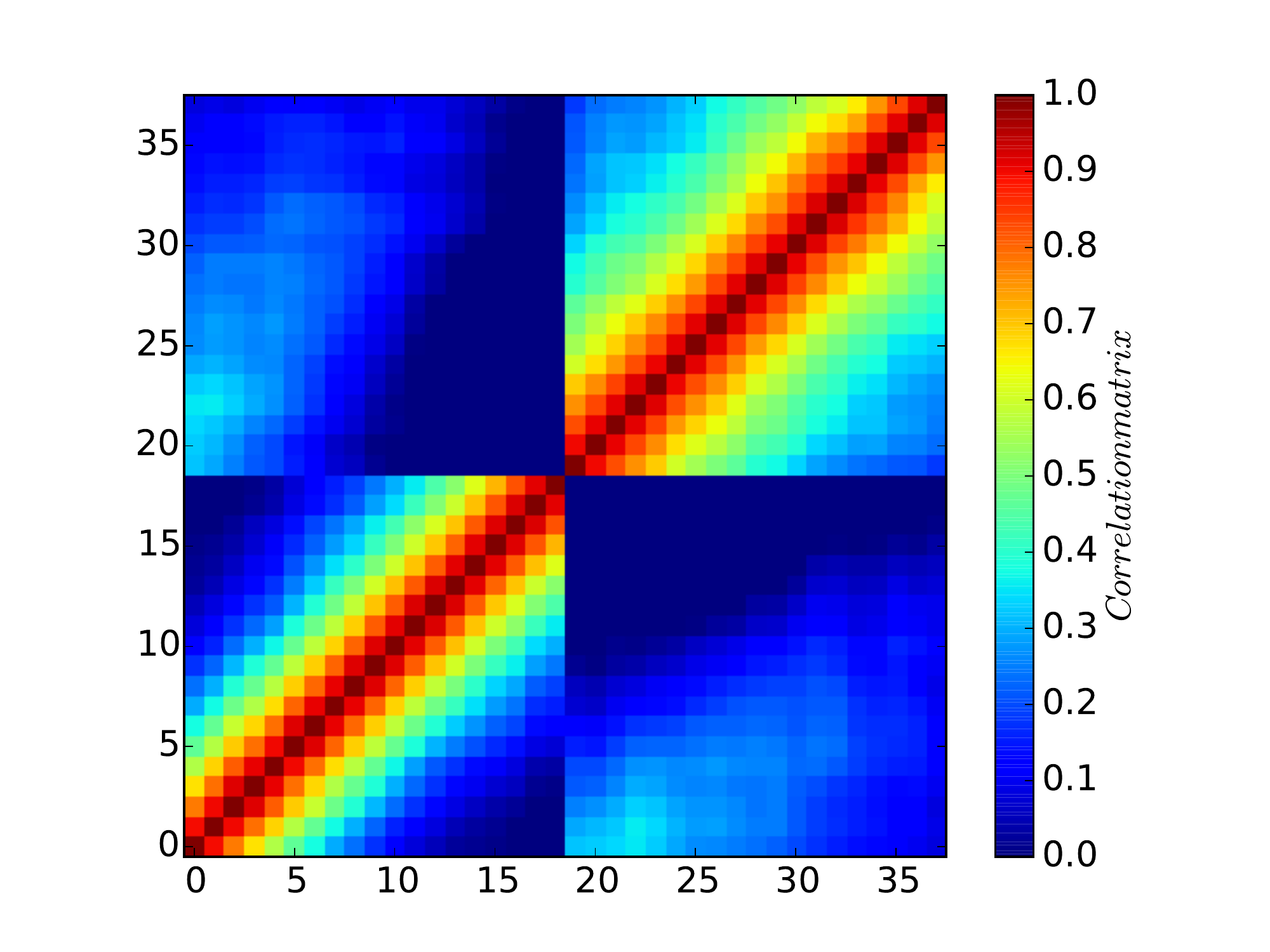}
    \includegraphics[width=2.0in]{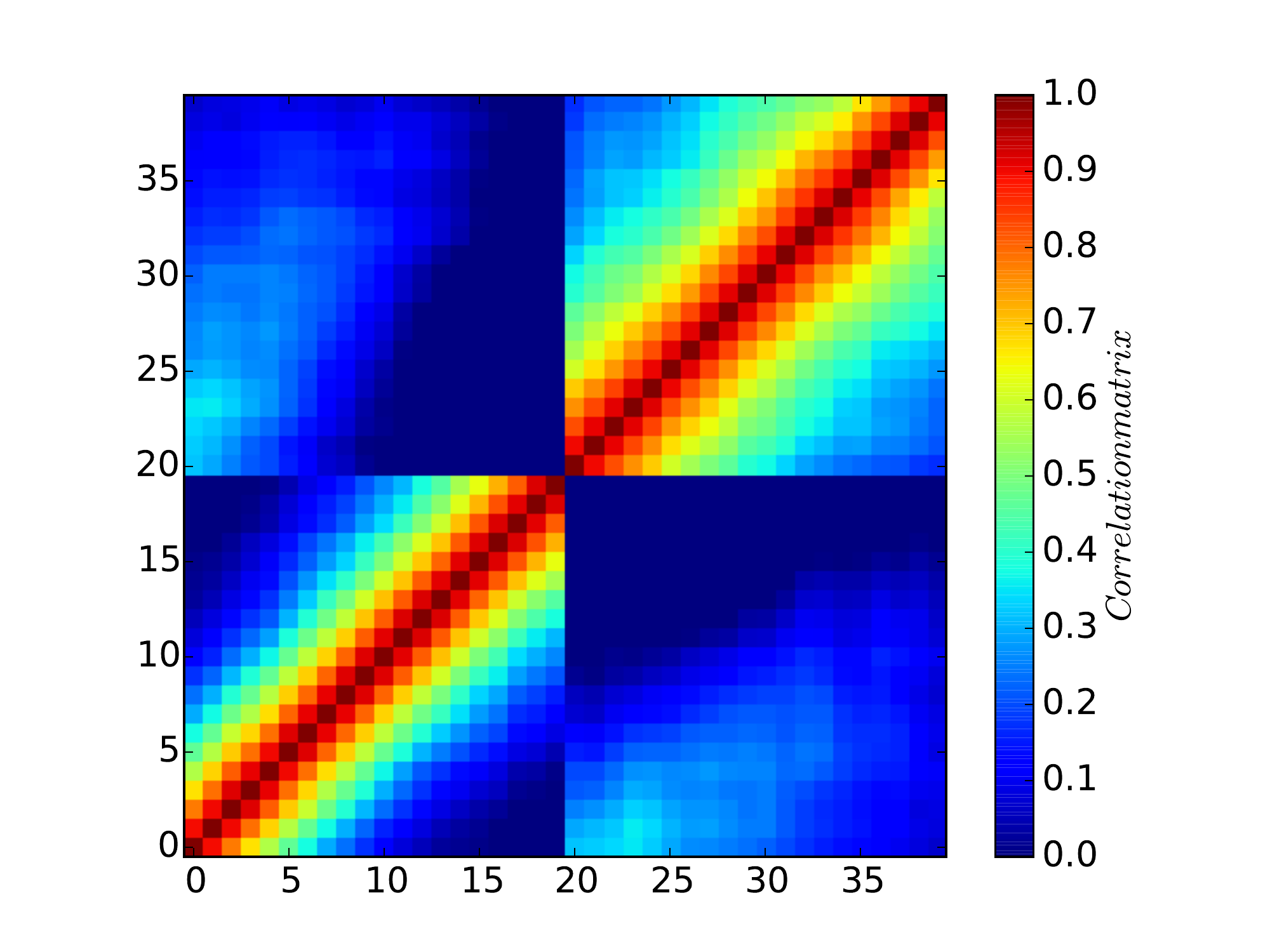}
    \includegraphics[width=2.0in]{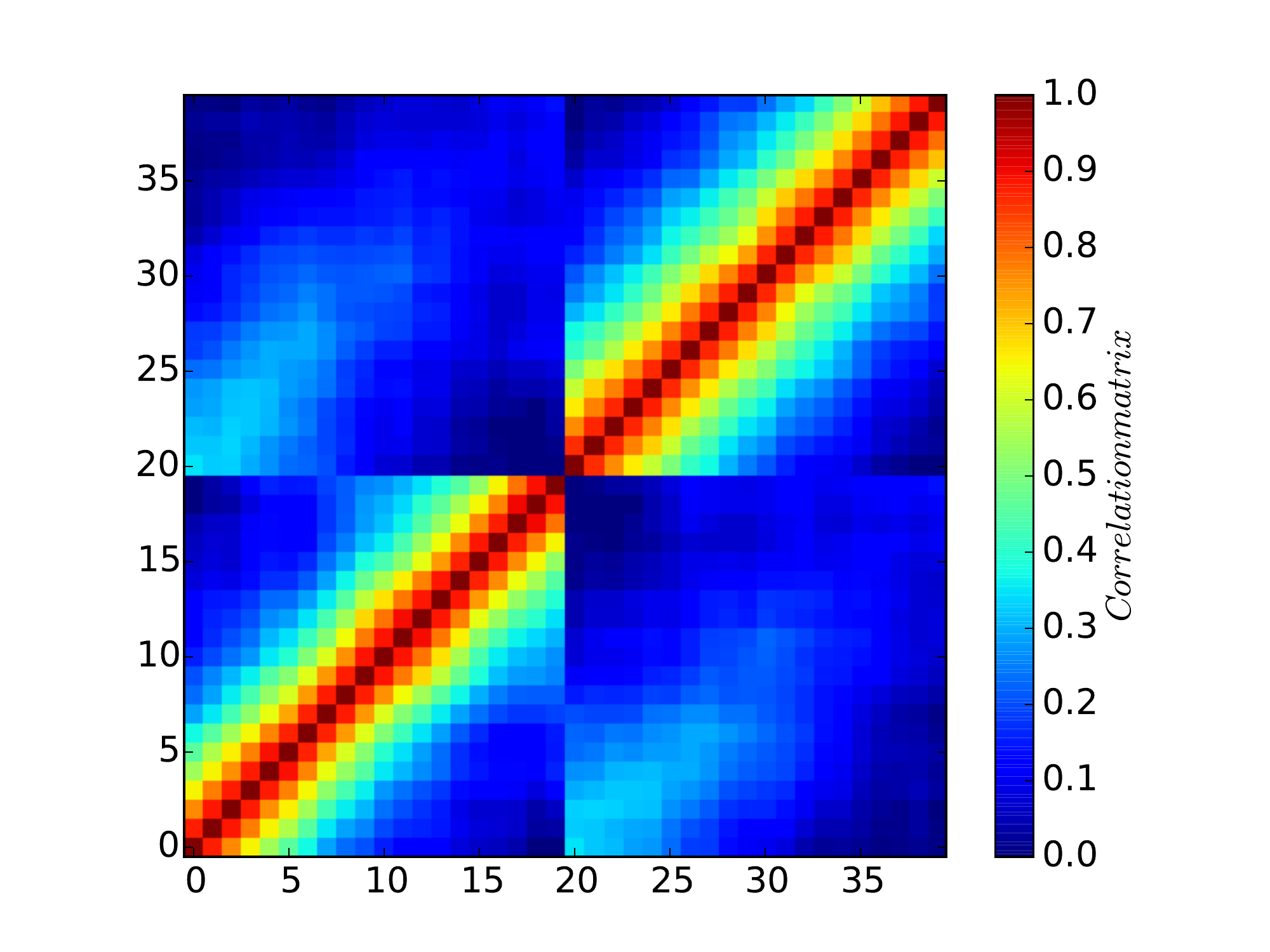}
    \includegraphics[width=2.0in]{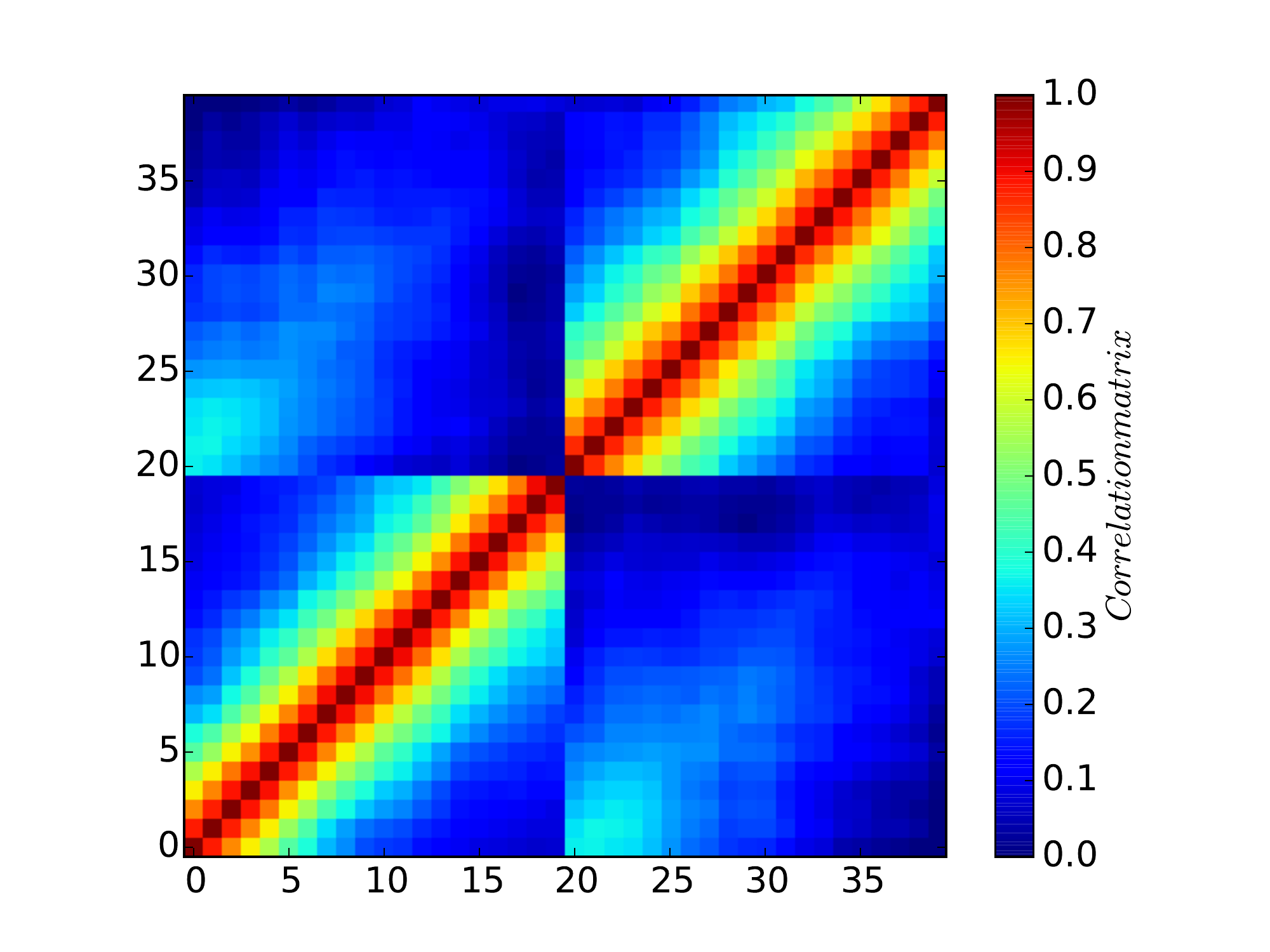}
     \includegraphics[width=2.0in]{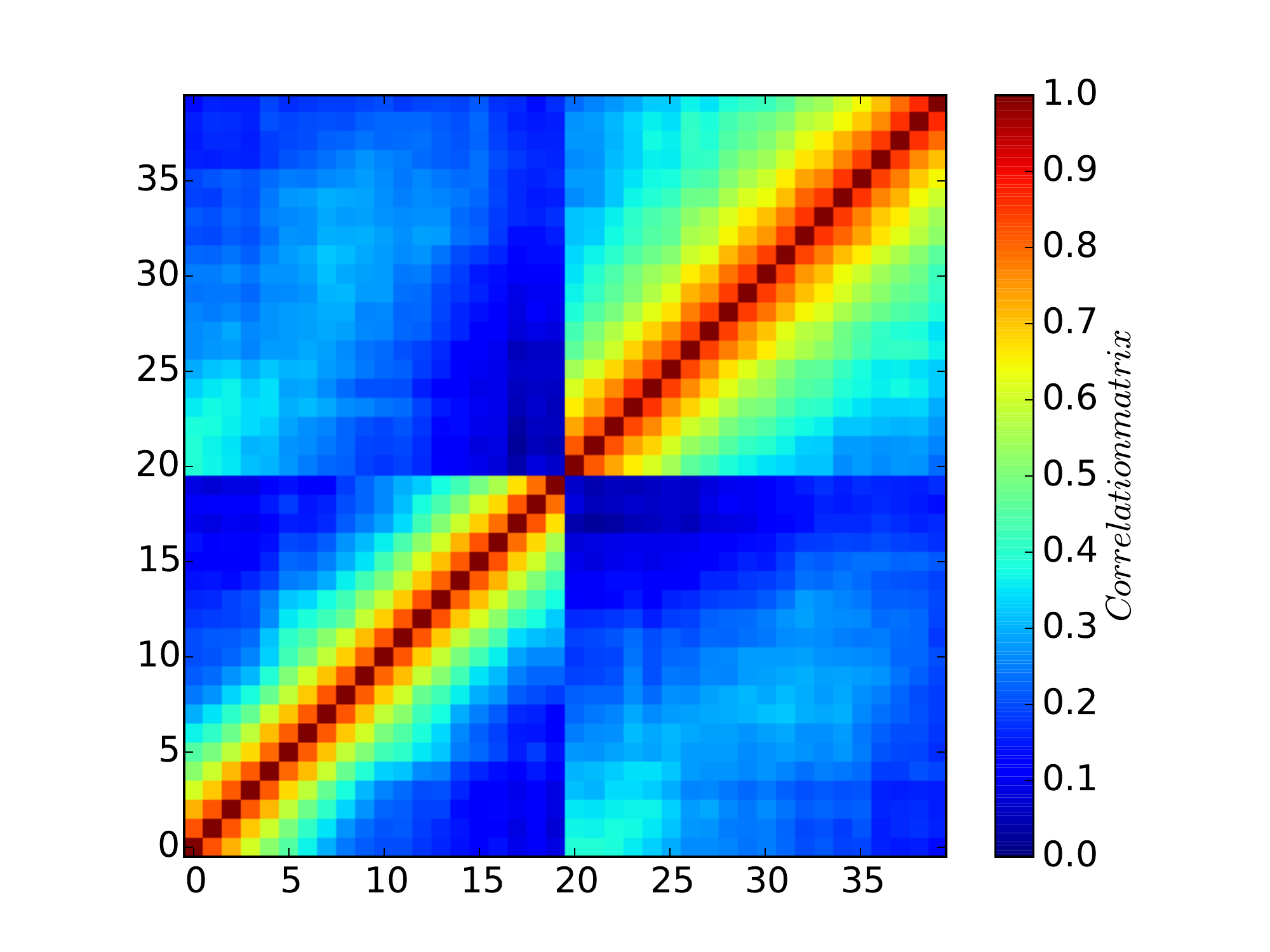}
%
    \includegraphics[width=2.0in]{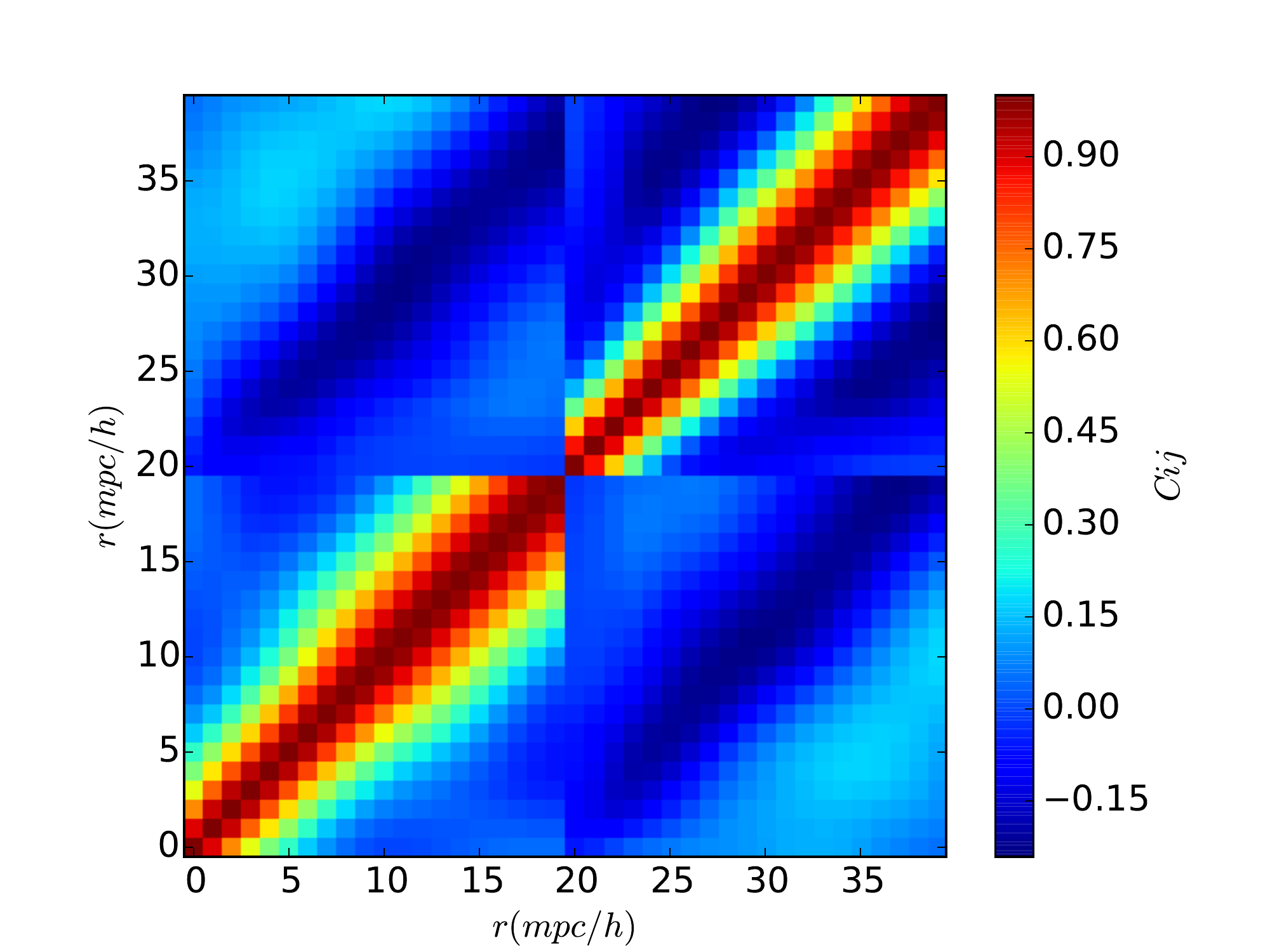}
    \includegraphics[width=2.0in]{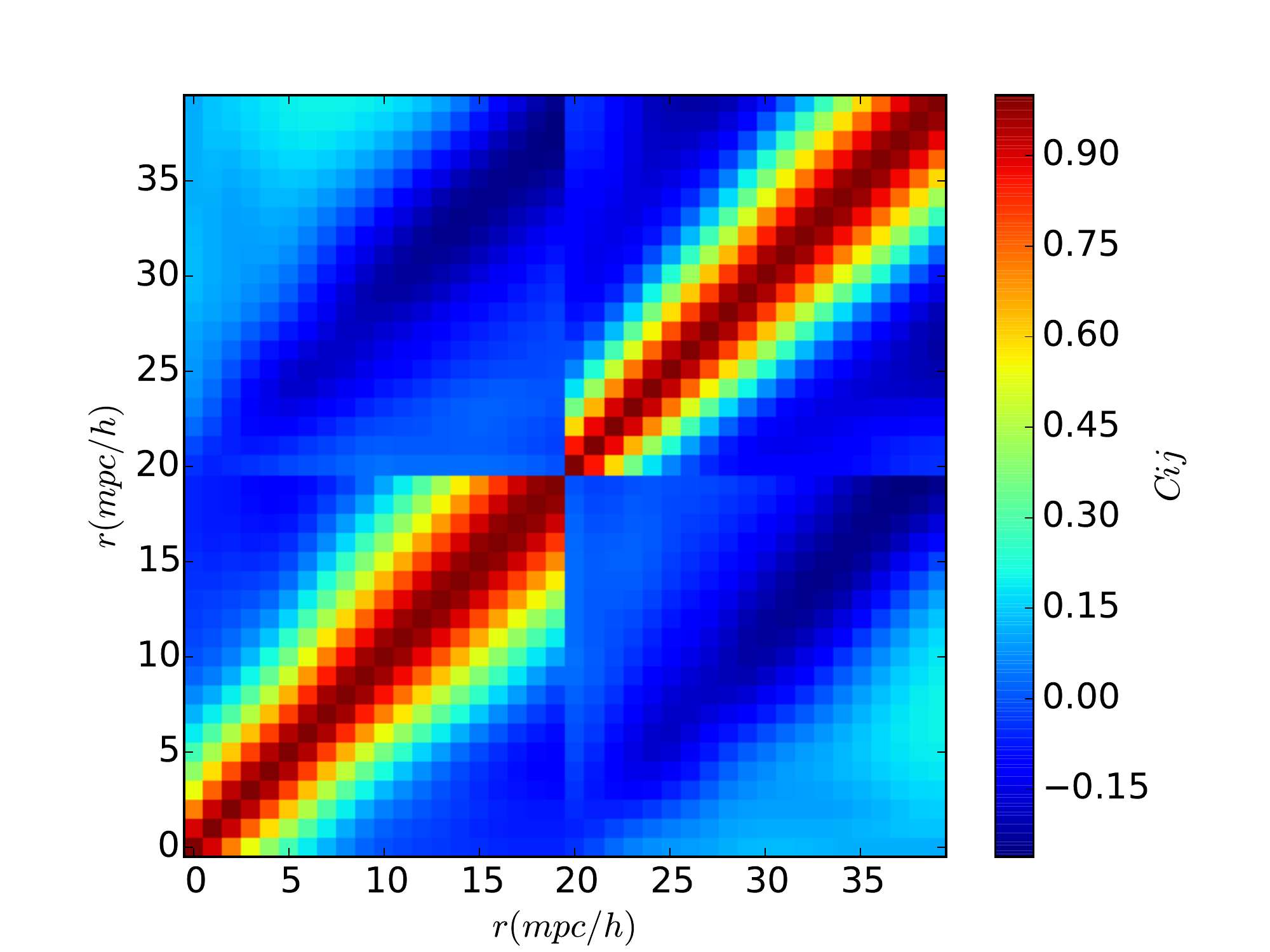}
    \includegraphics[width=2.0in]{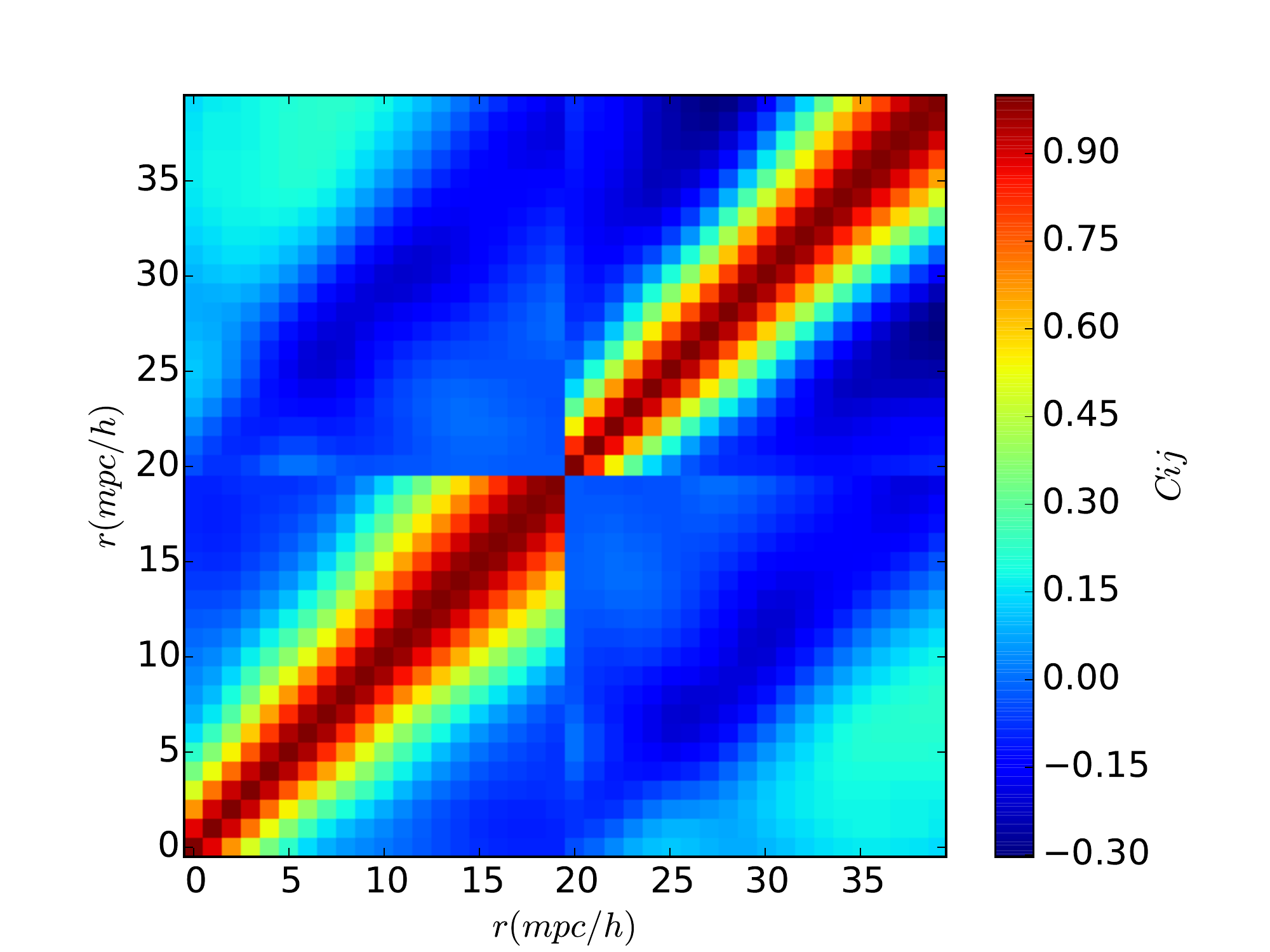}
    \includegraphics[width=2.0in]{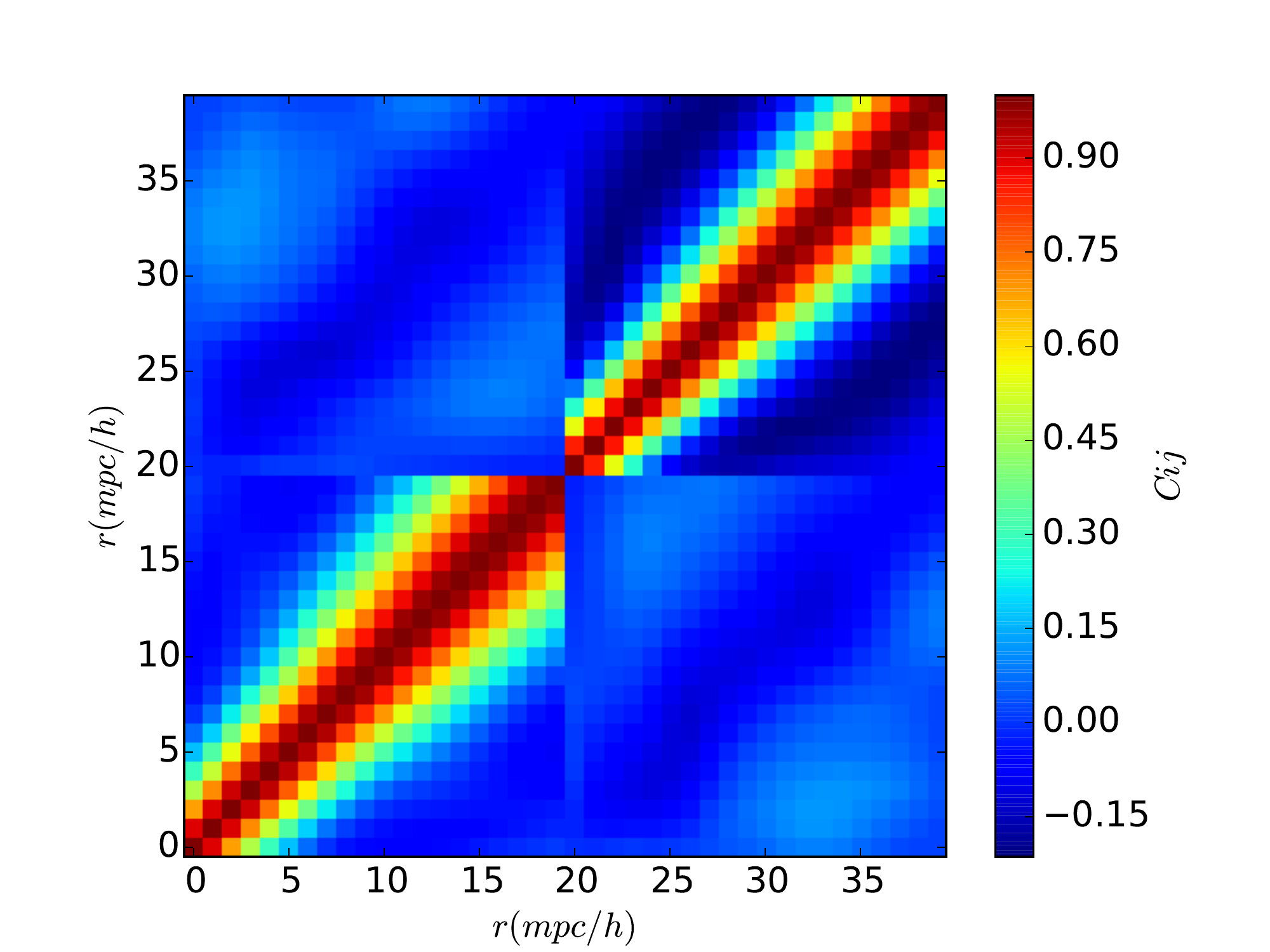}
     \includegraphics[width=2.0in]{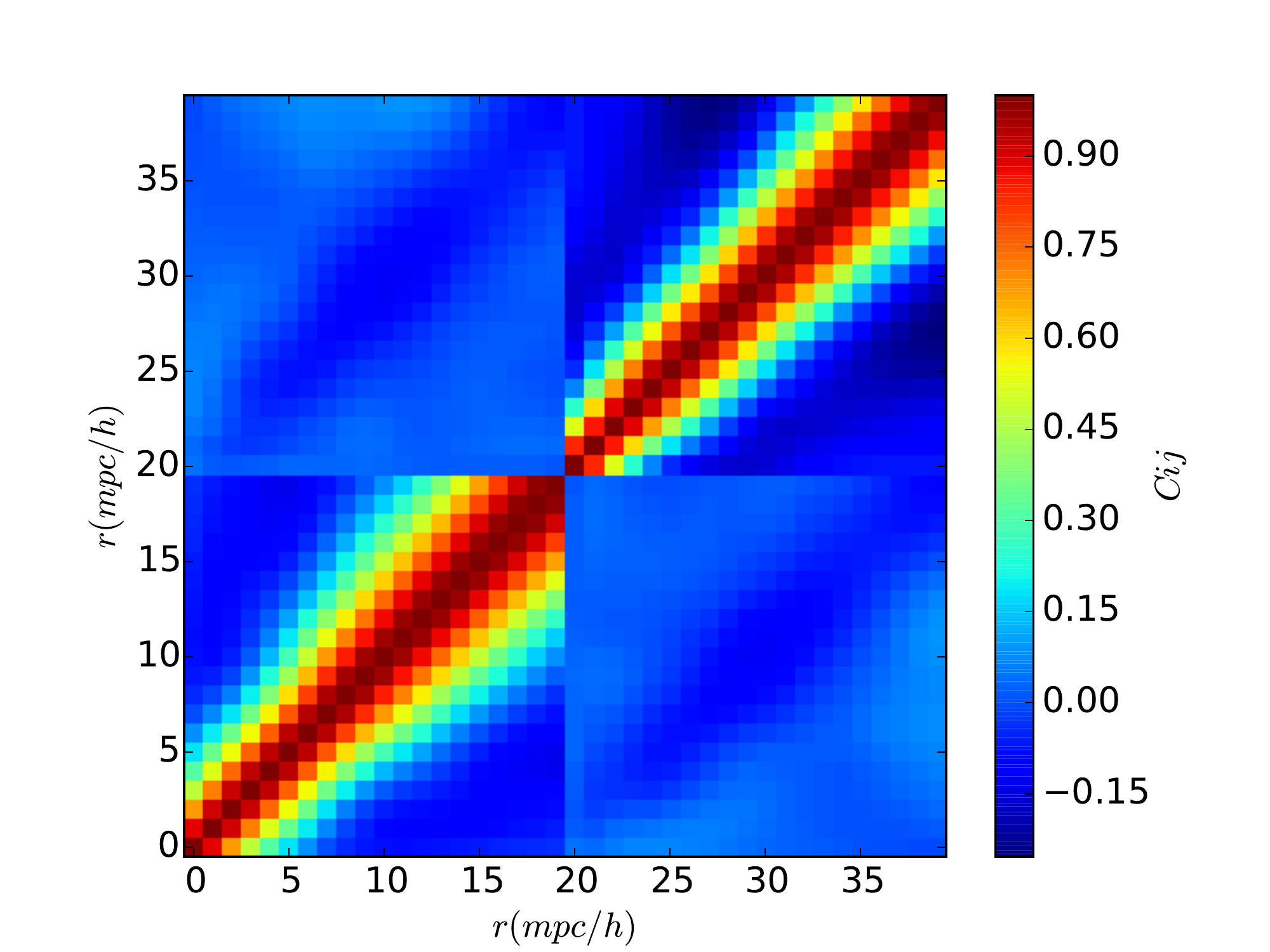}
     \includegraphics[width=2.0in]{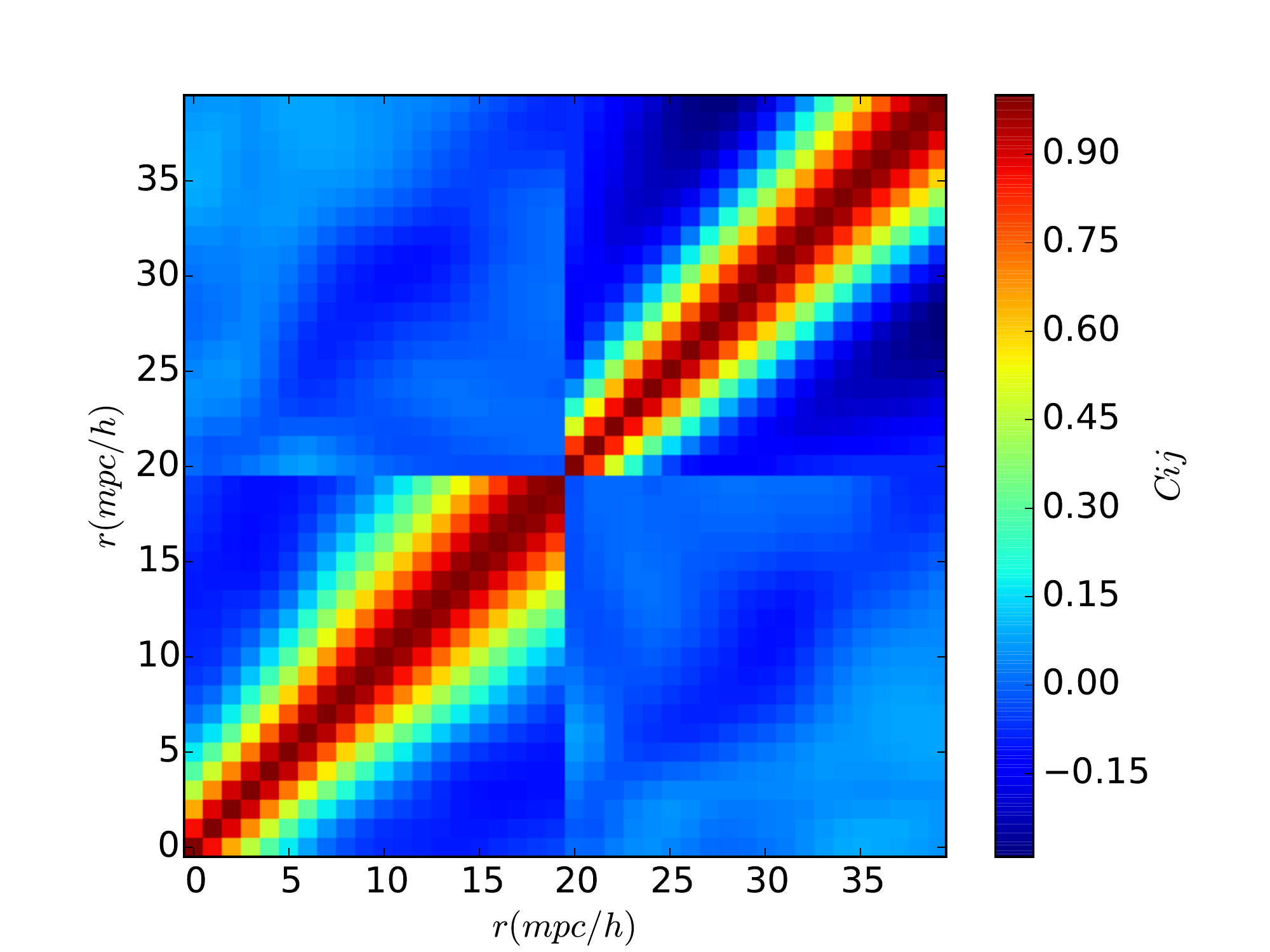}
     \caption{  Correlation matrix pre-reconstruction (top) and post-reconstruction (bottom) for MD-PATCHY for 1000 for the 3 redshift bins.Top panels: Multipoles Clustering Estimator including Hexadecapole.
     Intermediate panels: Wedges clustering estimator. 
   Bottom panels. $\omega-$Clustering Estimator. 
   \commentH{Check plots omega}\commentH{The values in x and y axis are not in Mpc/h, also it should be explained how $l=0$ and $l=2$ are arranged in this matrix}}
      \label{fig:covs}
\end{figure*}
\label{lastpage}

\begin{thebibliography}{}
\makeatletter
\relax
\def\mn@urlcharsother{\let\do\@makeother \do\$\do\&\do\#\do\^\do\_\do\%\do\~}
\def\mn@doi{\begingroup\mn@urlcharsother \@ifnextchar [ {\mn@doi@}
  {\mn@doi@[]}}
\def\mn@doi@[#1]#2{\def\@tempa{#1}\ifx\@tempa\@empty \href
  {http://dx.doi.org/#2} {doi:#2}\else \href {http://dx.doi.org/#2} {#1}\fi
  \endgroup}
\def\mn@eprint#1#2{\mn@eprint@#1:#2::\@nil}
\def\mn@eprint@arXiv#1{\href {http://arxiv.org/abs/#1} {{\tt arXiv:#1}}}
\def\mn@eprint@dblp#1{\href {http://dblp.uni-trier.de/rec/bibtex/#1.xml}
  {dblp:#1}}
\def\mn@eprint@#1:#2:#3:#4\@nil{\def\@tempa {#1}\def\@tempb {#2}\def\@tempc
  {#3}\ifx \@tempc \@empty \let \@tempc \@tempb \let \@tempb \@tempa \fi \ifx
  \@tempb \@empty \def\@tempb {arXiv}\fi \@ifundefined
  {mn@eprint@\@tempb}{\@tempb:\@tempc}{\expandafter \expandafter \csname
  mn@eprint@\@tempb\endcsname \expandafter{\@tempc}}}

\makeatother
\end{thebibliography}


\begin{thebibliography}{99}

\bibitem[{{Planck Collaboration} {et~al}\mbox{.}(2015{\natexlab{b}}){Planck
  Collaboration}, {Ade}, {Aghanim}, {Arnaud}, {Ashdown}, {Aumont},
  {Baccigalupi}, {Banday}, {Barreiro}, {Bartlett}, \& et~al.}]{Planck2015Cosmo}
 {Planck Collaboration} {et~al.}, 2015{\natexlab{b}}, ArXiv e-prints, arXiv:1502.01589

\bibitem[{{Alam} {et~al}\mbox{.}(2015){Alam}, {Albareti}, {Allende Prieto},
  {Anders}, {Anderson}, {Anderton}, {Andrews}, {Armengaud}, {Aubourg},
  {Bailey}, \& et~al.}]{DR12}
 {Alam} S. {et~al.}, 2015, \apjs, 219, 12

\bibitem[{{Alam et al.}(2016)}]{Acacia16}
  Alam S., et al., 2016, arXiv:1607.03155

\bibitem[{{Alcock \& Paczynski}(1979)}]{AP}
  Alcock C., Paczynski B., 1979, Nature, 281, 358.

\bibitem[{{Anderson et al.}(2012)}]{And12}
  Anderson L., et al., 2012, MNRAS, 427, 3435 

\bibitem[{{Anderson et al.}(2013)}]{And13}
  Anderson L., et al., 2014, MNRAS, 439, 83 

\bibitem[{{Anderson et al.}(2014)}]{And14}
  Anderson L., et al., 2014, MNRAS, 441, 24

\bibitem[{{Beutler et al.}(2016a)}]{Beutler16a}
  Beutler F., et al., 2016, arXiv:1607.03149 (BAO)

\bibitem[{{Beutler et al.}(2016b)}]{Beutler16b}
  Beutler F., et al., 2016, arXiv:1607.03150 (RSD)

\bibitem[{{Bolton et al.}(2012)}]{Bolton12}
  Bolton A., et al., 2012, AJ, 144, 144 

\bibitem[{{Bull} {et~al}\mbox{.}(2015){Bull}, {Ferreira}, {Patel}, \&
  {Santos}}]{SKA}
 {Bull} P., {Ferreira} P.~G., {Patel} P., {Santos} M.~G., 2015, \apj, 803, 21

\bibitem[{{Burden et al.}(2014)}]{bur14}
  Burden A., et al., 2014, MNRAS, 445, 3.

\bibitem[{{Burden et al.}(2015)}]{bur15}
  Burden A., Percival W. \& Howlett C., 2015, MNRAS, 453, 456 

\bibitem[{{Cuesta} {et~al}\mbox{.}(2016){Cuesta}, {Vargas-Maga\~na}, {Beutler}}]{Cuesta16}
 {Cuesta} A.~J., {Vargas-Maga\~na} M., {Beutler} F., et al., 2016, \mnras, 457, 1770

\bibitem[{{Dawson et al.}(2012)}]{Daw12}
  Dawson K., et al., 2013, AJ, 145, 10

\bibitem[{{Dawson et al.}(2016)}]{eBOSS}
  Dawson K.~S., et al., 2016, AJ, 151, 44

\bibitem[{{Eisenstein \& Hu}(1998)}]{Eis98}
  Eisenstein D.~J., Hu W., 1998, ApJ, 496, 605

\bibitem[{{Eisenstein et al.}(2007a)}]{Eis07a}
  Eisenstein D.~J., Seo H.-J., Sirko E., Spergel D.~N., 2007a, ApJ, 664, 675

\bibitem[{{Eisenstein et al.}(2007b)}]{EisSeoWhi07}
  Eisenstein D.~J., Seo H.-J., White M., 2007b, ApJ, 664, 660
  
  \bibitem[{{Eisenstein et al.}(2011)}]{Eis11}
  Eisenstein D.~J., et al., 2011, AJ, 142.
  
\bibitem[{{Fisher et al.}(1994)}]{Fisher94}
  Fisher K.~B., Scharf C.~A., Lahav O., 1994, MNRAS, 266, 219
  
  \bibitem[{{Fukugita et al.}(1996)}]{Fuk96}
  Fukugita M., Ichikawa T., Gunn J.~E., Doi M., Shimasaku K., Schneider D.~P., 1996, AJ, 111, 1748

\bibitem[{{Gil-Mar\'in et al.}(2016)}]{Gil15}
  Gil-Mar\'in H., Percival W.~J., Cuesta A.~J., et al., 2016, \mnras, 460, 4210

\bibitem[{{Grieb et al.}(2016)}]{Grieb16}
  Grieb M. et al. 2016, in prep (RSD wedges FS).

\bibitem[{{Gunn et al.}(1998)}]{Gun98}
  Gunn J.~E., et al., 1998, AJ, 116, 3040.

\bibitem[{{Gunn et al.}(2006)}]{Gun06}
  Gunn J.~E., et al., 2006, AJ, 131, 2332

\bibitem[{{Hartlap et al.}(2007)}]{Har07}
Hartlap J., Simon P., \& Schneider P. 2007, A\&A, 464, 399

\bibitem[{{Kazin et al.}(2012)}]{Kaz12}
  Kazin E., Sanchez A.~G., Blanton M.~R., 2012, MNRAS, 419, 3223

\bibitem[{{Kazin et al.}(2013)}]{Kaz13}
  Kazin E., et al., 2013, MNRAS, 435, 64

\bibitem[{{Kitaura} {et~al}\mbox{.}(2015){Kitaura}, {Rodriguez-Torres},
  {Chuang}, {Zhao}, {Prada}, {Gil-Marin}, {Guo}, {Yepes}, {Klypin}, {Scoccola},
  {Tinker}, {McBride}, {Reid}, {Sanchez}, {Salazar-Albornoz}, {Niklas Grieb},
  {Vargas-Magana}, {Cuesta}, {Neyrinck}, {Beutler}, {Comparat}, {Percival}, \&
  {Ross}}]{PATCHY}
  {Kitaura} F.-S. {et~al.}, 2016, MNRAS, 456, 4156

\bibitem[{{Kitaura}, {Yepes} \& {Prada}(2014){Kitaura}, {Yepes}, \&
  {Prada}}]{PATCHYtheory}
  {Kitaura} F.-S., {Yepes} G., {Prada} F., 2014, \mnras, 439, L21

\bibitem[{{Klypin}, {Yepes}, {Gottlober}, {Prada}, \& {Hess}(2016)}]{Multidark}
   Klypin A., et al., 2016, MNRAS, 457, 4340
   
\bibitem[{{Landy \& Szalay}(1993)}]{LanSza93}
  Landy S.~D., Szalay A.~S., 1993, ApJ, 412, 64

\bibitem[{{Laureijs}(2009)}]{EuclidAssesment}
 {Laureijs} R., 2009, ArXiv e-prints, arXiv:0912.0914

\bibitem[{{Laureijs} {et~al}\mbox{.}(2011){Laureijs}, {Amiaux}, {Arduini},
  {Augu{\`e}res}, {Brinchmann}, {Cole}, {Cropper}, {Dabin}, {Duvet}, {Ealet},
  \& et~al.}]{EuclidDefinition}
 {Laureijs} R. {et~al.}, 2011, ArXiv e-prints, arXiv:1110.3193

\bibitem[{{Levi} {et~al}\mbox{.}(2013){Levi}, {Bebek}, {Beers}, {Blum}, {Cahn},
  {Eisenstein}, {Flaugher}, {Honscheid}, {Kron}, {Lahav}, {McDonald}, {Roe},
  {Schlegel}, \& {representing the DESI collaboration}}]{DesiWhitePaper}
 {Levi} M. {et~al.}, 2013, ArXiv e-prints, arXiv:1308.0847

\bibitem[{{Lewis et al.}(2000)}]{lewis00}
  Lewis A., Challinor A., Lasenby A., 2000, ApJ, 538, 473

 \bibitem[{{LSST Dark Energy Science Collaboration}(2012)}]{LSST}
{LSST Dark Energy Science Collaboration}, 2012, ArXiv e-prints, arXiv:1211.0310.

\bibitem[{{Manera et al.}(2013a)}]{Man12}
  Manera M., Scoccimarro R., Percival W.~J., et al.\ 2013, MNRAS, 428, 1036

\bibitem[{{Matsubara et al.}(2000)}]{Mat00}
  Matsubara T., Szalay, A.~S. \& Landy, S.~D., The Astrophysical Journal, Volume 535, Issue 1, pp. L1-L4.

\bibitem[{{Mostek} {et~al}\mbox{.}(2012){Mostek}, {Barbary}, {Bebek}, {Dey},
  {Edelstein}, {Jelinsky}, {Kim}, {Lampton}, {Levi}, {McDonald}, {Poppett},
  {Roe}, {Schlegel}, \& {Sholl}}]{DESIOverview}
 {Mostek} N. {et~al.}, 2012, in Society of Photo-Optical Instrumentation
  Engineers (SPIE) Conference Series, Vol. 8446, Society of Photo-Optical
  Instrumentation Engineers (SPIE) Conference Series, p.~0

\bibitem[{{Okumara et al.}(2008)}]{Oku08}
  Okumura T., et al., 2008, ApJ, 676, 889.

\bibitem[{{O'Connell et al.}(2015)}]{Oco15}
  O'Connell R., Eisenstein D., Vargas M., Ho S., Padmanabhan N., 2015, arXiv:1510.01740  

\bibitem[{{Padmanabhan et al.}(2007)}]{Pad07}
  Padmanabhan N., et al., 2007, MNRAS, 378, 852

\bibitem[{{Padmanabhan \& White}(2009)}]{PadWhi09}
  Padmanabhan N., White M., 2009, Phys. Rev. D, 80, 063508

\bibitem[{{Padmanabhan et al.}(2012)}]{Pad12} 
  Padmanabhan N., et al., 2012, \mnras, 427, 2132

\bibitem[{{Paz \& S\'anchez}(2015)}]{Paz15}
  Paz D.~J., S\'anchez A.~G., 2015, \mnras, 454, 4326

  \bibitem[{{Percival et al.}(2014)}]{Per14}
  Percival W.~J., et al., MNRAS, 439, 2531.
%
\bibitem[{{Pope \& Szapudi}(2008)}]{Pope08}
  Pope A.~C.; Szapudi I., 2008, \mnras, 389, 766
 
  \bibitem[{{Reid} {et~al}\mbox{.}(2016){Reid}, {Ho}, {Padmanabhan}, {Percival},
  {Tinker}, {Tojeiro}, {White}, {Eisenstein}, {Maraston}, {Ross},
  {S{\'a}nchez}, {Schlegel}, {Sheldon}, {Strauss}, {Thomas}, {Wake}, {Beutler},
  {Bizyaev}, {Bolton}, {Brownstein}, {Chuang}, {Dawson}, {Harding}, {Kitaura},
  {Leauthaud}, {Masters}, {McBride}, {More}, {Olmstead}, {Oravetz}, {Nuza},
  {Pan}, {Parejko}, {Pforr}, {Prada}, {Rodr{\'{\i}}guez-Torres},
  {Salazar-Albornoz}, {Samushia}, {Schneider}, {Sc{\'o}ccola}, {Simmons}, \&
  {Vargas-Magana}}]{Reid2015}
 {Reid} B., {et~al.}, 2016, \mnras, 455, 1553

\bibitem[{{Ross et al.}(2011)}]{Ross11}
  Ross A.~J., et al., 2011, MNRAS, 417, 1350

\bibitem[Ross et al.(2012)]{Ross12}
  Ross A.~J., Percival W.~J., S{\'a}nchez A.~G., et al., 2012, MNRAS, 424, 564.
    
\bibitem[Ross et al.(2015)]{Ross15}    
  Ross A.~J., Percival W.~J., \& Manera M., 2015 MNRAS, 451, 1331

\bibitem[{{Ross et al.}(2016)}]{Ross16}
  Ross A.~J., et al., 2016, arXiv:1607.03145, MNRAS in press.

\bibitem[{{Satpathy et al.}(2016)}]{Satpathy16}
 Satpathy S., et al., 2016, arXiv:1607.03148

\bibitem[{{S\'anchez et al.}(2013a)}]{Sanchez13}
  S\'anchez A.~G., et al., 2013a, MNRAS, 433, 1201
  
\bibitem[{{S\'anchez et al.}(2016a)}]{Sanchez16a}
  S\'anchez A.~G., et al., 2016a, arXiv:1607.03147, MNRAS in press.

\bibitem[{{S\'anchez et al.}(2016b)}]{Sanchez16b}
  S\'anchez A.~G., et al., 2016b, arXiv:1607.03146, MNRAS in press.

\bibitem[{{Schneider et al.}(2011)}]{Schneider11}
  Schneider M.~D., Cole S., Frenk C.~S., Szapudi I., 2011, \apj, 737, 11


\bibitem[{{Scoccimarro \& Sheth}(2002)}]{ScoShe02}
  Scoccimarro R., Sheth R.~K., 2002, MNRAS, 329, 629

\bibitem[{{Seo et al.}(2008)}]{Seo08}
Seo H.-J., Siegel E.~R., Eisenstein D.~J., White M., 2008, ApJ, 686, 13.

\bibitem[{{Seo et al.}(2010)}]{Seo10}
  Seo H.-J., et al., 2010, ApJ, 720, 1650
  
\bibitem[{{Seo et al.}(2016)}]{Seo15}
  Seo H.-J., et al., 2016, \mnras, 460, 2453

\bibitem[{{Smee et al.}(2013)}]{Smee13}
  Smee S., et al., 2013, AJ, 146, 32

\bibitem[{{Spergel} {et~al}\mbox{.}(2013{\natexlab{a}}){Spergel}, {Gehrels},
  {Breckinridge}, {Donahue}, {Dressler}, {Gaudi}, {Greene}, {Guyon}, {Hirata},
  {Kalirai}, {Kasdin}, {Moos}, {Perlmutter}, {Postman}, {Rauscher}, {Rhodes},
  {Wang}, {Weinberg}, {Centrella}, {Traub}, {Baltay}, {Colbert}, {Bennett},
  {Kiessling}, {Macintosh}, {Merten}, {Mortonson}, {Penny}, {Rozo},
  {Savransky}, {Stapelfeldt}, {Zu}, {Baker}, {Cheng}, {Content}, {Dooley},
  {Foote}, {Goullioud}, {Grady}, {Jackson}, {Kruk}, {Levine}, {Melton},
  {Peddie}, {Ruffa}, \& {Shaklan}}]{WFIRST}
 {Spergel} D. {et~al.}, 2013{\natexlab{a}}, ArXiv e-prints, arXiv:1305.5425

\bibitem[{{Spergel} {et~al}\mbox{.}(2013{\natexlab{b}}){Spergel}, {Gehrels},
  {Breckinridge}, {Donahue}, {Dressler}, {Gaudi}, {Greene}, {Guyon}, {Hirata},
  {Kalirai}, {Kasdin}, {Moos}, {Perlmutter}, {Postman}, {Rauscher}, {Rhodes},
  {Wang}, {Weinberg}, {Centrella}, {Traub}, {Baltay}, {Colbert}, {Bennett},
  {Kiessling}, {Macintosh}, {Merten}, {Mortonson}, {Penny}, {Rozo},
  {Savransky}, {Stapelfeldt}, {Zu}, {Baker}, {Cheng}, {Content}, {Dooley},
  {Foote}, {Goullioud}, {Grady}, {Jackson}, {Kruk}, {Levine}, {Melton},
  {Peddie}, {Ruffa}, \& {Shaklan}}]{WFIRSTReport}
 {Spergel} D. {et~al.}, 2013{\natexlab{b}}, ArXiv e-prints, arXiv:1305.5422

\bibitem[{{de la Torre \& Peacock}(2013)}]{delaTorre13}
  de la Torre S., Peacock J.~A., 2013, \mnras, 435, 743

\bibitem[{{Vargas-Magana et al.}(2014)}]{Vargas14}
  Vargas-Magana M., et al., 2014, MNRAS, 445, 2
  
\bibitem[{{Vargas-Magana et al.}(2015)}]{Vargas15}
  Vargas-Magana M., Ho S., Cuesta A.~J., Fromenteau S., 2015, ArXiv e-prints, arXiv:1509.06384.

\bibitem[{{White}, {Tinker} \& {McBride}(2014){White}, {Tinker}, \&
  {McBride}}]{QPM}
 {White} M., {Tinker} J.~L., {McBride} C.~K., 2014, \mnras, 437, 2594

\bibitem[{{Xu et al.}(2010)}]{Xu10}
  Xu X., et al., 2010, ApJ, 718, 1224.

\bibitem[{{Xu et al.}(2012a)}]{Xeaip}
  Xu X., et al., 2012a, MNRAS, 427, 2146

\bibitem[{{Xu et al.}(2012b)}]{Xu12b}
  Xu X., et al., 2012b, MNRAS, 431, 2834

\bibitem[{{Zhao, Kitaura, Chuang, Prada, Yepes, \& Tao }(2015)}]{HADRON}
  Zhao C., et al., 2015, MNRAS 451, 4266

\bibliographystyle{mnras}
\end{thebibliography}
\end{document}